\documentclass[
  journal=pasa,
  manuscript=research-paper, 
  year=2025,
  volume=37
]{cup-journal}

\usepackage{microtype,siunitx,booktabs}

\DeclareRobustCommand{\VAN}[3]{#2}
\let\VANthebibliography\thebibliography
\def\thebibliography{\DeclareRobustCommand{\VAN}[3]{##3}\VANthebibliography}

\usepackage{graphicx}	
\usepackage{amsmath}	
\usepackage{amssymb}	
\usepackage{xspace}
\usepackage{upgreek}
\usepackage{hyperref}
\usepackage{pdflscape}

\newcommand\ion[2]{\text{#1\,\textsc{\lowercase{#2}}}}	

\newcommand{\Teff}{$T_\mathrm{eff}$\xspace}
\newcommand{\logg}{$\log g$\xspace}
\newcommand{\feh}{$\mathrm{[Fe/H]}$\xspace}

\newcommand{\vmic}{$v_\mathrm{mic}$\xspace}
\newcommand{\vsini}{$v \sin i$\xspace}
\newcommand{\vrad}{$v_\mathrm{rad}$\xspace}

\newcommand{\allstarnumber}{917\,588\xspace}
\newcommand{\allspecnumber}{1\,085\,520\xspace}

\newcommand{\sme}{\textsc{sme}\xspace}
\newcommand{\marcs}{\textsc{marcs}\xspace}
\newcommand{\Gaia}{\textit{Gaia}\xspace}
\newcommand{\TLF}{\Teff, \logg, and \feh}
\newcommand{\breidablik}{\textsc{breidablik}\xspace}
\newcommand{\stagger}{\textsc{stagger}\xspace}

\newcommand{\dex}{\,\mathrm{dex}}	
\newcommand{\K}{\,\mathrm{K}}	
\newcommand{\Angstroem}{\,\text{\AA}}	
\newcommand{\kms}{\,\mathrm{km\,s^{-1}}}	

\title{The GALAH Survey: Data Release 4}

\author{S.~Buder}
\affiliation{Research School of Astronomy and Astrophysics, Australian National University, Canberra, ACT 2611, Australia}
\alsoaffiliation{ARC Centre of Excellence for All Sky Astrophysics in 3 Dimensions (ASTRO 3D), Australia}
\alsoaffiliation{ACCESS-NRI, Australian National University, Canberra, ACT2601, Australia}
\email[S. Buder]{sven.buder@anu.edu.au}

\author{J.~Kos}
\affiliation{Faculty of Mathematics \& Physics, University of Ljubljana, Jadranska 19, 1000 Ljubljana, Slovenia}

\author{E.~X.~Wang}
\affiliation{Research School of Astronomy and Astrophysics, Australian National University, Canberra, ACT 2611, Australia}
\alsoaffiliation{ARC Centre of Excellence for All Sky Astrophysics in 3 Dimensions (ASTRO 3D), Australia}

\author{M.~McKenzie}
\affiliation{Research School of Astronomy and Astrophysics, Australian National University, Canberra, ACT 2611, Australia}
\alsoaffiliation{ARC Centre of Excellence for All Sky Astrophysics in 3 Dimensions (ASTRO 3D), Australia}

\author{M.~Howell}
\affiliation{School of Physics and Astronomy, Monash University, Clayton, VIC 3800, Australia}
\alsoaffiliation{ARC Centre of Excellence for All Sky Astrophysics in 3 Dimensions (ASTRO 3D), Australia}

\author{S.~L.~Martell}
\affiliation{School of Physics, University of New South Wales, Sydney, NSW 2052, Australia}
\alsoaffiliation{ARC Centre of Excellence for All Sky Astrophysics in 3 Dimensions (ASTRO 3D), Australia}

\author{M.~R.~Hayden}
\affiliation{Homer L. Dodge Department of Physics \& Astronomy, University of Oklahoma, 440 W. Brooks St., Norman, OK 73019, USA}
\alsoaffiliation{School of Physics, UNSW, Sydney, NSW 2052, Australia}
\alsoaffiliation{Sydney Institute for Astronomy, School of Physics, A28, The University of Sydney, NSW 2006, Australia}
\alsoaffiliation{ARC Centre of Excellence for All Sky Astrophysics in 3 Dimensions (ASTRO 3D), Australia}

\author{D.~B.~Zucker}
\affiliation{School of Mathematical and Physical Sciences, Macquarie University, Balaclava Road, Sydney, NSW 2109, Australia}
\alsoaffiliation{Astrophysics and Space Technologies Research Centre, Macquarie University, Balaclava Road, Sydney, NSW 2109, Australia}
\alsoaffiliation{ARC Centre of Excellence for All Sky Astrophysics in 3 Dimensions (ASTRO 3D), Australia}

\author{T.~Nordlander}
\affiliation{Research School of Astronomy and Astrophysics, Australian National University, Canberra, ACT 2611, Australia}
\alsoaffiliation{Department of Physics and Astronomy, Uppsala University, Box 516, 751 20 Uppsala, Sweden}
\alsoaffiliation{ARC Centre of Excellence for All Sky Astrophysics in 3 Dimensions (ASTRO 3D), Australia}

\author{B.~T.~Montet}
\affiliation{School of Physics, University of New South Wales, Sydney, NSW 2052, Australia}
\alsoaffiliation{UNSW Data Science Hub, University of New South Wales, Sydney, NSW 2052, Australia}

\author{G.~Traven}
\affiliation{Faculty of Mathematics \& Physics, University of Ljubljana, Jadranska 19, 1000 Ljubljana, Slovenia}

\author{J.~Bland-Hawthorn}
\affiliation{Sydney Institute for Astronomy, School of Physics, A28, The University of Sydney, NSW 2006, Australia}
\alsoaffiliation{ARC Centre of Excellence for All Sky Astrophysics in 3 Dimensions (ASTRO 3D), Australia}

\author{G.~M.~De~Silva}
\affiliation{School of Mathematical and Physical Sciences, Macquarie University, Balaclava Road, Sydney, NSW 2109, Australia}
\alsoaffiliation{Astrophysics and Space Technologies Research Centre, Macquarie University, Balaclava Road, Sydney, NSW 2109, Australia}
\alsoaffiliation{ARC Centre of Excellence for All Sky Astrophysics in 3 Dimensions (ASTRO 3D), Australia}

\author{K.~C.~Freeman}
\affiliation{Research School of Astronomy and Astrophysics, Australian National University, Canberra, ACT 2611, Australia}
\alsoaffiliation{ARC Centre of Excellence for All Sky Astrophysics in 3 Dimensions (ASTRO 3D), Australia}

\author{G.~F.~Lewis}
\affiliation{Sydney Institute for Astronomy, School of Physics, A28, The University of Sydney, NSW 2006, Australia}

\author{K.~Lind}
\affiliation{Department of Astronomy, Stockholm University, AlbaNova University Centre, SE-106 91 Stockholm, Sweden}

\author{S.~Sharma}
\affiliation{Space Telescope Science Institute, 3700 San Martin Drive, Baltimore, MD, 21218, USA}

\author{J.~D.~Simpson}
\affiliation{School of Physics, UNSW, Sydney, NSW 2052, Australia}
\alsoaffiliation{ARC Centre of Excellence for All Sky Astrophysics in 3 Dimensions (ASTRO 3D), Australia}

\author{D.~Stello}
\affiliation{School of Physics, UNSW, Sydney, NSW 2052, Australia}
\alsoaffiliation{Sydney Institute for Astronomy, School of Physics, A28, The University of Sydney, NSW 2006, Australia}
\alsoaffiliation{ARC Centre of Excellence for All Sky Astrophysics in 3 Dimensions (ASTRO 3D), Australia}
\alsoaffiliation{Stellar Astrophysics Centre, Aarhus University, Ny Munkegade 120, DK-8000 Aarhus C, Denmark}

\author{T.~Zwitter}
\affiliation{Faculty of Mathematics \& Physics, University of Ljubljana, Jadranska 19, 1000 Ljubljana, Slovenia}

\author{A.~M.~Amarsi}
\affiliation{Department of Physics and Astronomy, Uppsala University, Box 516, 751 20 Uppsala, Sweden}

\author{J.~J.~Armstrong}
\affiliation{
Department of Space, Earth \& Environment, Chalmers University of Technology, SE-412 96 Gothenburg, Sweden}

\author{K.~Banks}
\affiliation{School of Physics, UNSW, Sydney, NSW 2052, Australia}
\alsoaffiliation{ARC Centre of Excellence for All Sky Astrophysics in 3 Dimensions (ASTRO 3D), Australia}

\author{M.~A.~Beavis}
\affiliation{Centre for Astrophysics, University of Southern Queensland, West Street, Toowoomba, QLD 4350, Australia}

\author{K.~Beeson}
\affiliation{Faculty of Mathematics \& Physics, University of Ljubljana, Jadranska 19, 1000 Ljubljana, Slovenia}

\author{B.~Chen}
\affiliation{Research School of Astronomy and Astrophysics, Australian National University, Canberra, ACT 2611, Australia}
\alsoaffiliation{ARC Centre of Excellence for All Sky Astrophysics in 3 Dimensions (ASTRO 3D), Australia}

\author{I.~Ciuc{\u{a}}}
\affiliation{Research School of Astronomy and Astrophysics, Australian National University, Canberra, ACT 2611, Australia}
\alsoaffiliation{ARC Centre of Excellence for All Sky Astrophysics in 3 Dimensions (ASTRO 3D), Australia}

\author{G.~S.~Da~Costa}
\affiliation{Research School of Astronomy and Astrophysics, Australian National University, Canberra, ACT 2611, Australia}
\alsoaffiliation{ARC Centre of Excellence for All Sky Astrophysics in 3 Dimensions (ASTRO 3D), Australia}

\author{R.~de~Grijs}
\affiliation{School of Mathematical and Physical Sciences, Macquarie University, Balaclava Road, Sydney, NSW 2109, Australia}
\alsoaffiliation{Astrophysics and Space Technologies Research Centre, Macquarie University, Balaclava Road, Sydney, NSW 2109, Australia}
\alsoaffiliation{International Space Science Institute--Beijing, 1 Nanertiao, Zhongguancun, Beijing 100190, China}

\author{B.~Martin}
\affiliation{Research School of Astronomy and Astrophysics, Australian National University, Canberra, ACT 2611, Australia}

\author{D.~M.~Nataf} 
\affiliation{Department of Physics \& Astronomy, University of Iowa, Iowa City, IA 52242, USA}

\author{M.~K.~Ness}
\affiliation{Research School of Astronomy and Astrophysics, Australian National University, Canberra, ACT 2611, Australia}
\alsoaffiliation{ARC Centre of Excellence for All Sky Astrophysics in 3 Dimensions (ASTRO 3D), Australia}

\author{A.~D.~Rains}
\affiliation{Department of Physics and Astronomy, Uppsala University, Box 516, 751 20 Uppsala, Sweden}

\author{T.~Scarr}
\affiliation{Research School of Astronomy and Astrophysics, Australian National University, Canberra, ACT 2611, Australia}

\author{R.~Vogrin{\v{c}}i{\v{c}}}
\affiliation{Faculty of Mathematics \& Physics, University of Ljubljana, Jadranska 19, 1000 Ljubljana, Slovenia}

\author{Z.~Wang}
\affiliation{Sydney Institute for Astronomy, School of Physics, A28, The University of Sydney, NSW 2006, Australia}
\alsoaffiliation{ARC Centre of Excellence for All Sky Astrophysics in 3 Dimensions (ASTRO 3D), Australia}
\alsoaffiliation{Department of Physics and Astronomy, University of Utah, Salt Lake City, UT 84112, USA}

\author{R.~A.~Wittenmyer}
\affiliation{Centre for Astrophysics, University of Southern Queensland, West Street, Toowoomba, QLD 4350, Australia}

\author{Y.~Xie}
\affiliation{Research School of Astronomy and Astrophysics, Australian National University, Canberra, ACT 2611, Australia}
\alsoaffiliation{ARC Centre of Excellence for All Sky Astrophysics in 3 Dimensions (ASTRO 3D), Australia}

\author{The GALAH Collaboration}
\affiliation{https://www.galah-survey.org}

\doi{10.1017/pasa.2020.32}

\received{01 October 2024}
\revised{06 January 2025}
\accepted{10 March 2025}
\published{10 March 2025}
\begin{document}

\begin{abstract}
The stars of the Milky Way carry the chemical history of our Galaxy in their atmospheres as they journey through its vast expanse. Like barcodes, we can extract the chemical fingerprints of stars from high-resolution spectroscopy. The fourth data release (DR4) of the Galactic Archaeology with HERMES (GALAH) Survey, based on a decade of observations, provides the chemical abundances of up to 32 elements for 917\,588 stars that also have exquisite astrometric data from the $Gaia$ satellite. For the first time, these elements include life-essential nitrogen to complement carbon, and oxygen as well as more measurements of rare-earth elements critical to modern-life electronics, offering unparalleled insights into the chemical composition of the Milky Way.

For this release, we use neural networks to simultaneously fit stellar parameters and abundances across the whole wavelength range, leveraging synthetic grids computed with Spectroscopy Made Easy. These grids account for atomic line formation in non-local thermodynamic equilibrium for 14 elements. In a two-iteration process, we first fit stellar labels to all 1\,085\,520 spectra, then co-add repeated observations and refine these labels using astrometric data from $Gaia$ and 2MASS photometry, improving the accuracy and precision of stellar parameters and abundances. Our validation thoroughly assesses the reliability of spectroscopic measurements and highlights key caveats.

GALAH DR4 represents yet another milestone in Galactic archaeology, combining detailed chemical compositions from multiple nucleosynthetic channels with kinematic information and age estimates. The resulting dataset, covering nearly a million stars, opens new avenues for understanding not only the chemical and dynamical history of the Milky Way, but also the broader questions of the origin of elements and the evolution of planets, stars, and galaxies.
\end{abstract}

\section{INTRODUCTION AND WORKFLOW}
\label{sec:introduction}

\subsection{Motivation} \label{sec:motivation}

The history of our Milky Way galaxy is written in starlight. By capturing and analysing the light from millions of stars, which are now millions or billions of years old, we can uncover the chemical compositions embedded in their atmospheres since birth and use stars as time capsules into the past evolution of the Milky Way. The light of stars can thus guide us to explore and map our environment and Country, just as it has guided Aboriginal and Torres Strait Islander peoples and their astronomers for tens of thousands of years.

With this fourth data release (DR4) from the Galactic Archaeology with HERMES (GALAH) Survey, we are proudly publishing the next set of measurements of stellar chemical abundances for almost a third of the elements in the periodic table that are created by stars. The initial motivation for measuring so many elemental abundances was laid out by \citet{DeSilva2015} and included the major motivation - chemical tagging -- with the aim to trace back stars that were born together through their (expected) similar chemical compositions. The recent and ongoing efforts of GALAH and other surveys like the SDSS/APOGEE surveys \citep[e.g.][]{SDSSDR17, Kollmeier2017}, LAMOST \citep{Zhao2012}, \Gaia-ESO \citep{Gilmore2022,Hourihane2023}, RAVE \citep{Steinmetz2020a}, and \Gaia RVS \citep{RecioBlanco2023} have taught us that the chemical evolution of our Galaxy and stars is complex and it is difficult to recover stellar siblings on a large scale due to limitations in our observations, analysis methods, and intrinsic changes to chemical composition due to stellar evolution. New observations and innovations in the analysis that are presented in this data release will allow us to make significant progress towards chemical tagging.

The unique observational setup of GALAH allows us to deliver chemical abundance information for a powerful and substantial set of stars: those which have exquisite astrometric information from the revolutionary \Gaia satellite \citep{Gaia-Collaboration2016} and for which we can estimate stellar ages either from empirical or theoretical models, like stellar isochrones or mass- and age-dependent relations of chemical compositions. By combining stellar ages, orbits, and chemistry, we have made major advances in the understanding of our Galaxy. In particular, the discovery of the major merger of the Milky Way with another slightly less massive galaxy between 8 and $10\,\mathrm{Gyr}$ ago \citep{Belokurov2018, Helmi2018} was paradigm shifting and motivated a new rush to collect more (and more diverse) information about the stars in our Milky Way.

GALAH DR4 presents two major improvements over the previous data releases. We have increased the quantity as well as quality of observations and we have implemented a hybrid spectrum synthesis approach that allows us to fit 95\% of the spectrum, including broad molecular absorption features from $\mathrm{C_2}$ and CN. This allows us to now infer up to 32 elements\footnote{
Li, C, N, O, Na, Mg, Al, Si, K, Ca,
Sc, Ti, V, Cr, Mn, Fe, Co, Ni, Cu, Zn,
Rb, Sr, Y, Zr, Mo, Ru, Ba, La, Ce, Nd,
Sm, and Eu}, including N, with unprecedented precision for a larger number of stars. GALAH DR4 naturally continues both the observing program aimed at acquiring spectra of 1 million stars \citep{DeSilva2015}, and our ongoing efforts to improve the spectrum reduction and analysis pipelines, including the novel and more accurate line modelling with non-local thermodynamics equilibrium. In GALAH DR1 and DR2 \citep{Martell2017, Buder2018}, we developed a novel, data-driven pipeline using the interpolation and fitting code \textit{The Cannon} \citep{Ness2015}. However, for DR3 \citep{Buder2021}, we reverted to the more computationally expensive method of spectrum synthesis, applying it to a limited wavelength range to confirm the accuracy of our data-driven approach. In this data release, we are now implementing a hybrid approach. We create a training set of synthetic spectra across the full wavelength range using the same synthesis code as DR3, then train a neural network to interpolate the spectra efficiently in a high-dimensional space with up to 36 dimensions. By using neural networks, we can model the entire wavelength range, including broad molecular absorption features from $\mathrm{C_2}$ and CN, rather than focusing on narrow atomic line windows. This approach allows us to simultaneously model all stellar labels—global parameters and elemental abundances. Additionally, we can infer the shape of the interstellar spectrum from the differences between observed and synthetic spectra, while also incorporating non-spectroscopic information during the optimisation process.

In the following section, we outline our workflow and provide detailed explanations of our methodology throughout this manuscript, offering insights that upcoming surveys like WEAVE \citep{Dalton2014}, SDSS-V \citep{Kollmeier2017}, and 4MOST \citep{4MOST2019} can readily utilise.

\begin{figure}[ht]
 \centering
 \includegraphics[width=0.95\textwidth]{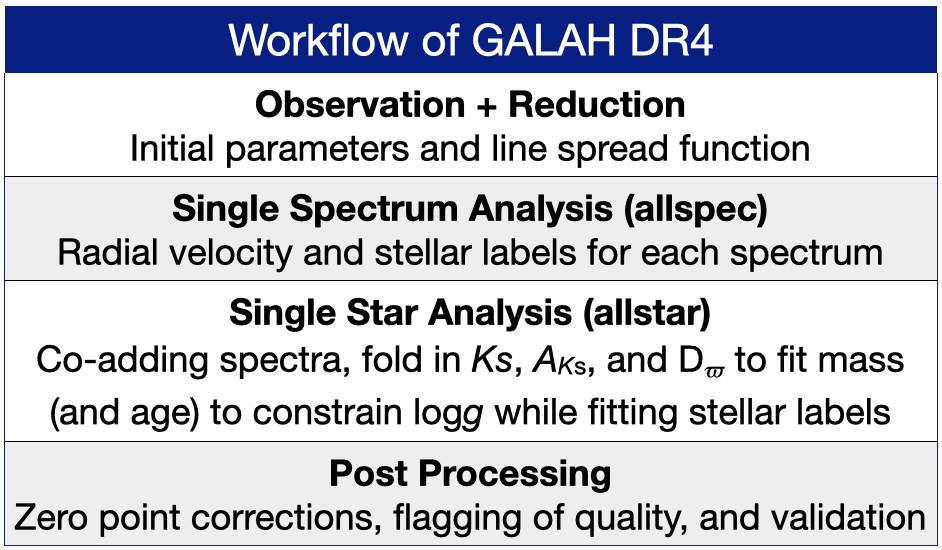}
 \caption{\textbf{Workflow of GALAH DR4.}}
 \label{fig:workflow_galah_dr4}
\end{figure}

\subsection{Workflow} \label{sec:workflow}

The workflow of GALAH DR4 is depicted in Figure~\ref{fig:workflow_galah_dr4} and will serve as a guideline for this manuscript: We first describe the collection of data in Section~\ref{sec:data}, most notably the observation of HERMES spectra. We explain how we create synthetic stellar spectra to compare with the observed ones in Section~\ref{sec:synthetic_spectra}. This comparison is done in two consecutive steps. In Section~\ref{sec:allspec_analysis}, we explain how we extract stellar labels from individual observations (without non-spectroscopic information folded into the optimisation), while Section~\ref{sec:allstar_analysis} describes how we co-add repeated observations and fold in non-spectroscopic information for each star. We describe the post-processing and validation of our data in Section~\ref{sec:post_processing}. The data products of this data release are explained in Section~\ref{sec:catalogues_release_products}. We describe identified caveats in Section~\ref{sec:caveats} and make suggestions for minimising them in the future, before concluding this manuscript in Section~\ref{sec:conclusion}.

\begin{figure*}[ht]
 \centering
 \includegraphics[width=0.91\textwidth]{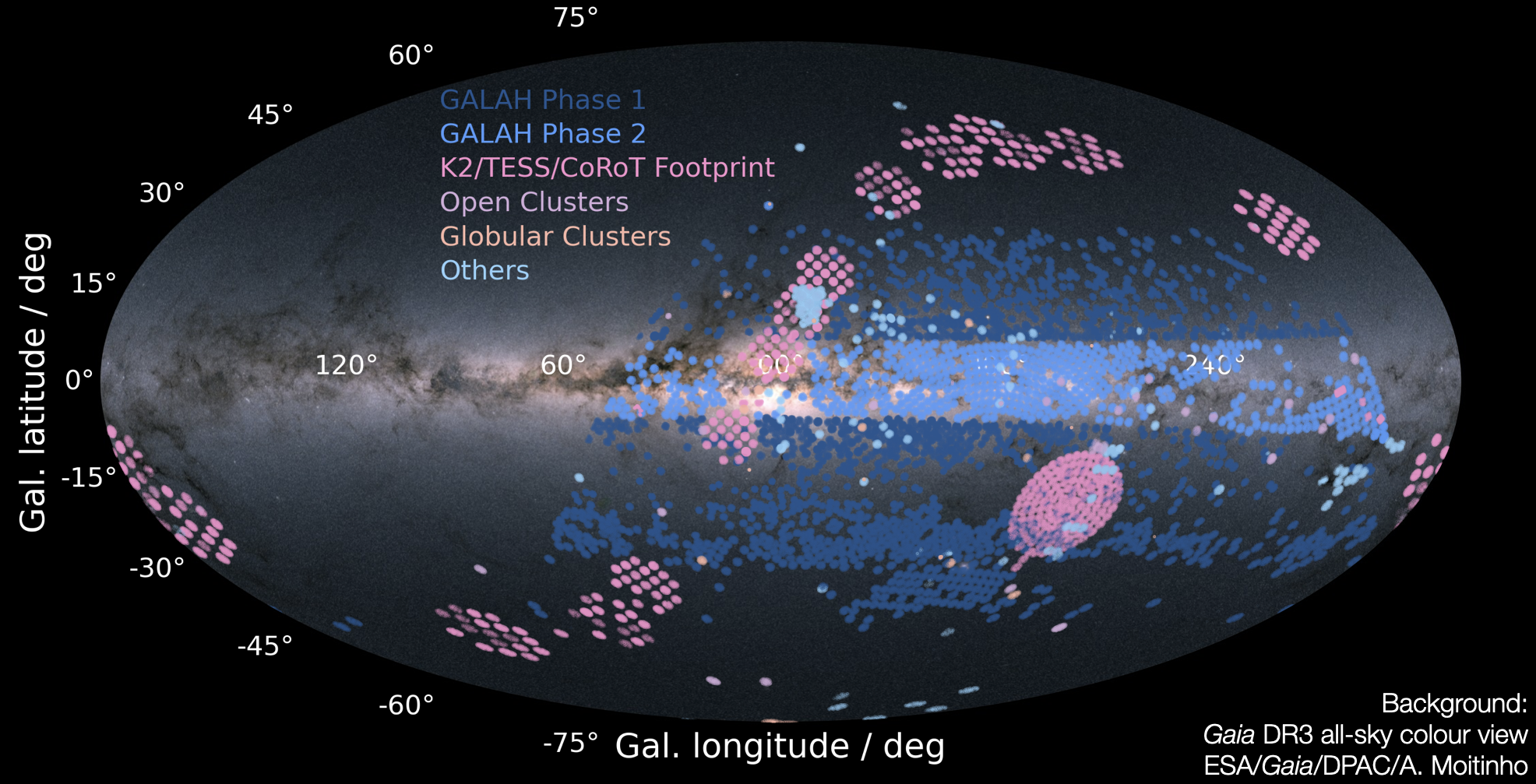}
 \caption{\textbf{Overview of the distribution of stars included in this fourth GALAH data release in Galactic coordinates with the centre of the Galaxy at the origin and the \Gaia DR3 all-sky colour view \citep{GaiaDR3} as background.}
Shown are the targets of GALAH Phase 1 (dark blue) and Phase 2 (medium blue), the targets of the K2-HERMES follow-up along the ecliptic and TESS-HERMES in the TESS Southern Continuous Viewing Zone as well as CoRoT fields (pink). Both open and globular cluster points are shown in purple and orange, respectively. All other targets are shown in in light blue across the Southern sky.
}
 \label{fig:galah_dr4_skymap_gaiadr3}
\end{figure*}

\section{DATA}
\label{sec:data}

The GALAH Survey uses the 3.9-metre Anglo-Australian Telescope at Siding Spring Observatory on Gamilaraay Country and its Two-Degree Field positioning system (2dF) top end \citep{Lewis2002}. 2dF magnetically places up to 400 fibre buttons on one of two metal field plates, which can be tumbled to allow observing with one set of fibres while configuring the other. Light is delivered through the fibres to the High Efficiency and Resolution Multi-Element Spectrograph (HERMES) spectrograph \citep{Barden2010, Brzeski2011, Heijmans2012, Farrell2014, Sheinis2015} and dispersed into four non-contiguous wavelength bands in the optical that cover $\sim 1000\,\text{\AA}$ in the range of $4713-4903$ (blue CCD or CCD1), $5648-5873$ (green / CCD2), $6478-6737$ (red / CCD3), and $7585-7887\,\text{\AA}$ (infrared IR / CCD4). The data used in this data release is primarily based on observations of stars with this setup, but also makes use of auxiliary photometric and astrometric information for the stars where available.

In this Section, we describe which stars we have targeted as part of configured fields \citep{Miszalski2006} and observed with the 2dF-HERMES setup (Section~\ref{sec:target_selection_observations}), including the first description of the second phase of GALAH observations (GALAH Phase 2) which has a sharper focus on main-sequence turn-off stars to estimate more precise ages. In Section~\ref{sec:spectroscopic_data_from_galah_observations}, we briefly summarise the properties of the spectroscopic data and how they were reduced to one-dimensional spectra. We also point out major changes in the observations and reductions with respect to the previous (third) data release \citep{Buder2021}. We further elaborate on the auxiliary information that was used for the analysis in Section~\ref{sec:non-spec_data}.

\subsection{Target selection and observational setup} \label{sec:target_selection_observations}

GALAH DR4 is a combination of the main GALAH survey and additional projects to observe asteroseismic targets from the K2 \citep{Howell2014} and TESS \citep{Ricker2015} missions, that is, K2-HERMES \citep{Sharma2019} and TESS-HERMES \citep{Sharma2018}, as well as numerous smaller programs and public HERMES data. Additional proposals with 2dF-HERMES have contributed targeted observations of globular cluster members (PI M. McKenzie and PI M. Howell), open clusters (PI G. De Silva and PI J. Kos), young stellar associations (PI J. Kos and J. Armstrong), and halo stars (PI S. Buder) in addition to their observation through the main surveys. The column \texttt{survey\_name} in our catalogues denotes the origin. An all-sky view of GALAH DR4 is shown in Figure~\ref{fig:galah_dr4_skymap_gaiadr3}.

\subsubsection{Target selection for GALAH Phase 1 and 2}

For GALAH Phase 1 (DR1-DR3) and in the absence of a precise and volume-complete survey in the optical, we used the 2MASS photometric survey \citep{Skrutskie2006} with its $J$ and $Ks$ filters as a precise and nearly volume-complete parent sample from which we selected stars based on approximated \citep{DeSilva2015} visual magnitudes
\begin{equation}
V_{JK_S} = K_S+2(J-K_S+0.14)+0.382e^{((J-K_S-0.2)/0.5)}.
\end{equation}

For GALAH Phase 1, a tiling pattern (with unique \texttt{field\_id} entries) with $2\,\mathrm{deg}$ fields of view below declination $\delta \leq +10\,\mathrm{deg}$ was created for regions with Galactic latitude $\vert b \vert \geq 10\,\mathrm{deg}$ to avoid crowding and strong extinction. For each tile, a selection of 400 stars within magnitudes $9 \leq V_{JK_S} \leq 12$ for a bright magnitude cut and $12 \leq V_{JK_S} \leq 14$ for the nominal magnitude cut is randomly selected from the complete parent sample of 2MASS. Of those, typically 350 stars are actually observed with around 2/3 main-sequence and turn-off stars and 1/3 evolved stars.

For GALAH Phase 2, a stronger focus on turn-off stars was implemented with the photometric and astrometric information of \Gaia data release 2 as a parent sample. For each field, we therefore first allocate fibres to stars with absolute \Gaia magnitude in the range of $2 \leq M_G \leq 4$, where
\begin{equation}
M_G = G + 5 \cdot \log_{10} \left( \frac{\varpi}{100\,\mathrm{mas}} \right)
\end{equation}
with apparent magnitude ($G~/~\mathrm{mag} \equiv \texttt{phot\_g\_mean\_mag}$) and parallax measurements ($\varpi~/~\mathrm{mas} \equiv \texttt{parallax}$) from \Gaia DR2 \citep{Brown2018, Evans2018, Lindegren2018}. Remaining fibres are filled with targets as done with the Phase 1 selection function. This leads to a different selection function for each phase. For science cases in which selection functions matter, we thus recommend to use the \texttt{survey\_name} (Table~\ref{tab:field_ids}) for a clean selection of phase and selection function.

\subsubsection{Observational setup}

\begin{figure*}[ht]
 \centering
 \includegraphics[width=0.96\textwidth]{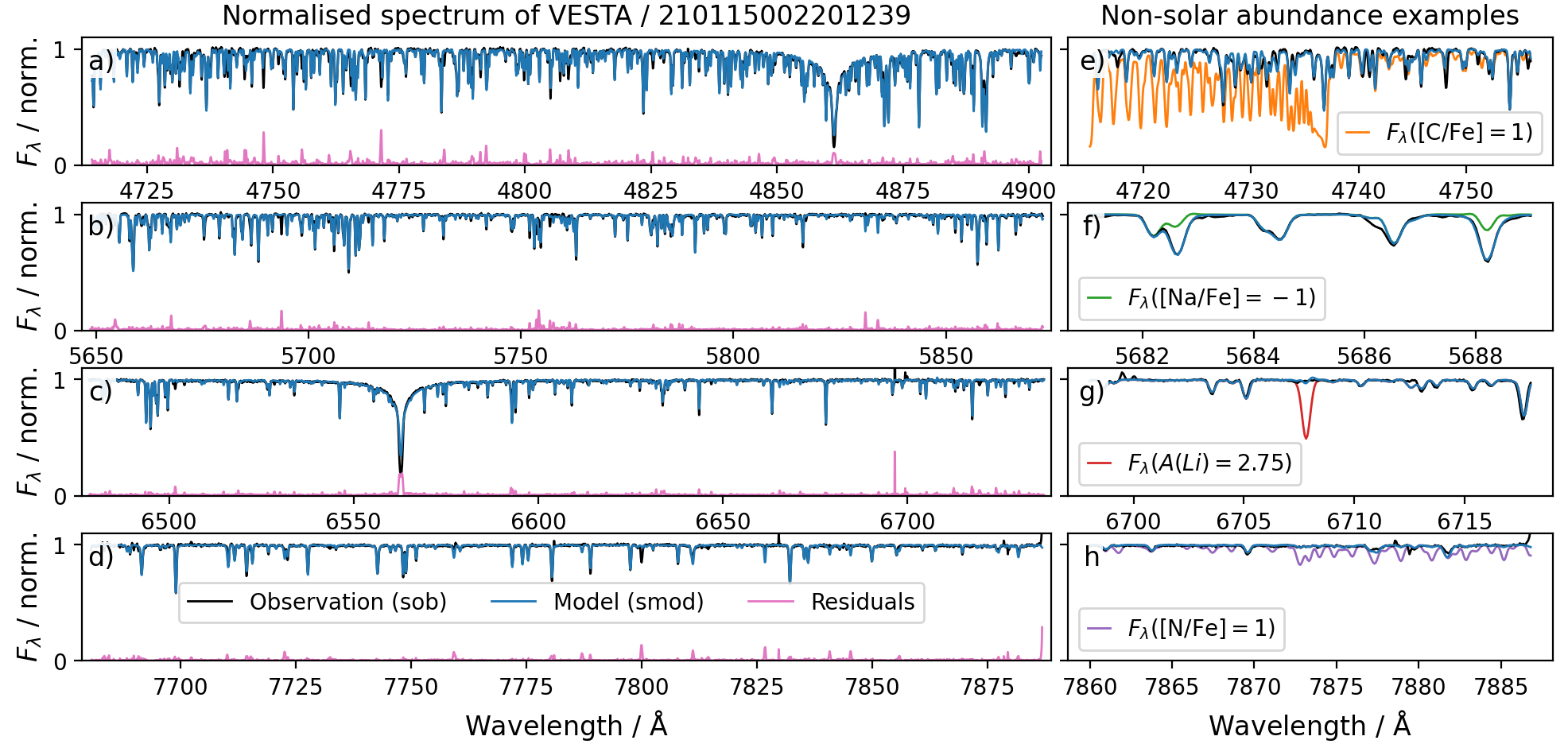}
 \caption{\textbf{Comparison of normalised observed (black) and synthetic spectra (blue) of the asteroid Vesta with solar composition as well as examples of synthetic spectra with non-solar abundances.}
 \textbf{Panels a-d)} show the observed and best-fitting synthetic spectrum as well as their absolute residual (pink) for the four wavelength channels of the HERMES spectrograph.
 \textbf{Panel e)} shows the beginning of the blue CCD 1 (left most part of panel a) with an additional synthetic spectrum with ten times higher [C/Fe] in orange, for which the $\mathrm{C}_2$ Swan bands are prominent.
 \textbf{Panel f)} shows the beginning of the green CCD 2 (left most part of panel b) and exemplifies with a synthetic spectrum in green that also has a ten times lower [Na/Fe] abundance (for example, in accreted stars) can still be reliably detected. 
 \textbf{Panel g)} shows the end of the red CCD 3 with a synthetic spectrum of primordial Li abundance of $\mathrm{A(Li)} = 2.75$ in red. While this abundance could be detected, the line for the Solar value $\mathrm{A(Li)} = 1.05$ is barely detectable.
 \textbf{Panel h)} shows the end of the infrared CCD 4, which would show strong molecular absorption features of the CN molecule for $\mathrm{[N/Fe]} = +1\,\mathrm{dex}$ (purple).
 }
 \label{fig:210115002201239_abundance_examples}
\end{figure*}

We list the observations under various sub-programs in Table~\ref{tab:field_ids}. Except for 2\,935 spectroscopic observations with the high-resolution mode of HERMES ($R \sim 42\,000$) on 7, 8, 10, 11 and 12 February 2014, all observations were made in the low-resolution mode ($R \sim 28\,000$) with different total exposure times chosen for different programs, but typically between 60 and 90 minutes. Under sufficient conditions (no clouds and seeing below $2\,\mathrm{arcsec}$), GALAH Phase 1 and TESS-HERMES observed 3 exposures for 6 minutes for bright targets ($9 \leq V_{JK_S} \leq 12$) and 3 exposures for 20 minutes for the majority of targets ($12 \leq V_{JK_S} \leq 14$).

\begin{table}
\centering
 \caption{\textbf{Overview of stars observed for the programs included in GALAH DR4.} Numbers of open and globular cluster observations were estimated after observations as described in Section~\ref{sec:oc_gc}. We have observed 30 globular clusters (23 with $\geq$ 5 stars) and 361 open clusters (109 with $\geq$ 5 stars).}
\label{tab:field_ids}
\begin{tabular}{crcr}
\hline \hline
Program & No. Stars & Program & No. Stars \\
\hline
galah\_bright & 67\,680 & 
k2\_hermes & 117\,736\\
galah\_main & 434\,901 & 
tess\_hermes & 37\,129\\
galah\_faint & 33\,907 & 
globular clusters & 2\,509\\ 
galah\_phase2 & 172\,494 & 
open clusters & 3\,706\\ 
commissioning & 2\,625 & 
other & 44\,901\\
  \hline
 \end{tabular}
\end{table}

\begin{table}
    \centering
    \caption{Data product 1: FITS files of reduced spectra.}
    \label{tab:reduction_fits}
    \begin{tabular}{cc}
    \hline \hline
    FITS Ext. & Description \\
    \hline
    Ext. 0 & Un-normalised signal~/~counts \\
    Ext. 1 & Normalised signal (by reduction pipeline) \\
    Ext. 2 & Relative uncertainty of signal \\
    Ext. 3 & Subtracted sky signal~/~counts \\
    Ext. 4 & Applied telluric correction \\
    Ext. 5 & Scattered light~/~counts \\
    Ext. 6 & Cross-talk~/~counts \\
    Ext. 7 & Resolution profile~/~FWHM \\
    \hline
    \end{tabular}
\end{table}

GALAH Phase 2 extended these times to 3 exposures of 10 or 30 minutes, respectively, and included repeat observations of GALAH Phase 1 main targets with another 3 exposures for 15 minutes. K2-HERMES observations targeted stars with $13 \leq V_{JK_S}~/~\mathrm{mag} \leq 15$ or even $13 \leq V_{JK_S}~/~\mathrm{mag} \leq 15.8$ to complement the K2 Galactic Archaeology Program \citep{Stello2015}. These fields were observed for 2 hours, similar to most globular and open cluster stars. Worse seeing conditions leading to increasing full-width at half maxima or thin clouds triggered between one ($2 < \mathrm{seeing} \leq 2.5\,\mathrm{arcsec}$) and 3 ($2.5 < \mathrm{seeing} \leq 3\,\mathrm{arcsec}$) additional exposures. In addition to the science frames, quartz fibre flat and ThXe arc observations were taken directly before or after each set of science exposures, and bias frames were taken at the beginning or end of each observing night.

\subsection{Spectroscopic data from GALAH observations}
\label{sec:spectroscopic_data_from_galah_observations}

Since the commissioning of the HERMES spectrograph in late 2013 until 6 August 2023, the GALAH collaboration and its partners have observed and successfully reduced \allspecnumber spectra of \allstarnumber stars. Each single observation is given a unique \texttt{sobject\_id} YYMMDDRRR01FFF that is based on its year (YY), month (MM), and day (DD) of observations, its exposure run number (RRRR), and the used fibre (FFF). A reduced example spectrum of the asteroid Vesta (observed on 15 January 2014 during run 22 through fibre 239 with \texttt{sobject\_id} 210115002201239) is shown in Figure~\ref{fig:210115002201239_abundance_examples} and used as a reference for a Solar spectrum. The reduction process to create FITS files of reduced spectra from two-dimensional images from the cameras employs an updated and publicly available \href{https://github.com/sheliak/galah_reduction/blob/master/extract6.0.py}{version 6} of the already well-tested reduction pipeline \citep{Kos2017}. The file extensions are listed in Table~\ref{tab:reduction_fits} and created as follows.

Science frames are corrected by removing the bias, dividing out the different gains (provided in the FITS headers) of the two readout amplifiers per CCD, flagging bad pixels, and dividing by master flat field frames, as well as removing cosmic rays and scattered light. Subsequently, apertures for each fibre trace are identified and used to extract the individual spectra. 

Wavelength calibrations are performed via Chebyshev polynomial functions based on the up to 62, 52, 41, or 31 emission lines within the ThXe arc frames of CCDs 1-4, with wavelengths reported in air, and the spectra are interpolated onto a linearly increasing wavelength grid. Typical root mean square values for the wavelength solutions of CCDs 1-4 are 0.010, 0.015, 0.019, and $0.028\,\mathrm{\AA}$, respectively. The starting wavelength \texttt{CRVAL1} and dispersion \texttt{CDELT1} are saved in the headers of each FITS file.

Finally, sky lines are subtracted and telluric features removed, before a barycentric correction is applied to create the `reduced' spectra that are saved in extension~0 of the reduction pipeline FITS files and used for the subsequent analysis. Reduction pipeline spectra are normalised by an eleventh order Legendre polynomial fit and saved in extension 1 of the reduction products.

Fractional noise/uncertainties are saved in extension~2 and calculated from the square root of the sum of squared counts, sky features (extension~3), scattered light (extension~5), and crosstalk (extension~6) measurements as well as the squared readout noise.

\begin{figure}[ht]
    \centering
    \includegraphics[width=\columnwidth]{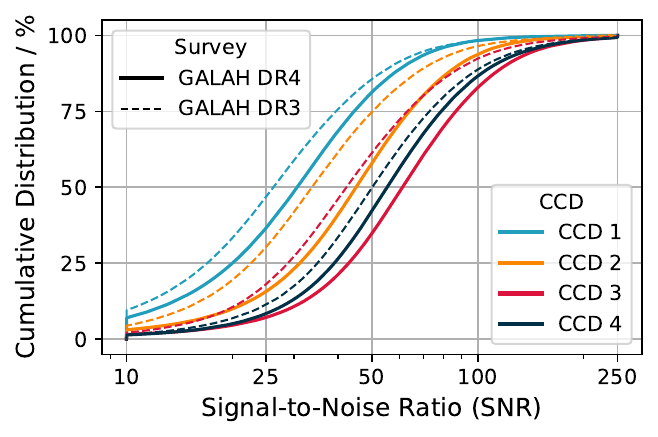}
    \caption{Cumulative Distribution Function (CDF) of the logarithmic Signal-to-Noise Ratio (SNR) per pixel for the 4 CCDs of the HERMES spectrograph comparing GALAH DR4 (solid lines) and GALAH DR3 (dashed lines).}
    \label{fig:snr_distribution}
\end{figure}

The wavelength dependent line spread functions (LSFs) are measured from the arc calibration frames for each spectrum and CCD by fitting modified Gaussian distributions with one boxiness parameter $b$ per CCD and full width half maxima \texttt{fwhm} for each wavelength point in the spectrum, that is
\begin{align}
    \exp \left(-0.693147 \cdot \vert 2 \cdot \textbf{\textit{x}}/\texttt{fwhm} \vert^\texttt{b}\right) \label{eq:lsf}
\end{align}
The array $\textbf{\textit{x}}$ then includes the pixels around each wavelength step that are used to apply the convolution from higher resolution to GALAH resolution spectra. The fitted values of \texttt{fwhm} are saved in extension~7 with $b$ saved in the headers.

The achieved Signal-to-Noise Ratio (SNR) per pixel of the individual exposures depends on the spectral types, reddening, and observational conditions. In particular the repeat observations of previous pointings have increased the SNR for co-added spectra with respect to GALAH DR3. This can be appreciated from Figure~\ref{fig:snr_distribution}, where we plot the cumulative distribution function for all stars of GALAH DR3 (dashed lines) and GALAH DR4 (solid lines) for the different CCDs. 

\subsection{Auxiliary data from \Gaia, 2MASS, and literature} \label{sec:non-spec_data}


To support our spectroscopic analysis, we make use of astrometric and photometric information from the \Gaia satellite \citep{Gaia-Collaboration2016} and 2MASS survey \citep{Skrutskie2006}, which is available for essentially all our targets. We further use the value-added catalogues, like distance estimates for field stars by \citet{BailerJones2021} as well as open and globular cluster membership probabilities from \citet{CantatGaudin2020} as well as \citet{Vasiliev2021} and \citet{Baumgardt2021}.

\subsubsection{\Gaia DR3}

We crossmatch our observations to the \Gaia DR3 catalogue \citep{Brown2021,GaiaDR3} using the 2MASS ID, via the nearest neighbour crossmatches provided as part of \Gaia DR3 \citep{Torra2021}. 
911\,754 (99.0\,\%) also have astrometric information \citep{Lindegren2021a} and 849\,867 (93.0\,\%) have radial velocity estimates \citep{Katz2023}. We apply the corrections to both photometric \citep{Riello2021} and astrometric \citep{Lindegren2021b} information. Where possible we prefer the photo-geometric distances over the geometric distances from \citet{BailerJones2021}. Where neither are available, we further try to find parallaxes from \cite{vanLeeuwen2007}. The average parallax uncertainty of the GALAH stars is $\sigma_{\varpi} / \varpi = 1.6_{-0.9}^{+2.6}\,\mathrm{\%}$. Only $2.3\,\%$ of GALAH stars have no parallax measurements\footnote{For stars without parallaxes, we only perform an analysis without astrometric information.} or parallax measurements beyond $20\%$ uncertainty, for which the priors adopted by \citet{BailerJones2021} start to dominate distance estimates.

\subsubsection{2MASS, WISE, and extinction}

In addition to the excellent infrared photometry for 99.9\,\% of our sources from the 2MASS survey \citep{Skrutskie2006}, 98.7\,\% of them have far-infrared measurements from the WISE mission \citep{Cutri2013}. We therefore can estimate the extinction in the $K_S$ band via the Rayleigh-Jeans colour excess (RJCE) method \citep{Majewski2011} $A_{K_S} = 0.917 \cdot \left( H - W2 - 0.08 \right)$ for most stars. We confirm this estimate by estimating the extinction in $K_S$ via the extrapolation of the colour extinction of $B-V$, that is, $A_{K_S} \sim 0.36 \cdot E(B-V)$ \citep{Cardelli1989}. We revert to this value if it is less than half the value of the RJCE estimate, or if either of the $H$ and $W2$ bands does not have an excellent quality flag `A'. For negative estimates by the RJCE method and very nearby stars ($<100\,\mathrm{pc}$) we null the value.

\subsubsection{Open and Globular Cluster members and distances} \label{sec:oc_gc}

We identify open cluster members using the membership catalogue from \citet{CantatGaudin2020} via crossmatch with the \Gaia \texttt{source\_id} and adjust their parallaxes and distance estimates to the average cluster values if the latter are more precise. We identify globular cluster members (with more than 70\% probability) via the membership catalogue from \citet{Vasiliev2021} by crossmatching with the \Gaia \texttt{source\_id}. We then adjust the parallaxes and distances for the member stars to the mean values listed by \citet{Baumgardt2021}.

\section{SYNTHETIC SPECTRA FOR 2DF-HERMES}
\label{sec:synthetic_spectra}

The goal of our spectroscopic analysis is to estimate the optimal set of stellar properties (labels) that influence a stellar spectrum by minimising the difference between observed stellar spectra and synthetic ones. In this data release, we push the analysis further by fitting up to 32 elemental abundances and stellar parameters simultaneously across the full GALAH wavelength range with the appropriate synthetic spectra.

To make this computationally feasible, we adopt a strategy inspired by \citet{Rix2016}, where we create flexible models for smaller regions of the parameter space, utilizing only a limited number of \textit{ab initio} synthetic spectra \citep[see also][]{Ting2016b}. These synthetic spectra are calculated using Spectroscopy Made Easy \citep[\sme;][]{Valenti1996,Piskunov2017}, covering the entire wavelength range and accounting for all visible atomic and molecular lines. The spectra are generated for random selections of elemental abundances and stellar parameters within the range allowed by \marcs atmospheric models \citep{Gustafsson2008}, at much higher resolution and sampling than our 2dF-HERMES spectra. From these, we select subsets of spectra corresponding to restricted regions of the parameter space defined by \Teff, \logg, and \feh. This method is analogous to using Solar twins \citep[see, e.g.,][]{Nissen2015} or performing differential abundance analysis of globular cluster stars \citep[e.g.,][]{Yong2013, Monty2023}. By reducing the impact of systematic uncertainties in atomic data and atmospheric models, these approaches have proven to be highly effective \citep{Nissen2018}.

For each parameter subset, we train a neural network to map stellar fluxes to their corresponding stellar parameters and abundances, similar to \textit{The Payne} \citep{Ting2019}. These models allow us to generate synthetic spectra across the full wavelength range for any combination of elemental abundances within the restricted parameter space in under a second—compared to the minutes or hours required by traditional physics-driven spectrum synthesis approaches.

Another key motivation for creating smaller training sets is the limited flexibility of interpolation methods when dealing with the full parameter space. Spectroscopic surveys like GALAH, RAVE, and APOGEE aim to fit all types of stellar spectra simultaneously, including Sun-like stars, red clump stars, metal-poor stars, evolved stars with strong molecular bands, and hot stars with shallow and broad lines. Attempting to model this vast range with a single model leads to systematic trends, particularly in extreme cases \citep{Casey2016, Buder2018, Ting2019}. To mitigate these issues, we deliberately limit the complexity of the models by creating smaller, more focused models. For example, the model for hot stars does not need to predict the strong molecular absorption features seen in cooler stars. The potential caveats and limitations of this approach are discussed in detail in Section~\ref{sec:caveats}.

In the following sections, we describe our approach to dividing the parameter space into smaller bins for training (Section~\ref{sec:higher_resolution_synthetic_spectra}) and explain how we generate high-resolution synthetic spectra for this parent sample (Section~\ref{sec:higher_resolution_synthetic_spectra}). We also outline how we train neural networks to rapidly interpolate these synthetic spectra (Section~\ref{sec:interpolating_synthetic_spectra_with_neural_networks}).

\begin{figure}[ht]
 \centering
 \includegraphics[width=\textwidth]{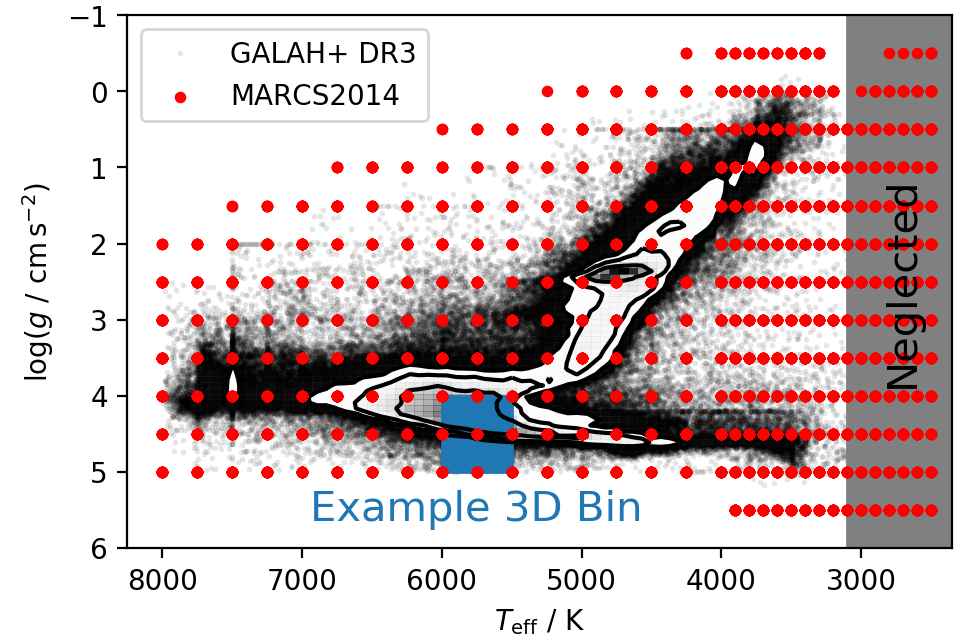}
 \caption{\textbf{Coverage in \Teff and \logg of the MARCS2014 grid (red) and GALAH DR3 (black, including density countours).} Shown is also an example of one of the 3D bins used to create stellar sibling models with each neural network. \marcs grid points \Teff$ < 3100\K$ or \feh$<-3\dex$ are neglected for GALAH DR4.}
 \label{fig:teff_logg_grid_coverage}
\end{figure}

\begin{figure*}[ht]
 \centering
 \includegraphics[width=\textwidth]{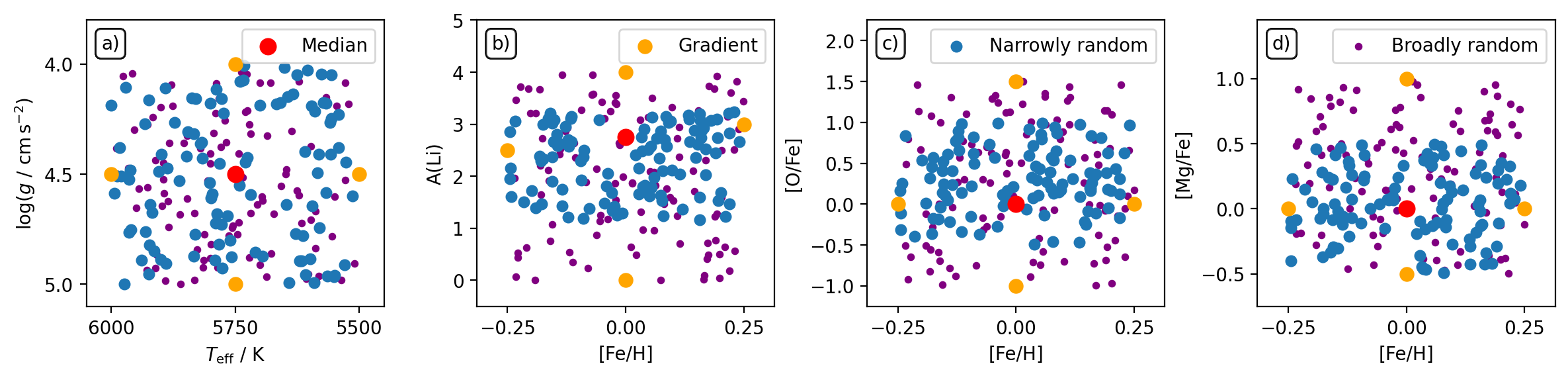}
 \caption{\textbf{Coverage of stellar parameters and abundances for one of the 3D bins.} Shown is the example of the Solar 3D bin ($T_\mathrm{eff}~/~\mathrm{K} = 5750$, $\log g~/~\mathrm{dex} = 4.5$, $\mathrm{[Fe/H]}~/~\mathrm{dex} = 0.0$). \textbf{Panel a):} \Teff and \logg, \textbf{Panel b):} [Fe/H] vs. A(Li), \textbf{Panel c):} [Fe/H] vs. [O/Fe], \textbf{Panel d):} [Fe/H] vs. [Mg/Fe]. While \Teff, \logg, and \feh are sampled randomly within the 3D bin, the abundances are sampled both narrowly (blue) and broadly (purple) within limits as described in the text. Red points indicate the median label values and orange points the adjusted label values (see Table~\ref{tab:sampling_xfe}) to test the gradient change of spectra with individual labels.}
 \label{fig:example_3d_bin_sample}
\end{figure*}

\begin{table*}[ht]
\centering
 \caption{Example of boundaries for the uniform sampling of synthetic spectrum labels (stellar parameters and elemental abundances) for the 3-dimensional bin of Solar siblings \texttt{5750\_4.50\_0.00}.}
\label{tab:sampling_xfe}
\begin{tabular}{lclclc}
\hline \hline
Parameter & Sampling & Element & Sampling Narrow & Element & Sampling Broad \\
\hline
$T_\text{eff}~/~\K$ & 5500..5750..6000 & $\mathrm{A(Li)}$ & {1.05..2.75..3.26} & $\mathrm{A(Li)}$ & {0.00..4.00} \\
$\log g~/~\dex$ & 4.0..4.5..5.0 &  $\mathrm{C,~N,~O}$ & {-0.5..0.0..1.0} & $\mathrm{C,~N,~O}$ & {-1.0..1.5} \\
$\mathrm{[Fe/H]}~/~\dex$ & {-0.25..0.0..0.25} & $\mathrm{Y,~Ba,~La,~Ce,~Nd}$ & {-0.5..0.0..1.0} & $\mathrm{Y,~Ba,~La,~Ce,~Nd}$ & {-1.0..1.5} \\
$v_\text{mic}~/~\kms$ & {0.5,1.5,4.0}, but see Equation~\ref{eq:vmic_initial} & $\mathrm{[X/Fe]~for~Mg,~Si,~Ti}$  &  {-0.5..0.0..0.5}& $\mathrm{[X/Fe]~for~Mg,~Si,~Ti}$ & {-0.5..1.0} \\
$v \sin i~/~\kms$ & 0.0\text{, but see Equation~\ref{eq:vsini}} & $\mathrm{[X/Fe]~for~all~other~elements}$ & {-0.5..0.0..0.5} & $\mathrm{[X/Fe]~for~all~other~elements}$ & {-1.0..1.0}  \\
\hline \hline
\end{tabular}
\end{table*}

\subsection{Stellar twin training sets rather than one-fits-all}
\label{sec:spectrum_grid}

The base grid for our training set computation is the \marcs grid \citep{Gustafsson2008}, which is shown with red points in Figure~\ref{fig:teff_logg_grid_coverage}. Following the aforementioned idea of restricting ourselves to stellar siblings, we create multiple 3-dimensional bins in \Teff, \logg, and \feh within $\pm 1$ grid points in \Teff (with either $\pm 250$ or $\pm 100\K$), \logg ($\pm 0.5\dex$), and \feh ($\pm 0.5$ or $\pm 0.25\dex $). An example box is shown for Solar siblings as a blue box in Figure~\ref{fig:teff_logg_grid_coverage}, which is centred on $T_\text{eff} = 5750\pm250\K$, $\log g = 4.5\pm0.5\dex$ and $\mathrm{[Fe/H]} = 0.0\pm0.25\dex$.

Within these bins we sample 280\footnote{This number is chosen to match the 28 CPUs of our computing nodes.} synthetic spectra with no rotational broadening, which are later broadened with different rotational velocities \vsini to create between 1680 and 2240 training set spectra for each bin. For clarity, we explain the parameter and abundance sampling for an example 3D bin centred on $T_\text{eff} = 5750\pm250\K$, $\log g = 4.5\pm0.5\dex$ and $\mathrm{[Fe/H]} = 0.0\pm0.25\dex$ (see blue box in Figure~\ref{fig:teff_logg_grid_coverage}.

Stellar parameters (\Teff, \logg, \feh, \vmic) and elemental abundances [X/Fe] of all 32 elements are randomly sampled within reasonable limits (see examples in Figure~\ref{fig:example_3d_bin_sample} and Table~\ref{tab:sampling_xfe}) and fed into \sme to create self-consistent synthetic spectra over the full HERMES wavelength range for \marcs atmospheres. 

Microturbulence velocity (\vmic) values are sampled uniformly between the upper and lower limits of the empirical relation from GALAH DR3 \citep[Eqs.~4 and 5 from][]{Buder2021} and an adjusted version of the relation of \citet{DutraFerreira2016}. The latter has been adjusted for $T_\text{eff}^\prime = T_\text{eff}~/~\mathrm{K} - 5500$ as well as $\log g^\prime = \log g~/~\mathrm{dex} - 4.0$ to return $v_\text{mic}~/~\mathrm{km\,s^{-1}}$:
\begin{align} 
v_\text{mic} = \begin{array}{l}
1.198 + 3.16 \cdot 10^{-4} \cdot T_\text{eff}^\prime - 0.253 \cdot \log g^\prime \\ - 2.86\cdot 10^{-4} \cdot T_\text{eff}^\prime \cdot \log g^\prime + 0.165 \cdot (\log g^\prime)^2
\end{array} \label{eq:vmic_initial}
\end{align}

\subsection{High-resolution synthetic spectra with \sme}
\label{sec:higher_resolution_synthetic_spectra}

\begin{figure*}[ht]
 \centering
 \includegraphics[width=\textwidth]{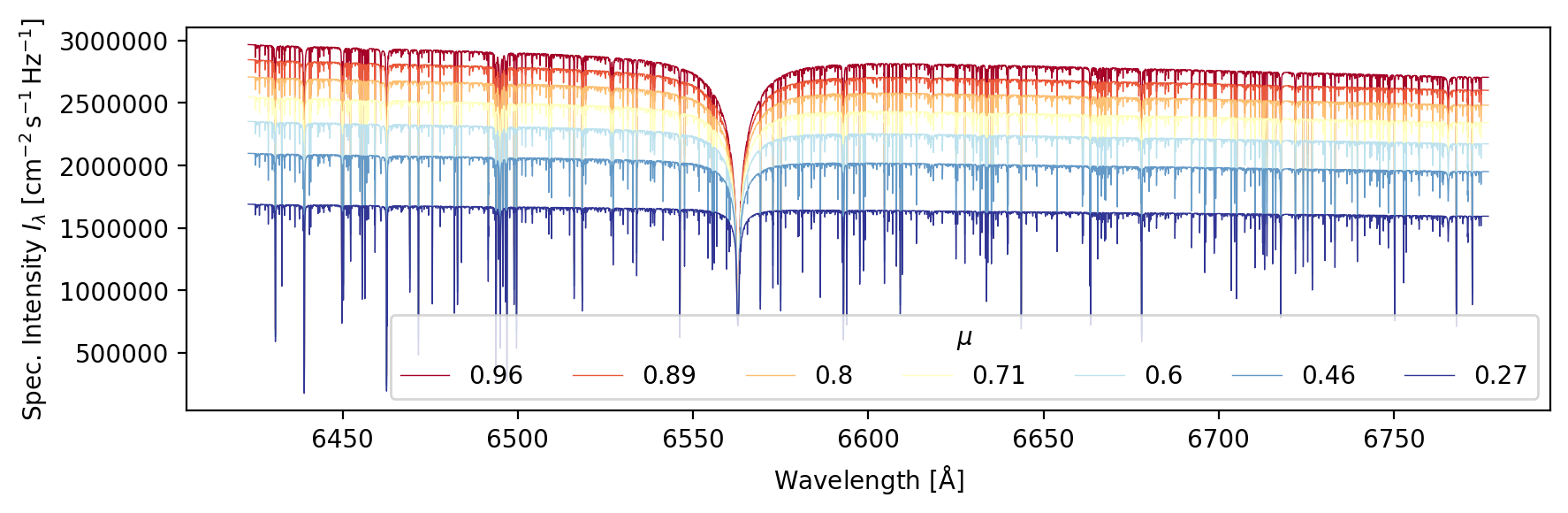}
 \caption{\textbf{Example output of \sme for a solar spectrum in HERMES CCD3 (around the Balmer $\mathrm{H}_\upalpha$ line).} Shown are the specific intensities (\texttt{sme.sint}) as a function of the viewing angle $\mu = \cos \theta$. }
 \label{fig:sme_mu_output}
\end{figure*}

We create training sets from high-resolution stellar spectra for each smaller 3D bin region of the parameter space. We compute oversampled synthetic intensity spectra at higher resolution and sampling than the typical GALAH resolution with \sme for seven equal-area angles (see Figure~\ref{fig:sme_mu_output}) of the plane-parallel or spherically symmetric stellar surfaces \citep{Gustafsson2008}.

For each spectrum, we first run a test on all available lines in the GALAH linelist. We use the same linelist as in GALAH DR3 \citep{Buder2021}. This linelist was adapted from the linelist of \citet{Heiter2021} and implements numerous updates to line data, such as updates or corrections of $\log gf$ values in the GALAH wavelength range. The test is used to restrict the myriad of possible molecular lines to the visible ones with \textsc{sme}.depth above 0.001, while keeping all atomic lines for the final synthesis.

Spectra are computed at a resolution of $R = 300\,000$ on a fine wavelength grid with 60,819 pixels spread over the extended wavelengths $4675.1-4949.9$, $5624.1-5900.9$, $6424.1-6775.9$, and $7549.1-7925.9 \Angstroem$. We note that these extend significantly beyond the actual GALAH wavelength range.

We use one-dimensional (1D) \marcs atmospheres from the \marcs grid \citep[][version 2014]{Gustafsson2008} with a trilinear interpolation for combinations of \Teff, \logg, and \feh. We use grids of non-LTE departure coefficients from \citet{Amarsi2020, Amarsi2022} for atomic lines of H, Li, C, N, O, Na, Mg, Al, Si, K, Ca, Mn, Fe, and Ba. For most elements, the non-LTE departure coefficient grids include isotropic and coherent scattering for lines from background atomic and ionic species \citep[see Equation 7 of ][]{Amarsi2020} as well as Thompson and Rayleigh scattering. The calculations for C include all background species in pure absorption \citep[Equation 6 of ][]{Amarsi2020}, whereas for Fe, Thompson and Rayleigh scattering were included but all background lines were treated in pure absorption.

Our synthetic grid explicitly includes C and N abundances. C was previously included in the analysis of GALAH DR3, but limited to the atomic C line. The analysis thus neglected the molecular absorption features of $\mathrm{C_2}$ and CN at the beginning of CCD1 and end of CCD4, respectively. With the new self-consistent grid, we can include these features, as they hold valuable information for both C and N, as well as several other features through the molecular equilibrium in stars \citep[see e.g.][]{Ting2018}.

To be able to test that the flux-label correlations found by our interpolation routine are limited to reasonable wavelength ranges, we also calculate one spectrum that is exactly in the middle of the parameter range and additional spectra, where we increase the value of one label at a time (e.g. increase [O/Fe] by $1\dex$) to test the response in the synthetic spectrum.

To save computational costs, we compute synthetic spectra with no rotational or macroturbulence broadening ($v_\text{mac} = v\sin i = 0\kms$), but save the continuum flux (\texttt{sme.cmod}) and the specific intensities (\texttt{sme.sint}) as a function of the equal-area midpoints of each equal-area annulus\footnote{$\mu \in [0.96, 0.89, 0.8, 0.71, 0.6, 0.46, 0.27]$} $\mu$ (see Figure~\ref{fig:sme_mu_output}). We then apply the broadening of spectra due to rotation (\vsini) with the flux integration code of the python-implementation \textsc{PySME} \citep{Wehrhahn2023} of \sme. Depending on the expected rotational velocities (increasing with temperature) we sample a range of
\begin{align} \label{eq:vsini}
    v \sin i~/~\kms \in \{ 1.5, 3, 6, 9, 12, 18, 24, 36\}.
\end{align}

Note that $v \sin i = 24 \kms$ is only included for bins with \Teff$\geq 5000\,\mathrm{K}$ and $v \sin i = 36 \kms$ for those with \Teff$\geq 6000\,\mathrm{K}$.


\subsection{Interpolating synthetic spectra with neural networks} \label{sec:interpolating_synthetic_spectra_with_neural_networks}

To allow a fast interpolation with new and different stellar labels, we use data-driven models to connect stellar fluxes at given pixels from a combination of stellar labels. This method is well established in stellar spectroscopy through the successful applications of quadratic models with \textit{The Cannon} \citep[see e.g.][]{Ness2015, Ness2016, Casey2016, Casey2017, Ho2017, Buder2018} as well as neural networks with \textit{The Payne} \citep[see e.g.][]{Ting2019, Xiang2019, Xiang2021}. Because of the needed flexibility to predict synthetic spectra with 36 stellar labels for a large parameter space \citep[for a detailed discussion of advantages of neural networks over quadratic models see][]{Ting2019}, we choose neural networks to interpolate between our synthetic spectra in this data release.

In this work, we utilise the neural network architecture and training algorithms from the spectrum interpolation framework of \textit{The Payne} \citep{Ting2019}. While we do not implement the full functionality of \textit{The Payne}, we specifically adopt its spectrum interpolation capabilities. Unlike the version originally published by \citet{Ting2019}, we use the architecture of the latest available version of \textit{The Payne}. This modified architecture connects $k$ stellar labels $\boldsymbol{\ell}$ with the flux $f$ at each wavelength pixel $\lambda$ via
\begin{equation}
f_\lambda = w \cdot \mathrm{lReLU} \bigg( \tilde{w}_\lambda^i \cdot \mathrm{lReLU}  \Big( w^k_{\lambda i} \ell_k + b_{\lambda i} \Big) + \tilde{b} \bigg) + \bar{f}_\lambda,
\label{eq:neural_network_function}
\end{equation}
which encapsulates the so-called layers of a neural network with $i = 300$ neurons with weights $w$ and biases $b$ as well as a leaky Rectified Linear Unit ($\mathrm{lReLU}$)
\begin{equation}
    \mathrm{lReLU} (x) =  \begin{cases}
        x \qquad &x \geq 0 \\
        0.01 x \qquad &x < 0.
    \end{cases}
\end{equation}

After optimising the mean absolute error loss function for $10^4$ steps, we consider the network trained with an optimised combination of three sets of weights and biases within the minimum and maximum ranges of each label. We discuss the performance and caveats of this particular neural network architecture and training setup in Section~\ref{sec:caveats_interpolation}. The trained networks can then be used with new input labels to quickly create synthetic spectra for the label optimisation. Computational resources could be conserved by training neural networks exclusively on spectra from non-rotating stars and subsequently applying broadening through convolution with a center-to-limb darkening law. This method, while less accurate, could enable the fitting of broader velocity ranges and enhance neural network performance by simplifying the spectral shapes they must learn. However, shifting the broadening process from training to post-processing does not necessarily guarantee a reduction in computational costs.

\section{SINGLE SPECTRUM ANALYSIS (ALLSPEC)}
\label{sec:allspec_analysis}

As outlined in Section~\ref{sec:introduction}, the workflow of GALAH DR4 includes a first analysis step of all observed spectra without including non-spectroscopic information for the optimisation. This allows us to identify shifts in radial velocity between separate spectroscopic observations of the same star\footnote{While repeat observations were only done for quality assurance in GALAH Phase 1, a significant number of repeat observations was performed as part of Phase 2.} and a better co-adding of spectra for the \texttt{allstar} analysis (see Section~\ref{sec:allstar_analysis}). Another motivation for this step is to get a first estimate of stellar labels without potentially biased photometric and astrometric information, for example for binary stars.

The optimisation of stellar labels thus aims to minimise the absolute difference between synthetic and observed spectra, weighted by their uncertainty. Starting from a set of initial labels (Section~\ref{sec:initial_stellar_labels}), we create high-resolution synthetic spectra and convolve them to the resolution and wavelength grid of each observed spectrum. We remind ourselves that in GALAH DR3, we used a repeated combination of spectrum normalisation followed by stellar parameter optimisation and a subsequent fit of individual elements with fixed stellar parameters. In the analysis of GALAH DR4, we perform an on-the-fly re-normalisation of the observed spectrum for every change of the simultaneously fitted stellar parameters and elemental abundances. This allows a more accurate comparison of synthetic and observed spectra (Section~\ref{sec:comparison_synthetic_spectra_to_observations}) and thus a more accurate stellar label optimisation (see Section~\ref{sec:stellar_label_optimisation}).

\subsection{Initial stellar labels}
\label{sec:initial_stellar_labels}


Initial values of all stellar labels are needed for creating a first synthetic spectrum. For \vrad, \Teff, \logg, and \vsini we use a combination of sources. Where possible, we use the previous estimates from GALAH DR3 \citep{Buder2021}, and otherwise use estimates from the GALAH DR4 reduction pipeline (Section~\ref{sec:spectroscopic_data_from_galah_observations}). Because of the limited accuracy of the latter for cool stars with $T_\text{eff} < 4000\,\mathrm{K}$ as well as the hot stars with $T_\text{eff} > 6500\,\mathrm{K}$, we perform a consistency check with photometric information from \Gaia DR3 \citep{Brown2021} and 2MASS \citep{Skrutskie2006}. For most of the aforementioned cool and hot stars, we therefore prefer the parameters from the \Gaia DR3 photometric pipeline GSP-Phot \citep{Andrae2023,Fouesneau2022} as initial values.

In selected cases, we further adjust the starting parameters toward reasonable limits. For example, hot stars are likely to be young and are adjusted to close to Solar metallicity. Furthermore, we recalculate the initial \vmic based on Equation~\ref{eq:vmic_initial} and limit rotational broadening values to $3 \leq v \sin i \leq 10\,\mathrm{km\,s^{-1}}$ for stars below $T_\text{eff} = 5500\,\mathrm{K}$ and $3 \leq v \sin i \leq 20\,\mathrm{km\,s^{-1}}$ for hotter stars. The explicit choices of starting values for \Teff, \logg, \feh, \vmic, and \vsini are described in our online repository\footnote{\href{https://github.com/svenbuder/GALAH_DR4/blob/main/spectrum_analysis/galah_dr4_initial_parameters.ipynb}{GALAH\_DR4/spectrum\_analysis/galah\_dr4\_initial\_parameters.ipynb}} and are depicted in Figure~A\ref{fig:initial_parameters}.

Based on the value of \feh we apply an offset to the $\upalpha$-process elements O, Mg, Si, Ca, and Ti. The initial value is 0.4 for $\mathrm{[Fe/H]} < -1$, 0.0 for $\mathrm{[Fe/H]} > 0$, and $-0.4 \cdot \mathrm{[Fe/H]}$ for $-1 \leq \mathrm{[Fe/H]} \leq 0$. All other abundances are initialised at $\mathrm{[X/Fe]} = 0$.

\subsection{Comparison of synthetic spectra to observations}
\label{sec:comparison_synthetic_spectra_to_observations}

The major aim of our spectroscopic analysis is to predict the best set of stellar labels by minimising the uncertainty-weighted difference between observed and synthetic spectra. In this section, we describe several important steps to enable the pixel-level comparison of the higher resolution, oversampled synthetic spectra created with the neural networks from Section~\ref{sec:interpolating_synthetic_spectra_with_neural_networks} and the observational data at actually measured resolution and sampling (presented in Section~\ref{sec:spectroscopic_data_from_galah_observations}).

\subsubsection{Downgrading synthetic spectra to observed resolution}

Because dedicated line-spread-function measurements are available for every spectrum (see Section~\ref{sec:spectroscopic_data_from_galah_observations}), we use this information to downgrade our synthetic spectrum with Gaussian kernels on an equidistant velocity grid to the measured resolution of each observation. We then interpolate the oversampled synthetic spectrum onto the observed wavelength grid.

\subsubsection{On-the-fly re-normalisation of observed spectra}

Measurements of the GALAH flux and flux uncertainty are reported in counts by the reduction pipeline. To compare with our synthetic spectra, which are normalised to the continuum, we fit an outlier-robust polynomial function to the ratio of observed and synthetic spectra and re-normalise our observed spectra and their uncertainties via this normalisation function.

This specific approach is similar to the internal routine of \sme and has the important advantage that no continuum points have to be defined. This is advantageous because we try to cover the full parameter range of FGKM stars for which positions of continuum points -- corresponding to 1 on a (pseudo-)continuum-normalised spectrum -- differ significantly or for which continuum points may not even be present, or will be a strong function of \Teff and \feh (as is the case for M stars).

We make two additional adjustments to the reduced spectra, which come in the form of counts and uncertainty per wavelength, $f_\lambda$ and $\sigma_{f,\lambda}$.

\begin{figure*}[ht]
\centering
\includegraphics[width=\textwidth]{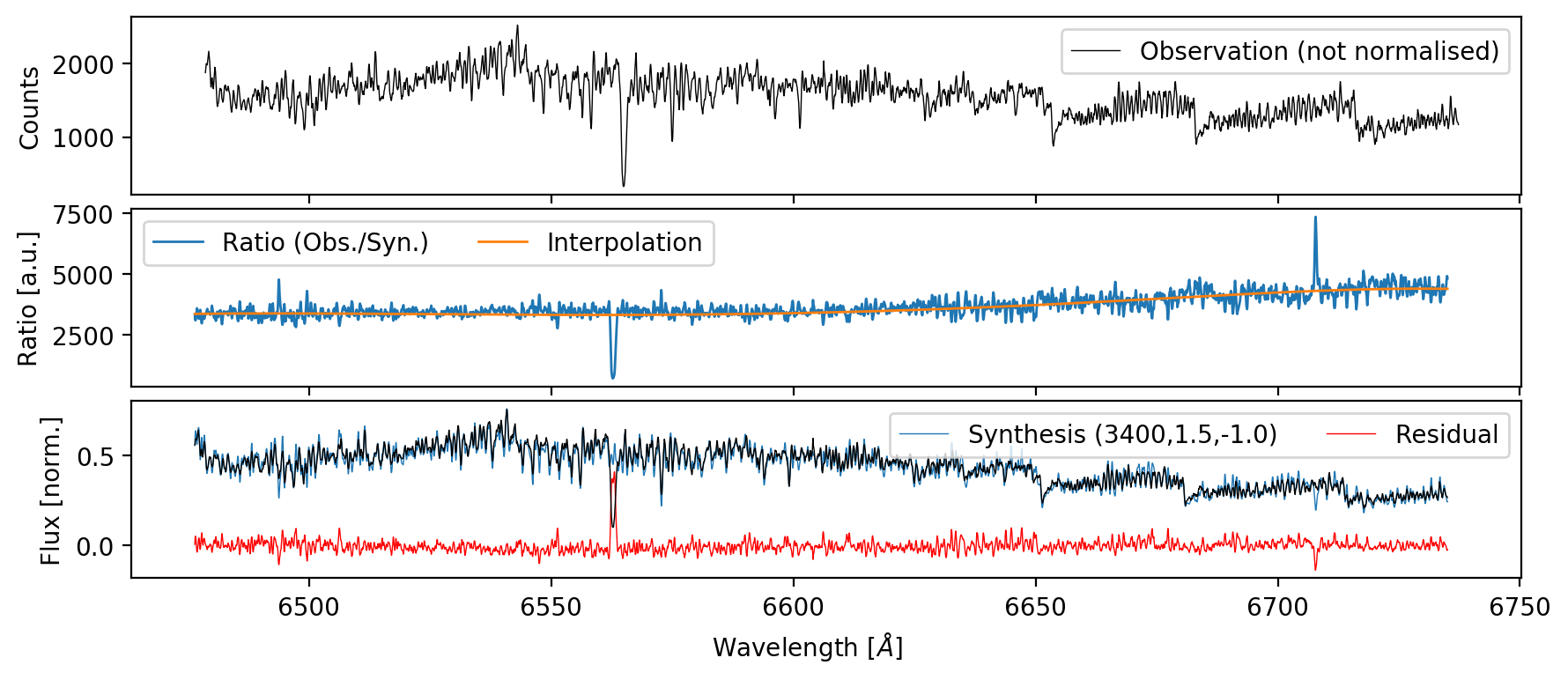}
\caption{
\textbf{Example of normalisation for GALAH DR4 for a model spectrum ($T_\mathrm{eff} = 3400\,\mathrm{K}$, $\log g = 1.5$, $\mathrm{[Fe/H]} = -1.0\,\mathrm{dexbest-fitting }$) that is selected during the label optimisation.}
\textbf{Panel (a):} Observed spectrum (counts).
\textbf{Panel (b):} Ratio (blue) of observed spectrum and model spectrum as well as Chebyshev polynomial fit (orange).
\textbf{Panel (c):} Normalised observed spectrum (black) compared to the model spectrum (blue). Residuals (red) can then be used as input for the likelihood function.
}
\label{fig:ratio_normalisation}
\end{figure*}

As we compare the observation to model spectra, we do not have to restrict ourselves to an \textit{a priori} normalisation, but can take into account the residual information on the continuum in parts of the spectrum. For each model spectrum that we compare to, we therefore perform a normalisation by fitting a fourth order Chebyshev polynomial with outlier clipping to the ratio of model and observation (see Figure~\ref{fig:ratio_normalisation}). This allows us to both overcome previous shortcomings of the synthetic analysis in GALAH+ DR3 \citep{Buder2021}, which had to be restricted to small wavelength segments and assumed a linear relation for those. Our new approach allows us to properly assess the structure of deep and steep molecular features that can dominate spectra of cool stars and carry significant information on \Teff, \vrad, as well as abundances \citep{Mann2012}.

\subsection{Stellar label optimisation}
\label{sec:stellar_label_optimisation}

In up to four major loops, we optimise the radial velocities and all other stellar labels and report a) their values, b) their co-variances, c) the best-fitting synthetic and re-normalised spectra along with d) their uncertainties and e) masks that indicate which pixels were used in the final optimisation.

\begin{figure*}[ht]
\centering
\includegraphics[width=0.75\textwidth]{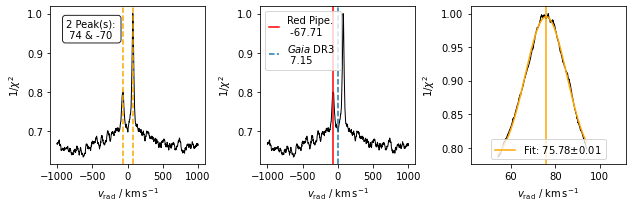}
\caption{\textbf{Output of the radial velocity fitting step.} \textbf{Panel a)} shows the initial broad search on a \vrad array of $-1000..(2)..1000\kms$. In the case of 2MASS J06084657\textminus7815235, two peaks (yellow, dashed) are visible for this double-lined spectroscopic binary. \textbf{Panel b)} shows the same plot, but overlaid with the GALAH DR4 reduction pipeline (red) and \Gaia DR3 (blue, dashed) estimates for \vrad. \textbf{Panel c)} shows the narrow window of $-20.00..(0.04)..20.00\kms$ around the highest peak and its Gaussian fit (yellow). Despite their low resolution (26 KB), these on-the-fly created diagnostic images already occupy 50GB in total.}
\label{fig:181221003101356_single_fit_rv}
\end{figure*}

Starting from the initial values, a first synthetic spectrum is computed and compared with the observation in order to assess the initial radial velocity. This is done by applying the \textsc{scipy.signal.find\_peaks} algorithm on the normalised inverse residuals of non-shifted observed and synthetic spectra, when the latter is shifted by $v_\text{rad} = -1000..(2)..1000\kms$ (see Figure~\ref{fig:181221003101356_single_fit_rv}a). If no peak is found, the initial \vrad value is used hereafter. If more than one peak is found (see Figure~\ref{fig:181221003101356_single_fit_rv} with \Gaia DR3 agreeing with the systemic radial velocity), the two strongest peaks are reported. For the purpose of the single star analysis, a narrower search is conducted around the highest peak with a \vrad shift of $-20.00..(0.04)..20.00\kms$ around said peak by fitting a Gaussian function to the inverse of the residuals that were normalised with the smallest residual values (see Figure~\ref{fig:181221003101356_single_fit_rv}c). The mean of this fit and its uncertainty are reported by the pipeline.

The centerpiece of our optimisation is the \textsc{scipy.optimize} module's \textsc{curve\_fit} function \citep{scipy}, which we call with counts and uncertainties (our absolute sigmas) as input for a placeholder function that self-consistently re-normalises the observed spectrum. We estimate the labels via the least squares optimisation within less than $10^4$ iterations and a desired relative error (\texttt{xtol}) below $10^{-4}$.

For each optimisation loop, a new, best-fitting 3D bin and neural network is identified via a grid search in the \TLF dimensions with \textsc{sklearn.cKDtree}. If the stellar labels that are being fitted have changed (for example if an element is deemed not detectable for the new 3D bin during the neural network training), the label and its value are either set to or initialised with $\mathrm{[X/Fe]} = 0$.

While the optimisation of the neural network selection has not converged (the final parameters \TLF are not within the current 3D bin), the optimisation is repeated, starting with the previous best-fitting parameters as starting guesses.

We measured the time taken for the individual steps in the \textsc{curve\_fit} function's execution to be approximately $80\,\mathrm{ms}$. The total fitting process for stellar labels, including input/output overheads, was timed at $89_{-29}^{+77}\,\mathrm{s}$ for the \texttt{allspec} module, and $125_{-33}^{+81}\,\mathrm{s}$ for the more complex \texttt{allstar} module.

\begin{figure*}[ht]
\centering  
\includegraphics[width=\textwidth]{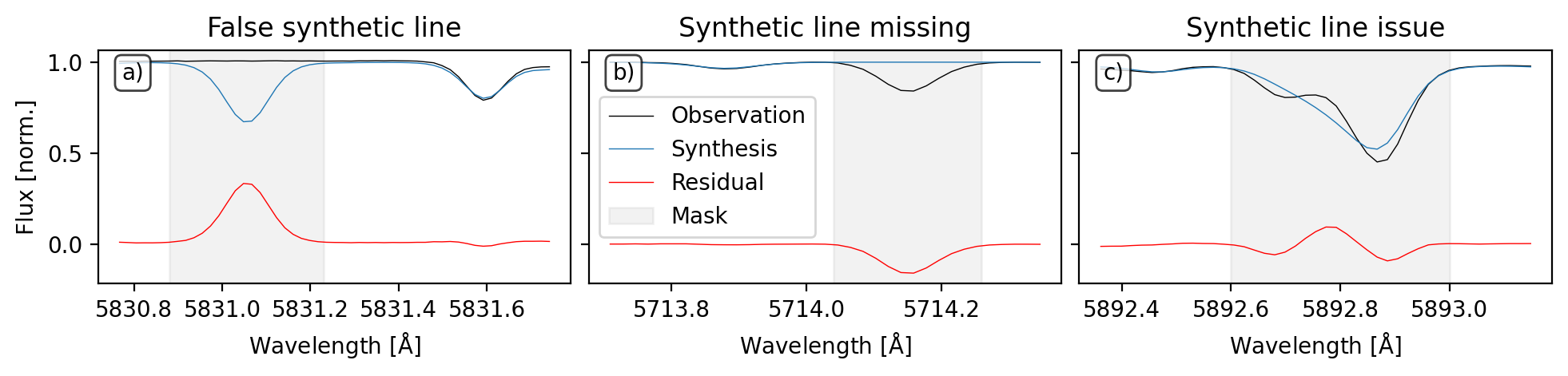}
\caption{\textbf{Examples of masks applied to unreliable pixels for the spectrum fitting, which is done by the minimisation of residuals (red) between observation (black) and synthesis (blue).} \textbf{Panel a)} shows a strong synthetic line, where no line is observed in the data. \textbf{Panel b)} shows an observed line without any line being synthesised. \textbf{Panel c)} shows significant disagreement between the two observed lines and the synthesis.} \label{fig:example_masking_sun}
\end{figure*}

\subsubsection{Which stellar labels are optimised?} \label{sec:which_labels_are_optimised}


As part of the synthetic grid computations, we have sampled each label of stellar parameters and elemental abundances individually between our chosen maximum and minimum ranges (see Section~\ref{sec:spectrum_grid}). This allows us to also judge which stellar labels to fit for each given star. We choose to fit a stellar label if either of these two cases applies to said label for the GALAH wavelength range when neglecting the cores of the Balmer lines: (i) The normalised spectrum between minimum and maximum label value at any pixel exceeds the threshold of 0.007 or (ii) The normalised spectrum between the minimum and maximum value changes by more than 0.005 for at least 25\% of the pixels. While the first case is constructed for atomic lines, such as \ion{Li}{i} 6708\,\AA, the second case addresses in particular molecular lines like the $\mathrm{C_2}$ and $\mathrm{CN}$ lines. The pipeline can handle missing arms, for example in the case of readout issues of a CCD, and will fix abundances to the scaled-Solar values for elements with absorption features solely in the missing arm, for example N, O, K, and Rb for CCD4.

Initial tests of the pipeline have revealed that in cases where the initial parameter estimates deviate significantly from the final values, several elemental abundance estimates were shifted towards their boundaries, leading to a masking of their elemental abundance lines by the masking module (Section~\ref{sec:masking_of_unreliable_wavelength_regions}) at the beginning of each optimisation loop. To minimise this effect, we therefore shift the interim abundance values towards the narrow label boundaries. In practice, we limit the initial and interim abundances to 1.05..3.26 for A(Li), $\mathrm{[X/Fe]} = -0.5..1.0$ for C, N, O, Y, Ba, La, Ce, and Nd, and $\mathrm{[X/Fe]} = -0.5..0.5$ for all other elements before optimising them again. For warm and hot stars ($T_\text{eff} > 6000\K$), this effect was seen to affect multiple abundances, such that we needed to implement a reset of all abundances except Li to their initial values for stars above $6000\K$, which would on average be expected to be young and have a Solar-like composition.

\subsubsection{Masking of unreliable wavelength regions} \label{sec:masking_of_unreliable_wavelength_regions}

Not all pixels of the observed or synthetic spectra might prove useful for estimating reliable stellar labels. Observations can include bad pixels/patterns and incorrect corrections (for example of telluric or sky lines). Flux predictions of synthetic spectra are only as good as the input physics (limited for example for specific lines via uncertain oscillator strengths).

To minimise the influence of inaccurate synthetic pixel predictions, we have compared a 2dF-HERMES observation of the asteroid 4~Vesta and a high-quality Solar spectrum from \citet{Hinkle2000} with the flux that would be predicted by our pipeline for a star with Solar labels ($T_\text{eff} = 5772\,\K$, $\log g = 4.438\dex$, $\mathrm{[Fe/H]} = 0.00\dex$, $v_\text{mic} = 1.06\,\kms$, $v \sin i = 1.6\,\kms$, $v_\text{mac} = 4.2\,\kms$ \citep{Prsa2016, Jofre2017}, and $\mathrm{[X/Fe]} = 0.00\dex$ for the default Solar abundance pattern for \marcs by \citet{Grevesse2007}).

We have identified all lines that showed differences of the normalised flux of more than $0.1$, lines where either a synthetic line or an observed one was completely missing, or lines that were significantly misaligned. Examples of masks\footnote{Example masks can be found in the GALAH DR4 repository  \href{https://github.com/svenbuder/GALAH_DR4/blob/main/spectrum_analysis/spectrum_masks}{here}.} are shown in Figure~\ref{fig:example_masking_sun}. To avoid the influence of bad spectrum regions with an observational origin, we mask pixels where the synthetic and re-normalised observed spectra differ by more than $5\sigma$ or a flux of 0.3 (0.4 before the initial optimisation). To avoid the masking of lines that are vital for our spectroscopic analysis, we have created a list\footnote{The list is available in the GALAH DR4 repository \href{https://github.com/svenbuder/GALAH_DR4/blob/main/spectrum_analysis/spectrum_masks/vital_lines.fits}{here}.}  with segments of such lines that is mainly based on the previous element masks from GALAH DR3 \citep{Buder2021}. The final mask of pixels to use for the optimisation then includes all vital line regions, as well as those wavelengths that do not show a too strong disagreement between observation and synthesis and are not deemed unreliable in their synthesis.

In addition to this default masking, we exclude pixels for each major iteration, for which the flux of observation and synthesis differ by more than $5 \sigma$ and 30\% of the normalised flux and by more than 100\% of the normalised flux for the vital line regions.

We further indirectly take into account the currently less constrained molecular data for cool stars in optical spectra, in particular towards the blue \citep[e.g.][]{Rains2021,Rains2024}. For presumably cool stars (with initial $T_\text{eff} < 4100\,\mathrm{K}$), we therefore double the observational uncertainty of the blue arm.

\section{SINGLE STAR ANALYSIS (ALLSTAR)}
\label{sec:allstar_analysis}

\begin{figure}[ht]
 \centering
 \includegraphics[width=\textwidth]{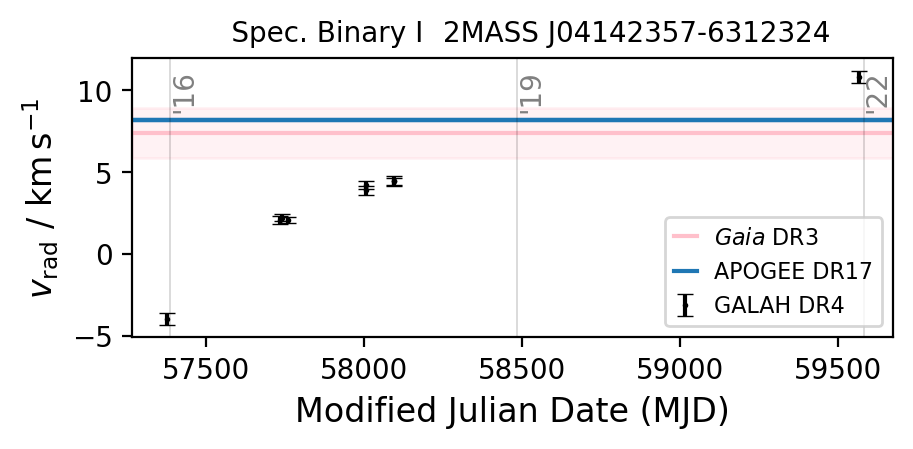}
 \caption{\textbf{Example of radial velocity evolution over modified Julian Date (vertical lines show the beginning of 2016, 2019, and 2022) for a single-lined spectroscopic binary (SB1)}.}
 \label{fig:examples_flag_sp_2}
\end{figure}

After the \texttt{allspec} module (Section~\ref{sec:allspec_analysis}) has been used to estimate spectroscopic labels for all spectra, we use the \texttt{allstar} module to co-add spectra and analyse one spectrum per star while taking into account photometric and astrometric information to constrain the surface gravities\footnote{In line with \citet{Nissen2015, Nissen2020}, we refer to these non-spectroscopically constrained surface gravities as photometric ones.}. The optimisation of stellar spectroscopic parameters with the help of non-spectroscopic information was successfully applied for GALAH DR3 \citep{Buder2021}, using \Gaia DR2 distances \citep{BailerJones2018} to overcome spectroscopic degeneracies. For the co-adding, we test whether the radial velocity estimates of individual exposures agree within $2\sigma$. Below this threshold, we apply no radial velocity correction and fit a global radial velocity. Above this threshold (which is useful for single-lined spectroscopic binaries as shown in Figure~\ref{fig:examples_flag_sp_2}), we apply a radial velocity correction before co-adding.

To speed up computation, we use the mean results of the \texttt{allspec} analyses as initial stellar labels for the \texttt{allstar} analysis. All other methodology of the comparison of synthetic spectra to observations (Section~\ref{sec:comparison_synthetic_spectra_to_observations}) and label optimisation (Section~\ref{sec:stellar_label_optimisation}) apply also to this module, with the exception of the optimisation of \logg. Contrary to the \texttt{allspec} approach, we do not fit \logg in this module, but estimate the logarithmic surface gravity $\log g$ using a combination of its definition ($g \propto \frac{\mathcal{M}}{\mathcal{R}^2}$) and the Stefan-Boltzmann law relative to the Solar values:
\begin{equation}
\log g = \log g_\odot + \log \frac{\mathcal{M}}{\mathcal{M_\odot}} + 4 \log \frac{T_\mathrm{eff}}{T_\mathrm{eff,\odot}} - \log \frac{L_\mathrm{bol}}{L_\mathrm{bol,\odot}} \label{eq:logg}
\end{equation}

While we can use our spectroscopically determined $T_\mathrm{eff}$ in Eq. \ref{eq:logg}, the other values have to be estimated through models or non-spectroscopic information. The logarithmic bolometric luminosity, $L_\mathrm{bol}$, can be estimated from the bolometric magnitude $M_\mathrm{bol}$, such that $\log \frac{L_\mathrm{bol}}{L_\mathrm{bol,\odot}} = -0.4 \cdot \left(M_\mathrm{bol} - M_\mathrm{bol,\odot} \right)$. The bolometric magnitude can be estimated from any given apparent magnitude, if we correct the latter for the distance modulus, bolometric correction, and extinction. Because essentially all stars in GALAH DR4 have high-quality infrared magnitudes available that suffer less from (uncertain) extinction corrections, we use $K_S$ as the magnitude to estimate our bolometric magnitudes and luminosities via
\begin{equation}
M_\mathrm{bol} = K_S - 5\cdot \log \frac{D_\varpi}{10} + BC(K_S) - A(K_S). \label{eq:mbol}
\end{equation}

While the values for $K_S$, curated distances $D_\varpi$ (rather than raw parallaxes $\varpi$), and $A(K_S)$ are readily available (see Section~\ref{sec:non-spec_data}), we need to estimate the bolometric correction from tabulated values using the routines provided by \citet{Casagrande2018}:
\begin{equation}
BC(K_S) = f(T_\mathrm{eff}, \log g, \mathrm{[Fe/H]})
\label{eq:bc_ks}
\end{equation}
We choose to assume an extinction value of $E(B-V) = 0\,\mathrm{mag}$ for this particular interpolation and post-correct the value by $A(K_S)$ based on the actual extinctions. The reason for this is that the latter values can exceed the maximum tabulated values of $E(B-V) = 0.72\,\mathrm{mag}$ of \citet{Casagrande2018}.

Because of the appearance of $\log g$ in Equation~\ref{eq:bc_ks}, we iterate the calculation of $BC(K_S)$ and subsequently $\log g$ up to four times or until the latter value changes less than $0.02\,\mathrm{dex}$ between iterations. Similarly, we need to estimate the stellar masses (and ages as a byproduct) from tabulated values, that is,
\begin{equation}
\mathcal{M}, \tau = f(T_\mathrm{eff}, \log g, \mathrm{[Fe/H]}, L_\mathrm{bol,\odot})
\label{eq:mass_age}
\end{equation}
For this on-the-fly estimate of masses and ages we use a likelihood-weighted estimate with default uncertainties of $100\,\mathrm{K}$, $0.25\,\mathrm{dex}$, $0.2\,\mathrm{dex}$, respectively, and an average uncertainty of $L_\mathrm{bol,\odot}$ from propagated uncertainties of Equation~\ref{eq:mbol}. We weigh the ages and masses via their likelihood of all isochrone grid points within these uncertainties of the {\sc parsec+colibri} isochrones \citep{Bressan2012, Marigo2017}, which cover the logarithmic ages of $\log (\tau~/~\mathrm{Gyr}) = 8.00..(0.01)..10.18$ by default and metallicities $\mathrm{[M/H]} = -2.75..(0.25)..-0.75$ as well as $\mathrm{[M/H]} = -0.6..(0.1)..0.7$. We exclude hot stars above $10\,000\,\mathrm{K}$ as well as extremely evolved white dwarf and extremely luminous giant stars ($\log g > 6\,\mathrm{dex}$ or $J - K_S > 2\,\mathrm{mag}$) as they fall far outside our spectroscopic pipeline range. We convert between the theoretical [M/H] and our measured [Fe/H] as well as an assumed $\mathrm{[\upalpha/Fe]}$ enhancement\footnote{We assume $\mathrm{[\upalpha/Fe]} = 0.4$ for $\mathrm{[Fe/H]} < -1$, $\mathrm{[\upalpha/Fe]} = 0.0$ for $\mathrm{[Fe/H]} > 0$ and linearly interpolate between these points for $-1 \leq \mathrm{[Fe/H]} \leq 0$.} via the correlation of \citet{Salaris2006}, $\mathrm{[M/H]} = \mathrm{[Fe/H]} + \log\left(10^{\mathrm{[\upalpha/Fe]}} \cdot 0.694 + 0.306 \right)$. For open clusters with age estimates below $1\,\mathrm{Gyr}$ as well as unevolved stars that are more luminous than expected from the oldest cool main-sequence isochrone with matching $\mathrm{[M/H]}$, we sample $\log (\tau~/~\mathrm{Gyr}) = 6.19..(0.01)..10.18$. For globular cluster stars identified in the crossmatch with \citet{Baumgardt2021}, we limit the isochrones to a minimum age of $4.5\,\mathrm{Gyr}$.

\begin{figure*}[ht]
\centering
\includegraphics[width=\textwidth]{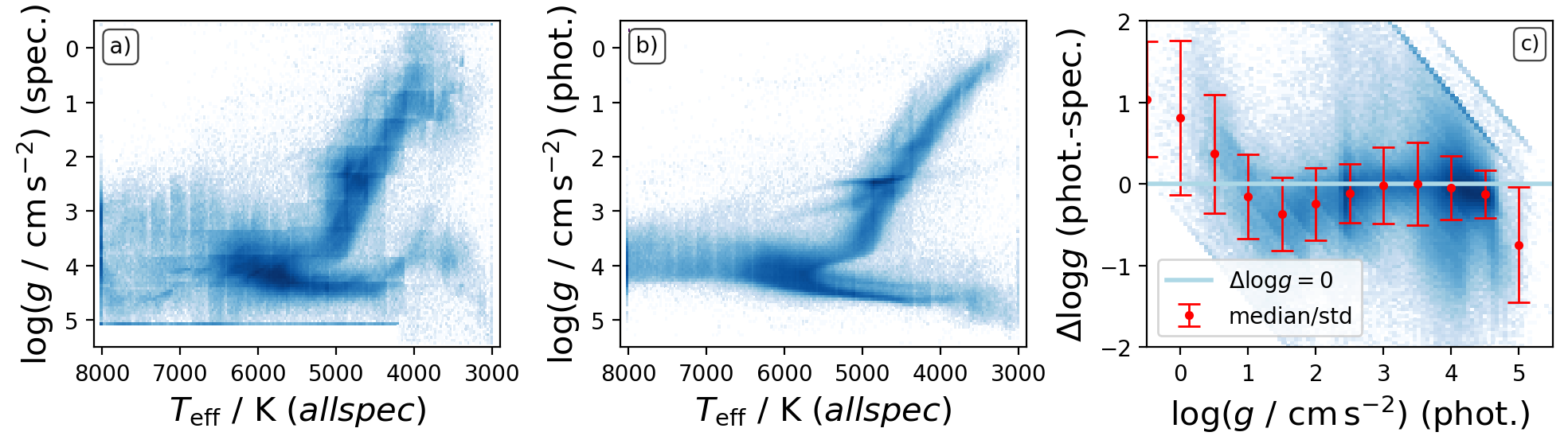}
\caption{\textbf{Comparison of spectroscopic and photometric \logg estimates in the \texttt{allspec} analysis.}
\textbf{Panel a)} shows the distribution of spectroscopic \logg and \Teff from the \texttt{allspec} module.
\textbf{Panel b)} shows the distribution of the same \Teff and photometric \logg.
\textbf{Panel c)} shows the difference of photometric \logg and spectroscopic \logg as a function of photometric \logg. Red error bars indicate the $1\sigma$ percentiles of this difference in $0.5\,\mathrm{dex}$ bins.}
\label{fig:dlogg_spec_plx}
\end{figure*}

\section{POST-PROCESSING}
\label{sec:post_processing}

After the \texttt{allspec} and  \texttt{allstar} modules have been run for a night's data (see Sections~\ref{sec:allspec_analysis} and \ref{sec:allstar_analysis}, respectively), a post-processing routine is used to estimate additional parameters from the residuals of the spectra (Section~\ref{sec:residual_analysis}), estimate and validate accuracy and precision uncertainties (Section~\ref{sec:uncertainty}), and perform quality assurance tests on a global scale (\texttt{flag\_sp}, see Section~\ref{sec:flag_sp}) as well as for the individual abundances of elements X (\texttt{flag\_X\_fe}, see Section~\ref{sec:flag_x_fe}).

\subsection{Analysis of spectral residuals} \label{sec:residual_analysis}

\subsubsection{Binary signatures} \label{sec:trigger_binary_module}

The residual spectrum of our best-fitting single star analysis can help us to identify a second flux contributor to the observed spectrum. In our case, there are two points in the analysis where we can identify such an influence. Firstly, the residuals are visible in the $\chi^2$ distribution as a function of radial velocity shifts (see Figure~\ref{fig:181221003101356_single_fit_rv}). While a single star would only show one peak (saved as \texttt{rv\_comp\_1}), a binary system like 2MASS J06084657-7815235 shows a second peak ($-70\,\mathrm{km\,s^{-1}}$ in addition to $74\,\mathrm{km\,s^{-1}}$) that is saved as \texttt{rv\_comp\_2}. Secondly, we perform an automatic search for reoccuring residuals as a function of radial velocity for a few selected lines. We chose the combination of strong lines in the spectra (Balmer lines, Fe lines at 4890 and $4891\,\text{\AA}$, Ni at $6644\,\text{\AA}$) as well as those with the largest expected wavelength shift in the infrared detector (O triplet at $7772-7775\,\text{\AA}$ as well as Mg at $7692\,\text{\AA}$). If we find several peaks with a reasonably similar radial velocity, the likely $X \in {16,50,84}^\text{th}$ percentiles of this radial velocity are saved in \texttt{sb2\_rv\_X}.

Because radial velocities from the \Gaia radial velocity spectrometer \citep{Katz2023} are reported in \Gaia DR3 for 94\% (774\,914) of the stars observed for GALAH DR4, we can also compare against those radial velocity estimates. For 6\% (50\,577) of our stars, we find a difference with respect to \Gaia DR3 larger than $10\,\mathrm{km\,s^{-1}}$. For these stars, we often noticed unrealistically high  \vmic and \vsini or negative velocities in our \texttt{allspec} analysis. We note that the \texttt{allspec} analysis was run without boundary conditions for global parameters and thus also resulted in negative velocities, which are later flagged and might indicate binarity (Section~\ref{sec:flag_sp}). \texttt{allstar}, however, was run with \vmic and \vsini forced to be above $0\,\mathrm{km\,s^{-1}}$.

\begin{figure*}[ht]
 \centering
 \includegraphics[width=\textwidth]{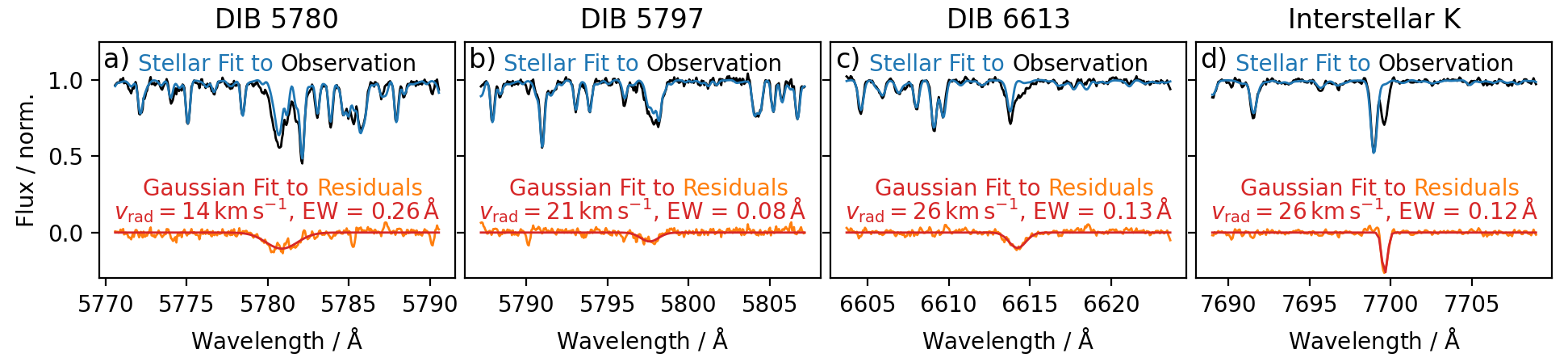}
 \caption{\textbf{Example of three diffuse interstellar bands (DIBs) and interstellar K absorption for 2MASS J06453479-0102137 with an $E(B-V) = 0.84\,\mathrm{mag}$ value from \citet{Schlegel1998}.} Shown are the observation (black) and stellar fit (blue) as well as a Gaussian fit (red) to the residual (orange), resulting in an estimate of the equivalent width (EW) as well as radial velocity.} 
 \label{fig:example_dibs_06453479-0102137}
\end{figure*}

\subsubsection{Post-correction of {log\textit{g}} for \texttt{allspec} results}

While we estimate logarithmic surface gravities \logg solely from spectra in the \texttt{allspec} results, we also perform a post-processing estimate where we employ the methodology of Section~\ref{sec:allstar_analysis} while fixing all other stellar parameters. The approach of only using spectroscopic information confirmed the previous conclusions of GALAH DR1-DR3 that the spectroscopic information in HERMES spectra to estimate \logg is not sufficient for the majority of the parameter space for the given SNR. We show the spectroscopic \logg in Figure~\ref{fig:dlogg_spec_plx}a and the photometric \logg and their difference in Figures~\ref{fig:dlogg_spec_plx}b and c, respectively.

We see an overall good agreement of both \logg estimates for stars between $4250 < T_\text{eff} < 6500\,\mathrm{K}$. Hotter stars show a strong dispersion of spectroscopic \logg due to limited information from fewer and shallower lines. Cooler stars show a significant trend towards much lower \logg for main-sequence stars and much higher \logg for cool evolved stars up to an order of $\Delta \log g$ of $1\,\mathrm{dex}$. This trend was previously seen in GALAH DR2 \citep{Buder2018} and is believed to be caused by the onset of molecular absorption features which suppress the continuum for almost the entire HERMES wavelength range (see for example Figure~\ref{fig:ratio_normalisation}), thus introducing several degeneracies. In addition, we can notice a significantly lower precision of the spectroscopic \logg in comparison to the excellent precision of photometric \logg, for example in the red clump stars.

On closer inspection, we notice several trends in Figure~\ref{fig:dlogg_spec_plx}a. Most notably, we see noding patterns along the \Teff and \logg grids where the \texttt{allspec} module switches between different neural network models. Our investigation of these noding effects is addressed in Section~\ref{sec:caveats}. In comparison to Figure~\ref{fig:dlogg_spec_plx}b, where a clear equal-mass binary sequence is visible just above the cool main-sequence, we do not see such a sequence in Figure~\ref{fig:dlogg_spec_plx}a. The difference between spectroscopic and photometric \logg will therefore be useful to identify photometric binaries with high quality spectra with \logg precisions below the single to binary system offset of up to $\Delta \log g = 0.3\,\mathrm{dex}$, as discussed in Section~\ref{sec:flag_sp}. We caution, however, that the use of stellar structure models for the estimation of surface gravities can introduce systematic trends, as we discuss in Section~\ref{sec:caveats_photospec}.

\subsubsection{Interstellar absorption} \label{sec:interstellar}

\begin{figure}[ht]
 \centering
 \includegraphics[width=\textwidth]{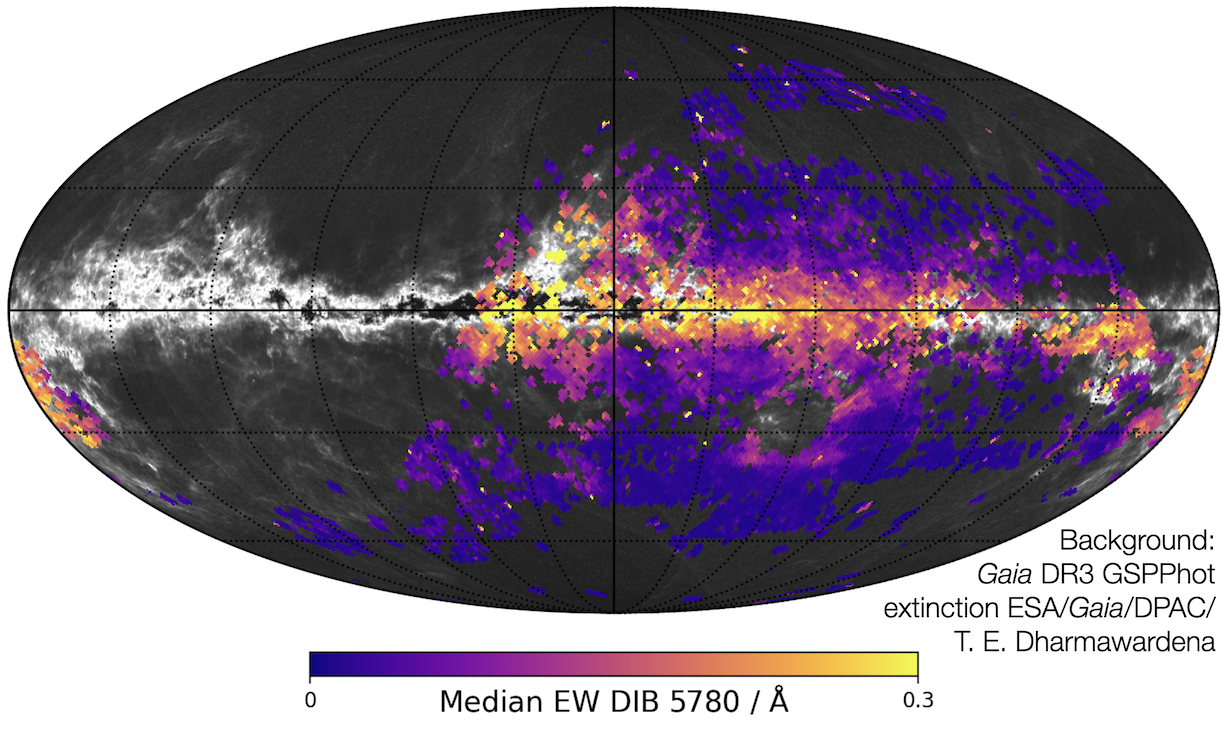}
 \caption{\textbf{All-sky map (l,b) of GALAH DR4 equivalent width measurements of the diffuse interstellar band around 5780\,\AA, with the GSPhot extinction by \citet{Andrae2023} in the background.}}
 \label{fig:galah_dr4_dibs_gaia_dr3_extinction}
\end{figure}

Because we can create synthetic stellar spectra for the full wavelength range, we can now also trace interstellar absorption in the residuals of observed spectra. By default, we try to calculate the equivalent width via Gaussian fits to the three strongest diffuse interstellar bands (DIBs; 5780.59, 5797.19, 6613.66\Angstroem) with central wavelengths identified by \citet{Vogrincic2023} as well as for interstellar K ($7698.9643\,\text{\AA}$), see Figure~\ref{fig:example_dibs_06453479-0102137}. We report the equivalent widths \texttt{eq\_x}, standard deviations \texttt{sigma\_x} and radial velocities \texttt{rv\_x}\footnote{In v240705, \texttt{rv\_comp\_1} has to be added to \texttt{rv\_k\_is} due to a bug.} for \texttt{x} in \texttt{k\_is} for interstellar K and \texttt{x} in \texttt{DIB\_5780}, \texttt{DIB\_5797}, and \texttt{DIB\_6613} for the DIBs. The coverage of interstellar material, estimated from \texttt{DIB\_5780}, within $D_\varpi < 5\,\mathrm{kpc}$ is shown in an all-sky map in Figure~\ref{fig:galah_dr4_dibs_gaia_dr3_extinction}, with the GSPhot extinction by \citet{Andrae2023} in the background.

\subsubsection{Emission estimates for the Balmer lines} \label{sec:emission}

\begin{figure}[ht]
 \centering
 \includegraphics[width=\textwidth]{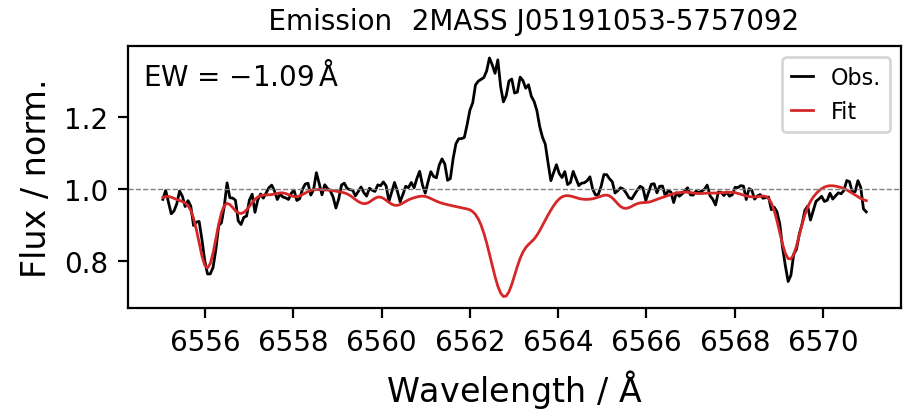}
 \caption{\textbf{Example of a flagged emission star with clear emission in the Balmer lines (here $\mathrm{H\upalpha}$).}}
 \label{fig:examples_flag_sp_1}
\end{figure}

The shape of the Balmer absorption lines holds valuable information on active stars as well as masses for evolved stars \citep{Bergemann2016} and possibly even information on unresolved binary systems \citep{Sayeed2024}. Although the cores of these lines suffer from inaccuracies in the synthesis, the residuals of synthetic and observed lines can be used in relative analyses. We therefore perform a trapezoidal integration around the Balmer lines of each normalised spectrum at $4861.3230$ and $6562.7970\,\text{\AA}$ whose values we report in \texttt{ew\_h\_beta} and \texttt{ew\_h\_alpha}. By default we integrate in a window of $\pm 0.75$ and $1.25\,\text{\AA}$ for $\text{H} \upbeta$ and $\text{H} \upalpha$, respectively, and increase this window to $5\,\text{\AA}$ if the average observed, normalised flux within $\pm 0.5\,\text{\AA}$ of the Balmer line core exceeds 1. An example of such a star is shown in Figure~\ref{fig:examples_flag_sp_1}, for which we measure a residual EW of $-1.09\,\text{\AA}$. Most emission line stars in the GALAH sample are found in the region of pre-main-sequence and hot stars (see Figure~A\ref{fig:flag_sp_overview_allstar}a). We conservatively only flag stars with a median normalised flux above 1 in $\text{H} \upbeta$ or $\text{H} \upalpha$ as emission line stars.

\begin{figure*}[ht]
 \centering
 \includegraphics[width=\textwidth]{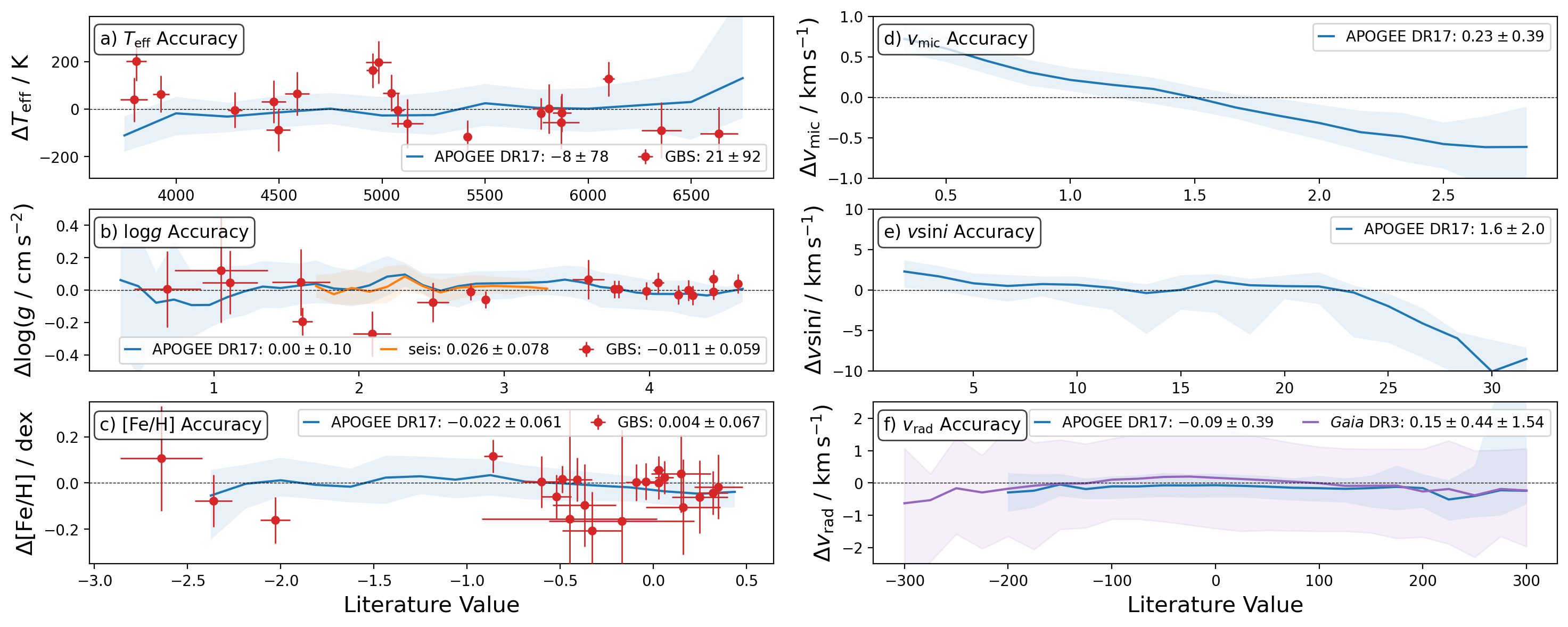}
 \caption{\textbf{Accuracy of the main stellar parameters \Teff, \logg, \feh, \vmic, \vsini, and \vrad for GALAH DR4.} Each panel shows the comparison to literature (DR4 - literature) with median values as lines and contours between 16th and 84th percentiles. Comparisons are performed for the \Gaia FGK Benchmark stars (red), APOGEE DR17 (blue), \logg inferred from asteroseismic measurements (orange) and \Gaia DR3 radial velocities (purple).}
 \label{fig:galah_dr4_validation_parameter_accuracy_allstar}
\end{figure*}

\subsection{Uncertainty estimation and validation}
\label{sec:uncertainty}

The uncertainties that we report for our spectroscopic data analysis are based on the comparison to literature measurements \citep[see also][]{Beeson2024} to estimate accuracy uncertainties and a combined precision uncertainty estimate from adjusted covariance estimates from the fitting process and the scatter of repeat observations. Formally, we estimate the total variance of measurements as a combination of the accuracy and precision variance

\begin{align} \label{eq:total_uncertainty}
    \sigma_\mathrm{total}^2 = \sigma_\mathrm{accuracy}^2 + \sigma_\mathrm{precision}^2
\end{align}

Representative values of accuracy and precision for our stellar parameters are listed in Table~\ref{tab:accuracy_precision}. We lay out how we estimate and validate accuracy and precision uncertainties in Sections~\ref{sec:uncertainty_accuracy} and \ref{sec:uncertainty_precision}, respectively.

\begin{table}[ht]
\centering
\caption{List of accuracy and representative precision uncertainties for stellar parameters in GALAH DR4. Accuracy values are estimated from comparisons with literature references (see Section~\ref{sec:uncertainty_accuracy}), whereas precision estimates are estimated from covariance uncertainties and repeat observations (Section~\ref{sec:uncertainty_precision}). Here, we list the median precision uncertainties for stars with $SNR = 50 \pm 10$ on CCD2 (see Figure~\ref{fig:galah_dr4_precision_parameters}).}
\label{tab:accuracy_precision}
\begin{tabular}{ccc}
\hline \hline
Parameter / Unit & Accuracy & Precision ($SNR = 50$)\\
\hline
$T_\text{eff}~/~\mathrm{K}$          & 66     & $23 \pm 5$ \\
$\log (g~/~\mathrm{cm\,s^{-2}})$     &  0.042 & -- \\
$\mathrm{[Fe/H]}~/~\mathrm{dex}$     &  0.051 & $0.025 \pm 0.004$ \\
$v_\text{mic}~/~\mathrm{km\,s^{-1}}$ &  0.28  & $0.05 \pm 0.03$ \\
$v \sin i~/~\mathrm{km\,s^{-1}}$     &  1.4   & $0.5 \pm 0.2$ \\
$v_\text{rad}~/~\mathrm{km\,s^{-1}}$ &  0.15  & $0.17 \pm 0.02$ \\
\hline
\end{tabular}
\end{table}

\subsubsection{Accuracy estimation and validation} \label{sec:uncertainty_accuracy}

Estimating the accuracy of spectroscopic measurements has always been a complicated endeavour, because there are no universal benchmark sets for all parameters across all stellar types. Subsequently, we describe the numerous comparisons that we have performed for both stellar parameters (\Teff, \logg, \feh, \vmic, \vsini, and \vrad) as well as the elemental abundance measurements. Consistent with GALAH DR3 \citep{Buder2021}, and caused by the limited coverage of benchmark literature, we continue to use a single accuracy estimate for each stellar parameter and ignore the possibly large accuracy uncertainties for the individual elemental abundances. In all cases, we estimate an overall bias with respect to literature values and then combine these estimates to a globally applied zero-point correction. Where not explicitly stated otherwise, we assume that the spread of stellar parameters residuals is indicative of the accuracy of either method and estimate our accuracy by dividing the parameter spread by $\sqrt{2}$. The applied shifts are listed in Table~\ref{tab:zeropoints}. We estimate the accuracy and bias correction for stellar parameters (including the iron abundance as a global parameter) and abundances separately.

Our primary reference source for parameter accuracy remains the \Gaia FGK benchmark stars \citep{Jofre2014, Jofre2015, Jofre2018, Heiter2015}. Additionally, we use asteroseismic estimates from the K2 and TESS photometry \citep{Zinn2020, Hon2021} to compare our surface gravities and perform a validation to higher quality observations of globular cluster stars with typically lower metallicities \citep{Carretta2009c, Carretta2009, Johnson2010}. Because the overlap with APOGEE DR17 \citep{SDSSDR17} has increased from 41\,941 stars in GALAH DR3 to 60\,046 stars with 92\,368 repeat observation matches in GALAH DR4, we also can assess systematic trends for a larger parameter space. For clarity, we discuss the stellar parameters separately, but show most accuracy estimates in a combined Figure~\ref{fig:galah_dr4_validation_parameter_accuracy_allstar}.

\Teff: The effective temperature estimates from GALAH DR4 show good agreement with the \Gaia FGK benchmark stars (Figure~\ref{fig:galah_dr4_validation_parameter_accuracy_allstar}a). Specifically, we find a mean difference of $\Delta T_\mathrm{eff} = 21 \pm 92\,\mathrm{K}$, indicating no significant bias between our temperatures and the benchmark values. Comparisons with APOGEE DR17 show an equally robust agreement, with $\Delta T_\mathrm{eff} = -8 \pm 78\,\mathrm{K}$. This small offset and uncertainty suggest that the GALAH DR4 \Teff estimates are highly reliable across a wide range of at least G- and K-, but possibly also F- and M-type stars. Here, we use $1/\sqrt{2}$ of the residual spread with respect to \Gaia benchmark stars as our accuracy estimate.

\logg: For surface gravity, we compared our \logg estimates to both the \Gaia FGK benchmark stars, asteroseismic measurements from \citet{Zinn2020} and \citet{Hon2021}, and APOGEE DR17. The asteroseismic \logg values are derived from $\nu_\mathrm{max}$ measurements for giant stars, and they show excellent agreement with our results, with a mean difference of $\Delta \log g = 0.026 \pm 0.078$. Both the asteroseismic comparison as well as the \Gaia benchmark star comparison ($\Delta \log g = -0.011 \pm 0.059$) and APOGEE DR17 ($\Delta \log g = 0.00 \pm 0.10$) agree well and show no trends across the \logg range. For the low metallicity regime, we compare GALAH \logg values with asteroseismically derived values from \cite{Howell2022} for the globular cluster M\,4 (NGC~6121). Stars from this cluster were observed as part of a dedicated survey (PI M. Howell) aimed at spectroscopically characterising their sample of stars observed by the K2 mission \citep{Howell2014}. Across the 75 overlapping targets, we find a $\Delta \log g = 0.056 \pm 0.128$. The comparison between independently derived light element abundance variations and asteroseismic masses will be presented in an upcoming paper (Howell et al., in preparation.). This is a significant improvement over GALAH DR3, where significant deviations were found for luminous giant stars - whose parameter estimates in GALAH DR3 suffered from less precise and systematically biased distance and thus \logg estimates. We find significant outliers, however, particularly for primary red clump stars, which were mistaken as secondary red clump stars, leading to larger deviations. We discuss this issue later in Section~\ref{sec:caveats_photospec}. Because this single group is driving the scatter in our disagreement with the asteroseismic estimates, we revert to the \Gaia benchmark stars to estimate the accuracy.

\begin{figure}[ht]
 \centering
 \includegraphics[width=\columnwidth]{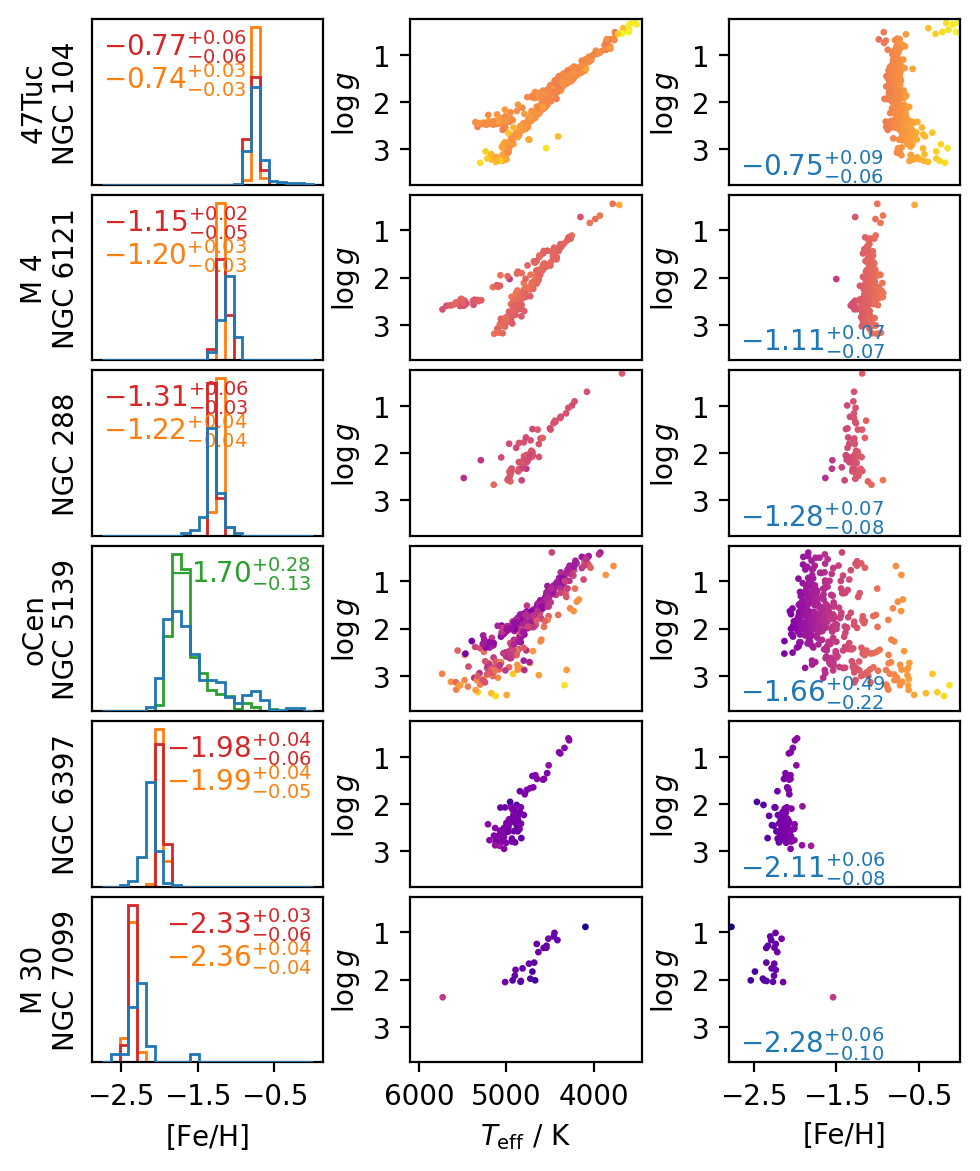}
 \caption{\textbf{Comparison of iron abundances (16th, 50th and 84th percentiles) and overview of spectroscopic and photometric properties of globular cluster stars in GALAH DR4.}
 \textbf{Left panels} show histograms of iron abundances from GALAH DR4 (blue) as well as literature estimates for the globular clusters from Giraffe (orange) and UVES (red) observations by \citep{Carretta2009c, Carretta2009} as well as observations from \cite{Johnson2010}.
 \textbf{Middle panels} show the spectroscopic \Teff-\logg diagrams coloured by iron abundance \feh.
 \textbf{Right panels} show the trend of GALAH DR4 [Fe/H] along the different \logg values.
}
 \label{fig:galah_dr4_allstar_globular_cluster_feh_comparison}
\end{figure}

\begin{figure*}[ht]
 \centering
 \includegraphics[width=0.93\textwidth]{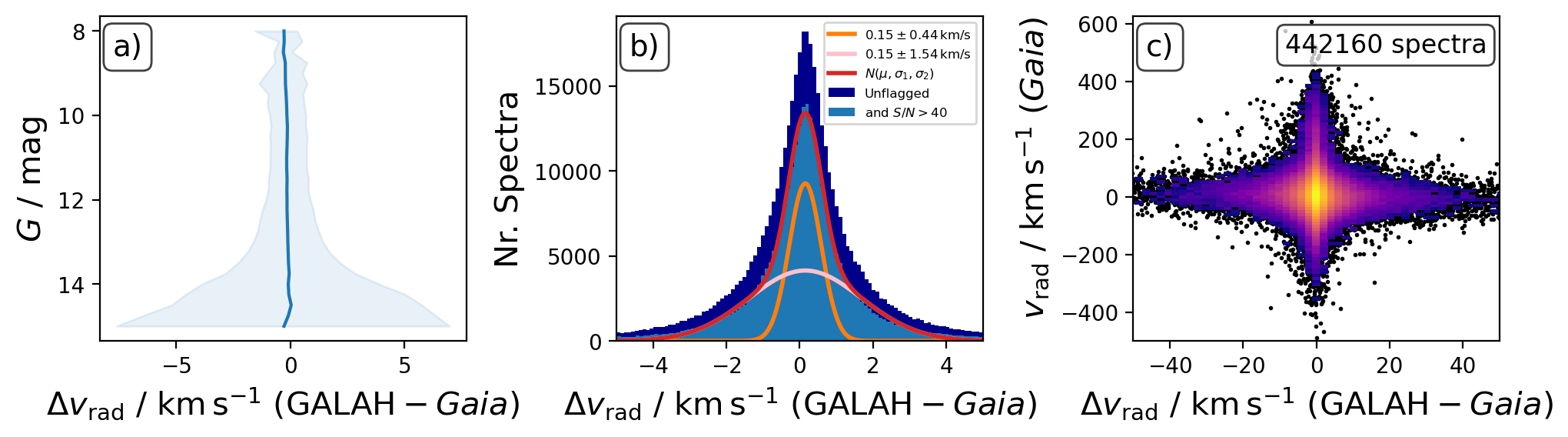}
 \caption{\textbf{Comparison of radial velocities between GALAH DR4 \texttt{allspec} and \Gaia DR3.}
 \textbf{Panel a)} shows the difference of radial velocities as function of \Gaia $G$ magnitude.
 \textbf{Panel b)} shows a histogram of the difference with two Gaussian distributions (with same mean) fitted to them to estimate a more robust, binary independent, radial velocity difference.
 \textbf{Panel c)} shows the difference of radial velocities as function of radial velocity, showing the systematic scatter introduced by binaries.
}
 \label{fig:galah_dr4_vrad_gaia_dr3}
\end{figure*}

\feh: The comparison of GALAH DR4 metallicities to the \Gaia FGK benchmark stars initially showed the similar bias of GALAH towards more metal-poor values at the $0.049\,\mathrm{dex}$ level. The application of a zero-point correction (see Table~\ref{tab:zeropoints}) yields an excellent agreement, with $\Delta \mathrm{[Fe/H]} = 0.004 \pm 0.067$ for the benchmark stars and $\Delta \mathrm{[Fe/H]} = -0.022 \pm 0.061$ for APOGEE DR17, confirming the reliability of the GALAH DR4 metallicity estimates across a large range of metallicities. For the metal-poor regime, benchmark estimates are still rare. Luckily, a dedicated observing program -- whose results are included in this data release -- was performed and an overview of globular cluster Kiel diagrams is appended in Figure~A\ref{fig:galah_dr4_gcs_teff_logg}. We therefore only perform a comparison with globular cluster stars -- often measured in 1D LTE -- to get a quantitative impression of the agreement. We restrict ourselves to a few studies, namely those by \citet{ Carretta2009c, Carretta2009} for NGC 104, 6121, 288, 6397, and 7099 as well as \citet{Johnson2010} for NGC 5139. In all cases, we find a good agreement of the metallicity distribution function for overlapping stars within the uncertainties (see Figure~\ref{fig:galah_dr4_allstar_globular_cluster_feh_comparison}). While this does not necessarily confirm our accuracy, it shows consistency within this uncertain parameter regime. We note however, a specific region in NGC~104, where the metallicity of the most luminous giants ($T_\mathrm{eff} < 3750\,\mathrm{K}$ and $\log g < 0.5$) is incorrectly estimated near the Solar value. We discuss this problem in detail as a caveat in Section~\ref{sec:caveats_fitting}, since we have not been able to systematically flag these stars. More generally, we note that the strong and unexpected abundance trends with \Teff or \logg in globular clusters from GALAH DR3 have decreased for most elements. However, we still urge users to take caution when using globular cluster abundances, and we discuss this further in Sec. \ref{sec:caveats_globulars}. A custom, by hand analysis of globular cluster abundances beyond [Fe/H] will be performed in a separate study (McKenzie et al., in preparation), as these observations have been part of a dedicated observing program (PIs M. McKenzie and M. Howell). Similarly, a dedicated verification of open cluster observations (PIs J. Kos and G. De Silva) will be performed in a separate study \citep{Kos2025}.

\vmic: Microturbulence velocities show a more complex pattern when compared to APOGEE DR17. We find a mean difference of $\Delta v_\mathrm{mic} = 0.23 \pm 0.39\,\mathrm{km\,s^{-1}}$. However, the comparison reveals a linear mismatch: APOGEE DR17 tends to measure lower \vmic values for stars with low microturbulence and larger \vmic values for stars with higher microturbulence compared to GALAH DR4. This systematic trend suggests that the \vmic calibration between the two surveys may differ slightly, particularly at the extremes of the parameter range. We note, however, that the surveys agree much better than for GALAH DR3, where a fixed quadratic relation was used that did not allow for deviations, for example for red clump stars. Adding \vmic as free parameter returned a similar pattern as the empirical relation by \citet{DutraFerreira2016} and shows a significantly different behaviour of \vmic for the hottest, coolest, and red clump stars (see Figure~A\ref{fig:initial_parameters}). This mismatch of \vmic could have indeed driven the metallicity mismatch of metal-rich red clump stars in GALAH DR2 and DR3 \citep{Buder2018, Buder2021}, since their metallicities are in agreement with other estimates now (e.g. APOGEE DR17).

\vsini: The rotational velocity estimates agree well with APOGEE DR17, with a mean difference of $\Delta v \sin i = 1.6 \pm 2.0\,\mathrm{km\,s^{-1}}$. However, at higher rotational velocities (above approximately $24\,\mathrm{km\,s^{-1}}$), our neural networks start to extrapolate, leading to an upper limit in the estimates and returning significantly lower \vsini values compared to APOGEE DR17. This issue highlights the limitations of the GALAH DR4 \vsini estimates for rapidly rotating stars.

\vrad: For radial velocity we compared our results to both APOGEE DR17 and \Gaia DR3. The comparison with APOGEE DR17 yields a small offset of $\Delta v_\mathrm{rad} = -0.09 \pm 0.39\,\mathrm{km\,s^{-1}}$, indicating excellent agreement between the two surveys. Accounting for the much lower SNR for faint \Gaia targets and unidentified binaries, we fit two Gaussian distributions to the overall difference of GALAH and \Gaia radial velocities (see Figure~\ref{fig:galah_dr4_vrad_gaia_dr3}). The comparison with \Gaia DR3 shows a slightly larger offset of $\Delta v_\mathrm{rad} = 0.15 \pm 0.44 \pm 1.54\,\mathrm{km\,s^{-1}}$, which is expected due to the lower precision of the \Gaia DR3 radial velocities \citep{Katz2023}. We use the median residual of $0.15\,\mathrm{km\,s^{-1}}$ with respect to \Gaia DR3 rather than the spread as our accuracy estimate.

\begin{figure*}[ht]
 \centering
 \includegraphics[width=\textwidth]{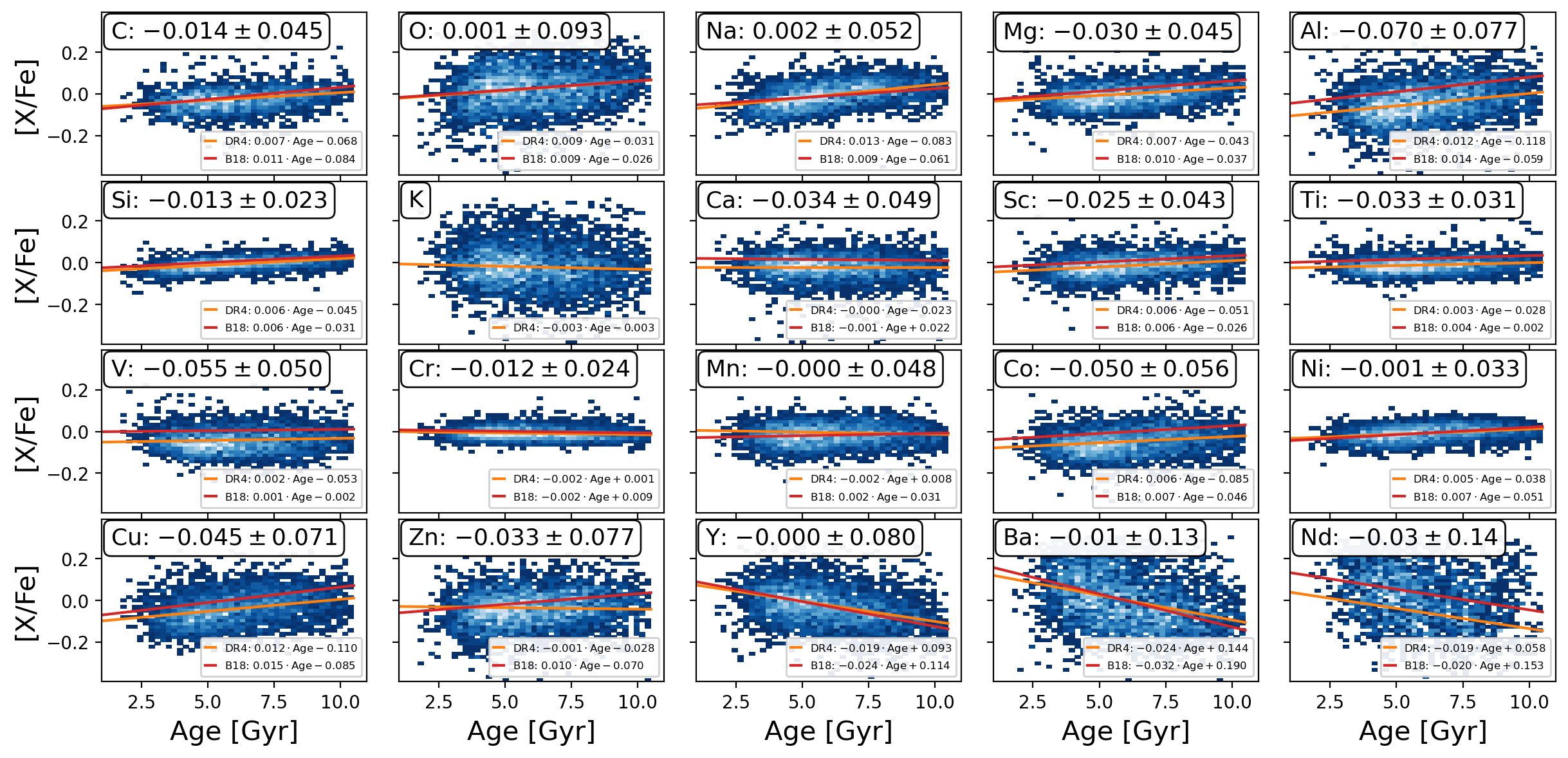}
 \caption{\textbf{Chemical abundances [X/Fe] of Solar twin stars as a function of ages that were estimated as part of the mass and age estimation of the \texttt{allstar} spectrum analysis.} We overplot linear fits to our age-abundance relations for Solar twins in orange and literature values from \citet{Bedell2018} in red. Panels also indicate the median and standard deviation with respect to \citet{Bedell2018} when assuming a correct age.}
 \label{fig:galah_dr4_age_xfe_trends_solar_twins_allstar}
\end{figure*}

Elemental abundances [X/Fe]: While there is no model-independent benchmark for absolute abundance accuracy, we continually perform comparisons with literature values to assess the consistency of our results with other studies. In GALAH DR4, we evaluate the abundance zero-points using up to five different reference estimates (see Figure~A\ref{fig:galah_dr4_zeropoint_checks_allstar}): (1) the spectroscopic analysis of a Solar-composition spectrum of the asteroid Vesta (\texttt{sobject\_id} 210115002201239), (2) abundance estimates for Solar twins corresponding to a Solar age of 4.5 Gyr (see Figure~\ref{fig:galah_dr4_age_xfe_trends_solar_twins_allstar}), (3) abundances of \Gaia FGK benchmark stars \citep{Jofre2015, Jofre2018}, (4) stars with Solar-like metallicity $-0.1 < \mathrm{[Fe/H]} < 0.1$ within $500\,\mathrm{pc}$ of the Sun \citep[a method introduced by][]{Joensson2020}, and (5) differences in abundance estimates for stars overlapping with the high-resolution, large-scale spectroscopic APOGEE DR17 survey \citep{SDSSDR17}. 

It is important to note that our abundance corrections, and consequently the Solar abundances presented in Table~A\ref{tab:zeropoints}, are determined within the framework of 1D LTE or 1D NLTE models and are not intended to represent the most accurate Solar abundances. Instead, they reflect our best effort to minimise discrepancies across different comparison cases. Given the differences in line modelling, such as those between \citet{Grevesse2007} (who used 3D atmospheres) and our 1D models, and possible deviations in the reference abundance from our Vesta spectrum, we refer to these adjusted values as \textit{zero-points}. For certain scientific applications, adjusting these abundance zero-points might be necessary to ensure consistency with other datasets.

While we are not able to include all of our validation plots, we refer the interested reader to the publicly available code in our code repository\footnote{\url{https://github.com/svenbuder/GALAH_DR4/tree/main/validation}}. Generally speaking, we have found that a large number of systematic trends of abundances with temperature and surface gravity has decreased with respect to GALAH DR3, as can be appreciated from dedicated validation plots (\href{https://github.com/svenbuder/GALAH_DR4/blob/main/validation/galah_dr4_validation_globular_clusters.ipynb}{online here}) -- with a similar appearance as the right hand panels of Figure~\ref{fig:galah_dr4_allstar_globular_cluster_feh_comparison}.

\begin{figure}[ht]
 \centering
 \includegraphics[width=\textwidth]{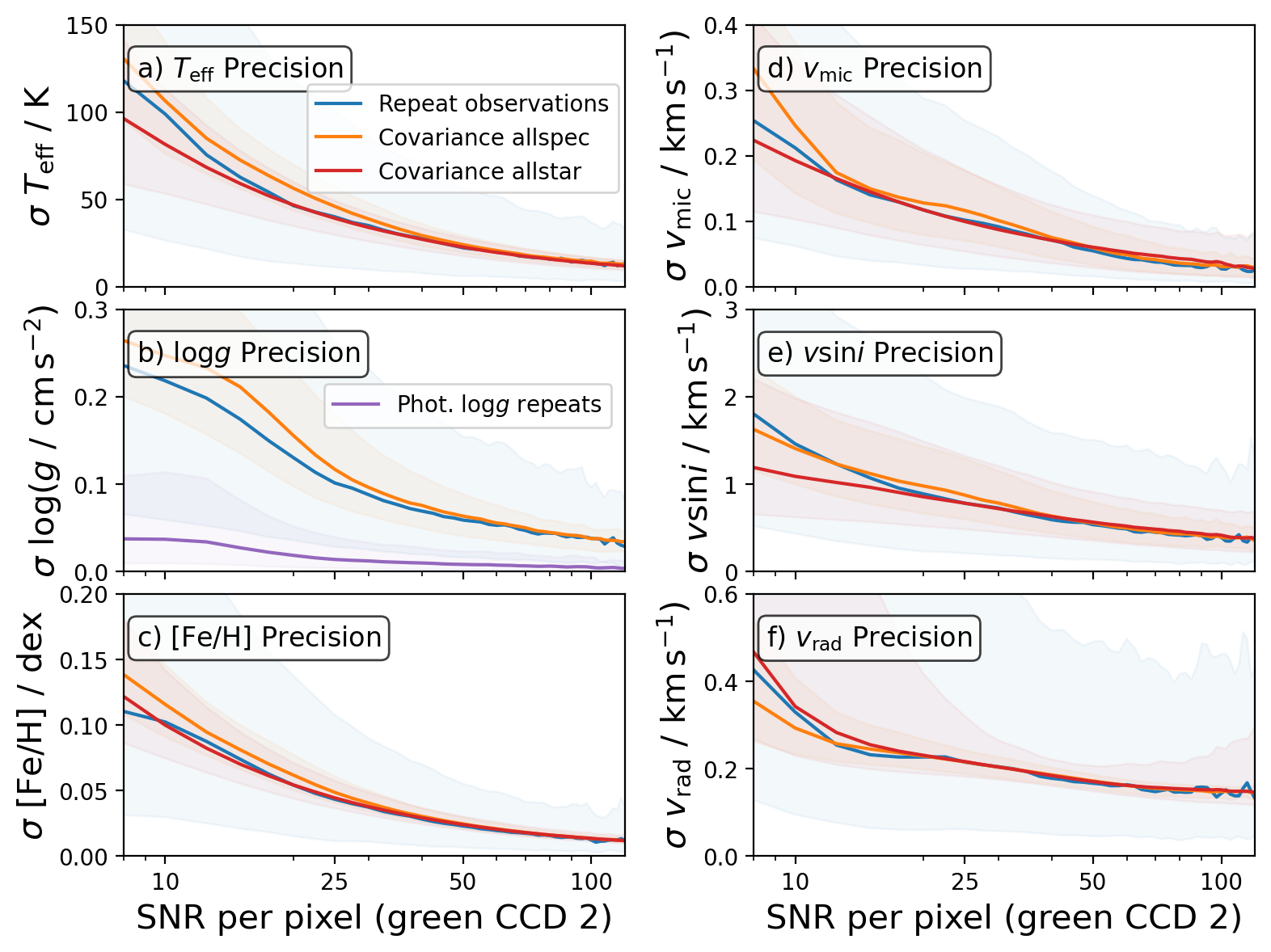}
 \caption{\textbf{Precision monitoring (with a median line and standard deviation shading) of stellar parameters as a function of SNR for the green CCD2 across GALAH DR4.} Each panel shows the behaviour for bins of width 10 for the scatter of repeat observations of the \texttt{allspec} runs (blue), covariance uncertainties of \texttt{allspec} (orange) and \texttt{allstar} (red) setups as well as scatter of photometric \logg from repeat observations (purple).}
 \label{fig:galah_dr4_precision_parameters}
\end{figure}

\subsubsection{Precision estimation and validation} \label{sec:uncertainty_precision}

In addition to the accuracy uncertainty, we estimate the total uncertainty through the additional precision uncertainty (Equation~\ref{eq:total_uncertainty}). For this purpose, we mainly rely on the fitting uncertainties of the \texttt{curve\_fit} function, which we rescale based on repeat observations.

While we report the raw fitting covariance matrix for each spectrum and module (see Figure~A\ref{fig:covariance_vesta_arcturus} for the covariance matrices of Vesta and Arcturus), their entries are not validated for reliability and have not been adjusted to incorporate a rescaling towards the final uncertainties. For the purposes of reporting stellar parameter and abundance fitting uncertainties, we restrict ourselves to the standard deviations of each feature, that is, the square root values of the diagonal covariance matrix entries.

Similar to GALAH DR3, we apply a precision adjustment of the fitting uncertainty towards consistency with the scatter of repeat observations only as a function of SNR of CCD2. Contrary to GALAH DR3, we have extended this rescaling function to be fitted in bins of SNR with both a constant, linear, and exponential term with \texttt{snr\_px\_ccd2} as the independent variable, that is, $c_1 + c2 \cdot SNR + c_3 \cdot \exp(c_4 \cdot SNR)$. We report the fitted constants for both \texttt{allspec} and \texttt{allstar} in the online repository\footnote{\texttt{*precision\_correction\_factors*} in \url{https://github.com/svenbuder/GALAH_DR4/tree/main/catalogs}.} for each stellar parameter and abundance. 

In Figures~\ref{fig:galah_dr4_precision_parameters} and A\ref{fig:galah_dr4_precision_abundances}, we then confirm that the overall trends of fitting uncertainties for \texttt{allspec} and \texttt{allstar} are consistent with the repeat observation scatter of the \texttt{allspec}. The latter has to be used as reference, because the \texttt{allstar} module uses co-added spectra of repeat observations rather than the repeat observations themselves. While this might actually overestimate the precision uncertainty of stellar parameters, we do not expect a too strong overprediction for abundances.

While the precision levels of stellar parameters have on average actually remained similar to the estimates of GALAH DR3, we see notable improvement of the precision for multiple elements, such as C, Mg, V, Cr, Co, Ni, La, Ce, Nd, and Sm. The precision of Eu, however, seems to have decreased.

Separately from this work, we performed an extensive analysis of the precision and accuracy of spectroscopic parameters from the observation of star clusters, 43 open clusters of all ages and 10 globular clusters \citep{Kos2025}. In this work, we compare \Teff, \logg, and stellar ages with the values obtained from cluster isochrone fitting. Ages show typical uncertainties of 10 to 50\%, depending on the stellar type. \Teff and \logg match well for stars hotter than $4000\,\mathrm{K}$ with a bias of $\Delta T_\mathrm{eff}=-68\,\mathrm{K}$ (GALAH - Isochrones), and $\Delta \log g = -0.03$. For stars cooler than stars $4000\,\mathrm{K}$, GALAH DR4 temperatures are overestimated by up to $250\,\mathrm{K}$ at $T_\mathrm{eff} = 3000\,\mathrm{K}$ and we find a complicated pattern in $\Delta \log g$, with \logg being sometimes severly overestimated for the coolest dwarf stars. 

Most interesting is the analysis of elemental abundances. Assuming that stellar clusters are chemically homogeneous, we can study the precision of the reported abundances over a large range of temperatures. We find that cold stars show consistent systematic trends, that can reach values of $0.5\,\mathrm{dex}$ for some elements. Dwarf stars are most affected at temperatures $T_\mathrm{eff} < 4600\,\mathrm{K}$, while giants show much smaller systematics with strong trends only at $T_\mathrm{eff} < 4000\,\mathrm{K}$. The results of this cluster validation \citep{Kos2025}, including a detrended set of elemental abundances, will be published as value-added-catalogues in DR4.

\begin{figure*}[ht]
 \centering
 \includegraphics[width=0.49\textwidth]{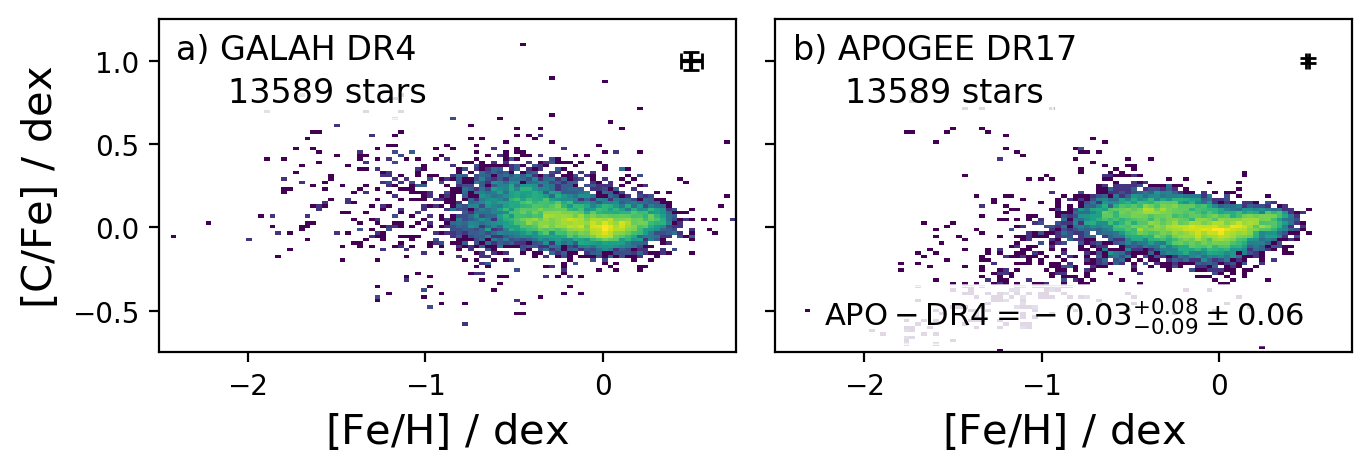}
 \includegraphics[width=0.49\textwidth]{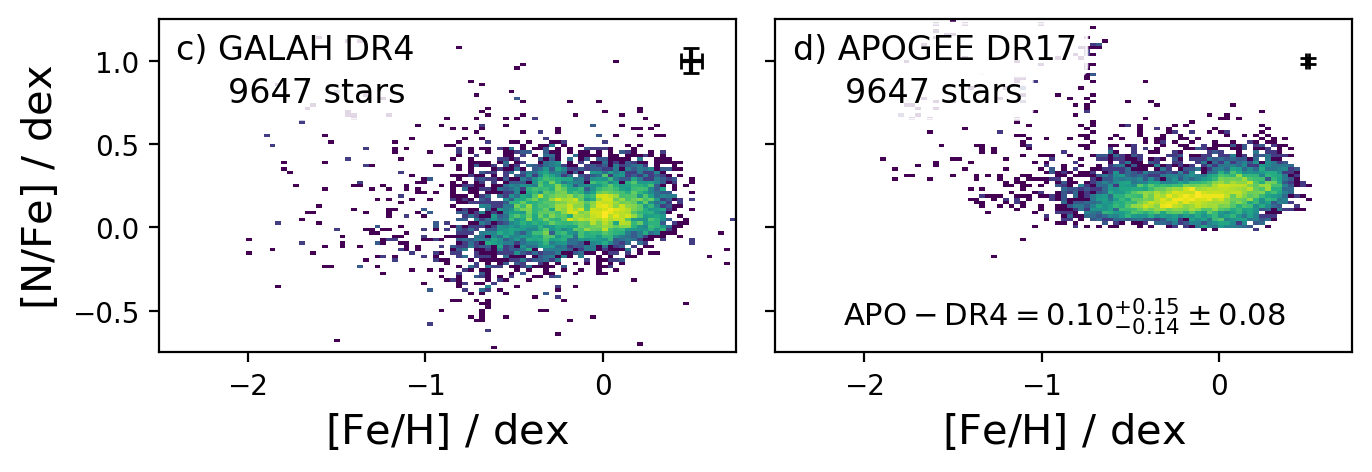}
 \caption{\textbf{Comparison of stars with available measurements in GALAH DR4 and APOGEE DR17 for [C/Fe] and [N/Fe]}.}
 \label{fig:comparison_dr4_apo17}
\end{figure*}

\begin{figure*}[ht]
 \centering
 \includegraphics[width=0.87\textwidth]{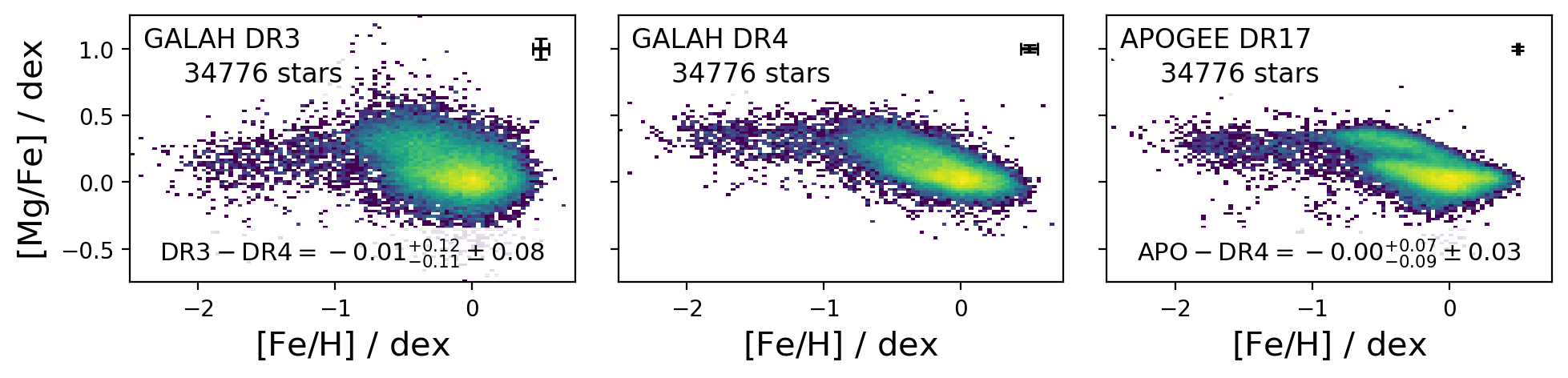}
 \includegraphics[width=0.87\textwidth]{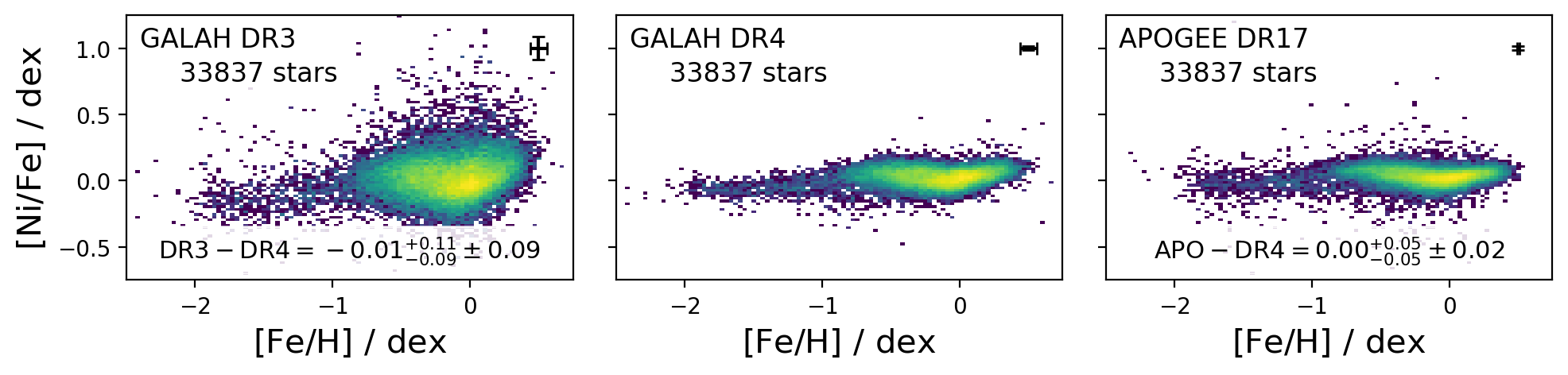}
 \caption{\textbf{Comparison of stars with available measurements in GALAH DR3 (left), GALAH DR4 (middle) as well as APOGEE DR17 (right) for [Mg/Fe] (top row) and [Ni/Fe] (bottom row)}.}
 \label{fig:comparison_dr4_dr3_apo17}
\end{figure*}

\subsubsection{Uncertainties in light of GALAH DR3 and APOGEE DR17}

To get a better idea of the actual improvement of accuracy and precision, we have performed more elaborate comparisons than those in Figure~\ref{fig:galah_dr4_validation_parameter_accuracy_allstar} and only showcase a few in this manuscript with reference to the online repository. We have found the comparison of GALAH DR4 with both GALAH DR3 and APOGEE DR17 highly informative.

Because GALAH DR3 did not include N measurements and only a limited amount of C measurements, our first comparison concerns the abundances of C and N between GALAH DR4 and APOGEE DR17 in Figure~\ref{fig:comparison_dr4_apo17}. We attach the comparisons for the other overlapping elements O, Na, Al, Si, K, Ca, Ti, V, Cr, Mn, Co, and Ce in Figures~A\ref{fig:comparison_dr3_dr4_apo17_rest} and A\ref{fig:comparison_dr3_dr4_apo17_rest2}. While we see a generally good agreement of the shapes, we notice biases of $-0.03\dex$ and $0.10\dex$ for C and N, respectively. These can be, however, explained by the lower precision of GALAH and might, in part, be driven by the slightly different trends of C and N towards lower metallicities. In particular, [C/Fe] decreases to sub-Solar level in APOGEE DR17 for metal-poor stars, whereas it is Solar or even enhanced in GALAH DR4. Enhanced levels would be expected for metal-poor disk stars, whereas sub-Solar levels are expected for accreted stars \citep{Amarsi2019c}, warranting a future population analysis to test the accuracy of either survey.

In addition to these novel abundances, we also showcase two previously measured abundances, namely [Mg/Fe] and [Ni/Fe] in Figure~\ref{fig:comparison_dr4_dr3_apo17}. The $\upalpha$-process element Mg has significant value for Galactic studies because it is predominantly produced by core-collapse supernovae \citep{Kobayashi2020}. In GALAH DR3, only the \ion{Mg}{i}~$5711\,\text{\AA}$ line was used, whereas we now use a combination of several lines. This has led to a significant improvement in precision, as can be appreciated from the comparison of Figures~\ref{fig:comparison_dr4_dr3_apo17}a and \ref{fig:comparison_dr4_dr3_apo17}b. Even more positive, we see an improved agreement of the [Fe/H] vs. [Mg/Fe] measurements between GALAH DR4 (Figure~\ref{fig:comparison_dr4_dr3_apo17}b) and APOGEE DR17 (Figure~\ref{fig:comparison_dr4_dr3_apo17}c), with no abundance bias. One of the elements with the most significant precision improvement is Ni. For this element, our move to fitting the full wavelength range has increased the number of lines from two very reliable lines to several dozen lines. Albeit less reliable in their line data and possibly blended, the sheer increase in flux information used has improved the precision almost to the level of APOGEE DR17 - with no bias and a standard deviation of only $0.05\,\mathrm{dex}$ between APOGEE DR17 and GALAH DR4 (see Figures~\ref{fig:comparison_dr4_dr3_apo17}c-f).

In addition to these instructive comparisons, we also return to the precision of Solar twins from Figure~\ref{fig:galah_dr4_age_xfe_trends_solar_twins_allstar}. Here we specifically highlight the significant improvement of precision from GALAH DR3 to GALAH DR4 with respect to the linear estimate from \citet{Bedell2018} for C (from 0.09 to 0.045), Si (from 0.04 to 0.023), Ca (0.07 to 0.049), Ti (0.05 to 0.031), V (0.13 to 0.050), Cr (0.06 to 0.024), Ni (0.07 to 0.033), and Y (0.12 to 0.080). This improvement of sometimes a factor of 2 is remarkable and most of our comparisons indicate that these values are representative of a precision improvement beyond the Solar twins, as shown for example for Ni in Figure~\ref{fig:comparison_dr4_dr3_apo17}.

\begin{figure*}[ht]
\centering
\includegraphics[width=0.99\textwidth]{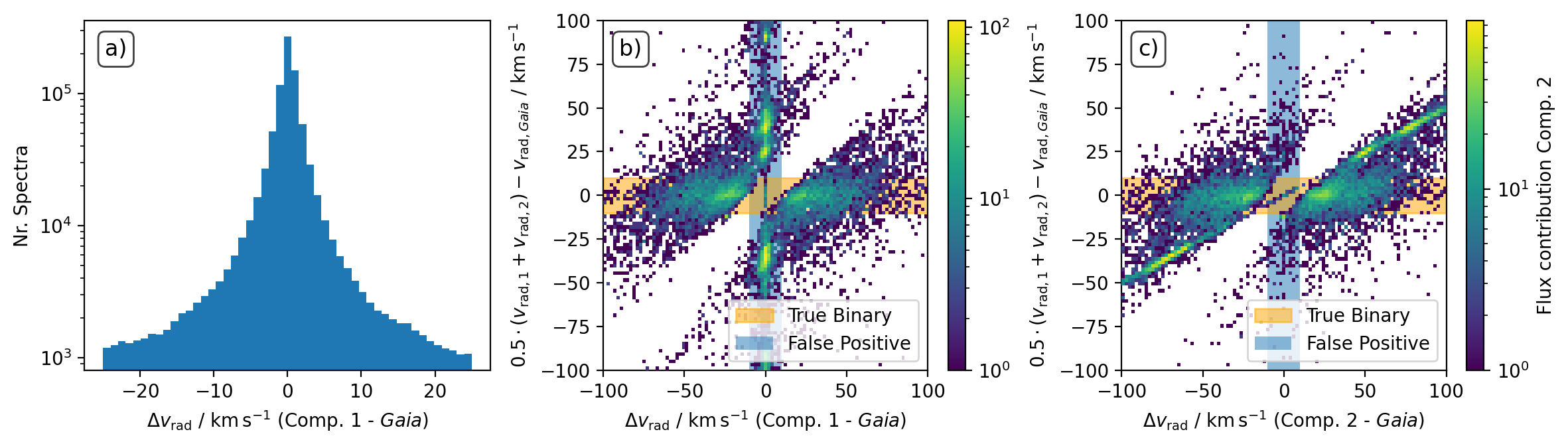}
\caption{\textbf{Comparison of radial velocity estimates of GALAH DR4 and \textit{Gaia} DR3.} \textbf{Panel a)} shows the difference of GALAH's primary component radial velocity with the mean \textit{Gaia} DR3. \textbf{Panels b) and c)} show stars for which two components were detected in GALAH DR4 and shows the difference between each component and \textit{Gaia} DR3 against the difference of mean (roughly systemic) radial velocities. The panels also include regions where actual binaries and false positive detections are expected.}
\label{fig:vrad_comparison_comp1_comp2_gaiadr3}
\end{figure*}

\subsection{Stellar parameter flags \texttt{flag\_sp}}
\label{sec:flag_sp}

We have implemented a series of post-processing routines to assess the quality of the stellar parameter determinations. These routines check for a variety of potential issues with the spectra and stellar label fitting, with each flag corresponding to a specific quality check. If any of these checks are not passed, the respective bit in the quality flag \texttt{flag\_sp} is raised. The description of the implemented bits/flags for \texttt{flag\_sp} and how often they were raised is listed in Table~\ref{tab:flag_sp} and distributions in the Kiel diagram (\Teff and \logg) are shown for each raised bit in Figure~A\ref{fig:flag_sp_overview_allstar} for the \texttt{allstar} catalogue. For examples of stars with raised flags, we refer back to the emission line star (\texttt{flag\_sp} = 1) of Figure~\ref{fig:examples_flag_sp_1} (Section~\ref{sec:emission}) and the clearly double-lined binary of Figure~\ref{fig:examples_flag_sp_2}.

Because quality cuts should be applied based on the specific science case at hand, we do not make a strict recommendation for which upper limit of \texttt{flag\_sp} should be applied. We note, however, that we have tried to implement flags that increase in concert. The first 8 bit masks (with values up to $2^9-1 = 511$) are therefore less problematic than those of 9  or higher ($\texttt{flag\_sp} \leq 512$).

\begin{table}[ht]
\centering
\caption{List of major quality flag \texttt{flag\_sp} listing the bit, description and how often the flag was raised for the \textit{allstar} and \textit{allspec} routines. Notes: Multiple bits can be raised for each of the 1\,085\,520 spectra of 917\,588 stars.}
\label{tab:flag_sp}
\begin{tabular}{ccccc}
\hline \hline
Raised Bit & Flag & Description & \textit{allspec} & \textit{allstar} \\
\hline
  & 0 & No flag & 700125 & 663075 \\ 
0 & 1 & Emission & 9568 & 7646 \\
1 & 2 & CCD missing & 70078 & 44344 \\
2 & 4 & Spectr. Binary 1 & 0 & 25538 \\
3 & 8 & Spectr. Binary 2 & 34833 & 32566 \\
4 & 16 & $\chi^2 > 3\sigma$ & 66859 & 20544 \\
5 & 32 & \vsini warning & 138317 & 95990 \\
6 & 64 & \vmic warning & 99692 & 78686 \\
7 & 128 & Triple Binary warning & 0 & 0 \\
8 & 256 & \Teff warning & 0 & 0 \\
9 & 512 & \logg warning & 19863 & 10900 \\
10 & 1024 & \feh warning & 0 & 0 \\
11 & 2048 & S/N low & 123736 & 71154 \\
12 & 4096 & Not converged & 32986 & 0 \\
13 & 8192 & Model extrapolated & 69613 & 5953 \\
14 & 16384 & No Results & 7400 & 10899 \\
\hline
\end{tabular}
\end{table}

While not intended to identify binaries, we believe that both the \vsini and \vmic flags are informative for binaries below $T_\mathrm{eff} < 6000\,\mathrm{K}$ (see their elevated position in Figures~A\ref{fig:flag_sp_overview_allstar}f and g). We have trained the stars of this region with a lower maximum \vsini range that would be reached for a spectrum that is broadened due to binarity. This region certainly overlaps with the one of identified single-lined and double-lined binaries with \texttt{flag\_sp} = 4 and 8, respectively (see Figures~A\ref{fig:flag_sp_overview_allstar}c and d). For the latter, we notice that especially cool giants are picked up by the automatic algorithm as well. This might be either due to strong extinction biasing our analysis or due to lines in the spectrum not being modelled properly and thus showing up as residual signal. While these stars are possibly flagged false-positively, we also find a remarkable amount of true binaries ($>41\%$ in orange area of Figure~\ref{fig:vrad_comparison_comp1_comp2_gaiadr3}b), for which the \Gaia DR3 radial velocity is likely the systemic radial velocity, as it is close to the mean radial velocity of both components identified in GALAH DR4. In Figure~\ref{fig:vrad_comparison_comp1_comp2_gaiadr3}, we visualise how one could use the radial velocities from GALAH DR4 and \Gaia DR3 to further assess the reliability of this flag.
To check if a particular bitmask flag (e.g. $2^3 = 8$) is raised, one can perform the check in \textsc{python} via
\begin{verbatim}
    flag_8_raised = (dr4['flag_sp'] & 8) != 0
\end{verbatim}

\subsection{Elemental abundance flags \texttt{flag\_X\_fe}}
\label{sec:flag_x_fe}

The quality of elemental abundance measurements is also captured through flags. When an element is reliably detected in the spectrum, no flag is raised. However, if the abundance of an element is estimated as an upper limit, often due to weak spectral lines or low SNR, an upper limit flag is triggered. If no measurement of the element is possible, a flag is raised to indicate that the relevant spectral features were too weak or the SNR too low to allow for an estimate. The list of bits and flags for elemental abundances, \texttt{flag\_X\_fe}, is shown in Table~\ref{tab:flag_x_fe}.

\begin{table}
\centering
\caption{List of elemental abundance quality flags \texttt{flag\_fe\_h} for \feh or \texttt{flag\_X\_fe} for element X.}
\label{tab:flag_x_fe}
\begin{tabular}{ccc}
\hline \hline
Raised Bit & Flag & Description \\
\hline
  & 0 & detection \\ 
0 & 1 & upper limit \\ 
1 & 2 & no measurement available\\
2 & 4 & no convergence\\
3 & 8 & measurement above limit\\
4 & 16 & measurement below limit\\
5 & 32 & measurement issue of CNO \\
6 & 64 & optimisation may have failed \\
\hline
\end{tabular}
\end{table}

By default, we recommend to only use significant detections ($\texttt{flag\_x\_fe} = 0$) for an element. Because of a bug in the flagging of the [Fe/H] detection (see discussion in Section~\ref{sec:caveats_flags}), we do not recommend to consider $\texttt{flag\_fe\_h}$ for quality cuts.

\subsection{Abundance detection or upper limit}
\label{sec:abundance_detection_or_upper_limit}

To assess whether the abundance estimates are a true detection or an upper limit for each element X, we compare a synthetic spectrum with the best-fitting parameters to a synthetic spectrum with the same parameters, except for element X, for which we use the lower limit abundance of the neural network. The residuals in units of $\sigma$ between the best-fitting spectrum and the spectrum with the lowest possible [X/Fe] or lowered \feh then allow us to identify a detection (with maximum differences beyond 3 $\sigma$) or upper limits, for which we raise the flag \texttt{flag\_x\_fe} by 1. Our initial test of overall detectability (Section~\ref{sec:which_labels_are_optimised}) allowed us to raise the flag \texttt{flag\_x\_fe} by 2 for elements for which not even an upper limit was expected.

\begin{landscape}
\begin{figure}[ht]
\includegraphics[width=0.975\columnwidth]{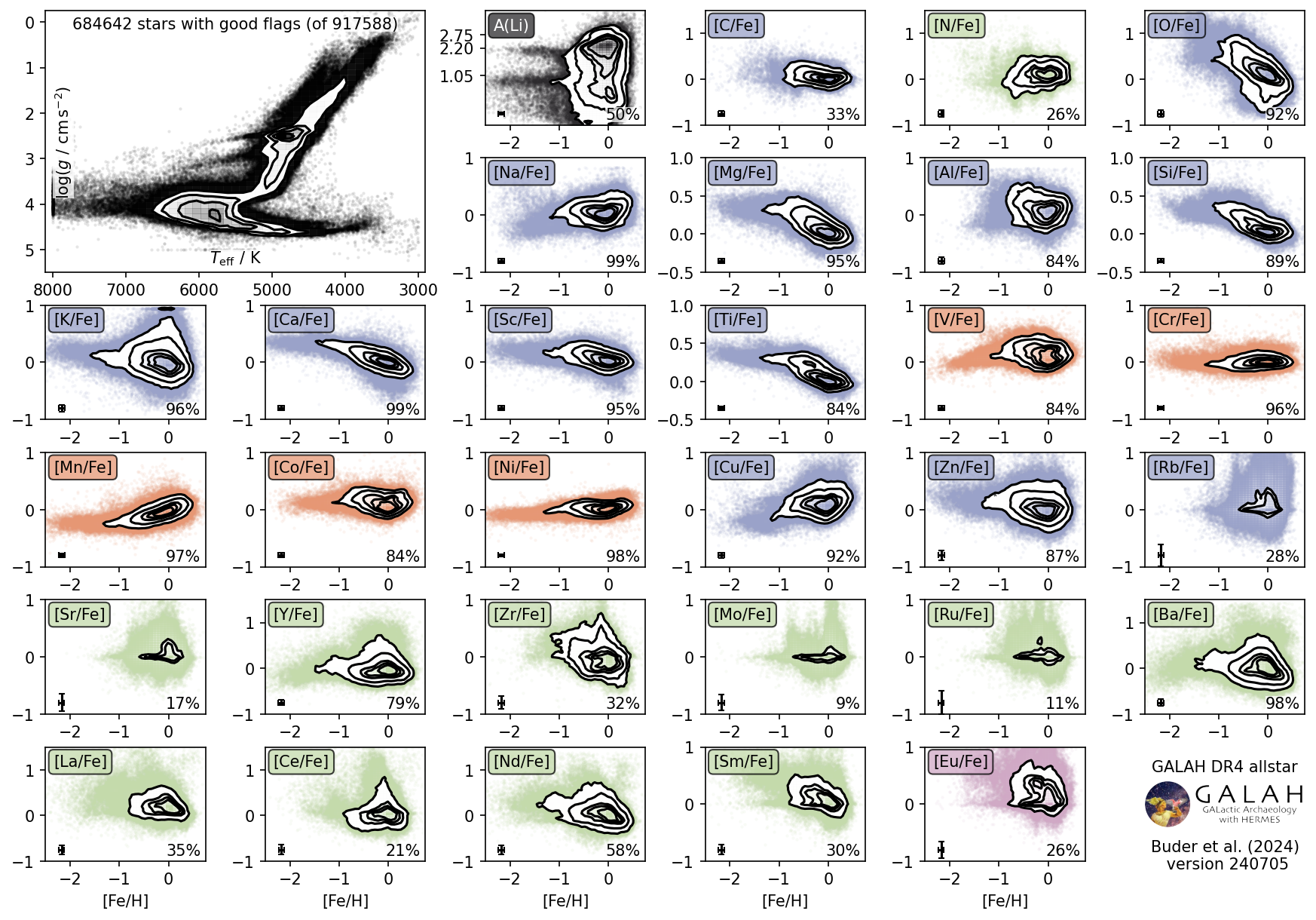}
\caption{x
\textbf{Overview of stellar parameters and elemental abundances for the \texttt{allstar} estimates of GALAH DR4.}
\textbf{The top left panel} shows the density distribution of stars in the Kiel diagram of \Teff and \logg.
\textbf{All other panels} show the logarithmic elemental abundances (for elements indicated in the top left of the panel) as a function of the logarithmic iron abundances \feh. Elements are coloured by different nucleosynthetic channels (black for big bang nucleosynthesis, blue for core-collapse supernovae, red for supernovae Type Ia, green for asymptotic giant branch star contributions and pink for the rapid neutron capture process with contributions from merging neutron stars) following the colour schema from \citet{Kobayashi2020}. Percentages indicate the fraction of detections of stars for each element.
}
\label{fig:galah_dr4_overview_allstar}
\end{figure}
\end{landscape}

We further raise a flag for \texttt{allspec} abundances, if the element was fit above (3) or below (4) the neural network training set range. For CNO, we have identified specific regions, in particular dwarfs, for which could not verify abundances and therefore caution their use (flag 5). We have further tried to identify abundances, for which the optimisation may have failed and flagged these with flag 6 (see Section~\ref{sec:caveats_fitting}).

\section{DATA RELEASE PRODUCTS}
\label{sec:catalogues_release_products}

GALAH DR4 encompasses a diverse range of data products. We describe the most important main catalogues in Section~\ref{sec:data_release_catalogues} and value-added catalogues in Section~\ref{sec:vacs}. We further explain the data products for each spectrum and star, that is, the reduced spectra (Section~\ref{sec:reduced_spectra}), \texttt{allspec} products (Section~\ref{sec:products_allspec}), and \texttt{allstar} products (Section~\ref{sec:products_allstar}).

The data products are provided directly on the AAO DataCentral website at \url{https://cloud.datacentral.org.au/teamdata/GALAH/public/GALAH_DR4/}. We further provide multiple ways to interact with the data release products, which are described in Section~\ref{sec:interactive}.

\subsection{Main Data Release Catalogues}
\label{sec:data_release_catalogues}

\begin{enumerate}
   \item \texttt{galah\_dr4\_allspec\_240705.fits}: analysis for each spectrum (including radial velocity estimation for each spectrum) based on a single spectrum.
   \item \texttt{galah\_dr4\_allstar\_240705.fits}: analysis for each star based on co-added spectra of each star and using non-spectroscopic information to constrain \logg.
\end{enumerate}

We present the main catalogue table schema in Table~\ref{tab:main_catalog_schema} (see also Figure~\ref{fig:galah_dr4_overview_allstar}), but refer the reader to the FITS headers of each catalogue for more detailed information.

\begin{figure*}[ht]
    \centering
    \includegraphics[width=\textwidth]{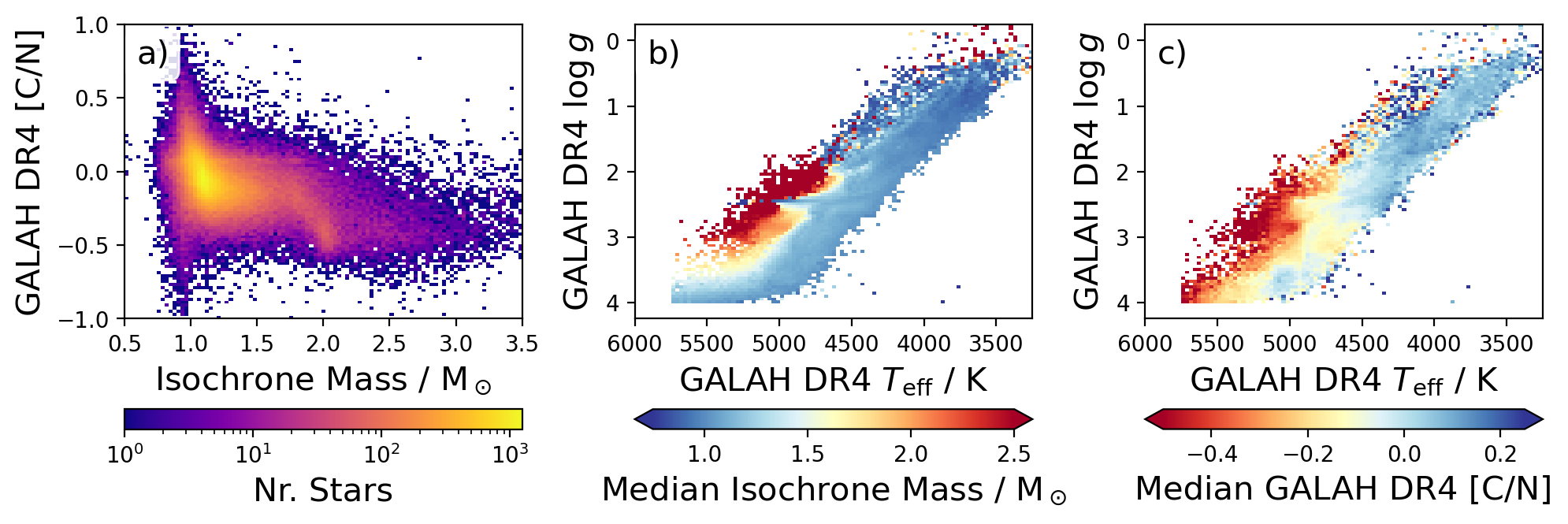}
    \caption{The ratio of [C/N] and isochrone masses in comparison (panel a), and as a function of \Teff and \logg in panels b) and c), respectively}
    \label{fig:cn_mass}
\end{figure*}

One of our greatest achievements as part of this data release is the extraction of C and N abundances for giant stars from molecular absorption features. In Figure~\ref{fig:cn_mass}, we show how stellar mass and [C/N] ratios are correlated in GALAH DR4, as is expected based on the pioneering work by \citet{Masseron2015}, \citet{Martig2016}, and \citet{Ness2016}. Our measurements demonstrate the potential of [C/N] abundances to better separate the core-helium burning from the red giant phase (around the blue areas of Figures~\ref{fig:cn_mass}b and \ref{fig:cn_mass}c) or at least better constrain stellar masses.

\subsection{Value-Added Catalogues (VAC)} \label{sec:vacs}

We provide several value-added catalogues, namely a crossmatch catalogue to all entries of the \Gaia DR3 main source catalogue and the most important entries from the 2MASS and WISE catalogues, a catalogue of stellar dynamics properties, a catalogue of 3D NLTE measurements of Li, and a catalogue with ages inferred via isochrone interpolation in a Bayesian framework.

\subsubsection{VAC of crossmatches with \Gaia DR3, 2MASS and WISE} \label{sec:vac_crossmatch}

The value-added catalogue of the crossmatch\footnote{\texttt{galah\_dr4\_vac\_wise\_tmass\_gaiadr3}} with the \Gaia DR3, 2MASS, and WISE catalogues as well as the distance catalogue of \citet{BailerJones2021} was calculated by performing an \texttt{OUTER JOIN} ADQL-query in the \Gaia archive. 

The query first performed an \texttt{INNER JOIN} with the 2MASS near-infrared photometry catalogue\footnote{\texttt{gaiadr1.tmass\_original\_valid}} via its \texttt{designation} and linked this match to the \Gaia DR3 catalogue via the best neighbour\footnote{ \texttt{gaiadr3.tmass\_psc\_xsc\_best\_neighbour}} and joined\footnote{\texttt{gaiadr3.tmass\_psc\_xsc\_join}} catalogues of 2MASS to \Gaia DR3 \citep{Torra2021}. When cross-matching between \Gaia DR3 and 2MASS, less than 1\% of stars were associated with multiple possible matches. To ensure the best match, the data were sorted from brightest to faintest \(G\)-band magnitude, and only the brightest match for each \texttt{sobject\_id} was retained.

The crossmatch to the WISE far-infrared photometry catalogue\footnote{\texttt{gaiadr1.allwise\_original\_valid}} \citep{Cutri2013} was performed via the \Gaia DR3's best neighbour catalogue\footnote{\texttt{gaiadr3.allwise\_best\_neighbour}} \citep{Torra2021}. The match to the distance catalogue\footnote{\texttt{external.gaiaedr3\_distance}} of \citet{BailerJones2021}  via the \Gaia DR3 \texttt{source\_id}.

The catalogue also includes uncertainties in the \Gaia DR3 photometric magnitudes (\(G\), \(G_{\rm BP}\), \(G_{\rm RP}\)) that were recalculated following the recommendations from the \Gaia Early Data Release 3 (EDR3) documentation \citep{Riello2021}. The total uncertainties were computed by combining the photon flux error with an additional systematic term.

We further corrected the \Gaia DR3 parallaxes for systematic zero-point errors by applying the correction model provided by \citet{Lindegren2021b}. This correction depends on several factors, including the \(G\)-band magnitude, effective wavenumber (\(\nu_{\rm eff}\)) used in astrometry, pseudocolour, latitude, and the astrometric solution type. The parallax zero-points and original parallaxes are reported as \texttt{plx\_zpt\_corr} and \texttt{parallax\_raw}, respectively.

Beyond the crossmatch with the \Gaia DR3 \texttt{gaia\_source} catalogue, multiple other crossmatches can easily be performed via the \texttt{gaiadr3\_source\_id} column. We have for example crossmatched the sources from GALAH DR4 with those from \Gaia DR3's variability catalogues \citep{Rimoldini2023}. We find 47\,493 stars in GALAH DR4 that overlap with the \textit{gaiadr3.vari\_classifier\_result} catalogue. In particular, we find 17\,256 SOLAR\_LIKE variables, 14\,477 stars in the {$\updelta$}Scuti, {$\upgamma$} Doradus, or SXPhoenicis category (DSCT/GDOR/SXPHE), 6\,247 LPV (long-period variables), 4\,074 ECL (eclipsing binaries), 3\,355 RS (RS Canum Venaticorum variables), 1\,096 YSO (young stellar objects), 401 RR (RR Lyrae types), and a large variety of other variables, including the white dwarf 2MASS J05005185-0930549 that was already found in GALAH data by \citet{Kawka2020}.

\begin{figure}[ht]
\includegraphics[width=\columnwidth]{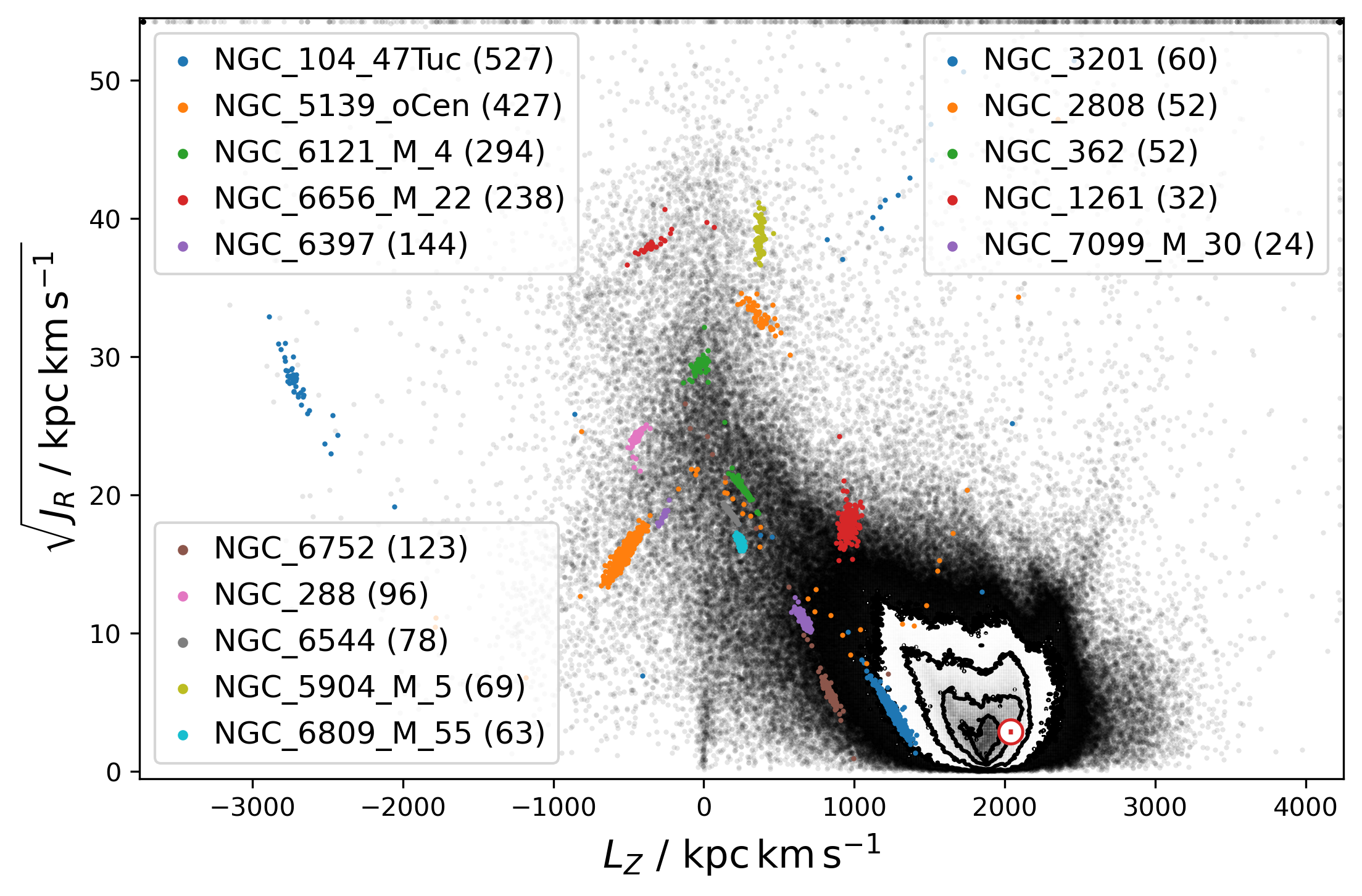}
\caption{
\textbf{Distribution of the dynamical properties of angular momentum $L_Z$ and radial action $J_R$ of stars in GALAH DR4 (black), with globular cluster members highlighted in colour.} Cluster members were selected as those with more than 70 percent membership probability according to \citet{Vasiliev2021}. The Sun is indicated with a red $\odot$ symbol.
}
\label{fig:galah_dr4_lz_jr_with_gcs}
\end{figure}

\begin{figure}[ht]
\includegraphics[width=\columnwidth]{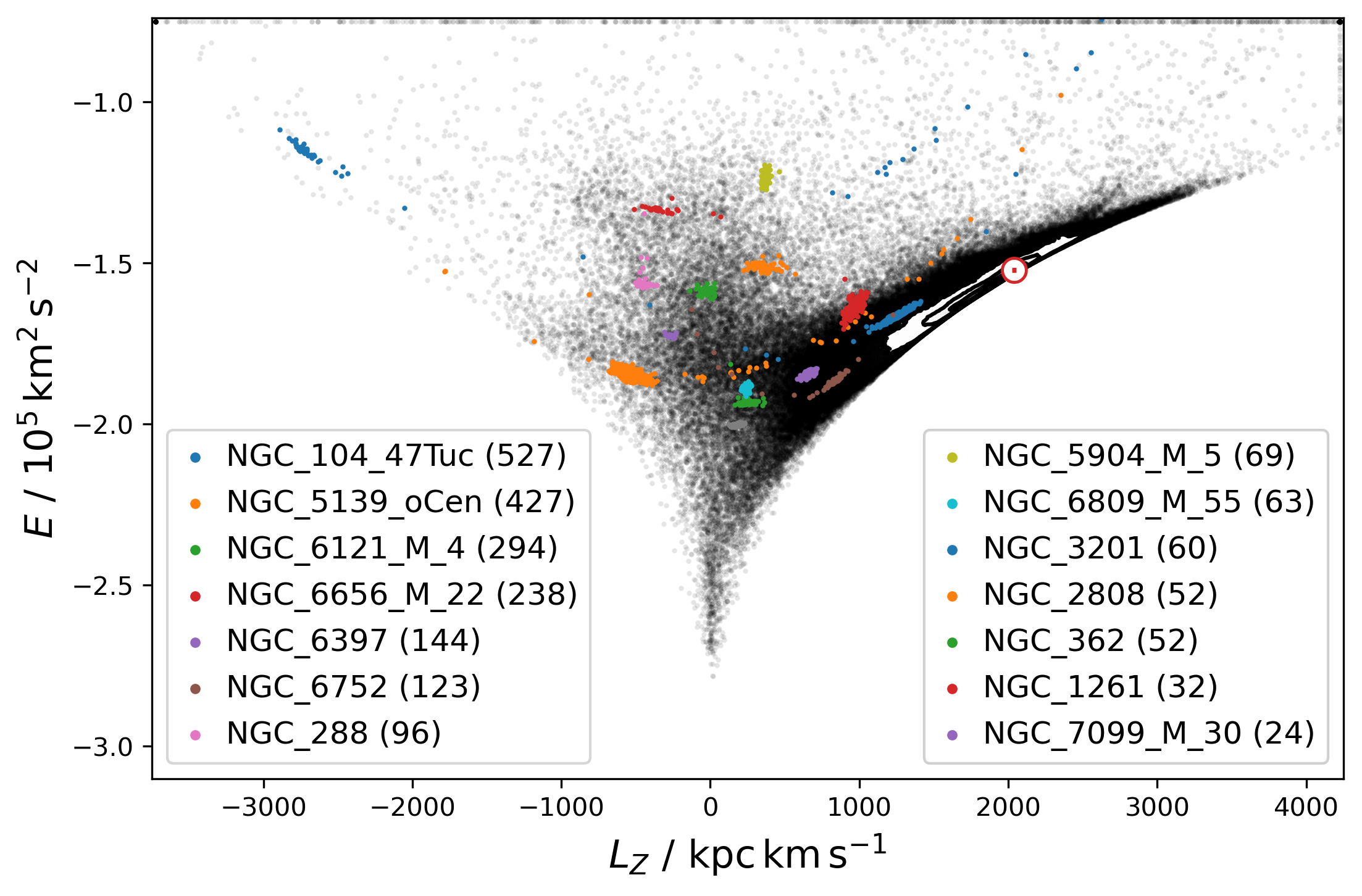}
\caption{
\textbf{Distribution of the dynamical properties of angular momentum $L_Z$ and orbital energy $E$ of stars in GALAH DR4 (black), with globular cluster members highlighted in colour.} Cluster members were selected as those with more than 70 percent membership probability according to \citet{Vasiliev2021}. The Sun is indicated with a red $\odot$ symbol.
}
\label{fig:galah_dr4_lz_e_with_gcs}
\end{figure}

\subsubsection{VAC of stellar dynamics}
\label{sec:vac_dynamics}

The value-added catalogue for stellar dynamics\footnote{\texttt{galah\_dr4\_vac\_dynamics}} includes the kinematic and dynamical properties for stars in the GALAH DR4 survey. The catalogue is created with a publicly available script\footnote{Accessible in the GALAH DR4 repository \href{https://github.com/svenbuder/GALAH_DR4/blob/main/catalogs/create_galah_dr4_vac_dynamics.ipynb}{here}.} as part of GALAH DR4. 
We define the position of the Sun in our Galactic reference frame as $R_\mathrm{GC} = 8.21\,\mathrm{kpc}$ \citep{McMillan2017}, $\varphi_\mathrm{GC} = 0\,\mathrm{rad}$, and $z_\mathrm{GC} = 25\,\mathrm{pc}$ \citep{BlandHawthorn_Gerhard2016}. We then combine the total velocity in $V$ of the Sun at $R_\mathrm{GC}$ based on the proper motion measurement of $6.379\pm0.024\,\mathrm{mas\,yr^{-1}}$ by \citep{Reid2004}, that is, $V_\odot = 248.27\,\mathrm{km\,s^{-1}}$ with the circular velocity of $V_\mathrm{circ} = 233.10\,\mathrm{km\,s^{-1}}$ from \citet{McMillan2017} to estimate a peculiar velocity of the Sun with respect to the local standard of rest of  $15.17\,\mathrm{km\,s^{-1}}$. For the other two components, we use the estimate by \citet{Schoenrich2010}, leading to a peculiar velocity of the Sun of $(U,V,W) = (11.1, 15.17, 7.25)\,\mathrm{km\,s^{-1}}$.

Starting from the crossmatch of GALAH DR4 with the \Gaia DR3 (see Section~\ref{sec:vac_crossmatch}), we use the \textsc{galpy.orbit} module by \citet{Bovy2015} to estimate heliocentric Cartesian coordinates $(X,Y,Z)$ and velocities $(U,V,W)$ as well as Galactocentric cylindrical coordinates $(R, \varphi, Z)$ and velocities ($v_R, v_\varphi, v_Z$). We approximate the orbit actions $J_R, J_\varphi = L_Z, J_Z$ and frequencies $\omega_i$ with the \textsc{galpy.actionAngle.actionAngleStaeckel} function with a focal length of the confocal coordinate system $\texttt{delta} = 0.45$ in the Milky Way potential by \citet{McMillan2017}. We further use the Staeckel approximation \citep{Binney2012} to calculate eccentricity, maximum orbit Galactocentric height, and apocentre/pericentre radii with \textsc{galpy}'s \textsc{EccZmaxRperiRap} \citep{Mackereth2018}. Our assumption of a time-invariant, axisymmetric potential further allows us to extract the orbit energy via \textsc{galpy.Orbit.E}.

In particular the dedicated observing programs of GALAH towards low angular momentum stars (PI S. Buder) and globular clusters (PI M. McKenzie and PI M. Howell) have increased the number of spectroscopic observations for stars on halo-like orbits. This is showcased by both the action-action diagram of angular momentum $L_Z$ versus radial action $\sqrt{J_R}$ (Figure~\ref{fig:galah_dr4_lz_jr_with_gcs}) and angular momentum $L_Z$ versus orbit energy $E$ (Figure~\ref{fig:galah_dr4_lz_e_with_gcs}) and visualises the potential of GALAH DR4 observations to complement Galactic dynamics studies and enable Galactic chemodynamic studies.

\subsubsection{VAC of 3D NLTE lithium abundances}
\label{sec:VAC_li}

In this value-added catalogue\footnote{\texttt{galah\_dr4\_vac\_3dnlte\_a\_li}}, we use spectrum fitting to infer 3D non-local thermodynamic equilibrium (NLTE) lithium abundances. For each spectrum, the Li line is modelled with a 3D NLTE \breidablik line profile \citep{Wang2021}. In cases where the Li line is blended with nearby lines such as Fe and CN, we model blending lines as Gaussian absorption profiles. From this model, we measure the equivalent width (EW) and errors in EW of the Li line using \textsc{UltraNest} \citep{Buchner2021}, a Monte Carlo nested sampling algorithm. The Li abundance, A(Li), is then inferred from the measured EW using \breidablik and the stellar parameters from GALAH DR4's \texttt{allstar}. See \citet{Wang2024} for a detailed description of the methodology. 

\citet{Wang2024} measured a local line width for the Li region and fit the width of the Li line separately from other lines. For this work, we use the GALAH instrumental profile convolved with the rotational velocity as the width of our blending lines and set the width of the Li line based on this convolved kernel, better constraining the line widths. 
Whilst we still measure a local radial velocity due to a lack of ThXe arc lines for CCD3 (see Section~\ref{sec:caveats_reduction}), we apply the GALAH radial velocity for poorly constrained Li depleted stars where we cannot measure the local radial velocity. In addition, the sampled EW posterior is now modelled using a first order boundary corrected kernel density estimator from \citet{Lewis2019}, which has better convergence than histograms. Lastly, GALAH DR4 analyses stars down to 3000\,K, but the \stagger model atmospheres only reach 4000\,K, therefore, we provide an additional column of 1D NLTE\footnote{Note that these 1D NLTE Li abundances are different from the 1D NLTE Li abundances published in \texttt{allstar}.} A(Li) inferred through our measured EWs using a new interpolator. Similar to the existing EW interpolators in \breidablik \citep{Wang2021}, we train a feedforward neural network on NLTE Li abundances synthesised using the 1D \marcs model atmospheres. We use a 2-layer architecture with the ReLU activation function and find best hyperparameters: $i = 900$ neurons, and $\alpha = 0.1$ L2 penalty. This model is included in the \breidablik package.
Using the updated methodology it takes $\sim$2 minutes per star in comparison to $0.5$--$2$ hours per star reported in \citet{Wang2024}, with the main speed up coming from fixing the Li line width. 

\begin{figure}[ht]
\includegraphics[width=\columnwidth]{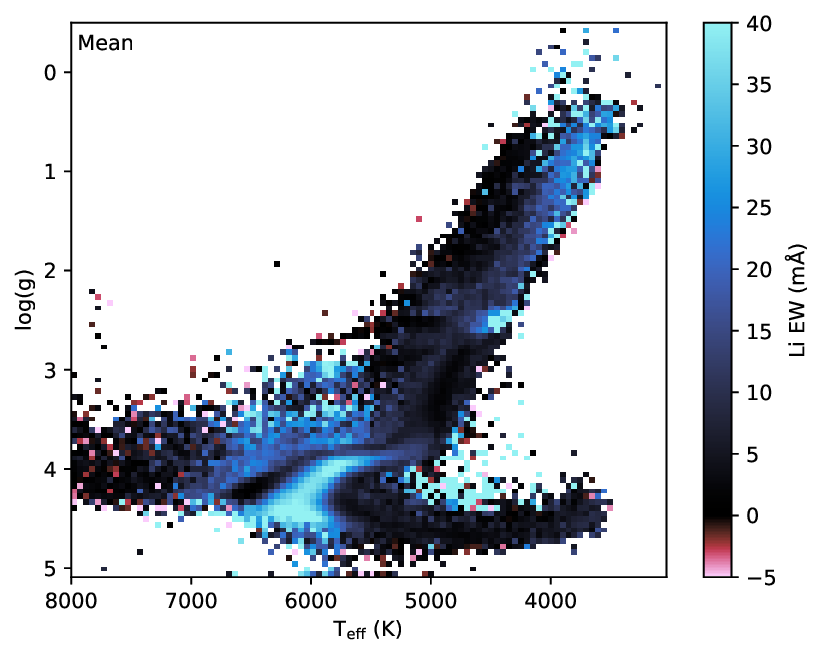}
\caption{
Mean EW binned in \Teff and \logg. The Li-dip can be seen at \Teff$\approx 6500$\,K and \logg$\approx 4.2$. At \logg$\approx 2.5$, red clump stars have a higher mean Li EW whilst horizontal branch stars have a lower mean Li EW compared to surrounding stars. The mean Li EW increases going up the red giant branch. 
}
\label{fig:mean_ew}
\end{figure}

EWs are measured for 892\,223 stars (97\% of GALAH DR4), with 3D NLTE A(Li) detections reported for 417\,825 stars (46\%) and upper limits for 474\,398 stars (52\%). Figure~\ref{fig:mean_ew} shows the mean EW over \Teff and \logg. The Li-dip can be seen on the main-sequence turn-off at \Teff$\approx 6500$\,K and \logg$\approx 4.2$ and extends up the subgiant branch. There is a Li enhanced population of stars at \logg$\approx 2.5$ in the red clump whilst the horizontal branch is depleted in Li. Although the secondary red clump appears to be depleted in Li, these stars have an overestimated \logg driven by incorrectly inferred masses (see Section~\ref{sec:caveats_photospec}), and should be primary red clump stars. The increase of mean Li EW up the giant branch is due to a large proportion of stars with EW$\approx 100$\,m\AA. Features of this figure will be studied in follow-up papers. 

A quality flag (\texttt{flag\_ALi}) is raised by 1 for upper limits, 2 or more to indicate other quality issues, such as stellar parameters falling outside of the model atmosphere grid \citep[see][]{Wang2024} for more details on the bitmask flag). We recommend \texttt{flag\_ALi < 2} when using the 3D NLTE A(Li), and \texttt{flag\_ALi < 4} when using Li EWs. A similar quality flag \texttt{flag\_ALi\_1D} is provided corresponding to the 1D NLTE A(Li) included in the VAC. 

For convenience, we have included the most important columns of this catalogue in the \texttt{allstar} catalogue (see Table~\ref{tab:main_catalog_schema}), as we recommend to use them instead of the less accurate 1D NLTE abundances estimated with an imperfect neural network interpolation, which we indicate with \texttt{nn\_li*}.

\subsubsection{Ages}

A value-added-catalogue for stellar ages and masses from BSTEP \citep{Sharma2018} is currently in preparation. In the meantime, users can rely on the on-the-fly \texttt{age} and \texttt{mass} estimates already provided in the \texttt{allstar} and \texttt{allspec} catalogues from the pipeline.

\subsection{Data products for each spectrum and star}
\label{sec:data_products_for_each_spectrum}

We provide individual data products in an orderly fashion that allow users to create links to these products based solely on the \texttt{sobject\_id}. To download data products for individual stars we recommend creating a url string and using \textsc{wget} or similar commands. For bulk downloads of the advanced data products of this section, we recommend contacting the GALAH collaboration or using the bulk download interfaces of AAO DataCentral.

\subsubsection{Reduced spectra} \label{sec:reduced_spectra}

The reduced spectra of each night are provided in the observations directory\footnote{\texttt{observations/YYMMDD/spectra/com/sobject\_id*.fits}} and sorted into directories with four spectra - one for each of the four CCDs. These spectra are produced by the reduction pipeline (see Section~\ref{sec:spectroscopic_data_from_galah_observations}) and include several extensions as outlined in Table~\ref{tab:reduction_fits}, with wavelength information stored in the fits headers with starting wavelength \texttt{CRVAL1} in \AA\xspace and linear pixel scale \texttt{CDELT1} in \AA/px, and the number of pixels \texttt{NAXIS1}. The reduced spectra are only provided per exposure and not in a co-added manner, since the co-adding was performed as part of the \texttt{allstar} module (see Section~\ref{sec:products_allstar} for co-added spectra). We note that not all files might be available for a given exposure due to the rare failure of CCD readouts.

\subsubsection{Additional products of the \texttt{allspec} module} \label{sec:products_allspec}

The \texttt{allspec} analysis product directory\footnote{\texttt{analysis\_products\_single/YYMMDD/sobject\_id/}} provides the files that were produced by the \texttt{allspec} module. These include the on-the-fly assessment of the radial velocity fit \texttt{*rv.png} (similar to Figure~\ref{fig:181221003101356_single_fit_rv}), the raw fitting results \texttt{*results.fits} and their covariance matrices \texttt{*covariances.npz} (similar to the entries used to produce Figure~A\ref{fig:covariance_vesta_arcturus}). We also provide a combined \texttt{*spectrum.fits} file (concatenated over the four bands) that includes the wavelength, flux, and flux uncertainty of the velocity-corrected and re-normalised observed spectrum as well as the best-fitting model spectrum interpolated onto the same wavelength. Finally, we provide a \texttt{*comparison.pdf} (similar to Figure~A\ref{fig:210115002201239_allstar_fit_comparison}) which displays the fit results, comparison of observed and model spectrum, masked wavelength regions, and wavelengths of the most important element lines. If the module did not run to completion, for example because the SNR of the spectra was below the threshold of $\mathrm{SNR} = 10$ for any CCD to even attempt a fit, not all products are available for a spectrum.

\subsubsection{Additional products of the \texttt{allstar} module} \label{sec:products_allstar}

The \texttt{allstar} analysis product directory\footnote{\texttt{analysis\_products\_allstar/YYMMDD/sobject\_id/}} also includes the radial velocity monitoring \texttt{*rv.png}, results files  \texttt{*results.fits}, combined spectra \texttt{*spectrum.fits} and \texttt{*comparison.pdf} overview, similar to the ones described in Section~\ref{sec:products_allspec}. In addition, each directory also includes a \texttt{*sobject\_ids.txt} file that lists all individual spectra that were co-added to create the observed spectrum and its uncertainty in \texttt{*spectrum.fits}.


\subsection{Interactive access via AAO DataCentral}
\label{sec:interactive}

In collaboration with the AAO Data Central, a number of interactive ways are provided to explore the data of this release. Data Central provides both Simple Spectral Access and Single Object Viewer services. In addition, we recommend to download files or easily crossmatch user catalogues with the TAP server \url{https://datacentral.org.au/vo/tap} in \textsc{TOPCAT} \citep{Taylor2005}. \url{https://apps.datacentral.org.au/galah/spectra} also provides an interactive plotting application to show normalised or un-normalised spectra of different repeat observations. As these tools are under active development, we refer to the latest documentation on both the DataCentral and the main Survey website \url{https://www.galah-survey.org}.

\section{CAVEATS AND FUTURE IMPROVEMENTS} \label{sec:caveats}

In this section, we attempt a detailed discussion of caveats at different steps of our analysis, while also giving suggestions for future improvements - both for GALAH and other surveys. We first discuss caveats of the spectrum reduction (Section~\ref{sec:caveats_reduction}), before extensively discussing the spectrum synthesis (Section~\ref{sec:caveats_synthesis}) and spectrum interpolation (Section~\ref{sec:caveats_interpolation}). We discuss possible problems arising from the use of photometric information (Section~\ref{sec:caveats_photospec}), in particular for stars that could be binaries (Section~\ref{sec:caveats_binaries}). We elaborate on caveats regarding globular clusters in Section~\ref{sec:caveats_globulars} and the fitting iteslf in Section~\ref{sec:caveats_fitting}. Finally, we point out caveats regarding the flags in Section~\ref{sec:caveats_flags} as well as a bug and its correction in the reported radial velocity of interstellar K in Section~\ref{sec:caveats_interstellar}. We summarise the most important caveats in Section~\ref{sec:caveats_summary}.

\subsection{Spectrum reduction}  \label{sec:caveats_reduction}

Although a significant amount of work was spent on improving the spectrum reduction, several persistent issues remain, which are summarised below.

\subsubsection{Wavelength solutions}

For each CCD, the reduction pipeline estimates the most suitable wavelength solution, linking pixels with actual wavelengths based on the ThXe arc lines. In GALAH DR3 \citep{Buder2021} we identified several issues for spectra where not enough ThXe lines could be used to constrain the wavelength solution. Improvements have been made for the new reduction version to improve the number of useful ThXe lines and restrict the flexibility of wavelength solutions to move them closer to previous results. This has helped us to decrease the number of problematic wavelength solutions towards the red end of CCD3 which includes the used absorption features of Li and Eu. We have decreased bad wavelength solutions for this CCD from initially 7.9\% of the spectra to roughly 1\% bad solutions, that is, similar to the other CCDs.

\subsubsection{Holistic spectrum extraction}

Although much work has been spent on improving telluric and sky lines in the reduction step, most reduction steps are currently run sequentially rather than in parallel. Using the information of stellar spectra when modelling the wavelength solution would certainly help to overcome the limited information in ThXe calibration spectra in the absence of laser combs \citep{Kos2018b}. Multiple steps in this direction have been taken \citep{Saydjari2023} and should be rolled out in future spectrum analysis. This would especially help to mitigate imperfect telluric and sky line removal while simultaneously improving the wavelength solution -- among many other effects.

\subsection{Imperfect spectrum synthesis} \label{sec:caveats_synthesis}

\subsubsection{Spectrum synthesis}

The GALAH survey's success relies heavily on the ability to accurately model stellar spectra to infer accurate stellar properties. The survey has seen significant improvements in moving from the approximation of 1D LTE towards 1D NLTE \citep{Amarsi2020}. This includes the use of 1D NLTE synthesis for atomic lines using the 3D NLTE code \textsc{balder} \citep{Amarsi2018}, a custom version of Multi3D \citep{Botnen1999, Leenaarts2009}. The code employs model atoms for H  \citep{Amarsi2018}, Li \citep{Lind2009, Wang2021}, C \citep{Amarsi2019}, N \citep{Amarsi2020b}, O \citep{Amarsi2018b}, Na \citep{Lind2011}, Mg \citep{Osorio2015}, Al \citep{Nordlander2017}, Si \citep{Amarsi2017}, K \citep{Reggiani2019}, Ca \citep{Osorio2019}, Mn \citep{Bergemann2019b}, Fe \citep{Amarsi2018, Amarsi2022}, and Ba \citep{Gallagher2020} over the {\sc marcs} model atmosphere grid. The work by \citet{Wang2024} also enables us to present measurements of Li in 3D NLTE as part of this release.

All of these advances contrast with the lack of a proper way of modelling molecular features appropriately. This could explain the significant mismatch of oxygen abundances between the optical and infrared \citep[compare e.g.][]{Bensby2014, SDSSDR17}. It can, however, also lead to mismatches in the GALAH wavelength range, where atomic features, such as \ion{C}{I}, can be modelled in 1D NLTE, whereas much stronger molecular features of $\mathrm{C}_2$ and CN have to be modelled in 1D LTE and linelists of molecules, such as TiO, might be incomplete \citep{Hoeijmakers2015, McKemmish2019}.

For our synthesis, we have employed version 580 of the IDL-based code Spectroscopy Made Easy \citep{Valenti1996, Piskunov2017}. As part of the continuing improvement of this code, several bugs have been identified and fixed. We also note that a Python-based version of \textsc{SME}, \textsc{pySME} \citep{Wehrhahn2023}, has become available. In addition, the spectrum synthesis code \textsc{KORG} \citep{Wheeler2023, Wheeler2024} has been published in Julia with a Python interface. It offers a faster alternative to \textsc{SME} once 1D NLTE synthesis is implemented, which is essential for applying to many NLTE-sensitive lines, such as O and K, in the GALAH wavelength range. \textsc{KORG} already internally adjusts the metallicity that is used to interpolate atmospheres based on the overall chemical abundances, whereas this would need to be adjusted in \textsc{SME} by hand, since atmospheres are interpolated with the \textsc{sme.feh} entry that is independent of the chemical composition \textsc{sme.abund}. Because we have not performed said adjustment, we note that the spectrum synthesis for chemical compositions far from scaled-Solar may have used an mismatched atmosphere in the synthesis in \textsc{SME} for GALAH DR4.

\subsubsection{Mismatch of atmosphere and spectrum chemistry}

For several of our synthetic spectra, the chosen chemical composition deviates significantly from the scaled-Solar pattern of the \marcs model atmospheres, particularly for $\upalpha$-process elements such as O and Mg, as well as C and N. These elements can substantially affect opacity and energy transport, and therefore, their abundances must be adjusted to match observed spectra more accurately. For instance, $\upalpha$-enhancements in stars with non-Solar abundance patterns can shift line strengths and depths significantly \citep{Asplund2005, VandenBerg2012}. Likewise, variations in C and N abundances, particularly in cooler stars, can impact molecular equilibrium, altering CO and CN molecular line strengths significantly \citep{Tsuji1976, Smith2013}. Dedicated \marcs atmospheres with modified $\upalpha$ and C abundances \citep{Meszaros2012, Joensson2020}, such as those used in modelling APOGEE spectra \citep{SDSSDR17}, or more flexible interpolation schemes by \citet{WestendorpPlaza2023}, address this mismatch. However, the NLTE grids would also need to be expanded to cover all grid points of the extended \marcs models to ensure consistency.

\subsection{Spectrum interpolation with neural networks} \label{sec:caveats_interpolation}

\subsubsection{Training set selection}

Before the neural networks are computed, it should actually be tested what the abundance zero-points are. In the case of several elements like Na and Al they are significant, on the order of $0.2\,\mathrm{dex}$. When this occurs, stars with actual high abundances of $0.7-0.8\,\mathrm{dex}$, for example in old stars and especially in globular clusters \citep[see e.g.][]{Carretta2009}, are not sufficiently covered.

One of the primary challenges in creating an optimal training set for spectrum interpolation lies in the choice of parameter sampling. A common caveat is the use of randomised, uncorrelated parameter sampling, which can lead to unrealistic combinations of elemental abundances. Elements that share a similar nucleosynthesis channel often exhibit correlated behaviour, for instance, stars with high abundances of Mg are typically also enhanced in Si, Ca, and Ti, while Na and Al tend to be elevated together. Similarly, neutron-capture elements like Y and Ba often follow similar trends \citep[e.g.][]{Ting2012, Kobayashi2020, Buder2021}. To better capture this behaviour in the training set, the use of scaled linear functions or normalising flows could be advantageous. These approaches would help minimise the occurrence of unlikely parameter combinations and yield a more representative sample. 

For stars of the thin disk population, one could for example consider sampling from a noisy age-[X/Fe] relation to model chemical evolution \citep[see Figure~\ref{fig:galah_dr4_age_xfe_trends_solar_twins_allstar},][]{Nissen2015, Spina2016, Bedell2018}. This approach becomes more complicated when considering the thick disk, halo, and peculiar stars, where distinct nucleosynthesis histories introduce greater variability in elemental abundance trends.

\subsubsection{Masking of spectra} \label{sec:masking_spectra}

Because the correlation between spectral features, stellar parameters, and abundances is often complex, degeneracies can arise when two stellar properties influence similar pixels of a spectrum (e.g. C and N for CN, or \Teff and \feh for cool dwarfs) or two stellar properties tend to act in lockstep in actual stars (e.g. Mg, Si, and Ti as $\upalpha$-process elements). In GALAH DR2 \citep{Buder2018}, we attempted to overcome these issues by specifically masking the coefficients of spectrum interpolation, that is, effectively restricting the interpolation to only change smaller parts of the spectrum for a given stellar property.

In GALAH DR4, we have relaxed this restriction again, since we have trained on random abundance combinations in the hope of being able to break correlation degeneracies. We note, however, that too little information in spectra can again cause by-chance correlations (e.g. if neutron-capture lines are always very weak and the training set is not sufficiently large). We believe that this is the cause of the decrease in precision for Eu measurements from GALAH DR3 to GALAH DR4. The Eu abundance was mainly measured only from the weak Eu $6645\,\text{\AA}$ line in DR3, whereas the neural networks of DR4 are not restricted to this region.

\subsubsection{Flexibility of neural networks in general}

The decision to use a large set of neural networks, each covering a restricted region in the \TLF space, was motivated by the goal of reducing the complexity required of any single model. By dividing the parameter space into smaller subsets, each neural network can be specialised and therefore less flexible, which allows for more precise modelling within its specific region. This approach avoids the trade-off faced by a single, monolithic neural network, which would either lack sufficient flexibility across the entire parameter space or be computationally more expensive to train and evaluate. For this data release, we have fixed the chosen network architecture of a 2-layer perceptron with 300 neurons and specific learning rate. While we have tested other activation functions than leaky rectified linear units, namely sigmoid, tanh and exponential linear unit functions, we found the lowest root mean square errors for our chosen activation function. Given the found issues with model fluxes above 1, we also would recommend to test a sigmoid as last activation layer of the neural network to ensure that the neural network always predicts fluxes between 0 and 1, as is expected from modelled stellar spectra. We have further tested a larger number of neurons, but found the root mean square errors to stabilise around 300 neurons for our test cases. It has to be acknowledged that due to the limit of human power to properly train and test the neural networks, we have not been able to properly test all neural networks and explore more flexible architectures. For this data release, we have decided not to rerun these steps, but make the current results available to the community. In the future, the restriction to one or only a few network models is recommended. The latter could cover regions of cool dwarfs, main-sequence turn-off stars, hot stars, and giant stars with individual models -- and possibly explore the split in metal-poor and solar-like regimes. This would also decrease overhead, in particular for training and loading different models as well as possible noding effects between different models.

\subsubsection{Flexibility of neural networks for extreme abundances}

While this approach has proved to be powerful for all elements across their abundance ranges, we have noticed sinusoidal shapes for weak Li lines \citep[see also][]{Wang2021}. This is likely caused by the large dynamical range of $0 < \mathrm{A(Li)} < 4$ that has to be covered by the neural network. For Li, the more sophisticated approach is to fit Gaussian lines to multiple components in the wavelength range around $6708\,\text{\AA}$, measure EW(Li), which are then used to infer 3D-NLTE based A(Li) abundances. This inference is preferable to our 1D-NLTE based neural network estimates, as it is independent of the network flexibility and superior to our less accurate spectrum synthesis in 1D.

While several studies have identified that the abundances of stars in the Galactic disk are often very similar \citep[e.g.][]{Ness2019b}, the Galactic halo offers a more diverse picture. An example is 2MASS J22353100-6658174 (140707003601047), a turn-off star with extremely high s-process abundances and actually visible lines of La and Nd in addition to the usually visible Y and Ba. In this case, the fits to the La and Nd lines are significantly weaker than the observations. GALAH DR3 actually produced reasonable fits to this star with high abundances in [Y/Fe]=1.2, [Ba/Fe]=1.5, [La/Fe]=1.5, [Ce/Fe]=1.1, [Nd/Fe]=1.9, and [Sm/Fe]=1.2. A neural network that is not trained on such high abundances is likely to improperly extrapolate stellar spectra.

While we have tried to extract abundances of chemically peculiar stars, such as carbon-enhanced metal-poor stars, the significant effect of their molecular features onto the whole stellar spectrum is not to be underestimated and can in-itself pose a problem to the flexibility of neural networks.

\begin{figure*}
    \centering
    \includegraphics[width=\textwidth]{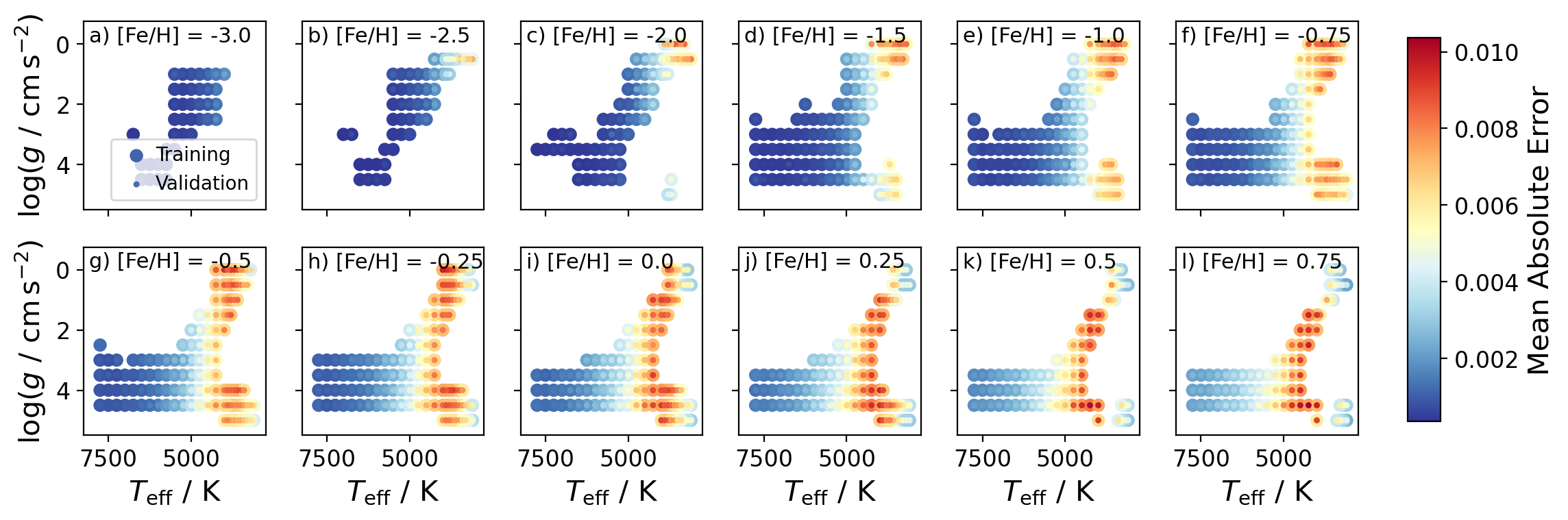}
    \caption{Neural network performance shown as a function of $T_\mathrm{eff}$ vs. $\log g$ with each panel showing a different range of [Fe/H]. Colours indicate the mean absolute errors of the training (large circles) and validation (small circles) for the neural networks.}
    \label{fig:loss_teff_logg_feh}
\end{figure*}

\subsubsection{Over- and underdensities at neural network edges}

While the use of one neural network to interpolate the high-dimensional spectrum space is preferable, in practice, different science cases may drive the decision to use several networks.
If the science case is to reach maximum precision, one neural network that is trained on the typical spectrum could be used at the expense of properly modelling peculiar spectra.
If the science case is to reach maximum accuracy, only the regions with reliable line data and spectrum synthesis might be preferable.
If the science case is to find peculiar stars, a larger coverage is needed to avoid the inaccurate extrapolation of stars with extreme abundances.
In practice, large collaborations likely unite all of these goals, and a compromise has to be struck among the different approaches.
For future analyses, a possible solution could therefore be to follow a two-step approach of first running one generic neural network for all spectra and then using optimised neural networks -- or full spectrum synthesis -- on smaller target samples of specific science cases.

\subsubsection{Quantitative performance of neural networks}

\begin{figure}
    \centering
    \includegraphics[width=0.93\columnwidth]{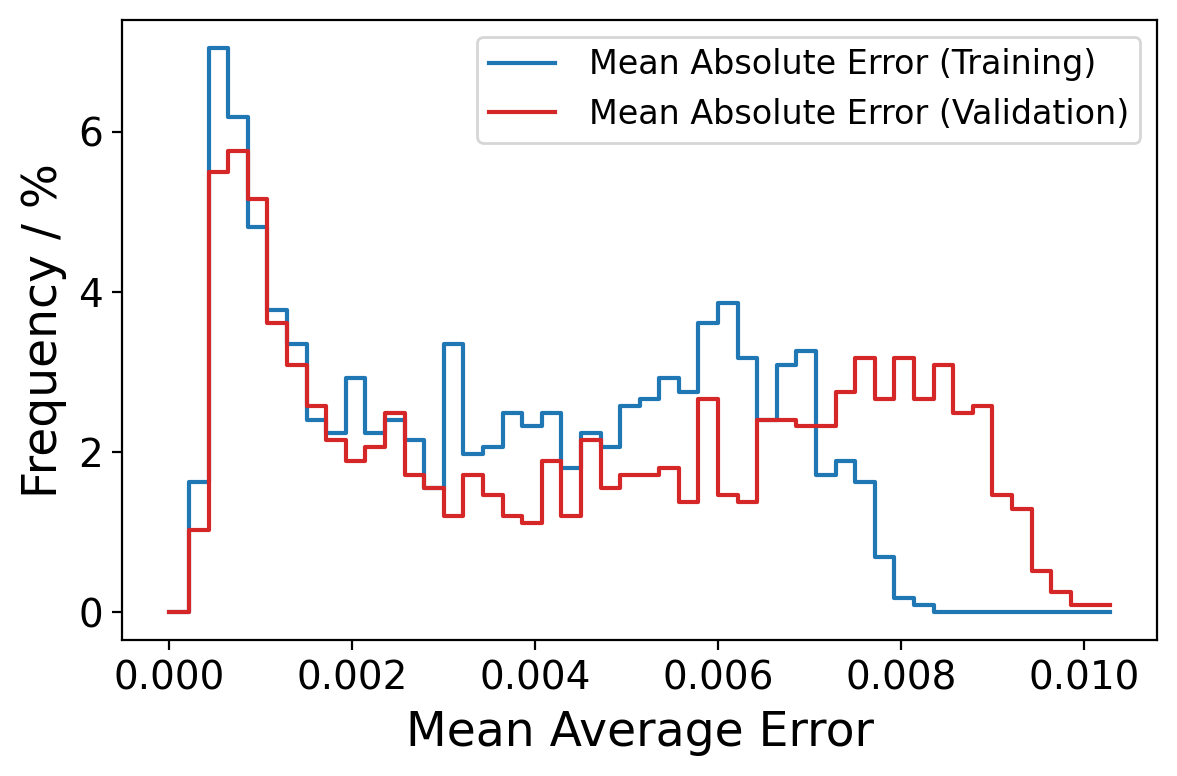}
    \caption{Histogram of the mean absolute errors for the neural networks. These were used as loss function during the training (blue) and validation (red) on seen and unseen spectra, respectively.}
    \label{fig:loss_histogram}
\end{figure}

Throughout the training of our neural networks, we optimised model parameters using a mean absolute error (MAE) loss function across the spectrum pixels. The MAE remains consistently below 0.01 for all neural networks, indicating high accuracy in model predictions, particularly for turn-off and most metal-poor stars where errors are typically below 0.001 (see Figures~\ref{fig:loss_teff_logg_feh} and \ref{fig:loss_histogram}).

Despite these low average error rates, the performance of neural networks can vary significantly across different spectral regions. Errors are minimal in continuum areas but tend to increase around strong or strongly changing absorption features, such as those of lithium, which are discussed in Section 8.3.4. The neural network architecture does not track uncertainty for each weight and bias, limiting our ability to generate perturbed models for assessing the impact of interpolation uncertainties on derived parameters and abundances. Additionally, retraining networks with varied initial conditions to evaluate prediction stability is computationally intensive.

To gauge the practical impact of these uncertainties, we compared the MAE against the noise levels in the GALAH spectra. Errors significantly lower than the noise levels for stars above $T_\mathrm{eff} > 5000\,\mathrm{K}$ suggest that the interpolation inaccuracies minimally impact our analysis. However, for cooler stars with MAE around 0.01, interpolation inaccuracies could potentially influence precise chemical abundance studies more substantially.

Despite the high degree of accuracy achieved, the limitations outlined necessitate careful interpretation of derived parameters, especially in regions with significant absorption features (see Fig~\ref{fig:loss_teff_logg_feh} and \ref{fig:loss_histogram}).

\subsection{Mismatch of spectroscopic and photometric information} \label{sec:caveats_photospec}

\subsubsection{Incorrect masses driving incorrect stellar parameters}

We estimate masses and ages through isochrone matching, where stellar parameters (validated against photometric estimates) are known for not being fully consistent with spectroscopic values.
We believe this leads to significant mismatches especially for stars close to the red clump. In this region, a small change in spectroscopic and photometric information can imply a significant change in inferred mass (e.g. from primary to secondary red clump, with the latter being 2 or more solar masses and thus significantly more than the usual $\sim$ 1 solar mass). This issue has only become noticeable after the production runs and we have therefore decided not to rerun this particular region of the parameter space for this data release. We have extensively tested the possible reasons and identified the mismatch of isochrones and actual stellar spectroscopic parameters as the cause. We have not been able to fully resolve this issue by either including a prior based on the initial mass function to weigh against massive stars \citep[see e.g.][]{Sharma2018} and move from likelihood-weighted to posterior mass estimates. Similarly, we have not been able to resolve these effects by artificially upscaling the spectroscopic uncertainties when calculating the likelihood-weighted masses. More work needs to be done to mitigate the current inconsistencies of theoretical isochrones and spectroscopic estimates.

Another solution for this particular region of the parameter space could be the use of chemical stellar evolution through the correlation of core and thus total mass with the ratio of [C/N] after the first dredge-up \citep{Masseron2015, Martig2016}, given that GALAH spectra also contain information on both elements. This could thus be used to better constrain high masses and counteract the information from isochrone-inferred masses. For this data release, the [C/N] information could at least serve as an indicator of how trustworthy high masses for giant stars are.

\subsubsection{To use or not to use non-spectroscopic information?}

The implementation of non-spectroscopic information, as done in our \texttt{allstar} module, has the advantage of overcoming spectroscopic degeneracies (as proven for the limited information on \logg in the HERMES wavelength range) as well as improving accuracy and precision also for the lowest quality spectra (because \logg is no longer solely dependent on the spectrum information).

However, this approach is only useful if the non-spectroscopic information is not biased (as it would be for astrometric and photometric information in the case of unresolved binarity). While the astrometric information for almost all GALAH targets is exquisite, this may not be the case for other surveys. The significant improvement from GALAH DR3 to GALAH DR4 has most definitely benefited from the improved astrometric information of \textit{Gaia} EDR3 \citep{GaiaEDR3, Lindegren2021a} and \textit{Gaia} DR3 \citep{GaiaDR3} with respect to \textit{Gaia} DR2 \citep{Brown2018, Lindegren2018}. Further improvement could be expected when also taking \textit{Gaia}'s photometric information into account, in addition to our use of 2MASS photometry.

\subsection{Binaries} \label{sec:caveats_binaries}

\begin{figure}[ht]
 \centering
 \includegraphics[width=\textwidth]{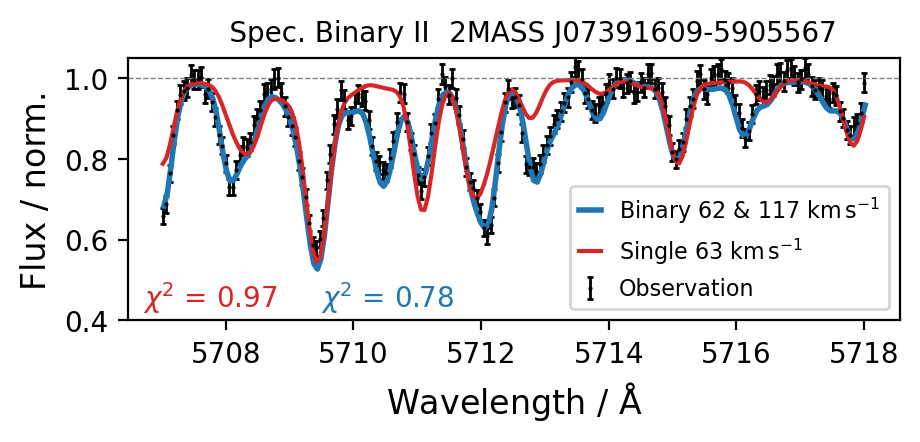}
 \caption{\textbf{Example spectrum for a double-lined spectroscopic binary star (SB2) that is better fitted with our binary fitting algorithm.}} \label{fig:examples_flag_sp_3}
\end{figure}

Although not part of this release, we have created an analysis module for spectroscopic binaries. The module will be presented in a separate work (Lach et al., in preparation) with a catalogue becoming a value-added catalogue of this release. The module is motivated by the extensive study of GALAH binary star spectra by \citet{Traven2020} and our ability to model the full spectrum via neural networks. We show a first analysis result of the module in Figure~\ref{fig:examples_flag_sp_3}, where the module was applied to a spectroscopic binary type 2 and resulted in a significantly better fit than the single star analysis.

\subsection{Globular clusters} \label{sec:caveats_globulars}

Globular clusters are well known for their light element anti-correlations (i.e., the Na-O or Mg-Al anti-correlations), though the underlying cause remains a subject of debate \citep[for recent reviews see][]{Bastian2018, Gratton2019, Milone2022}. It is widely accepted that one population is enhanced in elements including He, N, Na and Al, and depleted in O and C. Previous GALAH data releases have encountered issues in removing trends between abundances and stellar parameters (as discussed in Sec \ref{sec:uncertainty_accuracy}), and DR4 represents a marked decrease of scatter within the Kiel diagrams of the globular clusters (see \ref{sec:stellar_param_abund_valid}). Despite these improvements, light element abundance anti-correlations are still not well reproduced for DR4. We attribute this to two key factors: abundance zero points (Sec. \ref{sec:uncertainty_accuracy}) and the masking of spectra (Sec. \ref{sec:masking_spectra}).

Table~\ref{tab:zeropoints} illustrates that both Na and Al have some of the largest zero-point shifts (-0.171 and -0.185), meaning that for some clusters the full extent of the anti-correlations is not realised (particularly for the light element enhanced populations). Secondly, when inspecting the optimal synthesis for particular lines (e.g., the Na lines at $5682.6$ and $5688.2\,\text{\AA}$, or the O triplet around $7770\,\text{\AA}$), the fits are poorly constrained, leading to a more significant scatter in these critical elements than what has previously been reported in the literature. We expect this is related to the relaxed restrictions on the neural network. Based on the abundances in their current form, we do not recommend using these light element abundances to distinguish between the multiple populations in globular clusters. 

However, the 3D NLTE Li abundances discussed in Sec. \ref{sec:VAC_li} have effectively mitigated the above issues by adopting the GALAH stellar parameters and focusing exclusively on fitting the Li line. When analysing this Li data for the globular clusters, we can effectively reproduce the Li depletion patterns reported by \citet{Lind2009b} The large sample of clusters allows for a homogeneous study of Li depletion around the RGB bump, which will be detailed in McKenzie et al. (in preparation).

Dedicated observing programs have increased the average SNR for some clusters and expanded the sample to include additional clusters, such as M\,22 (PI: M. McKenzie) and M\,4 (PI: M. Howell). As discussed in Sec. \ref{sec:uncertainty_accuracy}, M\,4 will be used to spectroscopically confirm whether the stars with lower asteroseismic masses belong to the light-element-enhanced population. This confirmation will be achieved through the re-analysis of Na, O, Mg, and Al lines, following the approach used for Li, since we advise against relying on these current light element abundances for globular cluster stars.

The cluster M\,22, renowned for its bimodal \textit{s}-process population \citep{Marino2011, McKenzie2022, McKenzie2024}, was observed as a crucial test case to evaluate the GALAH pipeline's ability to detect \textit{s}-process abundance variations. While the pipeline successfully recovers the bimodal distribution, the scatter is larger than reported in previous studies. Additionally, as noted in Section~\ref{sec:uncertainty_precision}, the precision of Eu measurements appears to have decreased between DR3 and DR4, particularly within globular cluster populations. Therefore, we recommend against using Eu from DR4 in future globular cluster publications.

Due to their low metallicity, globular clusters are particularly susceptible to the bug in the \texttt{flag\_fe\_h} discussed in Section~\ref{sec:bug_in_fe_h}. If this condition is relaxed, we recommend that all spectra and corresponding fits be manually inspected for quality before being included in any publications. Again, we reiterate that a boutique, custom reanalysis aiming to address these caveats in the globular cluster data will be the focus of upcoming work from McKenzie et al. (in preparation).

\subsection{Fit optimisation} \label{sec:caveats_fitting}

As described in Section~\ref{sec:allspec_analysis}, we are using the \textsc{curve\_fit} function of \texttt{scipy.optimize} \citep{scipy} to fit synthetic spectra to observed spectra, whose optimisation can get stuck in local minima. We have tried to automatically identify regions of the parameter space where the \textsc{scipy.optimize.curve\_fit} function has become stuck. In particular for some red clump stars as well as cool giant stars with $T_\mathrm{eff} < 3750\,\mathrm{K}$ and $\log g < 0.5$ (see Section~\ref{sec:uncertainty_accuracy}), we have been able to recover a pattern of abundances that are stuck around their initial value. However, this pattern is not consistent enough to flag stars without a significant amount of false-positives. Because of the zero point corrections, these are shifted away from the usual initial guess of $0\,\mathrm{dex}$ depending on the element (see zero-points in Table~\ref{tab:zeropoints}).

Such a fitting failure would also be expected when applying \textit{The Payne} \citep{Ting2019} with its similar default setup that adopts parameter bounds for the fitting parameters and thus employs the \texttt{curve\_fit} function with the \textit{trust region reflective} (\texttt{trf}) method. Given the common use of \texttt{curve\_fit}, future pipelines should test a range of approaches to avoid this issue. Firstly, instead of using \texttt{trf}, the \textit{Dogbox} (\texttt{dogbox}) method, could be used. The method is potentially slower but more reliable for complex parameter spaces. It could be used to randomly check the convergence of the \texttt{trf} method or be applied only to regions where multiple local minima are expected.

Moving away from the \texttt{curve\_fit} function, the \textsc{leastsq}\footnote{\textsc{leastsq} is used for example by \textit{The Cannon} version by \citet{Casey2016}.}, \textsc{minimize} or the \textsc{differential\_evolution} function of \texttt{scipy}'s \texttt{optimize} module could be used to test options of a more expensive but more extensive optimisation. Finally, multiple randomised initial starting guesses could be applied for \texttt{curve\_fit}, but would multiply the computing costs linearly by the number of initial guesses.

The fitting optimisation and uncertainty estimation should be performed in a more sophisticated Bayesian framework that folds in photometric, astrometric, and asteroseismic information and their uncertainties. We have indeed implemented such a framework with a likelihood estimate from spectroscopic information and prior information based on photometric, astrometric, and asteroseismic estimates for test purposes. When implementing the resulting posterior into the Markov-Chain Monte-Carlo machinery of \textsc{emcee} \citep{ForemanMackey2013}, we have not been able to limit the computational time (when fitting all labels) to a competitive level with \texttt{curve\_fit} and thus not implemented this approach for the analysis of a million spectra. We note, however, that a future analysis should implement this approach - either with \textsc{emcee} or Monte Carlo nested sampling algorithms like \textsc{UltraNest} \citep{Buchner2021}. Furthermore, we suggest to either separate the likelihood and posterior estimation steps \citep[see e.g.][]{Gent2022} or limit the optimisation to only a few major stellar labels \citep[see e.g.][]{Traven2020}.

\subsection{Reliability of flags} \label{sec:caveats_flags}

We have tried to develop a quality assurance pipeline that automatically flags results and stars that may not be adequately analysed with our assumptions.

\subsubsection{Bug in \texttt{flag\_fe\_h}} \label{sec:bug_in_fe_h}

The quality flag for iron abundance, \texttt{flag\_fe\_h}, was computed similarly to the elemental abundances, that is, by comparing the best-fitting spectrum with a spectrum with the lowest grid value of the neural network subgrids. In the case of \feh, however, this is not the appropriate reference value. For example, for a star with $\mathrm{[Fe/H]} = -0.74\,\mathrm{dex}$, the spectrum will be compared to a reference with $\mathrm{[Fe/H]} = -0.75\,\mathrm{dex}$, which will appear essentially identical within the spectrum uncertainties, and the code concludes there are no spectral features that are significantly different. This has affected up to 34\% of stars -- most with detectable iron lines -- and we therefore do not recommend the use of this flag at all. In the future, such a test should be performed with respect to an actually low (undetectable) amount of iron, such as $\mathrm{[Fe/H]} = -4\,\mathrm{dex}$.

\subsubsection{Fitting machinery stuck in local minimum}

As laid out in Section~\ref{sec:caveats_fitting}, we have not been able to automatically flag all estimates for which our fitting machinery has become stuck in local minima, most notably at the initial value.

\subsubsection{Binary or fast rotating star?}

With the increasing number of turn-off stars as part of ongoing GALAH observations, we have tried to implement a more sensitive approach to identify binaries in this region. This may, however, mean that we have also introduced more false-positive detections of stars that are only fast rotating with higher \vsini, rather than being a binary system. We therefore suggest carefully considering using or neglecting the accompanying flag in GALAH DR4 (see Table~\ref{tab:flag_sp}).

\subsection{Bug of interstellar K velocity} \label{sec:caveats_interstellar}

As mentioned in Section~\ref{sec:interstellar}, \texttt{rv\_k\_is} in v240705 is reported relative to the stellar radial velocity. To compute the barycentric radial velocity of the measured interstellar K, \texttt{rv\_comp\_1} has to be added to \texttt{rv\_k\_is}.

\subsection{Summary of caveats} \label{sec:caveats_summary}

In summary, the most important caveats are:
\begin{itemize}
    \item Noding in \TLF around edges between neural networks: Our tests when switching between neural networks indicate that this effect for \TLF should stay within the precision uncertainties. A more problematic effect might be that some elements could be fitted as part of one neural network based on the detectability tests that were performed at the grid centres of each neural network.
    \item Mismatches of photometry and spectroscopy: Both imperfect isochrone and spectrum models can drive a mismatch in the estimation of spectroscopic parameters. This is most notable around the secondary red clump region and also expected for highly extincted regions.
    \item Imperfect synthesis leading to trends in cool stars: The unreliable line data in cool stars causes increasingly inaccurate models and inferred stellar properties towards the coolest stars \citep[see ][]{Kos2025}. The coolest giant stars ($T_\mathrm{eff} < 3750\,\mathrm{K}$ and $\log g < 0.5$) still have unreliable parameters.
    \item Lower precision for Eu due to missing masking of neural networks.
    \item The radial velocity of interstellar K has to be corrected from the stellar to barycentric frame by adding \texttt{rv\_comp\_1}.
\end{itemize}

These caveats are a by-product of our ambitious goal to enhance the accuracy and precision of stellar parameters and elemental abundances, while vastly expanding the number of stars for which we report measurements. Each region of the Hertzsprung–Russell diagram brings its own set of challenges, whether in the complex physics of evolved stars or the fine-tuned data analysis required for main-sequence stars. Yet, these efforts have culminated in the remarkable success of GALAH DR4, providing an incredibly rich and robust dataset for researchers. As we continue to explore the Galaxy, this data enables new discoveries and insights, but it is essential to take measurements and peculiar findings with a grain of salt—they may sometimes reflect the complexity of data analysis rather than intrinsic stellar properties. Despite these challenges, GALAH DR4 marks a significant leap forward, opening up exciting opportunities for the community to unravel the mysteries of our Galaxy.

\section{CONCLUSIONS}
\label{sec:conclusion}

The GALAH survey celebrates its 10th anniversary with the release of GALAH DR4, marking a decade of transformative contributions to our understanding of the Milky Way and the elemental composition of its stars. Over the years, GALAH has been pivotal in measuring and cataloguing the chemical fingerprints of stars, which serve as cosmic barcodes that reveal their formation histories, migration patterns, and the evolutionary processes that shaped our Galaxy.

With GALAH DR4, we have achieved notable advancements in the precision and accuracy of stellar parameters and elemental abundances for nearly a million stars. This release benefits from a decade of continuous development in spectroscopic techniques, calibration processes, and the adoption of cutting-edge models like 1D NLTE and even 3D NLTE synthesis for lithium abundances. The inclusion of photometric and astrometric information from \Gaia DR3 has enhanced the reliability of stellar parameters, particularly for surface gravities, helping to resolve degeneracies of spectroscopic data. The unique value of GALAH lies in its detailed mapping of elements crucial to the studies of exoplanets and life as we know it. By tracking the abundances of carbon, nitrogen, and oxygen (CNO), rock-forming elements (e.g., Mg, Si, and Fe), as well as rare heavy elements used in modern electronics (e.g., Ce, La, and Nd), GALAH has provided key insights into how the building blocks of planets, life, and technology were forged in the interiors of stars and distributed throughout the Milky Way over billions of years.

In the last decade, 321 research outputs (176 of them refereed) have mentioned GALAH in their abstract\footnote{A total of 1\,539 astronomical research outputs mentions (1\,193 refereed) mentioned GALAH throughout their manuscript.}. GALAH DR3 \citep{Buder2021}, the predecessor of this data release, was by far the most cited paper of the Monthly Notices of the Royal Astronomical Society in 2021 at the time when this manuscript was published. GALAH DR3 has made significant contributions across several major research fields. In stellar physics and evolution, GALAH has expanded our understanding of stellar structures, nucleosynthesis \citep{Sanders2021, Griffith2022}, and lithium enrichment \citep{Martell2021, Simpson2021, Bouma2021, Sayeed2024, Wang2024}. In galactic astronomy and archaeology, GALAH has mapped the Milky Way's chemical and kinematic properties \citep[e.g.][]{BlandHawthorn2019, Sharma2021, Sharma2022}, shedding light on its formation, dynamics, and past mergers \citep{Buder2022}. The survey has also influenced planetary formation by examining the chemical environments of exoplanet host stars \citep{Clark2021, SoaresFurtado2021, Spaargaren2023, Wang2024b}, while deepening our knowledge of the Galaxy's chemical evolution and complexity \citep{Kos2021}, especially regarding neutron-capture and r-process elements \citep{Matsuno2021, Aguado2021, Horta2022, Manea2024}. Additionally, GALAH has provided insights into open clusters and star formation across the Galactic disc \citep[e.g.][]{Spina2021}, with broader applications in extragalactic astronomy through a refined understanding of surviving structures of galaxy mergers and streams \citep{Myeong2022, Buder2022, Manea2022} through its innovative chemical tagging techniques \citep{Buder2022, Buder2024}. In addition to its scientific discoveries, GALAH's influence has always been mutually beneficial with both photometric \citep{Huang2021}, asteroseismic \citep{Zinn2022}, and spectroscopic analyses \citep{Nandakumar2022, Tsantaki2022, Soubiran2022}. GALAH information has aided the calibration and validation of surveys \citep{Casagrande2021, Katz2023, Fremat2023} as well as the improvement of stellar ages by exploring chemical abundances \citep{Hayden2022, Ratcliffe2024} and combining spectroscopic and other data \citep{Hayden2022, Sahlholdt2022, Queiroz2023}. GALAH's extensive observations also covered a range of rare or peculiar objects, such as variable stars \citep{Jayasinghe2021} or metal-poor stars \citep{DaCosta2023}.

The next decade holds tremendous potential for further breakthroughs as GALAH continues its mission to observe and analyse stars across the Milky Way. With a clear goal of surpassing the 1 million star milestone, GALAH not only refines its data reduction and spectral analysis techniques but also paves the way for other ambitious surveys, such as SDSS-V \citep{Kollmeier2017}, 4MOST \citep{4MOST2019}, and WEAVE \citep{Dalton2014} at similar spectral resolution or MSE \citep{MSE2019} and HRMOS \citep{HRMOS2023} at higher spectral resolution. As a trailblazer in the field of stellar spectroscopy, GALAH’s approach has set the standard for these upcoming surveys, and its legacy will be cemented by the release of one final data set that will address the caveats and challenges discussed in this fourth data release. GALAH will undoubtedly continue to influence not only planetary, stellar and galactic astronomy but also broaden our understanding of the cosmos and the elements that shape modern life.


\section*{Acknowledgements}

We acknowledge the traditional owners of the land on which the AAT and ANU stand, the Gamilaraay, the Ngunnawal and the Ngambri peoples. We pay our respects to Elders, past and present, and are proud to continue their tradition of surveying the night sky in the Southern hemisphere.

We extend our heartfelt thanks to the entire staff at Siding Spring Observatory, both past and present, for their dedicated maintenance of 2dF-HERMES and invaluable support during the decade of observations. This project would not have been possible without the collective efforts of the many individuals who have contributed their expertise, time, and hard work, including
Ashley Anderson, 
Paula Boubel, 
Rob Brookfield, 
James Cameron, 
Steve Chapman, 
Tony Farrell, 
Kristin Feigert, 
Andy Green, 
Doug Grey, 
Dionne Haynes, 
Steve Lee, 
Chris Lidman, 
Angel Lopez-Sanchez, 
Chris McCowage, 
Quentin Parker, 
Rob Patterson, 
Susan Patterson, 
Michael Andre Phillips, 
Chris Ramage, 
Murray Riding, 
Mike Sharrott, 
Andy Sheinis, 
Lee Spitler, 
Darren Stafford, 
Lew Waller, 
Fred Watson, 
Duncan Wright, 
as well as Tayyaba Zafar, 
and all those who played a part in making GALAH a success.

The operation of the AAT is funded by the AAT Consortium, which includes The Australian National University (operator), The University of New South Wales, The University of Sydney, Macquarie University, Western Sydney University, The University of Melbourne, Swinburne University, Monash University, The University of Queensland, The University of Southern Queensland and The University of Tasmania, with Astronomy Australia Limited (AAL) as Consortium Manager.

This work was supported by the Australian Research Council Centre of Excellence for All Sky Astrophysics in 3 Dimensions (ASTRO 3D), through project number CE170100013. SB acknowledges support from the Australian Research Council under grant number DE240100150. SLM, DBZ and GFL acknowledge support from the Australian Research Council through Discovery Program grant DP220102254. SLM, BTM and KB acknowledge support from the UNSW Scientia Program. JK, GT and TZ acknowledge financial support of the Slovenian Research Agency (research core funding No. P1-0188) and the European Space Agency (PRODEX Experiment Arrangements No. 4000142234 and No. 4000143450). AMA acknowledges support from the Swedish Research Council (VR 2020-03940) and from the Crafoord Foundation via the Royal Swedish Academy of Sciences (CR 2024-0015).

\section*{Facilities}

\textbf{AAT with 2dF-HERMES at Siding Spring Observatory:}
AAT observations for this data release were performed under programs {2013B/13}, {2014A/25}, {2015A/3}, {2015A/19}, {2015B/1}, {2015B/19}, {2016A/22}, {2016B/10}, {2016B/12}, {2017A/14}, {2017A/18}, {2017B/16}, {2018A/18}, {2018B/15}, {2019A/1}, {2019A/15}, {2020B/14}, {2020B/23}, {2022B/02}, {2022B/05}, {2023A/04}, {2023A/08}, {2023A/09}, {2023B/04}, and {2023B/05}.

\textbf{AAO Data Central:} This paper includes data that has been provided by AAO Data Central  (\url{datacentral.org.au}) and makes use of services and code that have been provided by AAO Data Central.

\textbf{\Gaia: } This work has made use of data from the European Space Agency (ESA) mission \Gaia (\url{http://www.cosmos.esa.int/gaia}), processed by the \Gaia Data Processing and Analysis Consortium (DPAC, \url{http://www.cosmos.esa.int/web/gaia/dpac/consortium}). Funding for the DPAC has been provided by national institutions, in particular the institutions participating in the \Gaia Multilateral Agreement. 

\textbf{Other facilities:} This publication makes use of data products from the Two Micron All Sky Survey \citep{Skrutskie2006} and the CDS VizieR catalogue access tool \citep{Vizier2000}. This research was supported by computational resources provided by the Australian Government through the National Computational Infrastructure (NCI) under the National Computational Merit Allocation Scheme and the ANU Merit Allocation Scheme (project y89) and HPCAI Talent Programme Scholarship (project hl99).

\section*{Software}

The research for this publication was coded in \textsc{python} (version 3.7.4) and included its packages
\textsc{astropy} \citep[v. 3.2.2;][]{Robitaille2013,PriceWhelan2018},
\textsc{astroquery} \citep[v. 0.4;][]{Ginsburg2019},
\textsc{corner} \citep[v. 2.0.1;][]{corner},
\textsc{galpy} \citep[version 1.6.0;][]{Bovy2015},
\textsc{IPython} \citep[v. 7.8.0;][]{ipython},
\textsc{matplotlib} \citep[v. 3.1.3;][]{matplotlib},
\textsc{NumPy} \citep[v. 1.17.2;][]{numpy},
\textsc{scipy} \citep[version 1.3.1;][]{scipy},
\textsc{sklearn} \citep[v. 0.21.3;][]{scikit-learn},
We further made use of \textsc{topcat} \citep[version 4.7;][]{Taylor2005};

\section*{Linelist}

Our linelist, as described in Section~\ref{sec:higher_resolution_synthetic_spectra}, makes use of the following work: References: 1982ApJ...260..395C: \cite{1982ApJ...260..395C},  1983MNRAS.204..883B|1989A\&A...208..157G: \cite{1983MNRAS.204..883B,1989A&A...208..157G},  1990JQSRT..43..207C: \cite{1990JQSRT..43..207C}, 
1992A\&A...255..457D: \cite{1992A&A...255..457D},  1993A\&AS...99..179H: \cite{1993A&AS...99..179H}, \\ 1993PhyS...48..297N: \cite{1993PhyS...48..297N},  1998PhRvA..57.1652Y: \cite{1998PhRvA..57.1652Y}, 
1999ApJS..122..557N: \cite{1999ApJS..122..557N},  2008JPCRD..37..709K: \cite{2008JPCRD..37..709K},  \\ 2009A\&A...497..611M: \cite{2009A&A...497..611M},  \\ 
2014ApJS..211...20W: \cite{2014ApJS..211...20W},  2014ApJS..215...20L: \cite{2014ApJS..215...20L},  2014ApJS..215...23D: \cite{2014ApJS..215...23D},  2014MNRAS.441.3127R: \cite{2014MNRAS.441.3127R},  \\
2015ApJS..220...13L: \cite{2015ApJS..220...13L},   \\ 2015ApJS..220...13L\_1982ApJ...260..395C: \cite{2015ApJS..220...13L,1982ApJ...260..395C},  2017MNRAS.471..532P: \cite{2017MNRAS.471..532P}, 
2017PhRvA..95e2507T: \cite{2017PhRvA..95e2507T},  BGHL: \cite{BGHL},  BIPS: \cite{BIPS},  BK: \cite{BK},  BK+BWL: \cite{BK,BWL},  BK+GESB82d+BWL: \cite{BK,GESB82d,BWL},  BKK: \cite{BKK}, 
BKK+GESB82c+BWL: \cite{BKK,GESB82c,BWL},  BLNP: \cite{BLNP},  BWL: \cite{BWL},   \\ BWL+2014MNRAS.441.3127R: \cite{BWL,2014MNRAS.441.3127R},  BWL+GESHRL14: \cite{BWL,GESHRL14},  CB: \cite{CB},  DLSSC: \cite{DLSSC}, 
FMW: \cite{FMW},  GARZ|BL: \cite{GARZ,BL},  GESB82c+BWL: \cite{GESB82c,BWL},  GESB86: \cite{GESB86},  GESB86+BWL: \cite{GESB86,BWL},  GESMCHF: \cite{GESMCHF},  Grevesse2015: \cite{Grevesse2015},  HLSC: \cite{HLSC}, 
K06: \cite{K06},  K07: \cite{K07},  K08: \cite{K08},  K09: \cite{K09},  K10: \cite{K10},  K13: \cite{K13},  K14: \cite{K14},  KL-astro: astrophysical,  KR|1989ZPhyD..11..287C: \cite{KR,1989ZPhyD..11..287C}, 
LBS: \cite{LBS},  LD: \cite{LD},  LD-HS: \cite{LD-HS},  LGWSC: \cite{LGWSC},  LSCI: \cite{LSCI},  LWHS: \cite{LWHS},  MA-astro: astrophysical,  MC: \cite{MC},  MFW: \cite{MFW},  MRW: \cite{MRW},  NIST: \cite{NIST10}, 
NWL: \cite{NWL},  PQWB: \cite{PQWB},  RU: \cite{RU},  S: \cite{S},  SLS: \cite{SLS},  SR: \cite{SR},  VGH: \cite{VGH},  WLSC: \cite{WLSC},  WSL: \cite{WSL}.

\bibliography{bib}

\begin{thebibliography}{}
\expandafter\ifx\csname natexlab\endcsname\relax\def\natexlab#1{#1}\fi

\bibitem[{{Abdurro'uf} {et~al.}(2022){Abdurro'uf}, {Accetta}, {Aerts}, {Silva
  Aguirre}, {Ahumada}, {Ajgaonkar}, {Filiz Ak}, {Alam}, {Allende Prieto},
  {Almeida}, {Anders}, {Anderson}, {Andrews}, {Anguiano}, {Aquino-Ort{\'\i}z},
  {Arag{\'o}n-Salamanca}, {Argudo-Fern{\'a}ndez}, {Ata}, {Aubert},
  {Avila-Reese}, {Badenes}, {Barb{\'a}}, {Barger}, {Barrera-Ballesteros},
  {Beaton}, {Beers}, {Belfiore}, {Bender}, {Bernardi}, {Bershady}, {Beutler},
  {Bidin}, {Bird}, {Bizyaev}, {Blanc}, {Blanton}, {Boardman}, {Bolton},
  {Boquien}, {Borissova}, {Bovy}, {Brandt}, {Brown}, {Brownstein}, {Brusa},
  {Buchner}, {Bundy}, {Burchett}, {Bureau}, {Burgasser}, {Cabang}, {Campbell},
  {Cappellari}, {Carlberg}, {Wanderley}, {Carrera}, {Cash}, {Chen}, {Chen},
  {Cherinka}, {Chiappini}, {Choi}, {Chojnowski}, {Chung}, {Clerc}, {Cohen},
  {Comerford}, {Comparat}, {da Costa}, {Covey}, {Crane}, {Cruz-Gonzalez},
  {Culhane}, {Cunha}, {Dai}, {Damke}, {Darling}, {Davidson}, {Davies},
  {Dawson}, {De Lee}, {Diamond-Stanic}, {Cano-D{\'\i}az}, {S{\'a}nchez},
  {Donor}, {Duckworth}, {Dwelly}, {Eisenstein}, {Elsworth}, {Emsellem},
  {Eracleous}, {Escoffier}, {Fan}, {Farr}, {Feng}, {Fern{\'a}ndez-Trincado},
  {Feuillet}, {Filipp}, {Fillingham}, {Frinchaboy}, {Fromenteau}, {Galbany},
  {Garc{\'\i}a}, {Garc{\'\i}a-Hern{\'a}ndez}, {Ge}, {Geisler}, {Gelfand},
  {G{\'e}ron}, {Gibson}, {Goddy}, {Godoy-Rivera}, {Grabowski}, {Green},
  {Greener}, {Grier}, {Griffith}, {Guo}, {Guy}, {Hadjara}, {Harding},
  {Hasselquist}, {Hayes}, {Hearty}, {Hern{\'a}ndez}, {Hill}, {Hogg},
  {Holtzman}, {Horta}, {Hsieh}, {Hsu}, {Hsu}, {Huber}, {Huertas-Company},
  {Hutchinson}, {Hwang}, {Ibarra-Medel}, {Chitham}, {Ilha}, {Imig}, {Jaekle},
  {Jayasinghe}, {Ji}, {Johnson}, {Jones}, {J{\"o}nsson}, {Katkov}, {Khalatyan},
  {Kinemuchi}, {Kisku}, {Knapen}, {Kneib}, {Kollmeier}, {Kong}, {Kounkel},
  {Kreckel}, {Krishnarao}, {Lacerna}, {Lane}, {Langgin}, {Lavender}, {Law},
  {Lazarz}, {Leung}, {Leung}, {Lewis}, {Li}, {Li}, {Lian}, {Liang}, {Lin},
  {Lin}, {Lin}, {Lintott}, {Long}, {Longa-Pe{\~n}a}, {L{\'o}pez-Cob{\'a}},
  {Lu}, {Lundgren}, {Luo}, {Mackereth}, {de la Macorra}, {Mahadevan},
  {Majewski}, {Manchado}, {Mandeville}, {Maraston}, {Margalef-Bentabol},
  {Masseron}, {Masters}, {Mathur}, {McDermid}, {Mckay}, {Merloni},
  {Merrifield}, {Meszaros}, {Miglio}, {Di Mille}, {Minniti}, {Minsley},
  {Monachesi}, {Moon}, {Mosser}, {Mulchaey}, {Muna}, {Mu{\~n}oz}, {Myers},
  {Myers}, {Nadathur}, {Nair}, {Nandra}, {Neumann}, {Newman}, {Nidever},
  {Nikakhtar}, {Nitschelm}, {O'Connell}, {Garma-Oehmichen}, {Luan Souza de
  Oliveira}, {Olney}, {Oravetz}, {Ortigoza-Urdaneta}, {Osorio}, {Otter},
  {Pace}, {Padilla}, {Pan}, {Pan}, {Parikh}, {Parker}, {Peirani}, {Pe{\~n}a
  Ram{\'\i}rez}, {Penny}, {Percival}, {Perez-Fournon}, {Pinsonneault},
  {Poidevin}, {Poovelil}, {Price-Whelan}, {B{\'a}rbara de Andrade Queiroz},
  {Raddick}, {Ray}, {Rembold}, {Riddle}, {Riffel}, {Riffel}, {Rix}, {Robin},
  {Rodr{\'\i}guez-Puebla}, {Roman-Lopes}, {Rom{\'a}n-Z{\'u}{\~n}iga}, {Rose},
  {Ross}, {Rossi}, {Rubin}, {Salvato}, {S{\'a}nchez}, {S{\'a}nchez-Gallego},
  {Sanderson}, {Santana Rojas}, {Sarceno}, {Sarmiento}, {Sayres}, {Sazonova},
  {Schaefer}, {Schiavon}, {Schlegel}, {Schneider}, {Schultheis}, {Schwope},
  {Serenelli}, {Serna}, {Shao}, {Shapiro}, {Sharma}, {Shen}, {Shetrone}, {Shu},
  {Simon}, {Skrutskie}, {Smethurst}, {Smith}, {Sobeck}, {Spoo}, {Sprague},
  {Stark}, {Stassun}, {Steinmetz}, {Stello}, {Stone-Martinez},
  {Storchi-Bergmann}, {Stringfellow}, {Stutz}, {Su}, {Taghizadeh-Popp},
  {Talbot}, {Tayar}, {Telles}, {Teske}, {Thakar}, {Theissen}, {Tkachenko},
  {Thomas}, {Tojeiro}, {Hernandez Toledo}, {Troup}, {Trump}, {Trussler},
  {Turner}, {Tuttle}, {Unda-Sanzana}, {V{\'a}zquez-Mata}, {Valentini},
  {Valenzuela}, {Vargas-Gonz{\'a}lez}, {Vargas-Maga{\~n}a}, {Alfaro},
  {Villanova}, {Vincenzo}, {Wake}, {Warfield}, {Washington}, {Weaver},
  {Weijmans}, {Weinberg}, {Weiss}, {Westfall}, {Wild}, {Wilde}, {Wilson},
  {Wilson}, {Wilson}, {Wolf}, {Wood-Vasey}, {Yan}, {Zamora}, {Zasowski},
  {Zhang}, {Zhao}, {Zheng}, {Zheng}, \& {Zhu}}]{SDSSDR17}
{Abdurro'uf}, {Accetta}, K., {Aerts}, C., {et~al.} 2022, \apjs, 259, 35

\bibitem[{{Aguado} {et~al.}(2021){Aguado}, {Belokurov}, {Myeong}, {Evans},
  {Kobayashi}, {Sbordone}, {Chanam{\'e}}, {Navarrete}, \&
  {Koposov}}]{Aguado2021}
{Aguado}, D.~S., {Belokurov}, V., {Myeong}, G.~C., {et~al.} 2021, \apjl, 908,
  L8

\bibitem[{{Amarsi} \& {Asplund}(2017)}]{Amarsi2017}
{Amarsi}, A.~M., \& {Asplund}, M. 2017, \mnras, 464, 264

\bibitem[{{Amarsi} {et~al.}(2018{\natexlab{a}}){Amarsi}, {Barklem}, {Asplund},
  {Collet}, \& {Zatsarinny}}]{Amarsi2018b}
{Amarsi}, A.~M., {Barklem}, P.~S., {Asplund}, M., {Collet}, R., \&
  {Zatsarinny}, O. 2018{\natexlab{a}}, \aap, 616, A89

\bibitem[{{Amarsi} {et~al.}(2019{\natexlab{a}}){Amarsi}, {Barklem}, {Collet},
  {Grevesse}, \& {Asplund}}]{Amarsi2019}
{Amarsi}, A.~M., {Barklem}, P.~S., {Collet}, R., {Grevesse}, N., \& {Asplund},
  M. 2019{\natexlab{a}}, \aap, 624, A111

\bibitem[{{Amarsi} {et~al.}(2020{\natexlab{a}}){Amarsi}, {Grevesse}, {Grumer},
  {Asplund}, {Barklem}, \& {Collet}}]{Amarsi2020b}
{Amarsi}, A.~M., {Grevesse}, N., {Grumer}, J., {et~al.} 2020{\natexlab{a}},
  \aap, 636, A120

\bibitem[{{Amarsi} {et~al.}(2022){Amarsi}, {Liljegren}, \&
  {Nissen}}]{Amarsi2022}
{Amarsi}, A.~M., {Liljegren}, S., \& {Nissen}, P.~E. 2022, \aap, 668, A68

\bibitem[{{Amarsi} {et~al.}(2019{\natexlab{b}}){Amarsi}, {Nissen}, \&
  {Sk{\'u}lad{\'o}ttir}}]{Amarsi2019c}
{Amarsi}, A.~M., {Nissen}, P.~E., \& {Sk{\'u}lad{\'o}ttir}, {\'A}.
  2019{\natexlab{b}}, \aap, 630, A104

\bibitem[{{Amarsi} {et~al.}(2018{\natexlab{b}}){Amarsi}, {Nordlander},
  {Barklem}, {Asplund}, {Collet}, \& {Lind}}]{Amarsi2018}
{Amarsi}, A.~M., {Nordlander}, T., {Barklem}, P.~S., {et~al.}
  2018{\natexlab{b}}, \aap, 615, A139

\bibitem[{{Amarsi} {et~al.}(2020{\natexlab{b}}){Amarsi}, {Lind}, {Osorio},
  {Nordlander}, {Bergemann}, {Reggiani}, {Wang}, {Buder}, {Asplund}, {Barklem},
  {Wehrhahn}, {Sk{\'u}lad{\'o}ttir}, {Kobayashi}, {Karakas}, {Gao},
  {Bland-Hawthorn}, {de Silva}, {Kos}, {Lewis}, {Martell}, {Sharma}, {Simpson},
  {Zucker}, {{\v{C}}otar}, {Horner}, \& {Galah Collaboration}}]{Amarsi2020}
{Amarsi}, A.~M., {Lind}, K., {Osorio}, Y., {et~al.} 2020{\natexlab{b}}, \aap,
  642, A62

\bibitem[{{Andrae} {et~al.}(2023){Andrae}, {Fouesneau}, {Sordo},
  {Bailer-Jones}, {Dharmawardena}, {Rybizki}, {De Angeli}, {Lindstr{\o}m},
  {Marshall}, {Drimmel}, {Korn}, {Soubiran}, {Brouillet}, {Casamiquela}, {Rix},
  {Abreu Aramburu}, {{\'A}lvarez}, {Bakker}, {Bellas-Velidis}, {Bijaoui},
  {Brugaletta}, {Burlacu}, {Carballo}, {Chaoul}, {Chiavassa}, {Contursi},
  {Cooper}, {Creevey}, {Dafonte}, {Dapergolas}, {de Laverny}, {Delchambre},
  {Demouchy}, {Edvardsson}, {Fr{\'e}mat}, {Garabato}, {Garc{\'\i}a-Lario},
  {Garc{\'\i}a-Torres}, {Gavel}, {Gomez}, {Gonz{\'a}lez-Santamar{\'\i}a},
  {Hatzidimitriou}, {Heiter}, {Jean-Antoine Piccolo}, {Kontizas}, {Kordopatis},
  {Lanzafame}, {Lebreton}, {Licata}, {Livanou}, {Lobel}, {Lorca}, {Magdaleno
  Romeo}, {Manteiga}, {Marocco}, {Mary}, {Nicolas}, {Ordenovic}, {Pailler},
  {Palicio}, {Pallas-Quintela}, {Panem}, {Pichon}, {Poggio}, {Recio-Blanco},
  {Riclet}, {Robin}, {Santove{\~n}a}, {Sarro}, {Schultheis}, {Segol},
  {Silvelo}, {Slezak}, {Smart}, {S{\"u}veges}, {Th{\'e}venin}, {Torralba
  Elipe}, {Ulla}, {Utrilla}, {Vallenari}, {van Dillen}, {Zhao}, \&
  {Zorec}}]{Andrae2023}
{Andrae}, R., {Fouesneau}, M., {Sordo}, R., {et~al.} 2023, \aap, 674, A27

\bibitem[{{Asplund}(2005)}]{Asplund2005}
{Asplund}, M. 2005, \araa, 43, 481

\bibitem[{{Astropy Collaboration} {et~al.}(2013){Astropy Collaboration},
  {Robitaille}, {Tollerud}, {Greenfield}, {Droettboom}, {Bray}, {Aldcroft},
  {Davis}, {Ginsburg}, {Price-Whelan}, {Kerzendorf}, {Conley}, {Crighton},
  {Barbary}, {Muna}, {Ferguson}, {Grollier}, {Parikh}, {Nair}, {Unther},
  {Deil}, {Woillez}, {Conseil}, {Kramer}, {Turner}, {Singer}, {Fox}, {Weaver},
  {Zabalza}, {Edwards}, {Azalee Bostroem}, {Burke}, {Casey}, {Crawford},
  {Dencheva}, {Ely}, {Jenness}, {Labrie}, {Lim}, {Pierfederici}, {Pontzen},
  {Ptak}, {Refsdal}, {Servillat}, \& {Streicher}}]{Robitaille2013}
{Astropy Collaboration}, {Robitaille}, T.~P., {Tollerud}, E.~J., {et~al.} 2013,
  \aap, 558, A33

\bibitem[{{Astropy Collaboration} {et~al.}(2018){Astropy Collaboration},
  {Price-Whelan}, {Sip{\H{o}}cz}, {G{\"u}nther}, {Lim}, {Crawford}, {Conseil},
  {Shupe}, {Craig}, {Dencheva}, {Ginsburg}, {Vand erPlas}, {Bradley},
  {P{\'e}rez-Su{\'a}rez}, {de Val-Borro}, {Aldcroft}, {Cruz}, {Robitaille},
  {Tollerud}, {Ardelean}, {Babej}, {Bach}, {Bachetti}, {Bakanov}, {Bamford},
  {Barentsen}, {Barmby}, {Baumbach}, {Berry}, {Biscani}, {Boquien}, {Bostroem},
  {Bouma}, {Brammer}, {Bray}, {Breytenbach}, {Buddelmeijer}, {Burke},
  {Calderone}, {Cano Rodr{\'\i}guez}, {Cara}, {Cardoso}, {Cheedella}, {Copin},
  {Corrales}, {Crichton}, {D'Avella}, {Deil}, {Depagne}, {Dietrich}, {Donath},
  {Droettboom}, {Earl}, {Erben}, {Fabbro}, {Ferreira}, {Finethy}, {Fox},
  {Garrison}, {Gibbons}, {Goldstein}, {Gommers}, {Greco}, {Greenfield},
  {Groener}, {Grollier}, {Hagen}, {Hirst}, {Homeier}, {Horton}, {Hosseinzadeh},
  {Hu}, {Hunkeler}, {Ivezi{\'c}}, {Jain}, {Jenness}, {Kanarek}, {Kendrew},
  {Kern}, {Kerzendorf}, {Khvalko}, {King}, {Kirkby}, {Kulkarni}, {Kumar},
  {Lee}, {Lenz}, {Littlefair}, {Ma}, {Macleod}, {Mastropietro}, {McCully},
  {Montagnac}, {Morris}, {Mueller}, {Mumford}, {Muna}, {Murphy}, {Nelson},
  {Nguyen}, {Ninan}, {N{\"o}the}, {Ogaz}, {Oh}, {Parejko}, {Parley}, {Pascual},
  {Patil}, {Patil}, {Plunkett}, {Prochaska}, {Rastogi}, {Reddy Janga},
  {Sabater}, {Sakurikar}, {Seifert}, {Sherbert}, {Sherwood-Taylor}, {Shih},
  {Sick}, {Silbiger}, {Singanamalla}, {Singer}, {Sladen}, {Sooley},
  {Sornarajah}, {Streicher}, {Teuben}, {Thomas}, {Tremblay}, {Turner},
  {Terr{\'o}n}, {van Kerkwijk}, {de la Vega}, {Watkins}, {Weaver}, {Whitmore},
  {Woillez}, {Zabalza}, \& {Astropy Contributors}}]{PriceWhelan2018}
{Astropy Collaboration}, {Price-Whelan}, A.~M., {Sip{\H{o}}cz}, B.~M., {et~al.}
  2018, \aj, 156, 123

\bibitem[{{Bailer-Jones} {et~al.}(2021){Bailer-Jones}, {Rybizki}, {Fouesneau},
  {Demleitner}, \& {Andrae}}]{BailerJones2021}
{Bailer-Jones}, C.~A.~L., {Rybizki}, J., {Fouesneau}, M., {Demleitner}, M., \&
  {Andrae}, R. 2021, \aj, 161, 147

\bibitem[{{Bailer-Jones} {et~al.}(2018){Bailer-Jones}, {Rybizki}, {Fouesneau},
  {Mantelet}, \& {Andrae}}]{BailerJones2018}
{Bailer-Jones}, C.~A.~L., {Rybizki}, J., {Fouesneau}, M., {Mantelet}, G., \&
  {Andrae}, R. 2018, \aj, 156, 58

\bibitem[{{Bard} {et~al.}(1991){Bard}, {Kock}, \& {Kock}}]{BKK}
{Bard}, A., {Kock}, A., \& {Kock}, M. 1991, \aap, 248, 315, (BKK)

\bibitem[{{Bard} \& {Kock}(1994)}]{BK}
{Bard}, A., \& {Kock}, M. 1994, \aap, 282, 1014, (BK)

\bibitem[{{Barden} {et~al.}(2010){Barden}, {Jones}, {Barnes}, {Heijmans},
  {Heng}, {Knight}, {Orr}, {Smith}, {Churilov}, {Brzeski}, {Waller},
  {Shortridge}, {Horton}, {Mayfield}, {Haynes}, {Haynes}, {Whittard},
  {Goodwin}, {Smedley}, {Saunders}, {Gillingham}, {Penny}, {Farrell}, {Vuong},
  {Heald}, {Lee}, {Muller}, {Freeman}, {Bland-Hawthorn}, {Zucker}, \& {De
  Silva}}]{Barden2010}
{Barden}, S.~C., {Jones}, D.~J., {Barnes}, S.~I., {et~al.} 2010, SPIE, 7735, 09

\bibitem[{{Bastian} \& {Lardo}(2018)}]{Bastian2018}
{Bastian}, N., \& {Lardo}, C. 2018, \araa, 56, 83

\bibitem[{{Baumgardt} \& {Vasiliev}(2021)}]{Baumgardt2021}
{Baumgardt}, H., \& {Vasiliev}, E. 2021, \mnras, 505, 5957

\bibitem[{{Bedell} {et~al.}(2018){Bedell}, {Bean}, {Mel{\'e}ndez}, {Spina},
  {Ram{\'\i}rez}, {Asplund}, {Alves-Brito}, {dos Santos}, {Dreizler}, {Yong},
  {Monroe}, \& {Casagrande}}]{Bedell2018}
{Bedell}, M., {Bean}, J.~L., {Mel{\'e}ndez}, J., {et~al.} 2018, \apj, 865, 68

\bibitem[{{Beeson} {et~al.}(2024){Beeson}, {Kos}, {de Grijs}, {Martell},
  {Buder}, {Traven}, {Lewis}, {Zafar}, {Bland-Hawthorn}, {Freeman}, {Hayden},
  {Sharma}, \& {De Silva}}]{Beeson2024}
{Beeson}, K.~L., {Kos}, J., {de Grijs}, R., {et~al.} 2024, \mnras, 529, 2483

\bibitem[{{Belokurov} {et~al.}(2018){Belokurov}, {Erkal}, {Evans}, {Koposov},
  \& {Deason}}]{Belokurov2018}
{Belokurov}, V., {Erkal}, D., {Evans}, N.~W., {Koposov}, S.~E., \& {Deason},
  A.~J. 2018, \mnras, 478, 611

\bibitem[{{Bensby} {et~al.}(2014){Bensby}, {Feltzing}, \& {Oey}}]{Bensby2014}
{Bensby}, T., {Feltzing}, S., \& {Oey}, M.~S. 2014, \aap, 562, A71

\bibitem[{{Bergemann} {et~al.}(2016){Bergemann}, {Serenelli}, {Sch{\"o}nrich},
  {Ruchti}, {Korn}, {Hekker}, {Kovalev}, {Mashonkina}, {Gilmore}, {Randich},
  {Asplund}, {Rix}, {Casey}, {Jofre}, {Pancino}, {Recio-Blanco}, {de Laverny},
  {Smiljanic}, {Tautvaisiene}, {Bayo}, {Lewis}, {Koposov}, {Hourihane},
  {Worley}, {Morbidelli}, {Franciosini}, {Sacco}, {Magrini}, {Damiani}, \&
  {Bestenlehner}}]{Bergemann2016}
{Bergemann}, M., {Serenelli}, A., {Sch{\"o}nrich}, R., {et~al.} 2016, \aap,
  594, A120

\bibitem[{{Bergemann} {et~al.}(2019){Bergemann}, {Gallagher}, {Eitner},
  {Bautista}, {Collet}, {Yakovleva}, {Mayriedl}, {Plez}, {Carlsson},
  {Leenaarts}, {Belyaev}, \& {Hansen}}]{Bergemann2019b}
{Bergemann}, M., {Gallagher}, A.~J., {Eitner}, P., {et~al.} 2019, \aap, 631,
  A80

\bibitem[{{Biemont} {et~al.}(1981){Biemont}, {Grevesse}, {Hannaford}, \&
  {Lowe}}]{BGHL}
{Biemont}, E., {Grevesse}, N., {Hannaford}, P., \& {Lowe}, R.~M. 1981, \apj,
  248, 867, (BGHL)

\bibitem[{{Binney}(2012)}]{Binney2012}
{Binney}, J. 2012, \mnras, 426, 1324

\bibitem[{{Blackwell} {et~al.}(1986){Blackwell}, {Booth}, {Menon}, \&
  {Petford}}]{GESB86}
{Blackwell}, D.~E., {Booth}, A.~J., {Menon}, S.~L.~R., \& {Petford}, A.~D.
  1986, \mnras, 220, 289

\bibitem[{{Blackwell} {et~al.}(1979){Blackwell}, {Ibbetson}, {Petford}, \&
  {Shallis}}]{BIPS}
{Blackwell}, D.~E., {Ibbetson}, P.~A., {Petford}, A.~D., \& {Shallis}, M.~J.
  1979, \mnras, 186, 633, (BIPS)

\bibitem[{{Blackwell} {et~al.}(1983){Blackwell}, {Menon}, \&
  {Petford}}]{1983MNRAS.204..883B}
{Blackwell}, D.~E., {Menon}, S.~L.~R., \& {Petford}, A.~D. 1983, \mnras, 204,
  883

\bibitem[{{Blackwell} {et~al.}(1982{\natexlab{a}}){Blackwell}, {Petford},
  {Shallis}, \& {Simmons}}]{GESB82c}
{Blackwell}, D.~E., {Petford}, A.~D., {Shallis}, M.~J., \& {Simmons}, G.~J.
  1982{\natexlab{a}}, \mnras, 199, 43

\bibitem[{{Blackwell} {et~al.}(1982{\natexlab{b}}){Blackwell}, {Petford}, \&
  {Simmons}}]{GESB82d}
{Blackwell}, D.~E., {Petford}, A.~D., \& {Simmons}, G.~J. 1982{\natexlab{b}},
  \mnras, 201, 595

\bibitem[{{Blackwell-Whitehead} {et~al.}(2006){Blackwell-Whitehead},
  {Lundberg}, {Nave}, {Pickering}, {Jones}, {Lyubchik}, {Pavlenko}, \&
  {Viti}}]{BLNP}
{Blackwell-Whitehead}, R.~J., {Lundberg}, H., {Nave}, G., {et~al.} 2006,
  \mnras, 373, 1603, (BLNP)

\bibitem[{{Bland-Hawthorn} \& {Gerhard}(2016)}]{BlandHawthorn_Gerhard2016}
{Bland-Hawthorn}, J., \& {Gerhard}, O. 2016, \araa, 54, 529

\bibitem[{{Bland-Hawthorn} {et~al.}(2019){Bland-Hawthorn}, {Sharma},
  {Tepper-Garcia}, {Binney}, {Freeman}, {Hayden}, {Kos}, {De Silva}, {Ellis},
  {Lewis}, {Asplund}, {Buder}, {Casey}, {D'Orazi}, {Duong}, {Khanna}, {Lin},
  {Lind}, {Martell}, {Ness}, {Simpson}, {Zucker}, {Zwitter}, {Kafle},
  {Quillen}, {Ting}, \& {Wyse}}]{BlandHawthorn2019}
{Bland-Hawthorn}, J., {Sharma}, S., {Tepper-Garcia}, T., {et~al.} 2019, \mnras,
  486, 1167

\bibitem[{{Botnen} \& {Carlsson}(1999)}]{Botnen1999}
{Botnen}, A., \& {Carlsson}, M. 1999, in Astrophysics and Space Science
  Library, Vol. 240, Numerical Astrophysics, ed. S.~M. {Miyama}, K.~{Tomisaka},
  \& T.~{Hanawa}, 379

\bibitem[{{Bouma} {et~al.}(2021){Bouma}, {Curtis}, {Hartman}, {Winn}, \&
  {Bakos}}]{Bouma2021}
{Bouma}, L.~G., {Curtis}, J.~L., {Hartman}, J.~D., {Winn}, J.~N., \& {Bakos},
  G.~{\'A}. 2021, \aj, 162, 197

\bibitem[{{Bovy}(2015)}]{Bovy2015}
{Bovy}, J. 2015, \apjs, 216, 29

\bibitem[{{Bressan} {et~al.}(2012){Bressan}, {Marigo}, {Girardi}, {Salasnich},
  {Dal Cero}, {Rubele}, \& {Nanni}}]{Bressan2012}
{Bressan}, A., {Marigo}, P., {Girardi}, L., {et~al.} 2012, \mnras, 427, 127

\bibitem[{{Brzeski} {et~al.}(2011){Brzeski}, {Case}, \& {Gers}}]{Brzeski2011}
{Brzeski}, J., {Case}, S., \& {Gers}, L. 2011, SPIE, 8125, 04

\bibitem[{{Buchner}(2021)}]{Buchner2021}
{Buchner}, J. 2021, The Journal of Open Source Software, 6, 3001

\bibitem[{{Buder} {et~al.}(2024){Buder}, {Mijnarends}, \& {Buck}}]{Buder2024}
{Buder}, S., {Mijnarends}, L., \& {Buck}, T. 2024, \mnras, 532, 1010

\bibitem[{{Buder} {et~al.}(2018){Buder}, {Asplund}, {Duong}, {Kos}, {Lind},
  {Ness}, {Sharma}, {Bland-Hawthorn}, {Casey}, {De Silva}, {D'Orazi},
  {Freeman}, {Lewis}, {Lin}, {Martell}, {Schlesinger}, {Simpson}, {Zucker},
  {Zwitter}, {Amarsi}, {Anguiano}, {Carollo}, {Casagrande}, {{\v C}otar},
  {Cottrell}, {Da Costa}, {Gao}, {Hayden}, {Horner}, {Ireland}, {Kafle},
  {Munari}, {Nataf}, {Nordlander}, {Stello}, {Ting}, {Traven}, {Watson},
  {Wittenmyer}, {Wyse}, {Yong}, {Zinn}, {{\v Z}erjal}, , \& {The GALAH
  Collaboration}}]{Buder2018}
{Buder}, S., {Asplund}, M., {Duong}, L., {et~al.} 2018, \mnras, 478, 4513

\bibitem[{{Buder} {et~al.}(2021){Buder}, {Sharma}, {Kos}, {Amarsi},
  {Nordlander}, {Lind}, {Martell}, {Asplund}, {Bland-Hawthorn}, {Casey}, {de
  Silva}, {D'Orazi}, {Freeman}, {Hayden}, {Lewis}, {Lin}, {Schlesinger},
  {Simpson}, {Stello}, {Zucker}, {Zwitter}, {Beeson}, {Buck}, {Casagrande},
  {Clark}, {{\v{C}}otar}, {da Costa}, {de Grijs}, {Feuillet}, {Horner},
  {Kafle}, {Khanna}, {Kobayashi}, {Liu}, {Montet}, {Nandakumar}, {Nataf},
  {Ness}, {Spina}, {Tepper-Garc{\'\i}a}, {Ting}, {Traven},
  {Vogrin{\v{c}}i{\v{c}}}, {Wittenmyer}, {Wyse}, {{\v{Z}}erjal},
  {{\v{Z}}erjal}, \& {Galah Collaboration}}]{Buder2021}
{Buder}, S., {Sharma}, S., {Kos}, J., {et~al.} 2021, \mnras, 506, 150

\bibitem[{{Buder} {et~al.}(2022){Buder}, {Lind}, {Ness}, {Feuillet}, {Horta},
  {Monty}, {Buck}, {Nordlander}, {Bland-Hawthorn}, {Casey}, {de Silva},
  {D'Orazi}, {Freeman}, {Hayden}, {Kos}, {Martell}, {Lewis}, {Lin},
  {Schlesinger}, {Sharma}, {Simpson}, {Stello}, {Zucker}, {Zwitter},
  {Ciuc{\u{a}}}, {Horner}, {Kobayashi}, {Ting}, {Wyse}, \& {Wyse}}]{Buder2022}
{Buder}, S., {Lind}, K., {Ness}, M.~K., {et~al.} 2022, \mnras, 510, 2407

\bibitem[{{Cantat-Gaudin} \& {Anders}(2020)}]{CantatGaudin2020}
{Cantat-Gaudin}, T., \& {Anders}, F. 2020, \aap, 633, A99

\bibitem[{{Cardelli} {et~al.}(1989){Cardelli}, {Clayton}, \&
  {Mathis}}]{Cardelli1989}
{Cardelli}, J.~A., {Clayton}, G.~C., \& {Mathis}, J.~S. 1989, \apj, 345, 245

\bibitem[{{Cardon} {et~al.}(1982){Cardon}, {Smith}, {Scalo}, {Testerman}, \&
  {Whaling}}]{1982ApJ...260..395C}
{Cardon}, B.~L., {Smith}, P.~L., {Scalo}, J.~M., {Testerman}, L., \& {Whaling},
  W. 1982, \apj, 260, 395

\bibitem[{{Carlsson} {et~al.}(1989){Carlsson}, {Sturesson}, \&
  {Svanberg}}]{1989ZPhyD..11..287C}
{Carlsson}, J., {Sturesson}, L., \& {Svanberg}, S. 1989, Zeitschrift fur Physik
  D Atoms Molecules Clusters, 11, 287

\bibitem[{{Carretta} {et~al.}(2009{\natexlab{a}}){Carretta}, {Bragaglia},
  {Gratton}, \& {Lucatello}}]{Carretta2009c}
{Carretta}, E., {Bragaglia}, A., {Gratton}, R., \& {Lucatello}, S.
  2009{\natexlab{a}}, \aap, 505, 139

\bibitem[{{Carretta} {et~al.}(2009{\natexlab{b}}){Carretta}, {Bragaglia},
  {Gratton}, {Lucatello}, {Catanzaro}, {Leone}, {Bellazzini}, {Claudi},
  {D'Orazi}, {Momany}, {Ortolani}, {Pancino}, {Piotto}, {Recio-Blanco}, \&
  {Sabbi}}]{Carretta2009}
{Carretta}, E., {Bragaglia}, A., {Gratton}, R.~G., {et~al.} 2009{\natexlab{b}},
  \aap, 505, 117

\bibitem[{{Casagrande} \& {VandenBerg}(2018)}]{Casagrande2018}
{Casagrande}, L., \& {VandenBerg}, D.~A. 2018, \mnras, 475, 5023

\bibitem[{{Casagrande} {et~al.}(2021){Casagrande}, {Lin}, {Rains}, {Liu},
  {Buder}, {Horner}, {Asplund}, {Lewis}, {Martell}, {Nordlander}, {Stello},
  {Ting}, {Wittenmyer}, {Bland-Hawthorn}, {Casey}, {De Silva}, {D'Orazi},
  {Freeman}, {Hayden}, {Kos}, {Lind}, {Schlesinger}, {Sharma}, {Simpson},
  {Zucker}, \& {Zwitter}}]{Casagrande2021}
{Casagrande}, L., {Lin}, J., {Rains}, A.~D., {et~al.} 2021, \mnras, 507, 2684

\bibitem[{{Casey} {et~al.}(2016){Casey}, {Hogg}, {Ness}, {Rix}, {Ho}, \&
  {Gilmore}}]{Casey2016}
{Casey}, A.~R., {Hogg}, D.~W., {Ness}, M., {et~al.} 2016, ArXiv e-prints,
  arXiv:1603.03040

\bibitem[{{Casey} {et~al.}(2017){Casey}, {Hawkins}, {Hogg}, {Ness}, {Rix},
  {Kordopatis}, {Kunder}, {Steinmetz}, {Koposov}, {Enke}, {Sanders}, {Gilmore},
  {Zwitter}, {Freeman}, {Casagrande}, {Matijevi{\v c}}, {Seabroke},
  {Bienaym{\'e}}, {Bland-Hawthorn}, {Gibson}, {Grebel}, {Helmi}, {Munari},
  {Navarro}, {Reid}, {Siebert}, \& {Wyse}}]{Casey2017}
{Casey}, A.~R., {Hawkins}, K., {Hogg}, D.~W., {et~al.} 2017, \apj, 840, 59

\bibitem[{{Chang} \& {Tang}(1990)}]{1990JQSRT..43..207C}
{Chang}, T.~N., \& {Tang}, X. 1990, \jqsrt, 43, 207

\bibitem[{{Clark} {et~al.}(2021){Clark}, {Clert{\'e}}, {Hinkel}, {Unterborn},
  {Wittenmyer}, {Horner}, {Wright}, {Carter}, {Morton}, {Spina}, {Asplund},
  {Buder}, {Bland-Hawthorn}, {Casey}, {De Silva}, {D'Orazi}, {Duong}, {Hayden},
  {Freeman}, {Kos}, {Lewis}, {Lin}, {Lind}, {Martell}, {Sharma}, {Simpson},
  {Zucker}, {Zwitter}, {Tinney}, {Ting}, {Nordlander}, \& {Amarsi}}]{Clark2021}
{Clark}, J.~T., {Clert{\'e}}, M., {Hinkel}, N.~R., {et~al.} 2021, \mnras, 504,
  4968

\bibitem[{{Corliss} \& {Bozman}(1962)}]{CB}
{Corliss}, C.~H., \& {Bozman}, W.~R. 1962, NBS Monograph, Vol.~53,
  {Experimental transition probabilities for spectral lines of seventy
  elements} (US Government Printing Office), (CB)

\bibitem[{{Cutri} {et~al.}(2014)}]{Cutri2013}
{Cutri}, R.~M., {et~al.} 2014, VizieR Online Data Catalog, 2328

\bibitem[{{Da Costa} {et~al.}(2023){Da Costa}, {Bessell}, {Nordlander},
  {Hughes}, {Buder}, {Mackey}, {Spitler}, \& {Zucker}}]{DaCosta2023}
{Da Costa}, G.~S., {Bessell}, M.~S., {Nordlander}, T., {et~al.} 2023, \mnras,
  520, 917

\bibitem[{{Dalton} {et~al.}(2014){Dalton}, {Trager}, {Abrams}, {Bonifacio},
  {L{\'o}pez Aguerri}, {Middleton}, {Benn}, {Dee}, {Say{\`e}de}, {Lewis},
  {Pragt}, {Pico}, {Walton}, {Rey}, {Allende Prieto}, {Pe{\~n}ate}, {Lhome},
  {Ag{\'o}cs}, {Alonso}, {Terrett}, {Brock}, {Gilbert}, {Ridings}, {Guinouard},
  {Verheijen}, {Tosh}, {Rogers}, {Steele}, {Stuik}, {Tromp}, {Jasko}, {Kragt},
  {Lesman}, {Mottram}, {Bates}, {Gribbin}, {Rodriguez}, {Delgado}, {Martin},
  {Cano}, {Navarro}, {Irwin}, {Lewis}, {Gonzalez Solares}, {O'Mahony},
  {Bianco}, {Zurita}, {ter Horst}, {Molinari}, {Lodi}, {Guerra}, {Vallenari},
  \& {Baruffolo}}]{Dalton2014}
{Dalton}, G., {Trager}, S., {Abrams}, D.~C., {et~al.} 2014, SPIE, 9147, 0L

\bibitem[{{Davidson} {et~al.}(1992){Davidson}, {Snoek}, {Volten}, \&
  {Doenszelmann}}]{1992A&A...255..457D}
{Davidson}, M.~D., {Snoek}, L.~C., {Volten}, H., \& {Doenszelmann}, A. 1992,
  \aap, 255, 457

\bibitem[{{de Jong} {et~al.}(2019){de Jong}, {Agertz}, {Berbel}, {Aird},
  {Alexander}, {Amarsi}, {Anders}, {Andrae}, {Ansarinejad}, {Ansorge},
  {Antilogus}, {Anwand -Heerwart}, {Arentsen}, {Arnadottir}, {Asplund},
  {Auger}, {Azais}, {Baade}, {Baker}, {Baker}, {Balbinot}, {Baldry}, {Banerji},
  {Barden}, {Barklem}, {Barth{\'e}l{\'e}my-Mazot}, {Battistini}, {Bauer},
  {Bell}, {Bellido-Tirado}, {Bellstedt}, {Belokurov}, {Bensby}, {Bergemann},
  {Bestenlehner}, {Bielby}, {Bilicki}, {Blake}, {Bland-Hawthorn}, {Boeche},
  {Boland}, {Boller}, {Bongard}, {Bongiorno}, {Bonifacio}, {Boudon}, {Brooks},
  {Brown}, {Brown}, {Br{\"u}ggen}, {Brynnel}, {Brzeski}, {Buchert},
  {Buschkamp}, {Caffau}, {Caillier}, {Carrick}, {Casagrande}, {Case}, {Casey},
  {Cesarini}, {Cescutti}, {Chapuis}, {Chiappini}, {Childress}, {Christlieb},
  {Church}, {Cioni}, {Cluver}, {Colless}, {Collett}, {Comparat}, {Cooper},
  {Couch}, {Courbin}, {Croom}, {Croton}, {Daguis{\'e}}, {Dalton}, {Davies},
  {Davis}, {de Laverny}, {Deason}, {Dionies}, {Disseau}, {Doel}, {D{\"o}scher},
  {Driver}, {Dwelly}, {Eckert}, {Edge}, {Edvardsson}, {Youssoufi}, {Elhaddad},
  {Enke}, {Erfanianfar}, {Farrell}, {Fechner}, {Feiz}, {Feltzing}, {Ferreras},
  {Feuerstein}, {Feuillet}, {Finoguenov}, {Ford}, {Fotopoulou}, {Fouesneau},
  {Frenk}, {Frey}, {Gaessler}, {Geier}, {Fusillo}, {Gerhard}, {Giannantonio},
  {Giannone}, {Gibson}, {Gillingham}, {Gonz{\'a}lez-Fern{\'a}ndez},
  {Gonzalez-Solares}, {Gottloeber}, {Gould}, {Grebel}, {Gueguen}, {Guiglion},
  {Haehnelt}, {Hahn}, {Hansen}, {Hartman}, {Hauptner}, {Hawkins}, {Haynes},
  {Haynes}, {Heiter}, {Helmi}, {Aguayo}, {Hewett}, {Hinton}, {Hobbs}, {Hoenig},
  {Hofman}, {Hook}, {Hopgood}, {Hopkins}, {Hourihane}, {Howes}, {Howlett},
  {Huet}, {Irwin}, {Iwert}, {Jablonka}, {Jahn}, {Jahnke}, {Jarno}, {Jin},
  {Jofre}, {Johl}, {Jones}, {J{\"o}nsson}, {Jordan}, {Karovicova}, {Khalatyan},
  {Kelz}, {Kennicutt}, {King}, {Kitaura}, {Klar}, {Klauser}, {Kneib}, {Koch},
  {Koposov}, {Kordopatis}, {Korn}, {Kosmalski}, {Kotak}, {Kovalev}, {Kreckel},
  {Kripak}, {Krumpe}, {Kuijken}, {Kunder}, {Kushniruk}, {Lam}, {Lamer},
  {Laurent}, {Lawrence}, {Lehmitz}, {Lemasle}, {Lewis}, {Li}, {Lidman}, {Lind},
  {Liske}, {Lizon}, {Loveday}, {Ludwig}, {McDermid}, {Maguire}, {Mainieri},
  {Mali}, {Mandel}, {Mandel}, {Mannering}, {Martell}, {Martinez Delgado},
  {Matijevic}, {McGregor}, {McMahon}, {McMillan}, {Mena}, {Merloni}, {Meyer},
  {Michel}, {Micheva}, {Migniau}, {Minchev}, {Monari}, {Muller}, {Murphy},
  {Muthukrishna}, {Nandra}, {Navarro}, {Ness}, {Nichani}, {Nichol}, {Nicklas},
  {Niederhofer}, {Norberg}, {Obreschkow}, {Oliver}, {Owers}, {Pai},
  {Pankratow}, {Parkinson}, {Paschke}, {Paterson}, {Pecontal}, {Parry},
  {Phillips}, {Pillepich}, {Pinard}, {Pirard}, {Piskunov}, {Plank},
  {Pl{\"u}schke}, {Pons}, {Popesso}, {Power}, {Pragt}, {Pramskiy}, {Pryer},
  {Quattri}, {Queiroz}, {Quirrenbach}, {Rahurkar}, {Raichoor}, {Ramstedt},
  {Rau}, {Recio-Blanco}, {Reiss}, {Renaud}, {Revaz}, {Rhode}, {Richard},
  {Richter}, {Rix}, {Robotham}, {Roelfsema}, {Romaniello}, {Rosario},
  {Rothmaier}, {Roukema}, {Ruchti}, {Rupprecht}, {Rybizki}, {Ryde}, {Saar},
  {Sadler}, {Sahl{\'e}n}, {Salvato}, {Sassolas}, {Saunders}, {Saviauk},
  {Sbordone}, {Schmidt}, {Schnurr}, {Scholz}, {Schwope}, {Seifert}, {Shanks},
  {Sheinis}, {Sivov}, {Sk{\'u}lad{\'o}ttir}, {Smartt}, {Smedley}, {Smith},
  {Smith}, {Sorce}, {Spitler}, {Starkenburg}, {Steinmetz}, {Stilz}, {Storm},
  {Sullivan}, {Sutherland}, {Swann}, {Tamone}, {Taylor}, {Teillon}, {Tempel},
  {ter Horst}, {Thi}, {Tolstoy}, {Trager}, {Traven}, {Tremblay}, {Tresse},
  {Valentini}, {van de Weygaert}, {van den Ancker}, {Veljanoski}, {Venkatesan},
  {Wagner}, {Wagner}, {Walcher}, {Waller}, {Walton}, {Wang}, {Winkler},
  {Wisotzki}, {Worley}, {Worseck}, {Xiang}, {Xu}, {Yong}, {Zhao}, {Zheng},
  {Zscheyge}, \& {Zucker}}]{4MOST2019}
{de Jong}, R.~S., {Agertz}, O., {Berbel}, A.~A., {et~al.} 2019, The Messenger,
  175, 3

\bibitem[{{De Silva} {et~al.}(2015){De Silva}, {Freeman}, {Bland-Hawthorn},
  {Martell}, {de Boer}, {Asplund}, {Keller}, {Sharma}, {Zucker}, {Zwitter},
  {Anguiano}, {Bacigalupo}, {Bayliss}, {Beavis}, {Bergemann}, {Campbell},
  {Cannon}, {Carollo}, {Casagrande}, {Casey}, {Da Costa}, {D'Orazi}, {Dotter},
  {Duong}, {Heger}, {Ireland}, {Kafle}, {Kos}, {Lattanzio}, {Lewis}, {Lin},
  {Lind}, {Munari}, {Nataf}, {O'Toole}, {Parker}, {Reid}, {Schlesinger},
  {Sheinis}, {Simpson}, {Stello}, {Ting}, {Traven}, {Watson}, {Wittenmyer},
  {Yong}, \& {{\v Z}erjal}}]{DeSilva2015}
{De Silva}, G.~M., {Freeman}, K.~C., {Bland-Hawthorn}, J., {et~al.} 2015,
  \mnras, 449, 2604

\bibitem[{{Den Hartog} {et~al.}(2003){Den Hartog}, {Lawler}, {Sneden}, \&
  {Cowan}}]{HLSC}
{Den Hartog}, E.~A., {Lawler}, J.~E., {Sneden}, C., \& {Cowan}, J.~J. 2003,
  Astrophys. J. Suppl. Ser., 148, 543, (HLSC)

\bibitem[{{Den Hartog} {et~al.}(2011){Den Hartog}, {Lawler}, {Sobeck},
  {Sneden}, \& {Cowan}}]{DLSSC}
{Den Hartog}, E.~A., {Lawler}, J.~E., {Sobeck}, J.~S., {Sneden}, C., \&
  {Cowan}, J.~J. 2011, \apjs, 194, 35, (DLSSC)

\bibitem[{{Den Hartog} {et~al.}(2014{\natexlab{a}}){Den Hartog}, {Ruffoni},
  {Lawler}, {Pickering}, {Lind}, \& {Brewer}}]{2014ApJS..215...23D}
{Den Hartog}, E.~A., {Ruffoni}, M.~P., {Lawler}, J.~E., {et~al.}
  2014{\natexlab{a}}, \apjs, 215, 23

\bibitem[{{Den Hartog} {et~al.}(2014{\natexlab{b}}){Den Hartog}, {Ruffoni},
  {Lawler}, {Pickering}, {Lind}, \& {Brewer}}]{GESHRL14}
---. 2014{\natexlab{b}}, \apjs, 215, 23

\bibitem[{{Dutra-Ferreira} {et~al.}(2016){Dutra-Ferreira}, {Pasquini},
  {Smiljanic}, {Porto de Mello}, \& {Steffen}}]{DutraFerreira2016}
{Dutra-Ferreira}, L., {Pasquini}, L., {Smiljanic}, R., {Porto de Mello}, G.~F.,
  \& {Steffen}, M. 2016, \aap, 585, A75

\bibitem[{{Evans} {et~al.}(2018){Evans}, {Riello}, {De Angeli}, {Carrasco},
  {Montegriffo}, {Fabricius}, {Jordi}, {Palaversa}, {Diener}, {Busso},
  {Cacciari}, {van Leeuwen}, {Burgess}, {Davidson}, {Harrison}, {Hodgkin},
  {Pancino}, {Richards}, {Altavilla}, {Balaguer-N{\'u}{\~n}ez}, {Barstow},
  {Bellazzini}, {Brown}, {Castellani}, {Cocozza}, {De Luise}, {Delgado},
  {Ducourant}, {Galleti}, {Gilmore}, {Giuffrida}, {Holl}, {Kewley}, {Koposov},
  {Marinoni}, {Marrese}, {Osborne}, {Piersimoni}, {Portell}, {Pulone},
  {Ragaini}, {Sanna}, {Terrett}, {Walton}, {Wevers}, \&
  {Wyrzykowski}}]{Evans2018}
{Evans}, D.~W., {Riello}, M., {De Angeli}, F., {et~al.} 2018, \aap, 616, A4

\bibitem[{{Farrell} {et~al.}(2014){Farrell}, {Birchall}, {Heald}, {Shortridge},
  {Vuong}, \& {Sheinis}}]{Farrell2014}
{Farrell}, T.~J., {Birchall}, M.~N., {Heald}, R.~W., {et~al.} 2014, SPIE, 9152,
  23

\bibitem[{Foreman-Mackey(2016)}]{corner}
Foreman-Mackey, D. 2016, The Journal of Open Source Software, 1, 24

\bibitem[{{Foreman-Mackey} {et~al.}(2013){Foreman-Mackey}, {Hogg}, {Lang}, \&
  {Goodman}}]{ForemanMackey2013}
{Foreman-Mackey}, D., {Hogg}, D.~W., {Lang}, D., \& {Goodman}, J. 2013, \pasp,
  125, 306

\bibitem[{{Fouesneau} {et~al.}(2023){Fouesneau}, {Fr{\'e}mat}, {Andrae},
  {Korn}, {Soubiran}, {Kordopatis}, {Vallenari}, {Heiter}, {Creevey}, {Sarro},
  {de Laverny}, {Lanzafame}, {Lobel}, {Sordo}, {Rybizki}, {Slezak},
  {{\'A}lvarez}, {Drimmel}, {Garabato}, {Delchambre}, {Bailer-Jones},
  {Hatzidimitriou}, {Lorca}, {Le Fustec}, {Pailler}, {Mary}, {Robin},
  {Utrilla}, {Abreu Aramburu}, {Bakker}, {Bellas-Velidis}, {Bijaoui}, {Blomme},
  {Bouret}, {Brouillet}, {Brugaletta}, {Burlacu}, {Carballo}, {Casamiquela},
  {Chaoul}, {Chiavassa}, {Contursi}, {Cooper}, {Dafonte}, {Demouchy},
  {Dharmawardena}, {Garc{\'\i}a-Lario}, {Garc{\'\i}a-Torres}, {Gomez},
  {Gonz{\'a}lez-Santamar{\'\i}a}, {Jean-Antoine Piccolo}, {Kontizas},
  {Lebreton}, {Licata}, {Lindstr{\o}m}, {Livanou}, {Magdaleno Romeo},
  {Manteiga}, {Marocco}, {Martayan}, {Marshall}, {Nicolas}, {Ordenovic},
  {Palicio}, {Pallas-Quintela}, {Pichon}, {Poggio}, {Recio-Blanco}, {Riclet},
  {Santove{\~n}a}, {Schultheis}, {Segol}, {Silvelo}, {Smart}, {S{\"u}veges},
  {Th{\'e}venin}, {Torralba Elipe}, {Ulla}, {van Dillen}, {Zhao}, \&
  {Zorec}}]{Fouesneau2022}
{Fouesneau}, M., {Fr{\'e}mat}, Y., {Andrae}, R., {et~al.} 2023, \aap, 674, A28

\bibitem[{{Fr{\'e}mat} {et~al.}(2023){Fr{\'e}mat}, {Royer}, {Marchal},
  {Blomme}, {Sartoretti}, {Guerrier}, {Panuzzo}, {Katz}, {Seabroke},
  {Th{\'e}venin}, {Cropper}, {Benson}, {Damerdji}, {Haigron}, {Lobel}, {Smith},
  {Baker}, {Chemin}, {David}, {Dolding}, {Gosset}, {Jan{\ss}en}, {Jasniewicz},
  {Plum}, {Samaras}, {Snaith}, {Soubiran}, {Vanel}, {Zorec}, {Zwitter},
  {Brouillet}, {Caffau}, {Crifo}, {Fabre}, {Fragkoudi}, {Huckle}, {Lasne},
  {Leclerc}, {Mastrobuono-Battisti}, {Jean-Antoine Piccolo}, \&
  {Viala}}]{Fremat2023}
{Fr{\'e}mat}, Y., {Royer}, F., {Marchal}, O., {et~al.} 2023, \aap, 674, A8

\bibitem[{{Froese Fischer} {et~al.}(2006){Froese Fischer}, {Tachiev}, \&
  {Irimia}}]{GESMCHF}
{Froese Fischer}, C., {Tachiev}, G., \& {Irimia}, A. 2006, Atomic Data and
  Nuclear Data Tables, 92, 607

\bibitem[{{Fuhr} {et~al.}(1988){Fuhr}, {Martin}, \& {Wiese}}]{FMW}
{Fuhr}, J.~R., {Martin}, G.~A., \& {Wiese}, W.~L. 1988, JPCRD, 17, (FMW)

\bibitem[{{Gaia Collaboration} {et~al.}(2016){Gaia Collaboration}, {Prusti},
  {de Bruijne}, {Brown}, {Vallenari}, {Babusiaux}, {Bailer-Jones}, {Bastian},
  {Biermann}, {Evans}, {Eyer}, {Jansen}, {Jordi}, {Klioner}, {Lammers},
  {Lindegren}, {Luri}, {Mignard}, {Milligan}, {Panem}, {Poinsignon},
  {Pourbaix}, {Randich}, {Sarri}, {Sartoretti}, {Siddiqui}, {Soubiran},
  {Valette}, {van Leeuwen}, {Walton}, {Aerts}, {Arenou}, {Cropper}, {Drimmel},
  {H{\o}g}, {Katz}, {Lattanzi}, {O'Mullane}, {Grebel}, {Holland}, {Huc},
  {Passot}, {Bramante}, {Cacciari}, {Casta{\~n}eda}, {Chaoul}, {Cheek}, {De
  Angeli}, {Fabricius}, {Guerra}, {Hern{\'a}ndez}, {Jean-Antoine-Piccolo},
  {Masana}, {Messineo}, {Mowlavi}, {Nienartowicz}, {Ord{\'o}{\~n}ez-Blanco},
  {Panuzzo}, {Portell}, {Richards}, {Riello}, {Seabroke}, {Tanga},
  {Th{\'e}venin}, {Torra}, {Els}, {Gracia-Abril}, {Comoretto},
  {Garcia-Reinaldos}, {Lock}, {Mercier}, {Altmann}, {Andrae}, {Astraatmadja},
  {Bellas-Velidis}, {Benson}, {Berthier}, {Blomme}, {Busso}, {Carry},
  {Cellino}, {Clementini}, {Cowell}, {Creevey}, {Cuypers}, {Davidson}, {De
  Ridder}, {de Torres}, {Delchambre}, {Dell'Oro}, {Ducourant}, {Fr{\'e}mat},
  {Garc{\'{\i}}a-Torres}, {Gosset}, {Halbwachs}, {Hambly}, {Harrison},
  {Hauser}, {Hestroffer}, {Hodgkin}, {Huckle}, {Hutton}, {Jasniewicz},
  {Jordan}, {Kontizas}, {Korn}, {Lanzafame}, {Manteiga}, {Moitinho},
  {Muinonen}, {Osinde}, {Pancino}, {Pauwels}, {Petit}, {Recio-Blanco}, {Robin},
  {Sarro}, {Siopis}, {Smith}, {Smith}, {Sozzetti}, {Thuillot}, {van Reeven},
  {Viala}, {Abbas}, {Abreu Aramburu}, {Accart}, {Aguado}, {Allan}, {Allasia},
  {Altavilla}, {{\'A}lvarez}, {Alves}, {Anderson}, {Andrei}, {Anglada Varela},
  {Antiche}, {Antoja}, {Ant{\'o}n}, {Arcay}, {Atzei}, {Ayache}, {Bach},
  {Baker}, {Balaguer-N{\'u}{\~n}ez}, {Barache}, {Barata}, {Barbier}, {Barblan},
  {Baroni}, {Barrado y Navascu{\'e}s}, {Barros}, {Barstow}, {Becciani},
  {Bellazzini}, {Bellei}, {Bello Garc{\'{\i}}a}, {Belokurov}, {Bendjoya},
  {Berihuete}, {Bianchi}, {Bienaym{\'e}}, {Billebaud}, {Blagorodnova},
  {Blanco-Cuaresma}, {Boch}, {Bombrun}, {Borrachero}, {Bouquillon}, {Bourda},
  {Bouy}, {Bragaglia}, {Breddels}, {Brouillet}, {Br{\"u}semeister},
  {Bucciarelli}, {Budnik}, {Burgess}, {Burgon}, {Burlacu}, {Busonero}, {Buzzi},
  {Caffau}, {Cambras}, {Campbell}, {Cancelliere}, {Cantat-Gaudin}, {Carlucci},
  {Carrasco}, {Castellani}, {Charlot}, {Charnas}, {Charvet}, {Chassat},
  {Chiavassa}, {Clotet}, {Cocozza}, {Collins}, {Collins}, {Costigan}, {Crifo},
  {Cross}, {Crosta}, {Crowley}, {Dafonte}, {Damerdji}, {Dapergolas}, {David},
  {David}, {De Cat}, {de Felice}, {de Laverny}, {De Luise}, {De March}, {de
  Martino}, {de Souza}, {Debosscher}, {del Pozo}, {Delbo}, {Delgado},
  {Delgado}, {di Marco}, {Di Matteo}, {Diakite}, {Distefano}, {Dolding}, {Dos
  Anjos}, {Drazinos}, {Dur{\'a}n}, {Dzigan}, {Ecale}, {Edvardsson}, {Enke},
  {Erdmann}, {Escolar}, {Espina}, {Evans}, {Eynard Bontemps}, {Fabre},
  {Fabrizio}, {Faigler}, {Falc{\~a}o}, {Farr{\`a}s Casas}, {Faye}, {Federici},
  {Fedorets}, {Fern{\'a}ndez-Hern{\'a}ndez}, {Fernique}, {Fienga}, {Figueras},
  {Filippi}, {Findeisen}, {Fonti}, {Fouesneau}, {Fraile}, {Fraser}, {Fuchs},
  {Furnell}, {Gai}, {Galleti}, {Galluccio}, {Garabato}, {Garc{\'{\i}}a-Sedano},
  {Gar{\'e}}, {Garofalo}, {Garralda}, {Gavras}, {Gerssen}, {Geyer}, {Gilmore},
  {Girona}, {Giuffrida}, {Gomes}, {Gonz{\'a}lez-Marcos},
  {Gonz{\'a}lez-N{\'u}{\~n}ez}, {Gonz{\'a}lez-Vidal}, {Granvik}, {Guerrier},
  {Guillout}, {Guiraud}, {G{\'u}rpide}, {Guti{\'e}rrez-S{\'a}nchez}, {Guy},
  {Haigron}, {Hatzidimitriou}, {Haywood}, {Heiter}, {Helmi}, {Hobbs},
  {Hofmann}, {Holl}, {Holland}, {Hunt}, {Hypki}, {Icardi}, {Irwin}, {Jevardat
  de Fombelle}, {Jofr{\'e}}, {Jonker}, {Jorissen}, {Julbe}, {Karampelas},
  {Kochoska}, {Kohley}, {Kolenberg}, {Kontizas}, {Koposov}, {Kordopatis},
  {Koubsky}, {Kowalczyk}, {Krone-Martins}, {Kudryashova}, {Kull}, {Bachchan},
  {Lacoste-Seris}, {Lanza}, {Lavigne}, {Le Poncin-Lafitte}, {Lebreton},
  {Lebzelter}, {Leccia}, {Leclerc}, {Lecoeur-Taibi}, {Lemaitre}, {Lenhardt},
  {Leroux}, {Liao}, {Licata}, {Lindstr{\o}m}, {Lister}, {Livanou}, {Lobel},
  {L{\"o}ffler}, {L{\'o}pez}, {Lopez-Lozano}, {Lorenz}, {Loureiro},
  {MacDonald}, {Magalh{\~a}es Fernandes}, {Managau}, {Mann}, {Mantelet},
  {Marchal}, {Marchant}, {Marconi}, {Marie}, {Marinoni}, {Marrese},
  {Marschalk{\'o}}, {Marshall}, {Mart{\'{\i}}n-Fleitas}, {Martino}, {Mary},
  {Matijevi{\v c}}, {Mazeh}, {McMillan}, {Messina}, {Mestre}, {Michalik},
  {Millar}, {Miranda}, {Molina}, {Molinaro}, {Molinaro}, {Moln{\'a}r},
  {Moniez}, {Montegriffo}, {Monteiro}, {Mor}, {Mora}, {Morbidelli}, {Morel},
  {Morgenthaler}, {Morley}, {Morris}, {Mulone}, {Muraveva}, {Musella},
  {Narbonne}, {Nelemans}, {Nicastro}, {Noval}, {Ord{\'e}novic},
  {Ordieres-Mer{\'e}}, {Osborne}, {Pagani}, {Pagano}, {Pailler}, {Palacin},
  {Palaversa}, {Parsons}, {Paulsen}, {Pecoraro}, {Pedrosa}, {Pentik{\"a}inen},
  {Pereira}, {Pichon}, {Piersimoni}, {Pineau}, {Plachy}, {Plum}, {Poujoulet},
  {Pr{\v s}a}, {Pulone}, {Ragaini}, {Rago}, {Rambaux}, {Ramos-Lerate},
  {Ranalli}, {Rauw}, {Read}, {Regibo}, {Renk}, {Reyl{\'e}}, {Ribeiro},
  {Rimoldini}, {Ripepi}, {Riva}, {Rixon}, {Roelens}, {Romero-G{\'o}mez},
  {Rowell}, {Royer}, {Rudolph}, {Ruiz-Dern}, {Sadowski}, {Sagrist{\`a}
  Sell{\'e}s}, {Sahlmann}, {Salgado}, {Salguero}, {Sarasso}, {Savietto},
  {Schnorhk}, {Schultheis}, {Sciacca}, {Segol}, {Segovia}, {Segransan},
  {Serpell}, {Shih}, {Smareglia}, {Smart}, {Smith}, {Solano}, {Solitro},
  {Sordo}, {Soria Nieto}, {Souchay}, {Spagna}, {Spoto}, {Stampa}, {Steele},
  {Steidelm{\"u}ller}, {Stephenson}, {Stoev}, {Suess}, {S{\"u}veges}, {Surdej},
  {Szabados}, {Szegedi-Elek}, {Tapiador}, {Taris}, {Tauran}, {Taylor},
  {Teixeira}, {Terrett}, {Tingley}, {Trager}, {Turon}, {Ulla}, {Utrilla},
  {Valentini}, {van Elteren}, {Van Hemelryck}, {van Leeuwen}, {Varadi},
  {Vecchiato}, {Veljanoski}, {Via}, {Vicente}, {Vogt}, {Voss}, {Votruba},
  {Voutsinas}, {Walmsley}, {Weiler}, {Weingrill}, {Werner}, {Wevers},
  {Whitehead}, {Wyrzykowski}, {Yoldas}, {{\v Z}erjal}, {Zucker}, {Zurbach},
  {Zwitter}, {Alecu}, {Allen}, {Allende Prieto}, {Amorim},
  {Anglada-Escud{\'e}}, {Arsenijevic}, {Azaz}, {Balm}, {Beck}, {Bernstein},
  {Bigot}, {Bijaoui}, {Blasco}, {Bonfigli}, {Bono}, {Boudreault}, {Bressan},
  {Brown}, {Brunet}, {Bunclark}, {Buonanno}, {Butkevich}, {Carret}, {Carrion},
  {Chemin}, {Ch{\'e}reau}, {Corcione}, {Darmigny}, {de Boer}, {de Teodoro}, {de
  Zeeuw}, {Delle Luche}, {Domingues}, {Dubath}, {Fodor}, {Fr{\'e}zouls},
  {Fries}, {Fustes}, {Fyfe}, {Gallardo}, {Gallegos}, {Gardiol}, {Gebran},
  {Gomboc}, {G{\'o}mez}, {Grux}, {Gueguen}, {Heyrovsky}, {Hoar}, {Iannicola},
  {Isasi Parache}, {Janotto}, {Joliet}, {Jonckheere}, {Keil}, {Kim},
  {Klagyivik}, {Klar}, {Knude}, {Kochukhov}, {Kolka}, {Kos}, {Kutka}, {Lainey},
  {LeBouquin}, {Liu}, {Loreggia}, {Makarov}, {Marseille}, {Martayan},
  {Martinez-Rubi}, {Massart}, {Meynadier}, {Mignot}, {Munari}, {Nguyen},
  {Nordlander}, {Ocvirk}, {O'Flaherty}, {Olias Sanz}, {Ortiz}, {Osorio},
  {Oszkiewicz}, {Ouzounis}, {Palmer}, {Park}, {Pasquato}, {Peltzer}, {Peralta},
  {P{\'e}turaud}, {Pieniluoma}, {Pigozzi}, {Poels}, {Prat}, {Prod'homme},
  {Raison}, {Rebordao}, {Risquez}, {Rocca-Volmerange}, {Rosen}, {Ruiz-Fuertes},
  {Russo}, {Sembay}, {Serraller Vizcaino}, {Short}, {Siebert}, {Silva},
  {Sinachopoulos}, {Slezak}, {Soffel}, {Sosnowska}, {Strai{\v z}ys}, {ter
  Linden}, {Terrell}, {Theil}, {Tiede}, {Troisi}, {Tsalmantza}, {Tur},
  {Vaccari}, {Vachier}, {Valles}, {Van Hamme}, {Veltz}, {Virtanen}, {Wallut},
  {Wichmann}, {Wilkinson}, {Ziaeepour}, \& {Zschocke}}]{Gaia-Collaboration2016}
{Gaia Collaboration}, {Prusti}, T., {de Bruijne}, J.~H.~J., {et~al.} 2016,
  \aap, 595, A1

\bibitem[{{Gaia Collaboration} {et~al.}(2018){Gaia Collaboration}, {Brown},
  {Vallenari}, {Prusti}, {de Bruijne}, {Babusiaux}, {Bailer-Jones}, {Biermann},
  {Evans}, {Eyer}, {Jansen}, {Jordi}, {Klioner}, {Lammers}, {Lindegren},
  {Luri}, {Mignard}, {Panem}, {Pourbaix}, {Randich}, {Sartoretti}, {Siddiqui},
  {Soubiran}, {van Leeuwen}, {Walton}, {Arenou}, {Bastian}, {Cropper},
  {Drimmel}, {Katz}, {Lattanzi}, {Bakker}, {Cacciari}, {Casta{\~n}eda},
  {Chaoul}, {Cheek}, {De Angeli}, {Fabricius}, {Guerra}, {Holl}, {Masana},
  {Messineo}, {Mowlavi}, {Nienartowicz}, {Panuzzo}, {Portell}, {Riello},
  {Seabroke}, {Tanga}, {Th{\'e}venin}, {Gracia-Abril}, {Comoretto},
  {Garcia-Reinaldos}, {Teyssier}, {Altmann}, {Andrae}, {Audard},
  {Bellas-Velidis}, {Benson}, {Berthier}, {Blomme}, {Burgess}, {Busso},
  {Carry}, {Cellino}, {Clementini}, {Clotet}, {Creevey}, {Davidson}, {De
  Ridder}, {Delchambre}, {Dell'Oro}, {Ducourant},
  {Fern{\'a}ndez-Hern{\'a}ndez}, {Fouesneau}, {Fr{\'e}mat}, {Galluccio},
  {Garc{\'{\i}}a-Torres}, {Gonz{\'a}lez-N{\'u}{\~n}ez}, {Gonz{\'a}lez-Vidal},
  {Gosset}, {Guy}, {Halbwachs}, {Hambly}, {Harrison}, {Hern{\'a}ndez},
  {Hestroffer}, {Hodgkin}, {Hutton}, {Jasniewicz}, {Jean-Antoine-Piccolo},
  {Jordan}, {Korn}, {Krone-Martins}, {Lanzafame}, {Lebzelter}, {L{\"o}ffler},
  {Manteiga}, {Marrese}, {Mart{\'{\i}}n-Fleitas}, {Moitinho}, {Mora},
  {Muinonen}, {Osinde}, {Pancino}, {Pauwels}, {Petit}, {Recio-Blanco},
  {Richards}, {Rimoldini}, {Robin}, {Sarro}, {Siopis}, {Smith}, {Sozzetti},
  {S{\"u}veges}, {Torra}, {van Reeven}, {Abbas}, {Abreu Aramburu}, {Accart},
  {Aerts}, {Altavilla}, {{\'A}lvarez}, {Alvarez}, {Alves}, {Anderson},
  {Andrei}, {Anglada Varela}, {Antiche}, {Antoja}, {Arcay}, {Astraatmadja},
  {Bach}, {Baker}, {Balaguer-N{\'u}{\~n}ez}, {Balm}, {Barache}, {Barata},
  {Barbato}, {Barblan}, {Barklem}, {Barrado}, {Barros}, {Barstow},
  {Bartholom{\'e} Mu{\~n}oz}, {Bassilana}, {Becciani}, {Bellazzini},
  {Berihuete}, {Bertone}, {Bianchi}, {Bienaym{\'e}}, {Blanco-Cuaresma}, {Boch},
  {Boeche}, {Bombrun}, {Borrachero}, {Bossini}, {Bouquillon}, {Bourda},
  {Bragaglia}, {Bramante}, {Breddels}, {Bressan}, {Brouillet},
  {Br{\"u}semeister}, {Brugaletta}, {Bucciarelli}, {Burlacu}, {Busonero},
  {Butkevich}, {Buzzi}, {Caffau}, {Cancelliere}, {Cannizzaro}, {Cantat-Gaudin},
  {Carballo}, {Carlucci}, {Carrasco}, {Casamiquela}, {Castellani},
  {Castro-Ginard}, {Charlot}, {Chemin}, {Chiavassa}, {Cocozza}, {Costigan},
  {Cowell}, {Crifo}, {Crosta}, {Crowley}, {Cuypers}, {Dafonte}, {Damerdji},
  {Dapergolas}, {David}, {David}, {de Laverny}, {De Luise}, {De March}, {de
  Martino}, {de Souza}, {de Torres}, {Debosscher}, {del Pozo}, {Delbo},
  {Delgado}, {Delgado}, {Di Matteo}, {Diakite}, {Diener}, {Distefano},
  {Dolding}, {Drazinos}, {Dur{\'a}n}, {Edvardsson}, {Enke}, {Eriksson},
  {Esquej}, {Eynard Bontemps}, {Fabre}, {Fabrizio}, {Faigler}, {Falc{\~a}o},
  {Farr{\`a}s Casas}, {Federici}, {Fedorets}, {Fernique}, {Figueras},
  {Filippi}, {Findeisen}, {Fonti}, {Fraile}, {Fraser}, {Fr{\'e}zouls}, {Gai},
  {Galleti}, {Garabato}, {Garc{\'{\i}}a-Sedano}, {Garofalo}, {Garralda},
  {Gavel}, {Gavras}, {Gerssen}, {Geyer}, {Giacobbe}, {Gilmore}, {Girona},
  {Giuffrida}, {Glass}, {Gomes}, {Granvik}, {Gueguen}, {Guerrier}, {Guiraud},
  {Guti{\'e}rrez-S{\'a}nchez}, {Haigron}, {Hatzidimitriou}, {Hauser},
  {Haywood}, {Heiter}, {Helmi}, {Heu}, {Hilger}, {Hobbs}, {Hofmann}, {Holland},
  {Huckle}, {Hypki}, {Icardi}, {Jan{\ss}en}, {Jevardat de Fombelle}, {Jonker},
  {Juh{\'a}sz}, {Julbe}, {Karampelas}, {Kewley}, {Klar}, {Kochoska}, {Kohley},
  {Kolenberg}, {Kontizas}, {Kontizas}, {Koposov}, {Kordopatis},
  {Kostrzewa-Rutkowska}, {Koubsky}, {Lambert}, {Lanza}, {Lasne}, {Lavigne}, {Le
  Fustec}, {Le Poncin-Lafitte}, {Lebreton}, {Leccia}, {Leclerc},
  {Lecoeur-Taibi}, {Lenhardt}, {Leroux}, {Liao}, {Licata}, {Lindstr{\o}m},
  {Lister}, {Livanou}, {Lobel}, {L{\'o}pez}, {Managau}, {Mann}, {Mantelet},
  {Marchal}, {Marchant}, {Marconi}, {Marinoni}, {Marschalk{\'o}}, {Marshall},
  {Martino}, {Marton}, {Mary}, {Massari}, {Matijevi{\v c}}, {Mazeh},
  {McMillan}, {Messina}, {Michalik}, {Millar}, {Molina}, {Molinaro},
  {Moln{\'a}r}, {Montegriffo}, {Mor}, {Morbidelli}, {Morel}, {Morris},
  {Mulone}, {Muraveva}, {Musella}, {Nelemans}, {Nicastro}, {Noval},
  {O'Mullane}, {Ord{\'e}novic}, {Ord{\'o}{\~n}ez-Blanco}, {Osborne}, {Pagani},
  {Pagano}, {Pailler}, {Palacin}, {Palaversa}, {Panahi}, {Pawlak},
  {Piersimoni}, {Pineau}, {Plachy}, {Plum}, {Poggio}, {Poujoulet}, {Pr{\v s}a},
  {Pulone}, {Racero}, {Ragaini}, {Rambaux}, {Ramos-Lerate}, {Regibo},
  {Reyl{\'e}}, {Riclet}, {Ripepi}, {Riva}, {Rivard}, {Rixon}, {Roegiers},
  {Roelens}, {Romero-G{\'o}mez}, {Rowell}, {Royer}, {Ruiz-Dern}, {Sadowski},
  {Sagrist{\`a} Sell{\'e}s}, {Sahlmann}, {Salgado}, {Salguero}, {Sanna},
  {Santana-Ros}, {Sarasso}, {Savietto}, {Schultheis}, {Sciacca}, {Segol},
  {Segovia}, {S{\'e}gransan}, {Shih}, {Siltala}, {Silva}, {Smart}, {Smith},
  {Solano}, {Solitro}, {Sordo}, {Soria Nieto}, {Souchay}, {Spagna}, {Spoto},
  {Stampa}, {Steele}, {Steidelm{\"u}ller}, {Stephenson}, {Stoev}, {Suess},
  {Surdej}, {Szabados}, {Szegedi-Elek}, {Tapiador}, {Taris}, {Tauran},
  {Taylor}, {Teixeira}, {Terrett}, {Teyssandier}, {Thuillot}, {Titarenko},
  {Torra Clotet}, {Turon}, {Ulla}, {Utrilla}, {Uzzi}, {Vaillant}, {Valentini},
  {Valette}, {van Elteren}, {Van Hemelryck}, {van Leeuwen}, {Vaschetto},
  {Vecchiato}, {Veljanoski}, {Viala}, {Vicente}, {Vogt}, {von Essen}, {Voss},
  {Votruba}, {Voutsinas}, {Walmsley}, {Weiler}, {Wertz}, {Wevers},
  {Wyrzykowski}, {Yoldas}, {{\v Z}erjal}, {Ziaeepour}, {Zorec}, {Zschocke},
  {Zucker}, {Zurbach}, \& {Zwitter}}]{Brown2018}
{Gaia Collaboration}, {Brown}, A.~G.~A., {Vallenari}, A., {et~al.} 2018, \aap,
  616, A1

\bibitem[{{Gaia Collaboration} {et~al.}(2021{\natexlab{a}}){Gaia
  Collaboration}, {Brown}, {Vallenari}, {Prusti}, {de Bruijne}, {Babusiaux},
  {Biermann}, {Creevey}, {Evans}, {Eyer}, {Hutton}, {Jansen}, {Jordi},
  {Klioner}, {Lammers}, {Lindegren}, {Luri}, {Mignard}, {Panem}, {Pourbaix},
  {Randich}, {Sartoretti}, {Soubiran}, {Walton}, {Arenou}, {Bailer-Jones},
  {Bastian}, {Cropper}, {Drimmel}, {Katz}, {Lattanzi}, {van Leeuwen}, {Bakker},
  {Cacciari}, {Casta{\~n}eda}, {De Angeli}, {Ducourant}, {Fabricius},
  {Fouesneau}, {Fr{\'e}mat}, {Guerra}, {Guerrier}, {Guiraud}, {Jean-Antoine
  Piccolo}, {Masana}, {Messineo}, {Mowlavi}, {Nicolas}, {Nienartowicz},
  {Pailler}, {Panuzzo}, {Riclet}, {Roux}, {Seabroke}, {Sordo}, {Tanga},
  {Th{\'e}venin}, {Gracia-Abril}, {Portell}, {Teyssier}, {Altmann}, {Andrae},
  {Bellas-Velidis}, {Benson}, {Berthier}, {Blomme}, {Brugaletta}, {Burgess},
  {Busso}, {Carry}, {Cellino}, {Cheek}, {Clementini}, {Damerdji}, {Davidson},
  {Delchambre}, {Dell'Oro}, {Fern{\'a}ndez-Hern{\'a}ndez}, {Galluccio},
  {Garc{\'\i}a-Lario}, {Garcia-Reinaldos}, {Gonz{\'a}lez-N{\'u}{\~n}ez},
  {Gosset}, {Haigron}, {Halbwachs}, {Hambly}, {Harrison}, {Hatzidimitriou},
  {Heiter}, {Hern{\'a}ndez}, {Hestroffer}, {Hodgkin}, {Holl}, {Jan{\ss}en},
  {Jevardat de Fombelle}, {Jordan}, {Krone-Martins}, {Lanzafame},
  {L{\"o}ffler}, {Lorca}, {Manteiga}, {Marchal}, {Marrese}, {Moitinho}, {Mora},
  {Muinonen}, {Osborne}, {Pancino}, {Pauwels}, {Petit}, {Recio-Blanco},
  {Richards}, {Riello}, {Rimoldini}, {Robin}, {Roegiers}, {Rybizki}, {Sarro},
  {Siopis}, {Smith}, {Sozzetti}, {Ulla}, {Utrilla}, {van Leeuwen}, {van
  Reeven}, {Abbas}, {Abreu Aramburu}, {Accart}, {Aerts}, {Aguado}, {Ajaj},
  {Altavilla}, {{\'A}lvarez}, {{\'A}lvarez Cid-Fuentes}, {Alves}, {Anderson},
  {Anglada Varela}, {Antoja}, {Audard}, {Baines}, {Baker},
  {Balaguer-N{\'u}{\~n}ez}, {Balbinot}, {Balog}, {Barache}, {Barbato},
  {Barros}, {Barstow}, {Bartolom{\'e}}, {Bassilana}, {Bauchet},
  {Baudesson-Stella}, {Becciani}, {Bellazzini}, {Bernet}, {Bertone}, {Bianchi},
  {Blanco-Cuaresma}, {Boch}, {Bombrun}, {Bossini}, {Bouquillon}, {Bragaglia},
  {Bramante}, {Breedt}, {Bressan}, {Brouillet}, {Bucciarelli}, {Burlacu},
  {Busonero}, {Butkevich}, {Buzzi}, {Caffau}, {Cancelliere}, {C{\'a}novas},
  {Cantat-Gaudin}, {Carballo}, {Carlucci}, {Carnerero}, {Carrasco},
  {Casamiquela}, {Castellani}, {Castro-Ginard}, {Castro Sampol}, {Chaoul},
  {Charlot}, {Chemin}, {Chiavassa}, {Cioni}, {Comoretto}, {Cooper}, {Cornez},
  {Cowell}, {Crifo}, {Crosta}, {Crowley}, {Dafonte}, {Dapergolas}, {David},
  {David}, {de Laverny}, {De Luise}, {De March}, {De Ridder}, {de Souza}, {de
  Teodoro}, {de Torres}, {del Peloso}, {del Pozo}, {Delbo}, {Delgado},
  {Delgado}, {Delisle}, {Di Matteo}, {Diakite}, {Diener}, {Distefano},
  {Dolding}, {Eappachen}, {Edvardsson}, {Enke}, {Esquej}, {Fabre}, {Fabrizio},
  {Faigler}, {Fedorets}, {Fernique}, {Fienga}, {Figueras}, {Fouron},
  {Fragkoudi}, {Fraile}, {Franke}, {Gai}, {Garabato}, {Garcia-Gutierrez},
  {Garc{\'\i}a-Torres}, {Garofalo}, {Gavras}, {Gerlach}, {Geyer}, {Giacobbe},
  {Gilmore}, {Girona}, {Giuffrida}, {Gomel}, {Gomez}, {Gonzalez-Santamaria},
  {Gonz{\'a}lez-Vidal}, {Granvik}, {Guti{\'e}rrez-S{\'a}nchez}, {Guy},
  {Hauser}, {Haywood}, {Helmi}, {Hidalgo}, {Hilger}, {H{\l}adczuk}, {Hobbs},
  {Holland}, {Huckle}, {Jasniewicz}, {Jonker}, {Juaristi Campillo}, {Julbe},
  {Karbevska}, {Kervella}, {Khanna}, {Kochoska}, {Kontizas}, {Kordopatis},
  {Korn}, {Kostrzewa-Rutkowska}, {Kruszy{\'n}ska}, {Lambert}, {Lanza}, {Lasne},
  {Le Campion}, {Le Fustec}, {Lebreton}, {Lebzelter}, {Leccia}, {Leclerc},
  {Lecoeur-Taibi}, {Liao}, {Licata}, {Lindstr{\o}m}, {Lister}, {Livanou},
  {Lobel}, {Madrero Pardo}, {Managau}, {Mann}, {Marchant}, {Marconi}, {Marcos
  Santos}, {Marinoni}, {Marocco}, {Marshall}, {Martin Polo},
  {Mart{\'\i}n-Fleitas}, {Masip}, {Massari}, {Mastrobuono-Battisti}, {Mazeh},
  {McMillan}, {Messina}, {Michalik}, {Millar}, {Mints}, {Molina}, {Molinaro},
  {Moln{\'a}r}, {Montegriffo}, {Mor}, {Morbidelli}, {Morel}, {Morris},
  {Mulone}, {Munoz}, {Muraveva}, {Murphy}, {Musella}, {Noval}, {Ord{\'e}novic},
  {Orr{\`u}}, {Osinde}, {Pagani}, {Pagano}, {Palaversa}, {Palicio}, {Panahi},
  {Pawlak}, {Pe{\~n}alosa Esteller}, {Penttil{\"a}}, {Piersimoni}, {Pineau},
  {Plachy}, {Plum}, {Poggio}, {Poretti}, {Poujoulet}, {Pr{\v{s}}a}, {Pulone},
  {Racero}, {Ragaini}, {Rainer}, {Raiteri}, {Rambaux}, {Ramos}, {Ramos-Lerate},
  {Re Fiorentin}, {Regibo}, {Reyl{\'e}}, {Ripepi}, {Riva}, {Rixon}, {Robichon},
  {Robin}, {Roelens}, {Rohrbasser}, {Romero-G{\'o}mez}, {Rowell}, {Royer},
  {Rybicki}, {Sadowski}, {Sagrist{\`a} Sell{\'e}s}, {Sahlmann}, {Salgado},
  {Salguero}, {Samaras}, {Sanchez Gimenez}, {Sanna}, {Santove{\~n}a},
  {Sarasso}, {Schultheis}, {Sciacca}, {Segol}, {Segovia}, {S{\'e}gransan},
  {Semeux}, {Shahaf}, {Siddiqui}, {Siebert}, {Siltala}, {Slezak}, {Smart},
  {Solano}, {Solitro}, {Souami}, {Souchay}, {Spagna}, {Spoto}, {Steele},
  {Steidelm{\"u}ller}, {Stephenson}, {S{\"u}veges}, {Szabados}, {Szegedi-Elek},
  {Taris}, {Tauran}, {Taylor}, {Teixeira}, {Thuillot}, {Tonello}, {Torra},
  {Torra}, {Turon}, {Unger}, {Vaillant}, {van Dillen}, {Vanel}, {Vecchiato},
  {Viala}, {Vicente}, {Voutsinas}, {Weiler}, {Wevers}, {Wyrzykowski}, {Yoldas},
  {Yvard}, {Zhao}, {Zorec}, {Zucker}, {Zurbach}, \& {Zwitter}}]{Brown2021}
---. 2021{\natexlab{a}}, \aap, 649, A1

\bibitem[{{Gaia Collaboration} {et~al.}(2021{\natexlab{b}}){Gaia
  Collaboration}, {Brown}, {Vallenari}, {Prusti}, {de Bruijne}, {Babusiaux},
  {Biermann}, {Creevey}, {Evans}, {Eyer}, {Hutton}, {Jansen}, {Jordi},
  {Klioner}, {Lammers}, {Lindegren}, {Luri}, {Mignard}, {Panem}, {Pourbaix},
  {Randich}, {Sartoretti}, {Soubiran}, {Walton}, {Arenou}, {Bailer-Jones},
  {Bastian}, {Cropper}, {Drimmel}, {Katz}, {Lattanzi}, {van Leeuwen}, {Bakker},
  {Cacciari}, {Casta{\~n}eda}, {De Angeli}, {Ducourant}, {Fabricius},
  {Fouesneau}, {Fr{\'e}mat}, {Guerra}, {Guerrier}, {Guiraud}, {Jean-Antoine
  Piccolo}, {Masana}, {Messineo}, {Mowlavi}, {Nicolas}, {Nienartowicz},
  {Pailler}, {Panuzzo}, {Riclet}, {Roux}, {Seabroke}, {Sordo}, {Tanga},
  {Th{\'e}venin}, {Gracia-Abril}, {Portell}, {Teyssier}, {Altmann}, {Andrae},
  {Bellas-Velidis}, {Benson}, {Berthier}, {Blomme}, {Brugaletta}, {Burgess},
  {Busso}, {Carry}, {Cellino}, {Cheek}, {Clementini}, {Damerdji}, {Davidson},
  {Delchambre}, {Dell'Oro}, {Fern{\'a}ndez-Hern{\'a}ndez}, {Galluccio},
  {Garc{\'\i}a-Lario}, {Garcia-Reinaldos}, {Gonz{\'a}lez-N{\'u}{\~n}ez},
  {Gosset}, {Haigron}, {Halbwachs}, {Hambly}, {Harrison}, {Hatzidimitriou},
  {Heiter}, {Hern{\'a}ndez}, {Hestroffer}, {Hodgkin}, {Holl}, {Jan{\ss}en},
  {Jevardat de Fombelle}, {Jordan}, {Krone-Martins}, {Lanzafame},
  {L{\"o}ffler}, {Lorca}, {Manteiga}, {Marchal}, {Marrese}, {Moitinho}, {Mora},
  {Muinonen}, {Osborne}, {Pancino}, {Pauwels}, {Petit}, {Recio-Blanco},
  {Richards}, {Riello}, {Rimoldini}, {Robin}, {Roegiers}, {Rybizki}, {Sarro},
  {Siopis}, {Smith}, {Sozzetti}, {Ulla}, {Utrilla}, {van Leeuwen}, {van
  Reeven}, {Abbas}, {Abreu Aramburu}, {Accart}, {Aerts}, {Aguado}, {Ajaj},
  {Altavilla}, {{\'A}lvarez}, {{\'A}lvarez Cid-Fuentes}, {Alves}, {Anderson},
  {Anglada Varela}, {Antoja}, {Audard}, {Baines}, {Baker},
  {Balaguer-N{\'u}{\~n}ez}, {Balbinot}, {Balog}, {Barache}, {Barbato},
  {Barros}, {Barstow}, {Bartolom{\'e}}, {Bassilana}, {Bauchet},
  {Baudesson-Stella}, {Becciani}, {Bellazzini}, {Bernet}, {Bertone}, {Bianchi},
  {Blanco-Cuaresma}, {Boch}, {Bombrun}, {Bossini}, {Bouquillon}, {Bragaglia},
  {Bramante}, {Breedt}, {Bressan}, {Brouillet}, {Bucciarelli}, {Burlacu},
  {Busonero}, {Butkevich}, {Buzzi}, {Caffau}, {Cancelliere}, {C{\'a}novas},
  {Cantat-Gaudin}, {Carballo}, {Carlucci}, {Carnerero}, {Carrasco},
  {Casamiquela}, {Castellani}, {Castro-Ginard}, {Castro Sampol}, {Chaoul},
  {Charlot}, {Chemin}, {Chiavassa}, {Cioni}, {Comoretto}, {Cooper}, {Cornez},
  {Cowell}, {Crifo}, {Crosta}, {Crowley}, {Dafonte}, {Dapergolas}, {David},
  {David}, {de Laverny}, {De Luise}, {De March}, {De Ridder}, {de Souza}, {de
  Teodoro}, {de Torres}, {del Peloso}, {del Pozo}, {Delbo}, {Delgado},
  {Delgado}, {Delisle}, {Di Matteo}, {Diakite}, {Diener}, {Distefano},
  {Dolding}, {Eappachen}, {Edvardsson}, {Enke}, {Esquej}, {Fabre}, {Fabrizio},
  {Faigler}, {Fedorets}, {Fernique}, {Fienga}, {Figueras}, {Fouron},
  {Fragkoudi}, {Fraile}, {Franke}, {Gai}, {Garabato}, {Garcia-Gutierrez},
  {Garc{\'\i}a-Torres}, {Garofalo}, {Gavras}, {Gerlach}, {Geyer}, {Giacobbe},
  {Gilmore}, {Girona}, {Giuffrida}, {Gomel}, {Gomez}, {Gonzalez-Santamaria},
  {Gonz{\'a}lez-Vidal}, {Granvik}, {Guti{\'e}rrez-S{\'a}nchez}, {Guy},
  {Hauser}, {Haywood}, {Helmi}, {Hidalgo}, {Hilger}, {H{\l}adczuk}, {Hobbs},
  {Holland}, {Huckle}, {Jasniewicz}, {Jonker}, {Juaristi Campillo}, {Julbe},
  {Karbevska}, {Kervella}, {Khanna}, {Kochoska}, {Kontizas}, {Kordopatis},
  {Korn}, {Kostrzewa-Rutkowska}, {Kruszy{\'n}ska}, {Lambert}, {Lanza}, {Lasne},
  {Le Campion}, {Le Fustec}, {Lebreton}, {Lebzelter}, {Leccia}, {Leclerc},
  {Lecoeur-Taibi}, {Liao}, {Licata}, {Lindstr{\o}m}, {Lister}, {Livanou},
  {Lobel}, {Madrero Pardo}, {Managau}, {Mann}, {Marchant}, {Marconi}, {Marcos
  Santos}, {Marinoni}, {Marocco}, {Marshall}, {Martin Polo},
  {Mart{\'\i}n-Fleitas}, {Masip}, {Massari}, {Mastrobuono-Battisti}, {Mazeh},
  {McMillan}, {Messina}, {Michalik}, {Millar}, {Mints}, {Molina}, {Molinaro},
  {Moln{\'a}r}, {Montegriffo}, {Mor}, {Morbidelli}, {Morel}, {Morris},
  {Mulone}, {Munoz}, {Muraveva}, {Murphy}, {Musella}, {Noval}, {Ord{\'e}novic},
  {Orr{\`u}}, {Osinde}, {Pagani}, {Pagano}, {Palaversa}, {Palicio}, {Panahi},
  {Pawlak}, {Pe{\~n}alosa Esteller}, {Penttil{\"a}}, {Piersimoni}, {Pineau},
  {Plachy}, {Plum}, {Poggio}, {Poretti}, {Poujoulet}, {Pr{\v{s}}a}, {Pulone},
  {Racero}, {Ragaini}, {Rainer}, {Raiteri}, {Rambaux}, {Ramos}, {Ramos-Lerate},
  {Re Fiorentin}, {Regibo}, {Reyl{\'e}}, {Ripepi}, {Riva}, {Rixon}, {Robichon},
  {Robin}, {Roelens}, {Rohrbasser}, {Romero-G{\'o}mez}, {Rowell}, {Royer},
  {Rybicki}, {Sadowski}, {Sagrist{\`a} Sell{\'e}s}, {Sahlmann}, {Salgado},
  {Salguero}, {Samaras}, {Sanchez Gimenez}, {Sanna}, {Santove{\~n}a},
  {Sarasso}, {Schultheis}, {Sciacca}, {Segol}, {Segovia}, {S{\'e}gransan},
  {Semeux}, {Shahaf}, {Siddiqui}, {Siebert}, {Siltala}, {Slezak}, {Smart},
  {Solano}, {Solitro}, {Souami}, {Souchay}, {Spagna}, {Spoto}, {Steele},
  {Steidelm{\"u}ller}, {Stephenson}, {S{\"u}veges}, {Szabados}, {Szegedi-Elek},
  {Taris}, {Tauran}, {Taylor}, {Teixeira}, {Thuillot}, {Tonello}, {Torra},
  {Torra}, {Turon}, {Unger}, {Vaillant}, {van Dillen}, {Vanel}, {Vecchiato},
  {Viala}, {Vicente}, {Voutsinas}, {Weiler}, {Wevers}, {Wyrzykowski}, {Yoldas},
  {Yvard}, {Zhao}, {Zorec}, {Zucker}, {Zurbach}, \& {Zwitter}}]{GaiaEDR3}
---. 2021{\natexlab{b}}, \aap, 649, A1

\bibitem[{{Gaia Collaboration} {et~al.}(2023){Gaia Collaboration}, {Vallenari},
  {Brown}, {Prusti}, {de Bruijne}, {Arenou}, {Babusiaux}, {Biermann},
  {Creevey}, {Ducourant}, {Evans}, {Eyer}, {Guerra}, {Hutton}, {Jordi},
  {Klioner}, {Lammers}, {Lindegren}, {Luri}, {Mignard}, {Panem}, {Pourbaix},
  {Randich}, {Sartoretti}, {Soubiran}, {Tanga}, {Walton}, {Bailer-Jones},
  {Bastian}, {Drimmel}, {Jansen}, {Katz}, {Lattanzi}, {van Leeuwen}, {Bakker},
  {Cacciari}, {Casta{\~n}eda}, {De Angeli}, {Fabricius}, {Fouesneau},
  {Fr{\'e}mat}, {Galluccio}, {Guerrier}, {Heiter}, {Masana}, {Messineo},
  {Mowlavi}, {Nicolas}, {Nienartowicz}, {Pailler}, {Panuzzo}, {Riclet}, {Roux},
  {Seabroke}, {Sordo}, {Th{\'e}venin}, {Gracia-Abril}, {Portell}, {Teyssier},
  {Altmann}, {Andrae}, {Audard}, {Bellas-Velidis}, {Benson}, {Berthier},
  {Blomme}, {Burgess}, {Busonero}, {Busso}, {C{\'a}novas}, {Carry}, {Cellino},
  {Cheek}, {Clementini}, {Damerdji}, {Davidson}, {de Teodoro}, {Nu{\~n}ez
  Campos}, {Delchambre}, {Dell'Oro}, {Esquej}, {Fern{\'a}ndez-Hern{\'a}ndez},
  {Fraile}, {Garabato}, {Garc{\'\i}a-Lario}, {Gosset}, {Haigron}, {Halbwachs},
  {Hambly}, {Harrison}, {Hern{\'a}ndez}, {Hestroffer}, {Hodgkin}, {Holl},
  {Jan{\ss}en}, {Jevardat de Fombelle}, {Jordan}, {Krone-Martins}, {Lanzafame},
  {L{\"o}ffler}, {Marchal}, {Marrese}, {Moitinho}, {Muinonen}, {Osborne},
  {Pancino}, {Pauwels}, {Recio-Blanco}, {Reyl{\'e}}, {Riello}, {Rimoldini},
  {Roegiers}, {Rybizki}, {Sarro}, {Siopis}, {Smith}, {Sozzetti}, {Utrilla},
  {van Leeuwen}, {Abbas}, {{\'A}brah{\'a}m}, {Abreu Aramburu}, {Aerts},
  {Aguado}, {Ajaj}, {Aldea-Montero}, {Altavilla}, {{\'A}lvarez}, {Alves},
  {Anders}, {Anderson}, {Anglada Varela}, {Antoja}, {Baines}, {Baker},
  {Balaguer-N{\'u}{\~n}ez}, {Balbinot}, {Balog}, {Barache}, {Barbato},
  {Barros}, {Barstow}, {Bartolom{\'e}}, {Bassilana}, {Bauchet}, {Becciani},
  {Bellazzini}, {Berihuete}, {Bernet}, {Bertone}, {Bianchi}, {Binnenfeld},
  {Blanco-Cuaresma}, {Blazere}, {Boch}, {Bombrun}, {Bossini}, {Bouquillon},
  {Bragaglia}, {Bramante}, {Breedt}, {Bressan}, {Brouillet}, {Brugaletta},
  {Bucciarelli}, {Burlacu}, {Butkevich}, {Buzzi}, {Caffau}, {Cancelliere},
  {Cantat-Gaudin}, {Carballo}, {Carlucci}, {Carnerero}, {Carrasco},
  {Casamiquela}, {Castellani}, {Castro-Ginard}, {Chaoul}, {Charlot}, {Chemin},
  {Chiaramida}, {Chiavassa}, {Chornay}, {Comoretto}, {Contursi}, {Cooper},
  {Cornez}, {Cowell}, {Crifo}, {Cropper}, {Crosta}, {Crowley}, {Dafonte},
  {Dapergolas}, {David}, {David}, {de Laverny}, {De Luise}, {De March}, {De
  Ridder}, {de Souza}, {de Torres}, {del Peloso}, {del Pozo}, {Delbo},
  {Delgado}, {Delisle}, {Demouchy}, {Dharmawardena}, {Di Matteo}, {Diakite},
  {Diener}, {Distefano}, {Dolding}, {Edvardsson}, {Enke}, {Fabre}, {Fabrizio},
  {Faigler}, {Fedorets}, {Fernique}, {Fienga}, {Figueras}, {Fournier},
  {Fouron}, {Fragkoudi}, {Gai}, {Garcia-Gutierrez}, {Garcia-Reinaldos},
  {Garc{\'\i}a-Torres}, {Garofalo}, {Gavel}, {Gavras}, {Gerlach}, {Geyer},
  {Giacobbe}, {Gilmore}, {Girona}, {Giuffrida}, {Gomel}, {Gomez},
  {Gonz{\'a}lez-N{\'u}{\~n}ez}, {Gonz{\'a}lez-Santamar{\'\i}a},
  {Gonz{\'a}lez-Vidal}, {Granvik}, {Guillout}, {Guiraud},
  {Guti{\'e}rrez-S{\'a}nchez}, {Guy}, {Hatzidimitriou}, {Hauser}, {Haywood},
  {Helmer}, {Helmi}, {Sarmiento}, {Hidalgo}, {Hilger}, {H{\l}adczuk}, {Hobbs},
  {Holland}, {Huckle}, {Jardine}, {Jasniewicz}, {Jean-Antoine Piccolo},
  {Jim{\'e}nez-Arranz}, {Jorissen}, {Juaristi Campillo}, {Julbe}, {Karbevska},
  {Kervella}, {Khanna}, {Kontizas}, {Kordopatis}, {Korn}, {K{\'o}sp{\'a}l},
  {Kostrzewa-Rutkowska}, {Kruszy{\'n}ska}, {Kun}, {Laizeau}, {Lambert},
  {Lanza}, {Lasne}, {Le Campion}, {Lebreton}, {Lebzelter}, {Leccia}, {Leclerc},
  {Lecoeur-Taibi}, {Liao}, {Licata}, {Lindstr{\o}m}, {Lister}, {Livanou},
  {Lobel}, {Lorca}, {Loup}, {Madrero Pardo}, {Magdaleno Romeo}, {Managau},
  {Mann}, {Manteiga}, {Marchant}, {Marconi}, {Marcos}, {Marcos Santos},
  {Mar{\'\i}n Pina}, {Marinoni}, {Marocco}, {Marshall}, {Martin Polo},
  {Mart{\'\i}n-Fleitas}, {Marton}, {Mary}, {Masip}, {Massari},
  {Mastrobuono-Battisti}, {Mazeh}, {McMillan}, {Messina}, {Michalik}, {Millar},
  {Mints}, {Molina}, {Molinaro}, {Moln{\'a}r}, {Monari}, {Mongui{\'o}},
  {Montegriffo}, {Montero}, {Mor}, {Mora}, {Morbidelli}, {Morel}, {Morris},
  {Muraveva}, {Murphy}, {Musella}, {Nagy}, {Noval}, {Oca{\~n}a}, {Ogden},
  {Ordenovic}, {Osinde}, {Pagani}, {Pagano}, {Palaversa}, {Palicio},
  {Pallas-Quintela}, {Panahi}, {Payne-Wardenaar}, {Pe{\~n}alosa Esteller},
  {Penttil{\"a}}, {Pichon}, {Piersimoni}, {Pineau}, {Plachy}, {Plum}, {Poggio},
  {Pr{\v{s}}a}, {Pulone}, {Racero}, {Ragaini}, {Rainer}, {Raiteri}, {Rambaux},
  {Ramos}, {Ramos-Lerate}, {Re Fiorentin}, {Regibo}, {Richards}, {Rios Diaz},
  {Ripepi}, {Riva}, {Rix}, {Rixon}, {Robichon}, {Robin}, {Robin}, {Roelens},
  {Rogues}, {Rohrbasser}, {Romero-G{\'o}mez}, {Rowell}, {Royer}, {Ruz Mieres},
  {Rybicki}, {Sadowski}, {S{\'a}ez N{\'u}{\~n}ez}, {Sagrist{\`a} Sell{\'e}s},
  {Sahlmann}, {Salguero}, {Samaras}, {Sanchez Gimenez}, {Sanna},
  {Santove{\~n}a}, {Sarasso}, {Schultheis}, {Sciacca}, {Segol}, {Segovia},
  {S{\'e}gransan}, {Semeux}, {Shahaf}, {Siddiqui}, {Siebert}, {Siltala},
  {Silvelo}, {Slezak}, {Slezak}, {Smart}, {Snaith}, {Solano}, {Solitro},
  {Souami}, {Souchay}, {Spagna}, {Spina}, {Spoto}, {Steele},
  {Steidelm{\"u}ller}, {Stephenson}, {S{\"u}veges}, {Surdej}, {Szabados},
  {Szegedi-Elek}, {Taris}, {Taylor}, {Teixeira}, {Tolomei}, {Tonello}, {Torra},
  {Torra}, {Torralba Elipe}, {Trabucchi}, {Tsounis}, {Turon}, {Ulla}, {Unger},
  {Vaillant}, {van Dillen}, {van Reeven}, {Vanel}, {Vecchiato}, {Viala},
  {Vicente}, {Voutsinas}, {Weiler}, {Wevers}, {Wyrzykowski}, {Yoldas}, {Yvard},
  {Zhao}, {Zorec}, {Zucker}, \& {Zwitter}}]{GaiaDR3}
{Gaia Collaboration}, {Vallenari}, A., {Brown}, A.~G.~A., {et~al.} 2023, \aap,
  674, A1

\bibitem[{{Gallagher} {et~al.}(2020){Gallagher}, {Bergemann}, {Collet}, {Plez},
  {Leenaarts}, {Carlsson}, {Yakovleva}, \& {Belyaev}}]{Gallagher2020}
{Gallagher}, A.~J., {Bergemann}, M., {Collet}, R., {et~al.} 2020, \aap, 634,
  A55

\bibitem[{{Garz}(1973)}]{GARZ}
{Garz}, T. 1973, \aap, 26, 471, (GARZ)

\bibitem[{{Gent} {et~al.}(2022){Gent}, {Bergemann}, {Serenelli}, {Casagrande},
  {Gerber}, {Heiter}, {Kovalev}, {Morel}, {Nardetto}, {Adibekyan}, {Silva
  Aguirre}, {Asplund}, {Belkacem}, {del Burgo}, {Bigot}, {Chiavassa},
  {Rodr{\'\i}guez D{\'\i}az}, {Goupil}, {Gonz{\'a}lez Hern{\'a}ndez},
  {Mourard}, {Merle}, {M{\'e}sz{\'a}ros}, {Marshall}, {Ouazzani}, {Plez},
  {Reese}, {Trampedach}, \& {Tsantaki}}]{Gent2022}
{Gent}, M.~R., {Bergemann}, M., {Serenelli}, A., {et~al.} 2022, \aap, 658, A147

\bibitem[{{Gilmore} {et~al.}(2022){Gilmore}, {Randich}, {Worley}, {Hourihane},
  {Gonneau}, {Sacco}, {Lewis}, {Magrini}, {Fran{\c{c}}ois}, {Jeffries},
  {Koposov}, {Bragaglia}, {Alfaro}, {Allende Prieto}, {Blomme}, {Korn},
  {Lanzafame}, {Pancino}, {Recio-Blanco}, {Smiljanic}, {Van Eck}, {Zwitter},
  {Bensby}, {Flaccomio}, {Irwin}, {Franciosini}, {Morbidelli}, {Damiani},
  {Bonito}, {Friel}, {Vink}, {Prisinzano}, {Abbas}, {Hatzidimitriou}, {Held},
  {Jordi}, {Paunzen}, {Spagna}, {Jackson}, {Ma{\'\i}z Apell{\'a}niz},
  {Asplund}, {Bonifacio}, {Feltzing}, {Binney}, {Drew}, {Ferguson}, {Micela},
  {Negueruela}, {Prusti}, {Rix}, {Vallenari}, {Bergemann}, {Casey}, {de
  Laverny}, {Frasca}, {Hill}, {Lind}, {Sbordone}, {Sousa}, {Adibekyan},
  {Caffau}, {Daflon}, {Feuillet}, {Gebran}, {Gonzalez Hernandez}, {Guiglion},
  {Herrero}, {Lobel}, {Merle}, {Mikolaitis}, {Montes}, {Morel}, {Ruchti},
  {Soubiran}, {Tabernero}, {Tautvai{\v{s}}ien{\.{e}}}, {Traven}, {Valentini},
  {Van der Swaelmen}, {Villanova}, {Viscasillas V{\'a}zquez}, {Bayo}, {Biazzo},
  {Carraro}, {Edvardsson}, {Heiter}, {Jofr{\'e}}, {Marconi}, {Martayan},
  {Masseron}, {Monaco}, {Walton}, {Zaggia}, {Aguirre B{\o}rsen-Koch}, {Alves},
  {Balaguer-Nunez}, {Barklem}, {Barrado}, {Bellazzini}, {Berlanas}, {Binks},
  {Bressan}, {Capuzzo-Dolcetta}, {Casagrande}, {Casamiquela}, {Collins},
  {D'Orazi}, {Dantas}, {Debattista}, {Delgado-Mena}, {Di Marcantonio},
  {Drazdauskas}, {Evans}, {Famaey}, {Franchini}, {Fr{\'e}mat}, {Fu}, {Geisler},
  {Gerhard}, {Gonz{\'a}lez Solares}, {Grebel}, {Guti{\'e}rrez Albarr{\'a}n},
  {Jim{\'e}nez-Esteban}, {J{\"o}nsson}, {Khachaturyants}, {Kordopatis}, {Kos},
  {Lagarde}, {Ludwig}, {Mahy}, {Mapelli}, {Marfil}, {Martell}, {Messina},
  {Miglio}, {Minchev}, {Moitinho}, {Montalban}, {Monteiro}, {Morossi},
  {Mowlavi}, {Mucciarelli}, {Murphy}, {Nardetto}, {Ortolani}, {Paletou},
  {Palou{\v{s}}}, {Pickering}, {Quirrenbach}, {Re Fiorentin}, {Read}, {Romano},
  {Ryde}, {Sanna}, {Santos}, {Seabroke}, {Spina}, {Steinmetz}, {Stonkut{\'e}},
  {Sutorius}, {Th{\'e}venin}, {Tosi}, {Tsantaki}, {Wright}, {Wyse}, {Zoccali},
  {Zorec}, \& {Zucker}}]{Gilmore2022}
{Gilmore}, G., {Randich}, S., {Worley}, C.~C., {et~al.} 2022, \aap, 666, A120

\bibitem[{{Ginsburg} {et~al.}(2019){Ginsburg}, {Sip{\H{o}}cz}, {Brasseur},
  {Cowperthwaite}, {Craig}, {Deil}, {Guillochon}, {Guzman}, {Liedtke}, {Lian
  Lim}, {Lockhart}, {Mommert}, {Morris}, {Norman}, {Parikh}, {Persson},
  {Robitaille}, {Segovia}, {Singer}, {Tollerud}, {de Val-Borro}, {Valtchanov},
  {Woillez}, {Astroquery Collaboration}, \& {a subset of astropy
  Collaboration}}]{Ginsburg2019}
{Ginsburg}, A., {Sip{\H{o}}cz}, B.~M., {Brasseur}, C.~E., {et~al.} 2019, \aj,
  157, 98

\bibitem[{{Gratton} {et~al.}(2019){Gratton}, {Bragaglia}, {Carretta},
  {D'Orazi}, {Lucatello}, \& {Sollima}}]{Gratton2019}
{Gratton}, R., {Bragaglia}, A., {Carretta}, E., {et~al.} 2019, \aapr, 27, 8

\bibitem[{{Grevesse} {et~al.}(2007){Grevesse}, {Asplund}, \&
  {Sauval}}]{Grevesse2007}
{Grevesse}, N., {Asplund}, M., \& {Sauval}, A.~J. 2007, \ssr, 130, 105

\bibitem[{{Grevesse} {et~al.}(1989){Grevesse}, {Blackwell}, \&
  {Petford}}]{1989A&A...208..157G}
{Grevesse}, N., {Blackwell}, D.~E., \& {Petford}, A.~D. 1989, \aap, 208, 157

\bibitem[{{Grevesse} {et~al.}(2015){Grevesse}, {Scott}, {Asplund}, \&
  {Sauval}}]{Grevesse2015}
{Grevesse}, N., {Scott}, P., {Asplund}, M., \& {Sauval}, A.~J. 2015, \aap, 573,
  A27

\bibitem[{{Griffith} {et~al.}(2022){Griffith}, {Weinberg}, {Buder}, {Johnson},
  {Johnson}, \& {Vincenzo}}]{Griffith2022}
{Griffith}, E.~J., {Weinberg}, D.~H., {Buder}, S., {et~al.} 2022, \apj, 931, 23

\bibitem[{{Gustafsson} {et~al.}(2008){Gustafsson}, {Edvardsson}, {Eriksson},
  {J{\o}rgensen}, {Nordlund}, \& {Plez}}]{Gustafsson2008}
{Gustafsson}, B., {Edvardsson}, B., {Eriksson}, K., {et~al.} 2008, \aap, 486,
  951

\bibitem[{{Hayden} {et~al.}(2022){Hayden}, {Sharma}, {Bland-Hawthorn}, {Spina},
  {Buder}, {Ciuc{\u{a}}}, {Asplund}, {Casey}, {De Silva}, {D'Orazi}, {Freeman},
  {Kos}, {Lewis}, {Lin}, {Lind}, {Martell}, {Schlesinger}, {Simpson}, {Zucker},
  {Zwitter}, {Chen}, {{\v{C}}otar}, {Feuillet}, {Horner}, {Joyce},
  {Nordlander}, {Stello}, {Tepper-Garcia}, {Ting}, {Wang}, {Wittenmyer}, \&
  {Wyse}}]{Hayden2022}
{Hayden}, M.~R., {Sharma}, S., {Bland-Hawthorn}, J., {et~al.} 2022, \mnras,
  517, 5325

\bibitem[{{Heijmans} {et~al.}(2012){Heijmans}, {Asplund}, {Barden}, {Birchall},
  {Carollo}, {Bland-Hawthorn}, {Brzeski}, {Case}, {Churilov}, {Colless},
  {Dean}, {De Silva}, {Farrell}, {Fiegert}, {Freeman}, {Gers}, {Goodwin},
  {Gray}, {Heald}, {Heng}, {Jones}, {Kobayashi}, {Klauser}, {Kondrat},
  {Lawrence}, {Lee}, {Mathews}, {Mayfield}, {Miziarski}, {Monnet}, {Muller},
  {Pai}, {Patterson}, {Penny}, {Orr}, {Sheinis}, {Shortridge}, {Smedley},
  {Smith}, {Stafford}, {Staszak}, {Vuong}, {Waller}, {Whittard}, {Wylie de
  Boer}, {Xavier}, {Zheng}, {Zhelem}, \& {Zucker}}]{Heijmans2012}
{Heijmans}, J., {Asplund}, M., {Barden}, S., {et~al.} 2012, SPIE, 8446, 0W

\bibitem[{{Heiter} {et~al.}(2015){Heiter}, {Jofr{\'e}}, {Gustafsson}, {Korn},
  {Soubiran}, \& {Th{\'e}venin}}]{Heiter2015}
{Heiter}, U., {Jofr{\'e}}, P., {Gustafsson}, B., {et~al.} 2015, \aap, 582, A49

\bibitem[{{Heiter} {et~al.}(2021){Heiter}, {Lind}, {Bergemann}, {Asplund},
  {Mikolaitis}, {Barklem}, {Masseron}, {de Laverny}, {Magrini}, {Edvardsson},
  {J{\"o}nsson}, {Pickering}, {Ryde}, {Bayo Ar{\'a}n}, {Bensby}, {Casey},
  {Feltzing}, {Jofr{\'e}}, {Korn}, {Pancino}, {Damiani}, {Lanzafame}, {Lardo},
  {Monaco}, {Morbidelli}, {Smiljanic}, {Worley}, {Zaggia}, {Randich}, \&
  {Gilmore}}]{Heiter2021}
{Heiter}, U., {Lind}, K., {Bergemann}, M., {et~al.} 2021, \aap, 645, A106

\bibitem[{{Helmi} {et~al.}(2018){Helmi}, {Babusiaux}, {Koppelman}, {Massari},
  {Veljanoski}, \& {Brown}}]{Helmi2018}
{Helmi}, A., {Babusiaux}, C., {Koppelman}, H.~H., {et~al.} 2018, \nat, 563, 85

\bibitem[{{Hibbert} {et~al.}(1993){Hibbert}, {Biemont}, {Godefroid}, \&
  {Vaeck}}]{1993A&AS...99..179H}
{Hibbert}, A., {Biemont}, E., {Godefroid}, M., \& {Vaeck}, N. 1993, \aaps, 99,
  179

\bibitem[{{Hinkle} {et~al.}(2000){Hinkle}, {Wallace}, {Valenti}, \&
  {Harmer}}]{Hinkle2000}
{Hinkle}, K., {Wallace}, L., {Valenti}, J., \& {Harmer}, D., eds. 2000, Visible
  and Near Infrared Atlas of the Arcturus Spectrum, 3727-9300 {\AA} (Astron.
  Soc. Pac.)

\bibitem[{{Ho} {et~al.}(2017){Ho}, {Ness}, {Hogg}, {Rix}, {Liu}, {Yang},
  {Zhang}, {Hou}, \& {Wang}}]{Ho2017}
{Ho}, A.~Y.~Q., {Ness}, M.~K., {Hogg}, D.~W., {et~al.} 2017, \apj, 836, 5

\bibitem[{{Hoeijmakers} {et~al.}(2015){Hoeijmakers}, {de Kok}, {Snellen},
  {Brogi}, {Birkby}, \& {Schwarz}}]{Hoeijmakers2015}
{Hoeijmakers}, H.~J., {de Kok}, R.~J., {Snellen}, I.~A.~G., {et~al.} 2015,
  \aap, 575, A20

\bibitem[{{Hon} {et~al.}(2021){Hon}, {Huber}, {Kuszlewicz}, {Stello}, {Sharma},
  {Tayar}, {Zinn}, {Vrard}, \& {Pinsonneault}}]{Hon2021}
{Hon}, M., {Huber}, D., {Kuszlewicz}, J.~S., {et~al.} 2021, \apj, 919, 131

\bibitem[{{Horta} {et~al.}(2022){Horta}, {Ness}, {Rybizki}, {Schiavon}, \&
  {Buder}}]{Horta2022}
{Horta}, D., {Ness}, M.~K., {Rybizki}, J., {Schiavon}, R.~P., \& {Buder}, S.
  2022, \mnras, 513, 5477

\bibitem[{{Hourihane} {et~al.}(2023){Hourihane}, {Fran{\c{c}}ois}, {Worley},
  {Magrini}, {Gonneau}, {Casey}, {Gilmore}, {Randich}, {Sacco}, {Recio-Blanco},
  {Korn}, {Allende Prieto}, {Smiljanic}, {Blomme}, {Bragaglia}, {Walton}, {Van
  Eck}, {Bensby}, {Lanzafame}, {Frasca}, {Franciosini}, {Damiani}, {Lind},
  {Bergemann}, {Bonifacio}, {Hill}, {Lobel}, {Montes}, {Feuillet},
  {Tautvai{\v{s}}ien{\.{e}}}, {Guiglion}, {Tabernero}, {Gonz{\'a}lez
  Hern{\'a}ndez}, {Gebran}, {Van der Swaelmen}, {Mikolaitis}, {Daflon},
  {Merle}, {Morel}, {Lewis}, {Gonz{\'a}lez Solares}, {Murphy}, {Jeffries},
  {Jackson}, {Feltzing}, {Prusti}, {Carraro}, {Biazzo}, {Prisinzano},
  {Jofr{\'e}}, {Zaggia}, {Drazdauskas}, {Stonkut{\'e}}, {Marfil},
  {Jim{\'e}nez-Esteban}, {Mahy}, {Guti{\'e}rrez Albarr{\'a}n}, {Berlanas},
  {Santos}, {Morbidelli}, {Spina}, \&
  {Minkevi{\v{c}}i{\={u}}t{\.{e}}}}]{Hourihane2023}
{Hourihane}, A., {Fran{\c{c}}ois}, P., {Worley}, C.~C., {et~al.} 2023, \aap,
  676, A129

\bibitem[{{Howell} {et~al.}(2022){Howell}, {Campbell}, {Stello}, \& {De
  Silva}}]{Howell2022}
{Howell}, M., {Campbell}, S.~W., {Stello}, D., \& {De Silva}, G.~M. 2022,
  \mnras, 515, 3184

\bibitem[{{Howell} {et~al.}(2014){Howell}, {Sobeck}, {Haas}, {Still},
  {Barclay}, {Mullally}, {Troeltzsch}, {Aigrain}, {Bryson}, {Caldwell},
  {Chaplin}, {Cochran}, {Huber}, {Marcy}, {Miglio}, {Najita}, {Smith},
  {Twicken}, \& {Fortney}}]{Howell2014}
{Howell}, S.~B., {Sobeck}, C., {Haas}, M., {et~al.} 2014, \pasp, 126, 398

\bibitem[{{Huang} {et~al.}(2021){Huang}, {Yuan}, {Li}, {Wolf}, {Onken},
  {Beers}, {Casagrande}, {Mackey}, {Da Costa}, {Bland-Hawthorn}, {Stello},
  {Nordlander}, {Ting}, {Buder}, {Sharma}, \& {Liu}}]{Huang2021}
{Huang}, Y., {Yuan}, H., {Li}, C., {et~al.} 2021, \apj, 907, 68

\bibitem[{Hunter(2007)}]{matplotlib}
Hunter, J.~D. 2007, Comput Sci Eng, 9, 90

\bibitem[{{Jayasinghe} {et~al.}(2021){Jayasinghe}, {Kochanek}, {Stanek},
  {Shappee}, {Holoien}, {Thompson}, {Prieto}, {Dong}, {Pawlak}, {Pejcha},
  {Pojmanski}, {Otero}, {Hurst}, \& {Will}}]{Jayasinghe2021}
{Jayasinghe}, T., {Kochanek}, C.~S., {Stanek}, K.~Z., {et~al.} 2021, \mnras,
  503, 200

\bibitem[{{Jofr{\'e}} {et~al.}(2018){Jofr{\'e}}, {Heiter}, {Tucci Maia},
  {Soubiran}, {Worley}, {Hawkins}, {Blanco-Cuaresma}, \& {Rodrigo}}]{Jofre2018}
{Jofr{\'e}}, P., {Heiter}, U., {Tucci Maia}, M., {et~al.} 2018, Research Notes
  of the American Astronomical Society, 2, 152

\bibitem[{{Jofr{\'e}} {et~al.}(2014){Jofr{\'e}}, {Heiter}, {Soubiran},
  {Blanco-Cuaresma}, {Worley}, {Pancino}, {Cantat-Gaudin}, {Magrini},
  {Bergemann}, {Gonz{\'a}lez Hern{\'a}ndez}, {Hill}, {Lardo}, {de Laverny},
  {Lind}, {Masseron}, {Montes}, {Mucciarelli}, {Nordlander}, {Recio Blanco},
  {Sobeck}, {Sordo}, {Sousa}, {Tabernero}, {Vallenari}, \& {Van
  Eck}}]{Jofre2014}
{Jofr{\'e}}, P., {Heiter}, U., {Soubiran}, C., {et~al.} 2014, \aap, 564, A133

\bibitem[{{Jofr{\'e}} {et~al.}(2015){Jofr{\'e}}, {Heiter}, {Soubiran},
  {Blanco-Cuaresma}, {Masseron}, {Nordlander}, {Chemin}, {Worley}, {Van Eck},
  {Hourihane}, {Gilmore}, {Adibekyan}, {Bergemann}, {Cantat-Gaudin},
  {Delgado-Mena}, {Gonz{\'a}lez Hern{\'a}ndez}, {Guiglion}, {Lardo}, {de
  Laverny}, {Lind}, {Magrini}, {Mikolaitis}, {Montes}, {Pancino},
  {Recio-Blanco}, {Sordo}, {Sousa}, {Tabernero}, \& {Vallenari}}]{Jofre2015}
---. 2015, \aap, 582, A81

\bibitem[{{Jofr{\'e}} {et~al.}(2017){Jofr{\'e}}, {Heiter}, {Worley},
  {Blanco-Cuaresma}, {Soubiran}, {Masseron}, {Hawkins}, {Adibekyan}, {Buder},
  {Casamiquela}, {Gilmore}, {Hourihane}, \& {Tabernero}}]{Jofre2017}
{Jofr{\'e}}, P., {Heiter}, U., {Worley}, C.~C., {et~al.} 2017, \aap, 601, A38

\bibitem[{{Johnson} \& {Pilachowski}(2010)}]{Johnson2010}
{Johnson}, C.~I., \& {Pilachowski}, C.~A. 2010, \apj, 722, 1373

\bibitem[{{J{\"o}nsson} {et~al.}(2020){J{\"o}nsson}, {Holtzman}, {Prieto},
  {Cunha}, {Garc{\'\i}a-Hern{\'a}ndez}, {Hasselquist}, {Masseron}, {Osorio},
  {Shetrone}, {Smith}, {Stringfellow}, {Bizyaev}, {Edvardsson}, {Majewski},
  {M{\'e}sz{\'a}ros}, {Souto}, {Zamora}, {Beaton}, {Bovy}, {Donor},
  {Pinsonneault}, {Poovelil}, \& {Sobeck}}]{Joensson2020}
{J{\"o}nsson}, H., {Holtzman}, J.~A., {Prieto}, C.~A., {et~al.} 2020, \aj, 160,
  120

\bibitem[{{Katz} {et~al.}(2023){Katz}, {Sartoretti}, {Guerrier}, {Panuzzo},
  {Seabroke}, {Th{\'e}venin}, {Cropper}, {Benson}, {Blomme}, {Haigron},
  {Marchal}, {Smith}, {Baker}, {Chemin}, {Damerdji}, {David}, {Dolding},
  {Fr{\'e}mat}, {Gosset}, {Jan{\ss}en}, {Jasniewicz}, {Lobel}, {Plum},
  {Samaras}, {Snaith}, {Soubiran}, {Vanel}, {Zwitter}, {Antoja}, {Arenou},
  {Babusiaux}, {Brouillet}, {Caffau}, {Di Matteo}, {Fabre}, {Fabricius},
  {Fragkoudi}, {Haywood}, {Huckle}, {Hottier}, {Lasne}, {Leclerc},
  {Mastrobuono-Battisti}, {Royer}, {Teyssier}, {Zorec}, {Crifo}, {Jean-Antoine
  Piccolo}, {Turon}, \& {Viala}}]{Katz2023}
{Katz}, D., {Sartoretti}, P., {Guerrier}, A., {et~al.} 2023, \aap, 674, A5

\bibitem[{{Kawka} {et~al.}(2020){Kawka}, {Simpson}, {Vennes}, {Bessell}, {Da
  Costa}, {Marino}, \& {Murphy}}]{Kawka2020}
{Kawka}, A., {Simpson}, J.~D., {Vennes}, S., {et~al.} 2020, \mnras, 495, L129

\bibitem[{{Kelleher} \& {Podobedova}(2008)}]{2008JPCRD..37..709K}
{Kelleher}, D.~E., \& {Podobedova}, L.~I. 2008, JPCRD, 37, 709

\bibitem[{{Kobayashi} {et~al.}(2020){Kobayashi}, {Karakas}, \&
  {Lugaro}}]{Kobayashi2020}
{Kobayashi}, C., {Karakas}, A.~I., \& {Lugaro}, M. 2020, \apj, 900, 179

\bibitem[{{Kock} \& {Richter}(1968)}]{KR}
{Kock}, M., \& {Richter}, J. 1968, \zap, 69, 180, (KR)

\bibitem[{{Kollmeier} {et~al.}(2017){Kollmeier}, {Zasowski}, {Rix}, {Johns},
  {Anderson}, {Drory}, {Johnson}, {Pogge}, {Bird}, {Blanc}, {Brownstein},
  {Crane}, {De Lee}, {Klaene}, {Kreckel}, {MacDonald}, {Merloni}, {Ness},
  {O'Brien}, {Sanchez-Gallego}, {Sayres}, {Shen}, {Thakar}, {Tkachenko},
  {Aerts}, {Blanton}, {Eisenstein}, {Holtzman}, {Maoz}, {Nandra}, {Rockosi},
  {Weinberg}, {Bovy}, {Casey}, {Chaname}, {Clerc}, {Conroy}, {Eracleous},
  {G{\"a}nsicke}, {Hekker}, {Horne}, {Kauffmann}, {McQuinn}, {Pellegrini},
  {Schinnerer}, {Schlafly}, {Schwope}, {Seibert}, {Teske}, \& {van
  Saders}}]{Kollmeier2017}
{Kollmeier}, J.~A., {Zasowski}, G., {Rix}, H.-W., {et~al.} 2017, arXiv
  e-prints, arXiv:1711.03234

\bibitem[{{Kos} {et~al.}(2017){Kos}, {Lin}, {Zwitter}, {{\v Z}erjal}, {Sharma},
  {Bland-Hawthorn}, {Asplund}, {Casey}, {De Silva}, {Freeman}, {Martell},
  {Simpson}, {Schlesinger}, {Zucker}, {Anguiano}, {Bacigalupo}, {Bedding},
  {Betters}, {Da Costa}, {Duong}, {Hyde}, {Ireland}, {Kafle}, {Leon-Saval},
  {Lewis}, {Munari}, {Nataf}, {Stello}, {Tinney}, {Traven}, {Watson}, \&
  {Wittenmyer}}]{Kos2017}
{Kos}, J., {Lin}, J., {Zwitter}, T., {et~al.} 2017, \mnras, 464, 1259

\bibitem[{{Kos} {et~al.}(2018){Kos}, {Bland-Hawthorn}, {Betters}, {Leon-Saval},
  {Asplund}, {Buder}, {Casey}, {D'Orazi}, {de Silva}, {Freeman}, {Lewis},
  {Lin}, {Martell}, {Schlesinger}, {Sharma}, {Simpson}, {Zucker}, {Zwitter},
  {Hayden}, {Horner}, {Nataf}, \& {Ting}}]{Kos2018b}
{Kos}, J., {Bland-Hawthorn}, J., {Betters}, C.~H., {et~al.} 2018, \mnras, 480,
  5475

\bibitem[{{Kos} {et~al.}(2021){Kos}, {Bland-Hawthorn}, {Buder}, {Nordlander},
  {Spina}, {Beeson}, {Lind}, {Asplund}, {Freeman}, {Hayden}, {Lewis},
  {Martell}, {Sharma}, {De Silva}, {Simpson}, {Zucker}, {Zwitter},
  {{\v{C}}otar}, {Horner}, {Ting}, \& {Traven}}]{Kos2021}
{Kos}, J., {Bland-Hawthorn}, J., {Buder}, S., {et~al.} 2021, \mnras, 506, 4232

\bibitem[{{Kos} {et~al.}(2025){Kos}, {Buder}, {Beeson}, {Bland-Hawthorn}, {De
  Silva}, {D'Orazi}, {Freeman}, {Hayden}, {Lewis}, {Lind}, {Martell}, {Sharma},
  {Zucker}, {Zwitter}, {Stello}, \& {de Grijs}}]{Kos2025}
{Kos}, J., {Buder}, S., {Beeson}, K.~L., {et~al.} 2025, arXiv e-prints,
  arXiv:2501.06140

\bibitem[{{Kurucz}(2006)}]{K06}
{Kurucz}, R.~L. 2006, Database of observed and predicted atomic transitions

\bibitem[{{Kurucz}(2007)}]{K07}
---. 2007, Database of observed and predicted atomic transitions

\bibitem[{{Kurucz}(2008)}]{K08}
---. 2008, Database of observed and predicted atomic transitions

\bibitem[{{Kurucz}(2009)}]{K09}
---. 2009, Database of observed and predicted atomic transitions

\bibitem[{{Kurucz}(2010)}]{K10}
---. 2010, Database of observed and predicted atomic transitions

\bibitem[{{Kurucz}(2013)}]{K13}
---. 2013, Database of observed and predicted atomic transitions

\bibitem[{{Kurucz}(2014)}]{K14}
---. 2014, Database of observed and predicted atomic transitions

\bibitem[{{Lawler} {et~al.}(2001{\natexlab{a}}){Lawler}, {Bonvallet}, \&
  {Sneden}}]{LBS}
{Lawler}, J.~E., {Bonvallet}, G., \& {Sneden}, C. 2001{\natexlab{a}},
  Astrophys. J., 556, 452, (LBS)

\bibitem[{{Lawler} \& {Dakin}(1989)}]{LD}
{Lawler}, J.~E., \& {Dakin}, J.~T. 1989, JOSA B Optical Physics, 6, 1457, (LD)

\bibitem[{{Lawler} {et~al.}(2006){Lawler}, {Den Hartog}, {Sneden}, \&
  {Cowan}}]{LD-HS}
{Lawler}, J.~E., {Den Hartog}, E.~A., {Sneden}, C., \& {Cowan}, J.~J. 2006,
  Astrophys. J. Suppl. Ser., 162, 227, (LD-HS)

\bibitem[{{Lawler} {et~al.}(2013){Lawler}, {Guzman}, {Wood}, {Sneden}, \&
  {Cowan}}]{LGWSC}
{Lawler}, J.~E., {Guzman}, A., {Wood}, M.~P., {Sneden}, C., \& {Cowan}, J.~J.
  2013, \apjs, 205, 11

\bibitem[{{Lawler} {et~al.}(2015){Lawler}, {Sneden}, \&
  {Cowan}}]{2015ApJS..220...13L}
{Lawler}, J.~E., {Sneden}, C., \& {Cowan}, J.~J. 2015, \apjs, 220, 13

\bibitem[{{Lawler} {et~al.}(2009){Lawler}, {Sneden}, {Cowan}, {Ivans}, \& {Den
  Hartog}}]{LSCI}
{Lawler}, J.~E., {Sneden}, C., {Cowan}, J.~J., {Ivans}, I.~I., \& {Den Hartog},
  E.~A. 2009, Astrophys. J. Suppl. Ser., 182, 51, (LSCI)

\bibitem[{{Lawler} {et~al.}(2001{\natexlab{b}}){Lawler}, {Wickliffe}, {den
  Hartog}, \& {Sneden}}]{LWHS}
{Lawler}, J.~E., {Wickliffe}, M.~E., {den Hartog}, E.~A., \& {Sneden}, C.
  2001{\natexlab{b}}, Astrophys. J., 563, 1075, (LWHS)

\bibitem[{{Lawler} {et~al.}(2014){Lawler}, {Wood}, {Den Hartog}, {Feigenson},
  {Sneden}, \& {Cowan}}]{2014ApJS..215...20L}
{Lawler}, J.~E., {Wood}, M.~P., {Den Hartog}, E.~A., {et~al.} 2014, \apjs, 215,
  20

\bibitem[{{Leenaarts} \& {Carlsson}(2009)}]{Leenaarts2009}
{Leenaarts}, J., \& {Carlsson}, M. 2009, \aspc, 415, 87

\bibitem[{{Lewis}(2019)}]{Lewis2019}
{Lewis}, A. 2019, arXiv e-prints, arXiv:1910.13970

\bibitem[{{Lewis} {et~al.}(2002){Lewis}, {Cannon}, {Taylor}, {Glazebrook},
  {Bailey}, {Baldry}, {Barton}, {Bridges}, {Dalton}, {Farrell}, {Gray},
  {Lankshear}, {McCowage}, {Parry}, {Sharples}, {Shortridge}, {Smith},
  {Stevenson}, {Straede}, {Waller}, {Whittard}, {Wilcox}, \&
  {Willis}}]{Lewis2002}
{Lewis}, I.~J., {Cannon}, R.~D., {Taylor}, K., {et~al.} 2002, \mnras, 333, 279

\bibitem[{{Lind} {et~al.}(2009{\natexlab{a}}){Lind}, {Asplund}, \&
  {Barklem}}]{Lind2009}
{Lind}, K., {Asplund}, M., \& {Barklem}, P.~S. 2009{\natexlab{a}}, \aap, 503,
  541

\bibitem[{{Lind} {et~al.}(2011){Lind}, {Asplund}, {Barklem}, \&
  {Belyaev}}]{Lind2011}
{Lind}, K., {Asplund}, M., {Barklem}, P.~S., \& {Belyaev}, A.~K. 2011, \aap,
  528, A103

\bibitem[{{Lind} {et~al.}(2009{\natexlab{b}}){Lind}, {Primas}, {Charbonnel},
  {Grundahl}, \& {Asplund}}]{Lind2009b}
{Lind}, K., {Primas}, F., {Charbonnel}, C., {Grundahl}, F., \& {Asplund}, M.
  2009{\natexlab{b}}, \aap, 503, 545

\bibitem[{{Lindegren} {et~al.}(2018){Lindegren}, {Hern{\'a}ndez}, {Bombrun},
  {Klioner}, {Bastian}, {Ramos-Lerate}, {de Torres}, {Steidelm{\"u}ller},
  {Stephenson}, {Hobbs}, {Lammers}, {Biermann}, {Geyer}, {Hilger}, {Michalik},
  {Stampa}, {McMillan}, {Casta{\~n}eda}, {Clotet}, {Comoretto}, {Davidson},
  {Fabricius}, {Gracia}, {Hambly}, {Hutton}, {Mora}, {Portell}, {van Leeuwen},
  {Abbas}, {Abreu}, {Altmann}, {Andrei}, {Anglada}, {Balaguer-N{\'u}{\~n}ez},
  {Barache}, {Becciani}, {Bertone}, {Bianchi}, {Bouquillon}, {Bourda},
  {Br{\"u}semeister}, {Bucciarelli}, {Busonero}, {Buzzi}, {Cancelliere},
  {Carlucci}, {Charlot}, {Cheek}, {Crosta}, {Crowley}, {de Bruijne}, {de
  Felice}, {Drimmel}, {Esquej}, {Fienga}, {Fraile}, {Gai}, {Garralda},
  {Gonz{\'a}lez-Vidal}, {Guerra}, {Hauser}, {Hofmann}, {Holl}, {Jordan},
  {Lattanzi}, {Lenhardt}, {Liao}, {Licata}, {Lister}, {L{\"o}ffler},
  {Marchant}, {Martin-Fleitas}, {Messineo}, {Mignard}, {Morbidelli}, {Poggio},
  {Riva}, {Rowell}, {Salguero}, {Sarasso}, {Sciacca}, {Siddiqui}, {Smart},
  {Spagna}, {Steele}, {Taris}, {Torra}, {van Elteren}, {van Reeven}, \&
  {Vecchiato}}]{Lindegren2018}
{Lindegren}, L., {Hern{\'a}ndez}, J., {Bombrun}, A., {et~al.} 2018, \aap, 616,
  A2

\bibitem[{{Lindegren} {et~al.}(2021{\natexlab{a}}){Lindegren}, {Bastian},
  {Biermann}, {Bombrun}, {de Torres}, {Gerlach}, {Geyer}, {Hern{\'a}ndez},
  {Hilger}, {Hobbs}, {Klioner}, {Lammers}, {McMillan}, {Ramos-Lerate},
  {Steidelm{\"u}ller}, {Stephenson}, \& {van Leeuwen}}]{Lindegren2021b}
{Lindegren}, L., {Bastian}, U., {Biermann}, M., {et~al.} 2021{\natexlab{a}},
  \aap, 649, A4

\bibitem[{{Lindegren} {et~al.}(2021{\natexlab{b}}){Lindegren}, {Klioner},
  {Hern{\'a}ndez}, {Bombrun}, {Ramos-Lerate}, {Steidelm{\"u}ller}, {Bastian},
  {Biermann}, {de Torres}, {Gerlach}, {Geyer}, {Hilger}, {Hobbs}, {Lammers},
  {McMillan}, {Stephenson}, {Casta{\~n}eda}, {Davidson}, {Fabricius},
  {Gracia-Abril}, {Portell}, {Rowell}, {Teyssier}, {Torra}, {Bartolom{\'e}},
  {Clotet}, {Garralda}, {Gonz{\'a}lez-Vidal}, {Torra}, {Abbas}, {Altmann},
  {Anglada Varela}, {Balaguer-N{\'u}{\~n}ez}, {Balog}, {Barache}, {Becciani},
  {Bernet}, {Bertone}, {Bianchi}, {Bouquillon}, {Brown}, {Bucciarelli},
  {Busonero}, {Butkevich}, {Buzzi}, {Cancelliere}, {Carlucci}, {Charlot},
  {Cioni}, {Crosta}, {Crowley}, {del Peloso}, {del Pozo}, {Drimmel}, {Esquej},
  {Fienga}, {Fraile}, {Gai}, {Garcia-Reinaldos}, {Guerra}, {Hambly}, {Hauser},
  {Jan{\ss}en}, {Jordan}, {Kostrzewa-Rutkowska}, {Lattanzi}, {Liao}, {Licata},
  {Lister}, {L{\"o}ffler}, {Marchant}, {Masip}, {Mignard}, {Mints}, {Molina},
  {Mora}, {Morbidelli}, {Murphy}, {Pagani}, {Panuzzo}, {Pe{\~n}alosa Esteller},
  {Poggio}, {Re Fiorentin}, {Riva}, {Sagrist{\`a} Sell{\'e}s}, {Sanchez
  Gimenez}, {Sarasso}, {Sciacca}, {Siddiqui}, {Smart}, {Souami}, {Spagna},
  {Steele}, {Taris}, {Utrilla}, {van Reeven}, \& {Vecchiato}}]{Lindegren2021a}
{Lindegren}, L., {Klioner}, S.~A., {Hern{\'a}ndez}, J., {et~al.}
  2021{\natexlab{b}}, \aap, 649, A2

\bibitem[{{Mackereth} \& {Bovy}(2018)}]{Mackereth2018}
{Mackereth}, J.~T., \& {Bovy}, J. 2018, \pasp, 130, 114501

\bibitem[{{Magrini} {et~al.}(2023){Magrini}, {Bensby}, {Brucalassi}, {Randich},
  {Jeffries}, {de Silva}, {Skuladottir}, {Smiljanic}, {Gonzalez}, {Hill},
  {Lagarde}, {Tolstoy}, {Arroyo-Polonio}, {Baratella}, {Barnes}, {Battaglia},
  {Baumgardt}, {Bellazzini}, {Biazzo}, {Bragaglia}, {Carter}, {Casali},
  {Cescutti}, {Danielski}, {Delgado Mena}, {Drazdauskas}, {Gieles},
  {Giribaldi}, {Hawkins}, {Hoeijmakers}, {Jablonka}, {Kamath}, {Louth},
  {Fabiola Marino}, {Martell}, {Merle}, {Montet}, {Murphy}, {Nisini},
  {Nordlander}, {D'Orazi}, {Pino}, {Romano}, {Sacco}, {Sandford}, {Sollima},
  {Spina}, {Tautvaisiene}, {Ting}, {Tozzi}, {Van der Swaelmen}, {Van Eck},
  {Watson}, {Worley}, \& {Zocchi}}]{HRMOS2023}
{Magrini}, L., {Bensby}, T., {Brucalassi}, A., {et~al.} 2023, arXiv e-prints,
  arXiv:2312.08270

\bibitem[{{Majewski} {et~al.}(2011){Majewski}, {Zasowski}, \&
  {Nidever}}]{Majewski2011}
{Majewski}, S.~R., {Zasowski}, G., \& {Nidever}, D.~L. 2011, \apj, 739, 25

\bibitem[{{Manea} {et~al.}(2022){Manea}, {Hawkins}, \& {Maas}}]{Manea2022}
{Manea}, C., {Hawkins}, K., \& {Maas}, Z.~G. 2022, \mnras, 511, 2829

\bibitem[{{Manea} {et~al.}(2024){Manea}, {Hawkins}, {Ness}, {Buder}, {Martell},
  \& {Zucker}}]{Manea2024}
{Manea}, C., {Hawkins}, K., {Ness}, M.~K., {et~al.} 2024, \apj, 972, 69

\bibitem[{{Mann} {et~al.}(2012){Mann}, {Gaidos}, {L{\'e}pine}, \&
  {Hilton}}]{Mann2012}
{Mann}, A.~W., {Gaidos}, E., {L{\'e}pine}, S., \& {Hilton}, E.~J. 2012, \apj,
  753, 90

\bibitem[{{Marigo} {et~al.}(2017){Marigo}, {Girardi}, {Bressan}, {Rosenfield},
  {Aringer}, {Chen}, {Dussin}, {Nanni}, {Pastorelli}, {Rodrigues}, {Trabucchi},
  {Bladh}, {Dalcanton}, {Groenewegen}, {Montalb{\'a}n}, \& {Wood}}]{Marigo2017}
{Marigo}, P., {Girardi}, L., {Bressan}, A., {et~al.} 2017, \apj, 835, 77

\bibitem[{{Marino} {et~al.}(2011){Marino}, {Sneden}, {Kraft}, {Wallerstein},
  {Norris}, {Da Costa}, {Milone}, {Ivans}, {Gonzalez}, {Fulbright}, {Hilker},
  {Piotto}, {Zoccali}, \& {Stetson}}]{Marino2011}
{Marino}, A.~F., {Sneden}, C., {Kraft}, R.~P., {et~al.} 2011, \aap, 532, A8

\bibitem[{{Martell} {et~al.}(2017){Martell}, {Sharma}, {Buder}, {Duong},
  {Schlesinger}, {Simpson}, {Lind}, {Ness}, {Marshall}, {Asplund},
  {Bland-Hawthorn}, {Casey}, {De Silva}, {Freeman}, {Kos}, {Lin}, {Zucker},
  {Zwitter}, {Anguiano}, {Bacigalupo}, {Carollo}, {Casagrande}, {Da Costa},
  {Horner}, {Huber}, {Hyde}, {Kafle}, {Lewis}, {Nataf}, {Navin}, {Stello},
  {Tinney}, {Watson}, \& {Wittenmyer}}]{Martell2017}
{Martell}, S.~L., {Sharma}, S., {Buder}, S., {et~al.} 2017, \mnras, 465, 3203

\bibitem[{{Martell} {et~al.}(2021){Martell}, {Simpson}, {Balasubramaniam},
  {Buder}, {Sharma}, {Hon}, {Stello}, {Ting}, {Asplund}, {Bland-Hawthorn}, {De
  Silva}, {Freeman}, {Hayden}, {Kos}, {Lewis}, {Lind}, {Zucker}, {Zwitter},
  {Campbell}, {{\v{C}}otar}, {Horner}, {Montet}, \& {Wittenmyer}}]{Martell2021}
{Martell}, S.~L., {Simpson}, J.~D., {Balasubramaniam}, A.~G., {et~al.} 2021,
  \mnras, 505, 5340

\bibitem[{{Martig} {et~al.}(2016){Martig}, {Fouesneau}, {Rix}, {Ness},
  {M{\'e}sz{\'a}ros}, {Garc{\'{\i}}a-Hern{\'a}ndez}, {Pinsonneault},
  {Serenelli}, {Silva Aguirre}, \& {Zamora}}]{Martig2016}
{Martig}, M., {Fouesneau}, M., {Rix}, H.-W., {et~al.} 2016, \mnras, 456, 3655

\bibitem[{{Martin} {et~al.}(1988){Martin}, {Fuhr}, \& {Wiese}}]{MFW}
{Martin}, G., {Fuhr}, J., \& {Wiese}, W. 1988, J. Phys. Chem. Ref. Data Suppl.,
  17

\bibitem[{{Masseron} \& {Gilmore}(2015)}]{Masseron2015}
{Masseron}, T., \& {Gilmore}, G. 2015, \mnras, 453, 1855

\bibitem[{{Matsuno} {et~al.}(2021){Matsuno}, {Hirai}, {Tarumi}, {Hotokezaka},
  {Tanaka}, \& {Helmi}}]{Matsuno2021}
{Matsuno}, T., {Hirai}, Y., {Tarumi}, Y., {et~al.} 2021, \aap, 650, A110

\bibitem[{{May} {et~al.}(1974){May}, {Richter}, \& {Wichelmann}}]{MRW}
{May}, M., {Richter}, J., \& {Wichelmann}, J. 1974, \aaps, 18, 405, (MRW)

\bibitem[{{McKemmish} {et~al.}(2019){McKemmish}, {Masseron}, {Hoeijmakers},
  {P{\'e}rez-Mesa}, {Grimm}, {Yurchenko}, \& {Tennyson}}]{McKemmish2019}
{McKemmish}, L.~K., {Masseron}, T., {Hoeijmakers}, H.~J., {et~al.} 2019,
  \mnras, 488, 2836

\bibitem[{{McKenzie} {et~al.}(2022){McKenzie}, {Yong}, {Marino}, {Monty},
  {Wang}, {Karakas}, {Milone}, {Legnardi}, {Roederer}, {Martell}, \&
  {Horta}}]{McKenzie2022}
{McKenzie}, M., {Yong}, D., {Marino}, A.~F., {et~al.} 2022, \mnras, 516, 3515

\bibitem[{{McKenzie} {et~al.}(2024){McKenzie}, {Yong}, {Karakas}, {Wang},
  {Monty}, {Marino}, {Milone}, {Nordlander}, {Mura-Guzm{\'a}n}, {Martell}, \&
  {Carlos}}]{McKenzie2024}
{McKenzie}, M., {Yong}, D., {Karakas}, A.~I., {et~al.} 2024, \mnras, 527, 7940

\bibitem[{{McMillan}(2017)}]{McMillan2017}
{McMillan}, P.~J. 2017, \mnras, 465, 76

\bibitem[{{Meggers} {et~al.}(1975){Meggers}, {Corliss}, \& {Scribner}}]{MC}
{Meggers}, W.~F., {Corliss}, C.~H., \& {Scribner}, B.~F. 1975, {Tables of
  spectral-line intensities. Part I, II\_- arranged by elements.} (NBS), (MC)

\bibitem[{{Mel{\'e}ndez} \& {Barbuy}(2009)}]{2009A&A...497..611M}
{Mel{\'e}ndez}, J., \& {Barbuy}, B. 2009, \aap, 497, 611

\bibitem[{{M{\'e}sz{\'a}ros} {et~al.}(2012){M{\'e}sz{\'a}ros}, {Allende
  Prieto}, {Edvardsson}, {Castelli}, {Garc{\'\i}a P{\'e}rez}, {Gustafsson},
  {Majewski}, {Plez}, {Schiavon}, {Shetrone}, \& {de Vicente}}]{Meszaros2012}
{M{\'e}sz{\'a}ros}, S., {Allende Prieto}, C., {Edvardsson}, B., {et~al.} 2012,
  \aj, 144, 120

\bibitem[{{Milone} \& {Marino}(2022)}]{Milone2022}
{Milone}, A.~P., \& {Marino}, A.~F. 2022, Universe, 8, 359

\bibitem[{{Miszalski} {et~al.}(2006){Miszalski}, {Shortridge}, {Saunders},
  {Parker}, \& {Croom}}]{Miszalski2006}
{Miszalski}, B., {Shortridge}, K., {Saunders}, W., {Parker}, Q.~A., \& {Croom},
  S.~M. 2006, \mnras, 371, 1537

\bibitem[{{Monty} {et~al.}(2023){Monty}, {Yong}, {Marino}, {Karakas},
  {McKenzie}, {Grundahl}, \& {Mura-Guzm{\'a}n}}]{Monty2023}
{Monty}, S., {Yong}, D., {Marino}, A.~F., {et~al.} 2023, \mnras, 518, 965

\bibitem[{{Myeong} {et~al.}(2022){Myeong}, {Belokurov}, {Aguado}, {Evans},
  {Caldwell}, \& {Bradley}}]{Myeong2022}
{Myeong}, G.~C., {Belokurov}, V., {Aguado}, D.~S., {et~al.} 2022, \apj, 938, 21

\bibitem[{{Nahar}(1993)}]{1993PhyS...48..297N}
{Nahar}, S.~N. 1993, \physscr, 48, 297

\bibitem[{{Nandakumar} {et~al.}(2022){Nandakumar}, {Hayden}, {Sharma}, {Buder},
  {Asplund}, {Bland-Hawthorn}, {De Silva}, {D'Orazi}, {Freeman}, {Kos},
  {Lewis}, {Martell}, {Schlesinger}, {Lin}, {Simpson}, {Zucker}, {Zwitter},
  {Nordlander}, {Casagrande}, {Lind}, {C{\^o}tar}, {Stello}, {Wittenmyer}, \&
  {Tepper-Garcia}}]{Nandakumar2022}
{Nandakumar}, G., {Hayden}, M.~R., {Sharma}, S., {et~al.} 2022, \mnras, 513,
  232

\bibitem[{{Ness} {et~al.}(2015){Ness}, {Hogg}, {Rix}, {Ho}, \&
  {Zasowski}}]{Ness2015}
{Ness}, M., {Hogg}, D.~W., {Rix}, H.-W., {Ho}, A.~Y.~Q., \& {Zasowski}, G.
  2015, \apj, 808, 16

\bibitem[{{Ness} {et~al.}(2016){Ness}, {Hogg}, {Rix}, {Martig}, {Pinsonneault},
  \& {Ho}}]{Ness2016}
{Ness}, M., {Hogg}, D.~W., {Rix}, H.-W., {et~al.} 2016, \apj, 823, 114

\bibitem[{{Ness} {et~al.}(2019){Ness}, {Johnston}, {Blancato}, {Rix}, {Beane},
  {Bird}, \& {Hawkins}}]{Ness2019b}
{Ness}, M.~K., {Johnston}, K.~V., {Blancato}, K., {et~al.} 2019, \apj, 883, 177

\bibitem[{{Nissen}(2015)}]{Nissen2015}
{Nissen}, P.~E. 2015, \aap, 579, A52

\bibitem[{{Nissen} {et~al.}(2020){Nissen}, {Christensen-Dalsgaard},
  {Mosumgaard}, {Silva Aguirre}, {Spitoni}, \& {Verma}}]{Nissen2020}
{Nissen}, P.~E., {Christensen-Dalsgaard}, J., {Mosumgaard}, J.~R., {et~al.}
  2020, \aap, 640, A81

\bibitem[{{Nissen} \& {Gustafsson}(2018)}]{Nissen2018}
{Nissen}, P.~E., \& {Gustafsson}, B. 2018, \aapr, 26, 6

\bibitem[{{Nitz} {et~al.}(1999){Nitz}, {Kunau}, {Wilson}, \&
  {Lentz}}]{1999ApJS..122..557N}
{Nitz}, D.~E., {Kunau}, A.~E., {Wilson}, K.~L., \& {Lentz}, L.~R. 1999, \apjs,
  122, 557

\bibitem[{{Nitz} {et~al.}(1998){Nitz}, {Wickliffe}, \& {Lawler}}]{NWL}
{Nitz}, D.~E., {Wickliffe}, M.~E., \& {Lawler}, J.~E. 1998, \apjs, 117, 313,
  (NWL)

\bibitem[{{Nordlander} \& {Lind}(2017)}]{Nordlander2017}
{Nordlander}, T., \& {Lind}, K. 2017, \aap, 607, A75

\bibitem[{{O'brian} \& {Lawler}(1991)}]{BL}
{O'brian}, T.~R., \& {Lawler}, J.~E. 1991, \pra, 44, 7134, (BL)

\bibitem[{{O'Brian} {et~al.}(1991){O'Brian}, {Wickliffe}, {Lawler}, {Whaling},
  \& {Brault}}]{BWL}
{O'Brian}, T.~R., {Wickliffe}, M.~E., {Lawler}, J.~E., {Whaling}, W., \&
  {Brault}, J.~W. 1991, JOSA B Optical Physics, 8, 1185, (BWL)

\bibitem[{{Ochsenbein} {et~al.}(2000){Ochsenbein}, {Bauer}, \&
  {Marcout}}]{Vizier2000}
{Ochsenbein}, F., {Bauer}, P., \& {Marcout}, J. 2000, \aaps, 143, 23

\bibitem[{{Osorio} {et~al.}(2015){Osorio}, {Barklem}, {Lind}, {Belyaev},
  {Spielfiedel}, {Guitou}, \& {Feautrier}}]{Osorio2015}
{Osorio}, Y., {Barklem}, P.~S., {Lind}, K., {et~al.} 2015, \aap, 579, A53

\bibitem[{{Osorio} {et~al.}(2019){Osorio}, {Lind}, {Barklem}, {Allende Prieto},
  \& {Zatsarinny}}]{Osorio2019}
{Osorio}, Y., {Lind}, K., {Barklem}, P.~S., {Allende Prieto}, C., \&
  {Zatsarinny}, O. 2019, \aap, 623, A103

\bibitem[{{Palmeri} {et~al.}(2017){Palmeri}, {Quinet}, {Lundberg},
  {Engstr{\"o}m}, {Nilsson}, \& {Hartman}}]{2017MNRAS.471..532P}
{Palmeri}, P., {Quinet}, P., {Lundberg}, H., {et~al.} 2017, \mnras, 471, 532

\bibitem[{{Palmeri} {et~al.}(2000){Palmeri}, {Quinet}, {Wyart}, \&
  {Bi{\'e}mont}}]{PQWB}
{Palmeri}, P., {Quinet}, P., {Wyart}, J., \& {Bi{\'e}mont}, E. 2000, Physica
  Scripta, 61, 323, (PQWB)

\bibitem[{Pedregosa {et~al.}(2011)Pedregosa, Varoquaux, Gramfort, Michel,
  Thirion, Grisel, Blondel, Prettenhofer, Weiss, Dubourg, Vanderplas, Passos,
  Cournapeau, Brucher, Perrot, \& Duchesnay}]{scikit-learn}
Pedregosa, F., Varoquaux, G., Gramfort, A., {et~al.} 2011, J Mach Learn Res,
  12, 2825

\bibitem[{P\'erez \& Granger(2007)}]{ipython}
P\'erez, F., \& Granger, B.~E. 2007, Comput Sci Eng, 9, 21

\bibitem[{{Piskunov} \& {Valenti}(2017)}]{Piskunov2017}
{Piskunov}, N., \& {Valenti}, J.~A. 2017, \aap, 597, A16

\bibitem[{{Pr{\v{s}}a} {et~al.}(2016){Pr{\v{s}}a}, {Harmanec}, {Torres},
  {Mamajek}, {Asplund}, {Capitaine}, {Christensen-Dalsgaard}, {Depagne},
  {Haberreiter}, {Hekker}, {Hilton}, {Kopp}, {Kostov}, {Kurtz}, {Laskar},
  {Mason}, {Milone}, {Montgomery}, {Richards}, {Schmutz}, {Schou}, \&
  {Stewart}}]{Prsa2016}
{Pr{\v{s}}a}, A., {Harmanec}, P., {Torres}, G., {et~al.} 2016, \aj, 152, 41

\bibitem[{{Queiroz} {et~al.}(2023){Queiroz}, {Anders}, {Chiappini},
  {Khalatyan}, {Santiago}, {Nepal}, {Steinmetz}, {Gallart}, {Valentini}, {Dal
  Ponte}, {Barbuy}, {P{\'e}rez-Villegas}, {Masseron}, {Fern{\'a}ndez-Trincado},
  {Khoperskov}, {Minchev}, {Fern{\'a}ndez-Alvar}, {Lane}, \&
  {Nitschelm}}]{Queiroz2023}
{Queiroz}, A.~B.~A., {Anders}, F., {Chiappini}, C., {et~al.} 2023, \aap, 673,
  A155

\bibitem[{{Raassen} \& {Uylings}(1998)}]{RU}
{Raassen}, A.~J.~J., \& {Uylings}, P.~H.~M. 1998, \aap, 340, 300, (RU)

\bibitem[{{Rains} {et~al.}(2021){Rains}, {{\v{Z}}erjal}, {Ireland},
  {Nordlander}, {Bessell}, {Casagrande}, {Onken}, {Joyce}, {Kammerer}, \&
  {Abbot}}]{Rains2021}
{Rains}, A.~D., {{\v{Z}}erjal}, M., {Ireland}, M.~J., {et~al.} 2021, \mnras,
  504, 5788

\bibitem[{{Rains} {et~al.}(2024){Rains}, {Nordlander}, {Monty}, {Casey},
  {Rojas-Ayala}, {{\v{Z}}erjal}, {Ireland}, {Casagrande}, \&
  {McKenzie}}]{Rains2024}
{Rains}, A.~D., {Nordlander}, T., {Monty}, S., {et~al.} 2024, \mnras, 529, 3171

\bibitem[{{Ralchenko} {et~al.}(2010){Ralchenko}, {Kramida}, {Reader}, \& {NIST
  ASD Team}}]{NIST10}
{Ralchenko}, Y., {Kramida}, A., {Reader}, J., \& {NIST ASD Team}. 2010, NIST
  Atomic Spectra Database (ver. 4.0.0), [Online].

\bibitem[{{Ratcliffe} {et~al.}(2024){Ratcliffe}, {Minchev}, {Cescutti},
  {Spitoni}, {J{\"o}nsson}, {Anders}, {Queiroz}, \&
  {Steinmetz}}]{Ratcliffe2024}
{Ratcliffe}, B., {Minchev}, I., {Cescutti}, G., {et~al.} 2024, \mnras, 528,
  3464

\bibitem[{{Recio-Blanco} {et~al.}(2023){Recio-Blanco}, {de Laverny}, {Palicio},
  {Kordopatis}, {{\'A}lvarez}, {Schultheis}, {Contursi}, {Zhao}, {Torralba
  Elipe}, {Ordenovic}, {Manteiga}, {Dafonte}, {Oreshina-Slezak}, {Bijaoui},
  {Fr{\'e}mat}, {Seabroke}, {Pailler}, {Spitoni}, {Poggio}, {Creevey}, {Abreu
  Aramburu}, {Accart}, {Andrae}, {Bailer-Jones}, {Bellas-Velidis}, {Brouillet},
  {Brugaletta}, {Burlacu}, {Carballo}, {Casamiquela}, {Chiavassa}, {Cooper},
  {Dapergolas}, {Delchambre}, {Dharmawardena}, {Drimmel}, {Edvardsson},
  {Fouesneau}, {Garabato}, {Garc{\'\i}a-Lario}, {Garc{\'\i}a-Torres}, {Gavel},
  {Gomez}, {Gonz{\'a}lez-Santamar{\'\i}a}, {Hatzidimitriou}, {Heiter},
  {Jean-Antoine Piccolo}, {Kontizas}, {Korn}, {Lanzafame}, {Lebreton}, {Le
  Fustec}, {Licata}, {Lindstr{\o}m}, {Livanou}, {Lobel}, {Lorca}, {Magdaleno
  Romeo}, {Marocco}, {Marshall}, {Mary}, {Nicolas}, {Pallas-Quintela}, {Panem},
  {Pichon}, {Riclet}, {Robin}, {Rybizki}, {Santove{\~n}a}, {Silvelo}, {Smart},
  {Sarro}, {Sordo}, {Soubiran}, {S{\"u}veges}, {Ulla}, {Vallenari}, {Zorec},
  {Utrilla}, \& {Bakker}}]{RecioBlanco2023}
{Recio-Blanco}, A., {de Laverny}, P., {Palicio}, P.~A., {et~al.} 2023, \aap,
  674, A29

\bibitem[{{Reggiani} {et~al.}(2019){Reggiani}, {Amarsi}, {Lind}, {Barklem},
  {Zatsarinny}, {Bartschat}, {Fursa}, {Bray}, {Spina}, \&
  {Mel{\'e}ndez}}]{Reggiani2019}
{Reggiani}, H., {Amarsi}, A.~M., {Lind}, K., {et~al.} 2019, \aap, 627, A177

\bibitem[{{Reid} \& {Brunthaler}(2004)}]{Reid2004}
{Reid}, M.~J., \& {Brunthaler}, A. 2004, \apj, 616, 872

\bibitem[{{Ricker} {et~al.}(2015){Ricker}, {Winn}, {Vanderspek}, {Latham},
  {Bakos}, {Bean}, {Berta-Thompson}, {Brown}, {Buchhave}, {Butler}, {Butler},
  {Chaplin}, {Charbonneau}, {Christensen-Dalsgaard}, {Clampin}, {Deming},
  {Doty}, {De Lee}, {Dressing}, {Dunham}, {Endl}, {Fressin}, {Ge}, {Henning},
  {Holman}, {Howard}, {Ida}, {Jenkins}, {Jernigan}, {Johnson}, {Kaltenegger},
  {Kawai}, {Kjeldsen}, {Laughlin}, {Levine}, {Lin}, {Lissauer}, {MacQueen},
  {Marcy}, {McCullough}, {Morton}, {Narita}, {Paegert}, {Palle}, {Pepe},
  {Pepper}, {Quirrenbach}, {Rinehart}, {Sasselov}, {Sato}, {Seager},
  {Sozzetti}, {Stassun}, {Sullivan}, {Szentgyorgyi}, {Torres}, {Udry}, \&
  {Villasenor}}]{Ricker2015}
{Ricker}, G.~R., {Winn}, J.~N., {Vanderspek}, R., {et~al.} 2015, Journal of
  Astronomical Telescopes, Instruments, and Systems, 1, 014003

\bibitem[{{Riello} {et~al.}(2021){Riello}, {De Angeli}, {Evans}, {Montegriffo},
  {Carrasco}, {Busso}, {Palaversa}, {Burgess}, {Diener}, {Davidson}, {Rowell},
  {Fabricius}, {Jordi}, {Bellazzini}, {Pancino}, {Harrison}, {Cacciari}, {van
  Leeuwen}, {Hambly}, {Hodgkin}, {Osborne}, {Altavilla}, {Barstow}, {Brown},
  {Castellani}, {Cowell}, {De Luise}, {Gilmore}, {Giuffrida}, {Hidalgo},
  {Holland}, {Marinoni}, {Pagani}, {Piersimoni}, {Pulone}, {Ragaini}, {Rainer},
  {Richards}, {Sanna}, {Walton}, {Weiler}, \& {Yoldas}}]{Riello2021}
{Riello}, M., {De Angeli}, F., {Evans}, D.~W., {et~al.} 2021, \aap, 649, A3

\bibitem[{{Rimoldini} {et~al.}(2023){Rimoldini}, {Holl}, {Gavras}, {Audard},
  {De Ridder}, {Mowlavi}, {Nienartowicz}, {Jevardat de Fombelle},
  {Lecoeur-Ta{\"\i}bi}, {Karbevska}, {Evans}, {{\'A}brah{\'a}m}, {Carnerero},
  {Clementini}, {Distefano}, {Garofalo}, {Garc{\'\i}a-Lario}, {Gomel},
  {Klioner}, {Kruszy{\'n}ska}, {Lanzafame}, {Lebzelter}, {Marton}, {Mazeh},
  {Molinaro}, {Panahi}, {Raiteri}, {Ripepi}, {Szabados}, {Teyssier},
  {Trabucchi}, {Wyrzykowski}, {Zucker}, \& {Eyer}}]{Rimoldini2023}
{Rimoldini}, L., {Holl}, B., {Gavras}, P., {et~al.} 2023, \aap, 674, A14

\bibitem[{{Rix} {et~al.}(2016){Rix}, {Ting}, {Conroy}, \& {Hogg}}]{Rix2016}
{Rix}, H.-W., {Ting}, Y.-S., {Conroy}, C., \& {Hogg}, D.~W. 2016, \apjl, 826,
  L25

\bibitem[{{Ruffoni} {et~al.}(2014){Ruffoni}, {Den Hartog}, {Lawler}, {Brewer},
  {Lind}, {Nave}, \& {Pickering}}]{2014MNRAS.441.3127R}
{Ruffoni}, M.~P., {Den Hartog}, E.~A., {Lawler}, J.~E., {et~al.} 2014, \mnras,
  441, 3127

\bibitem[{{Sahlholdt} {et~al.}(2022){Sahlholdt}, {Feltzing}, \&
  {Feuillet}}]{Sahlholdt2022}
{Sahlholdt}, C.~L., {Feltzing}, S., \& {Feuillet}, D.~K. 2022, \mnras, 510,
  4669

\bibitem[{{Salaris} \& {Cassisi}(2006)}]{Salaris2006}
{Salaris}, M., \& {Cassisi}, S. 2006, {Evolution of Stars and Stellar
  Populations} (J. Wiley)

\bibitem[{{Sanders} {et~al.}(2021){Sanders}, {Belokurov}, \&
  {Man}}]{Sanders2021}
{Sanders}, J.~L., {Belokurov}, V., \& {Man}, K. T.~F. 2021, \mnras, 506, 4321

\bibitem[{{Saydjari} {et~al.}(2023){Saydjari}, {Uzsoy}, {Zucker}, {Peek}, \&
  {Finkbeiner}}]{Saydjari2023}
{Saydjari}, A.~K., {Uzsoy}, A. S.~M., {Zucker}, C., {Peek}, J.~E.~G., \&
  {Finkbeiner}, D.~P. 2023, \apj, 954, 141

\bibitem[{{Sayeed} {et~al.}(2024){Sayeed}, {Ness}, {Montet}, {Cantiello},
  {Casey}, {Buder}, {Bedell}, {Breivik}, {Metzger}, {Martell}, \&
  {McGee-Gold}}]{Sayeed2024}
{Sayeed}, M., {Ness}, M.~K., {Montet}, B.~T., {et~al.} 2024, \apj, 964, 42

\bibitem[{{Schlegel} {et~al.}(1998){Schlegel}, {Finkbeiner}, \&
  {Davis}}]{Schlegel1998}
{Schlegel}, D.~J., {Finkbeiner}, D.~P., \& {Davis}, M. 1998, \apj, 500, 525

\bibitem[{{Sch{\"o}nrich} {et~al.}(2010){Sch{\"o}nrich}, {Binney}, \&
  {Dehnen}}]{Schoenrich2010}
{Sch{\"o}nrich}, R., {Binney}, J., \& {Dehnen}, W. 2010, \mnras, 403, 1829

\bibitem[{{Sharma} {et~al.}(2018){Sharma}, {Stello}, {Buder}, {Kos},
  {Bland-Hawthorn}, {Asplund}, {Duong}, {Lin}, {Lind}, {Ness}, {Huber},
  {Zwitter}, {Traven}, {Hon}, {Kafle}, {Khanna}, {Saddon}, {Anguiano}, {Casey},
  {Freeman}, {Martell}, {De Silva}, {Simpson}, {Wittenmyer}, \&
  {Zucker}}]{Sharma2018}
{Sharma}, S., {Stello}, D., {Buder}, S., {et~al.} 2018, \mnras, 473, 2004

\bibitem[{{Sharma} {et~al.}(2019){Sharma}, {Stello}, {Bland-Hawthorn},
  {Hayden}, {Zinn}, {Kallinger}, {Hon}, {Asplund}, {Buder}, {De Silva},
  {D'Orazi}, {Freeman}, {Kos}, {Lewis}, {Lin}, {Lind}, {Martell}, {Simpson},
  {Wittenmyer}, {Zucker}, {Zwitter}, {Bedding}, {Chen}, {Cotar}, {Esdaile},
  {Horner}, {Huber}, {Kafle}, {Khanna}, {Li}, {Ting}, {Nataf}, {Nordlander},
  {Saadon}, {Traven}, {Wright}, \& {Wyse}}]{Sharma2019}
{Sharma}, S., {Stello}, D., {Bland-Hawthorn}, J., {et~al.} 2019, \mnras, 490,
  5335

\bibitem[{{Sharma} {et~al.}(2021){Sharma}, {Hayden}, {Bland-Hawthorn},
  {Stello}, {Buder}, {Zinn}, {Kallinger}, {Asplund}, {De Silva}, {D'Orazi},
  {Freeman}, {Kos}, {Lewis}, {Lin}, {Lind}, {Martell}, {Simpson}, {Wittenmyer},
  {Zucker}, {Zwitter}, {Chen}, {Cotar}, {Esdaile}, {Hon}, {Horner}, {Huber},
  {Kafle}, {Khanna}, {Ting}, {Nataf}, {Nordlander}, {Saadon}, {Tepper-Garcia},
  {Tinney}, {Traven}, {Watson}, {Wright}, \& {Wyse}}]{Sharma2021}
{Sharma}, S., {Hayden}, M.~R., {Bland-Hawthorn}, J., {et~al.} 2021, \mnras,
  506, 1761

\bibitem[{{Sharma} {et~al.}(2022){Sharma}, {Hayden}, {Bland-Hawthorn},
  {Stello}, {Buder}, {Zinn}, {Spina}, {Kallinger}, {Asplund}, {De Silva},
  {D'Orazi}, {Freeman}, {Kos}, {Lewis}, {Lin}, {Lind}, {Martell},
  {Schlesinger}, {Simpson}, {Zucker}, {Zwitter}, {Chen}, {Cotar}, {Kafle},
  {Khanna}, {Tepper-Garcia}, {Wang}, \& {Wittenmyer}}]{Sharma2022}
---. 2022, \mnras, 510, 734

\bibitem[{{Sheinis} {et~al.}(2015){Sheinis}, {Anguiano}, {Asplund},
  {Bacigalupo}, {Barden}, {Birchall}, {Bland-Hawthorn}, {Brzeski}, {Cannon},
  {Carollo}, {Case}, {Casey}, {Churilov}, {Warrick}, {Dean}, {De Silva},
  {D'Orazi}, {Duong}, {Farrell}, {Fiegert}, {Freeman}, {Gabriella}, {Gers},
  {Goodwin}, {Gray}, {Green}, {Heald}, {Heijmans}, {Ireland}, {Jones}, {Kafle},
  {Keller}, {Klauser}, {Kondrat}, {Kos}, {Lawrence}, {Lee}, {Mali}, {Martell},
  {Mathews}, {Mayfield}, {Miziarski}, {Muller}, {Pai}, {Patterson}, {Penny},
  {Orr}, {Schlesinger}, {Sharma}, {Shortridge}, {Simpson}, {Smedley}, {Smith},
  {Stafford}, {Staszak}, {Vuong}, {Waller}, {de Boer}, {Xavier}, {Zheng},
  {Zhelem}, {Zucker}, \& {Zwitter}}]{Sheinis2015}
{Sheinis}, A., {Anguiano}, B., {Asplund}, M., {et~al.} 2015, J. Astron. Telesc.
  Instrum. Syst., 1, 035002

\bibitem[{{Simpson} {et~al.}(2021){Simpson}, {Martell}, {Buder},
  {Bland-Hawthorn}, {Casey}, {de Silva}, {D'Orazi}, {Freeman}, {Hayden}, {Kos},
  {Lewis}, {Lind}, {Schlesinger}, {Sharma}, {Stello}, {Zucker}, {Zwitter},
  {Asplund}, {da Costa}, {{\v{C}}otar}, {Tepper-Garc{\'\i}a}, {Horner},
  {Nordlander}, {Ting}, {Wyse}, {Wyse}, \& {The Galah
  Collaboration}}]{Simpson2021}
{Simpson}, J.~D., {Martell}, S.~L., {Buder}, S., {et~al.} 2021, \mnras, 507, 43

\bibitem[{Skrutskie {et~al.}(2006)Skrutskie, Cutri, Stiening, Weinberg,
  Schneider, Carpenter, Beichman, Capps, Chester, Elias, Huchra, Liebert,
  Lonsdale, Monet, Price, Seitzer, Jarrett, Kirkpatrick, Gizis, Howard, Evans,
  Fowler, Fullmer, Hurt, Light, Kopan, Marsh, McCallon, Tam, {Van Dyk}, \&
  Wheelock}]{Skrutskie2006}
Skrutskie, M.~F., Cutri, R.~M., Stiening, R., {et~al.} 2006, \aj, 131, 1163

\bibitem[{{Smith}(1988)}]{S}
{Smith}, G. 1988, JPhB, 21, 2827, (S)

\bibitem[{{Smith} \& {Raggett}(1981)}]{SR}
{Smith}, G., \& {Raggett}, D.~S.~J. 1981, JPhB Atomic Molecular Physics, 14,
  4015, (SR)

\bibitem[{{Smith} {et~al.}(2013){Smith}, {Cunha}, {Shetrone}, {Meszaros},
  {Allende Prieto}, {Bizyaev}, {Garc{\'\i}a P{\'e}rez}, {Majewski}, {Schiavon},
  {Holtzman}, \& {Johnson}}]{Smith2013}
{Smith}, V.~V., {Cunha}, K., {Shetrone}, M.~D., {et~al.} 2013, \apj, 765, 16

\bibitem[{{Soares-Furtado} {et~al.}(2021){Soares-Furtado}, {Cantiello},
  {MacLeod}, \& {Ness}}]{SoaresFurtado2021}
{Soares-Furtado}, M., {Cantiello}, M., {MacLeod}, M., \& {Ness}, M.~K. 2021,
  \aj, 162, 273

\bibitem[{{Sobeck} {et~al.}(2007){Sobeck}, {Lawler}, \& {Sneden}}]{SLS}
{Sobeck}, J.~S., {Lawler}, J.~E., \& {Sneden}, C. 2007, Astrophys. J., 667,
  1267, (SLS)

\bibitem[{{Soubiran} {et~al.}(2022){Soubiran}, {Brouillet}, \&
  {Casamiquela}}]{Soubiran2022}
{Soubiran}, C., {Brouillet}, N., \& {Casamiquela}, L. 2022, \aap, 663, A4

\bibitem[{{Spaargaren} {et~al.}(2023){Spaargaren}, {Wang}, {Mojzsis},
  {Ballmer}, \& {Tackley}}]{Spaargaren2023}
{Spaargaren}, R.~J., {Wang}, H.~S., {Mojzsis}, S.~J., {Ballmer}, M.~D., \&
  {Tackley}, P.~J. 2023, \apj, 948, 53

\bibitem[{{Spina} {et~al.}(2016){Spina}, {Mel{\'e}ndez}, {Karakas},
  {Ram{\'{\i}}rez}, {Monroe}, {Asplund}, \& {Yong}}]{Spina2016}
{Spina}, L., {Mel{\'e}ndez}, J., {Karakas}, A.~I., {et~al.} 2016, \aap, 593,
  A125

\bibitem[{{Spina} {et~al.}(2021){Spina}, {Ting}, {De Silva}, {Frankel},
  {Sharma}, {Cantat-Gaudin}, {Joyce}, {Stello}, {Karakas}, {Asplund},
  {Nordlander}, {Casagrande}, {D'Orazi}, {Casey}, {Cottrell},
  {Tepper-Garc{\'\i}a}, {Baratella}, {Kos}, {{\v{C}}otar}, {Bland-Hawthorn},
  {Buder}, {Freeman}, {Hayden}, {Lewis}, {Lin}, {Lind}, {Martell},
  {Schlesinger}, {Simpson}, {Zucker}, \& {Zwitter}}]{Spina2021}
{Spina}, L., {Ting}, Y.~S., {De Silva}, G.~M., {et~al.} 2021, \mnras, 503, 3279

\bibitem[{{Steinmetz} {et~al.}(2020){Steinmetz}, {Matijevi{\v{c}}}, {Enke},
  {Zwitter}, {Guiglion}, {McMillan}, {Kordopatis}, {Valentini}, {Chiappini},
  {Casagrande}, {Wojno}, {Anguiano}, {Bienaym{\'e}}, {Bijaoui}, {Binney},
  {Burton}, {Cass}, {de Laverny}, {Fiegert}, {Freeman}, {Fulbright}, {Gibson},
  {Gilmore}, {Grebel}, {Helmi}, {Kunder}, {Munari}, {Navarro}, {Parker},
  {Ruchti}, {Recio-Blanco}, {Reid}, {Seabroke}, {Siviero}, {Siebert}, {Stupar},
  {Watson}, {Williams}, {Wyse}, {Anders}, {Antoja}, {Birko}, {Bland-Hawthorn},
  {Bossini}, {Garc{\'\i}a}, {Carrillo}, {Chaplin}, {Elsworth}, {Famaey},
  {Gerhard}, {Jofre}, {Just}, {Mathur}, {Miglio}, {Minchev}, {Monari},
  {Mosser}, {Ritter}, {Rodrigues}, {Scholz}, {Sharma}, {Sysoliatina}, \& {RAVE
  Collaboration}}]{Steinmetz2020a}
{Steinmetz}, M., {Matijevi{\v{c}}}, G., {Enke}, H., {et~al.} 2020, \aj, 160, 82

\bibitem[{{Stello} {et~al.}(2015){Stello}, {Huber}, {Sharma}, {Johnson},
  {Lund}, {Handberg}, {Buzasi}, {Silva Aguirre}, {Chaplin}, {Miglio},
  {Pinsonneault}, {Basu}, {Bedding}, {Bland-Hawthorn}, {Casagrande}, {Davies},
  {Elsworth}, {Garcia}, {Mathur}, {Di Mauro}, {Mosser}, {Schneider},
  {Serenelli}, \& {Valentini}}]{Stello2015}
{Stello}, D., {Huber}, D., {Sharma}, S., {et~al.} 2015, \apjl, 809, L3

\bibitem[{{Taylor}(2005)}]{Taylor2005}
{Taylor}, M.~B. 2005, ASPC, 347, 29

\bibitem[{{The MSE Science Team} {et~al.}(2019){The MSE Science Team},
  {Babusiaux}, {Bergemann}, {Burgasser}, {Ellison}, {Haggard}, {Huber},
  {Kaplinghat}, {Li}, {Marshall}, {Martell}, {McConnachie}, {Percival},
  {Robotham}, {Shen}, {Thirupathi}, {Tran}, {Yeche}, {Yong}, {Adibekyan},
  {Silva Aguirre}, {Angelou}, {Asplund}, {Balogh}, {Banerjee}, {Bannister},
  {Barr{\'\i}a}, {Battaglia}, {Bayo}, {Bechtol}, {Beck}, {Beers}, {Bellinger},
  {Berg}, {Bestenlehner}, {Bilicki}, {Bitsch}, {Bland-Hawthorn}, {Bolton},
  {Boselli}, {Bovy}, {Bragaglia}, {Buzasi}, {Caffau}, {Cami}, {Carleton},
  {Casagrande}, {Cassisi}, {Catelan}, {Chang}, {Cortese}, {Damjanov}, {Davies},
  {de Grijs}, {de Rosa}, {Deason}, {di Matteo}, {Drlica-Wagner}, {Erkal},
  {Escorza}, {Ferrarese}, {Fleming}, {Font-Ribera}, {Freeman}, {G{\"a}nsicke},
  {Gabdeev}, {Gallagher}, {Gandolfi}, {Garc{\'\i}a}, {Gaulme}, {Geha},
  {Gennaro}, {Gieles}, {Gilbert}, {Gordon}, {Goswami}, {Greco}, {Grillmair},
  {Guiglion}, {H{\'e}nault-Brunet}, {Hall}, {Handler}, {Hansen}, {Hathi},
  {Hatzidimitriou}, {Haywood}, {Hern{\'a}ndez Santisteban}, {Hillenbrand},
  {Hopkins}, {Howlett}, {Hudson}, {Ibata}, {Ili{\'c}}, {Jablonka}, {Ji},
  {Jiang}, {Juneau}, {Karakas}, {Karinkuzhi}, {Kim}, {Kong}, {Konstantopoulos},
  {Krogager}, {Lagos}, {Lallement}, {Laporte}, {Lebreton}, {Lee}, {Lewis},
  {Lianou}, {Liu}, {Lodieu}, {Loveday}, {M{\'e}sz{\'a}ros}, {Makler}, {Mao},
  {Marchesini}, {Martin}, {Mateo}, {Melis}, {Merle}, {Miglio}, {Gohar
  Mohammad}, {Molaverdikhani}, {Monier}, {Morel}, {Mosser}, {Nataf}, {Necib},
  {Neilson}, {Newman}, {Nierenberg}, {Nord}, {Noterdaeme}, {O'Dea}, {Oshagh},
  {Pace}, {Palanque-Delabrouille}, {Pandey}, {Parker}, {Pawlowski}, {Peter},
  {Petitjean}, {Petric}, {Placco}, {Popovi{\'c}}, {Price-Whelan}, {Prsa},
  {Ravindranath}, {Rich}, {Ruan}, {Rybizki}, {Sakari}, {Sanderson}, {Schiavon},
  {Schimd}, {Serenelli}, {Siebert}, {Siudek}, {Smiljanic}, {Smith}, {Sobeck},
  {Starkenburg}, {Stello}, {Szab{\'o}}, {Szabo}, {Taylor}, {Thanjavur},
  {Thomas}, {Tollerud}, {Toonen}, {Tremblay}, {Tresse}, {Tsantaki},
  {Valentini}, {Van Eck}, {Variu}, {Venn}, {Villaver}, {Walker}, {Wang},
  {Wang}, {Wilson}, {Wright}, {Xu}, {Yildiz}, {Zhang}, {Zwintz}, {Anguiano},
  {Bedell}, {Chaplin}, {Collet}, {Cuillandre}, {Duc}, {Flagey}, {Hermes},
  {Hill}, {Kamath}, {Laychak}, {Ma{\l}ek}, {Marley}, {Sheinis}, {Simons},
  {Sousa}, {Szeto}, {Ting}, {Vegetti}, {Wells}, {Babas}, {Bauman}, {Bosselli},
  {C{\^o}t{\'e}}, {Colless}, {Comparat}, {Courtois}, {Crampton}, {Croom},
  {Davies}, {de Grijs}, {Denny}, {Devost}, {di Matteo}, {Driver},
  {Fernandez-Lorenzo}, {Guhathakurta}, {Han}, {Higgs}, {Hill}, {Ho}, {Hopkins},
  {Hudson}, {Ibata}, {Isani}, {Jarvis}, {Johnson}, {Jullo}, {Kaiser}, {Kneib},
  {Koda}, {Koshy}, {Mignot}, {Murowinski}, {Newman}, {Nusser}, {Pancoast},
  {Peng}, {Peroux}, {Pichon}, {Poggianti}, {Richard}, {Salmon}, {Seibert},
  {Shastri}, {Smith}, {Sutaria}, {Tao}, {Taylor}, {Tully}, {van Waerbeke},
  {Vermeulen}, {Walker}, {Willis}, {Willot}, \& {Withington}}]{MSE2019}
{The MSE Science Team}, {Babusiaux}, C., {Bergemann}, M., {et~al.} 2019, arXiv
  e-prints, arXiv:1904.04907

\bibitem[{{Ting} {et~al.}(2016){Ting}, {Conroy}, \& {Rix}}]{Ting2016b}
{Ting}, Y.-S., {Conroy}, C., \& {Rix}, H.-W. 2016, \apj, 826, 83

\bibitem[{{Ting} {et~al.}(2018){Ting}, {Conroy}, {Rix}, \&
  {Asplund}}]{Ting2018}
{Ting}, Y.-S., {Conroy}, C., {Rix}, H.-W., \& {Asplund}, M. 2018, \apj, 860,
  159

\bibitem[{{Ting} {et~al.}(2019){Ting}, {Conroy}, {Rix}, \&
  {Cargile}}]{Ting2019}
{Ting}, Y.-S., {Conroy}, C., {Rix}, H.-W., \& {Cargile}, P. 2019, \apj, 879, 69

\bibitem[{{Ting} {et~al.}(2012){Ting}, {Freeman}, {Kobayashi}, {De Silva}, \&
  {Bland-Hawthorn}}]{Ting2012}
{Ting}, Y.-S., {Freeman}, K.~C., {Kobayashi}, C., {De Silva}, G.~M., \&
  {Bland-Hawthorn}, J. 2012, \mnras, 421, 1231

\bibitem[{{Torra} {et~al.}(2021){Torra}, {Casta{\~n}eda}, {Fabricius},
  {Lindegren}, {Clotet}, {Gonz{\'a}lez-Vidal}, {Bartolom{\'e}}, {Bastian},
  {Bernet}, {Biermann}, {Garralda}, {G{\'u}rpide}, {Lammers}, {Portell}, \&
  {Torra}}]{Torra2021}
{Torra}, F., {Casta{\~n}eda}, J., {Fabricius}, C., {et~al.} 2021, \aap, 649,
  A10

\bibitem[{{Traven} {et~al.}(2020){Traven}, {Feltzing}, {Merle}, {Van der
  Swaelmen}, {{\v{C}}otar}, {Church}, {Zwitter}, {Ting}, {Sahlholdt},
  {Asplund}, {Bland-Hawthorn}, {De Silva}, {Freeman}, {Martell}, {Sharma},
  {Zucker}, {Buder}, {Casey}, {D'Orazi}, {Kos}, {Lewis}, {Lin}, {Lind},
  {Simpson}, {Stello}, {Munari}, \& {Wittenmyer}}]{Traven2020}
{Traven}, G., {Feltzing}, S., {Merle}, T., {et~al.} 2020, \aap, 638, A145

\bibitem[{{Trubko} {et~al.}(2017){Trubko}, {Gregoire}, {Holmgren}, \&
  {Cronin}}]{2017PhRvA..95e2507T}
{Trubko}, R., {Gregoire}, M.~D., {Holmgren}, W.~F., \& {Cronin}, A.~D. 2017,
  \pra, 95, 052507

\bibitem[{{Tsantaki} {et~al.}(2022){Tsantaki}, {Pancino}, {Marrese},
  {Marinoni}, {Rainer}, {Sanna}, {Turchi}, {Randich}, {Gallart}, {Battaglia},
  \& {Masseron}}]{Tsantaki2022}
{Tsantaki}, M., {Pancino}, E., {Marrese}, P., {et~al.} 2022, \aap, 659, A95

\bibitem[{{Tsuji}(1976)}]{Tsuji1976}
{Tsuji}, T. 1976, \pasj, 28, 543

\bibitem[{{Vaeck} {et~al.}(1988){Vaeck}, {Godefroid}, \& {Hansen}}]{VGH}
{Vaeck}, N., {Godefroid}, M., \& {Hansen}, J.~E. 1988, \pra, 38, 2830, (VGH)

\bibitem[{{Valenti} \& {Piskunov}(1996)}]{Valenti1996}
{Valenti}, J.~A., \& {Piskunov}, N. 1996, \aaps, 118, 595

\bibitem[{{van Leeuwen}(2007)}]{vanLeeuwen2007}
{van Leeuwen}, F. 2007, \aap, 474, 653

\bibitem[{{VandenBerg} {et~al.}(2012){VandenBerg}, {Bergbusch}, {Dotter},
  {Ferguson}, {Michaud}, {Richer}, \& {Proffitt}}]{VandenBerg2012}
{VandenBerg}, D.~A., {Bergbusch}, P.~A., {Dotter}, A., {et~al.} 2012, \apj,
  755, 15

\bibitem[{{Vasiliev} \& {Baumgardt}(2021)}]{Vasiliev2021}
{Vasiliev}, E., \& {Baumgardt}, H. 2021, \mnras, 505, 5978

\bibitem[{Virtanen {et~al.}(2020)Virtanen, Gommers, Oliphant, Haberland, Reddy,
  Cournapeau, Burovski, Peterson, Weckesser, Bright, {van der Walt}, Brett,
  Wilson, Millman, Mayorov, Nelson, Jones, Kern, Larson, Carey, Polat, Feng,
  Moore, {VanderPlas}, Laxalde, Perktold, Cimrman, Henriksen, Quintero, Harris,
  Archibald, Ribeiro, Pedregosa, {van Mulbregt}, \& {SciPy 1.0
  Contributors}}]{scipy}
Virtanen, P., Gommers, R., Oliphant, T.~E., {et~al.} 2020, Nature Methods, 17,
  261

\bibitem[{{Vogrin{\v{c}}i{\v{c}}} {et~al.}(2023){Vogrin{\v{c}}i{\v{c}}}, {Kos},
  {Zwitter}, {Traven}, {Beeson}, {{\v{C}}otar}, {Munari}, {Buder}, {Martell},
  {Lewis}, {De Silva}, {Hayden}, {Bland-Hawthorn}, \&
  {D'Orazi}}]{Vogrincic2023}
{Vogrin{\v{c}}i{\v{c}}}, R., {Kos}, J., {Zwitter}, T., {et~al.} 2023, \mnras,
  521, 3727

\bibitem[{Walt {et~al.}(2011)Walt, Colbert, \& Varoquaux}]{numpy}
Walt, S. v.~d., Colbert, S.~C., \& Varoquaux, G. 2011, Comput Sci Eng, 13, 22

\bibitem[{{Wang} {et~al.}(2021){Wang}, {Nordlander}, {Asplund}, {Amarsi},
  {Lind}, \& {Zhou}}]{Wang2021}
{Wang}, E.~X., {Nordlander}, T., {Asplund}, M., {et~al.} 2021, \mnras, 500,
  2159

\bibitem[{{Wang} {et~al.}(2024{\natexlab{a}}){Wang}, {Nordlander}, {Buder},
  {Ciuc{\u{a}}}, {Soen}, {Martell}, {Ness}, {Lind}, {McKenzie}, \&
  {Stello}}]{Wang2024}
{Wang}, E.~X., {Nordlander}, T., {Buder}, S., {et~al.} 2024{\natexlab{a}},
  \mnras, 528, 5394

\bibitem[{{Wang} {et~al.}(2024{\natexlab{b}}){Wang}, {Quanz}, {Mahadevan}, \&
  {Deal}}]{Wang2024b}
{Wang}, H.~S., {Quanz}, S.~P., {Mahadevan}, S., \& {Deal}, M.
  2024{\natexlab{b}}, \aap, 688, A225

\bibitem[{{Wehrhahn} {et~al.}(2023){Wehrhahn}, {Piskunov}, \&
  {Ryabchikova}}]{Wehrhahn2023}
{Wehrhahn}, A., {Piskunov}, N., \& {Ryabchikova}, T. 2023, \aap, 671, A171

\bibitem[{{Westendorp Plaza} {et~al.}(2023){Westendorp Plaza}, {Asensio Ramos},
  \& {Allende Prieto}}]{WestendorpPlaza2023}
{Westendorp Plaza}, C., {Asensio Ramos}, A., \& {Allende Prieto}, C. 2023,
  \aap, 675, A191

\bibitem[{{Wheeler} {et~al.}(2023){Wheeler}, {Abruzzo}, {Casey}, \&
  {Ness}}]{Wheeler2023}
{Wheeler}, A.~J., {Abruzzo}, M.~W., {Casey}, A.~R., \& {Ness}, M.~K. 2023, \aj,
  165, 68

\bibitem[{{Wheeler} {et~al.}(2024){Wheeler}, {Casey}, \&
  {Abruzzo}}]{Wheeler2024}
{Wheeler}, A.~J., {Casey}, A.~R., \& {Abruzzo}, M.~W. 2024, \aj, 167, 83

\bibitem[{{Wickliffe} {et~al.}(1994){Wickliffe}, {Salih}, \& {Lawler}}]{WSL}
{Wickliffe}, M.~E., {Salih}, S., \& {Lawler}, J.~E. 1994, \jqsrt, 51, 545,
  (WSL)

\bibitem[{{Wood} {et~al.}(2013){Wood}, {Lawler}, {Sneden}, \& {Cowan}}]{WLSC}
{Wood}, M.~P., {Lawler}, J.~E., {Sneden}, C., \& {Cowan}, J.~J. 2013, \apjs,
  208, 27

\bibitem[{{Wood} {et~al.}(2014){Wood}, {Lawler}, {Sneden}, \&
  {Cowan}}]{2014ApJS..211...20W}
---. 2014, \apjs, 211, 20

\bibitem[{{Xiang} {et~al.}(2019){Xiang}, {Ting}, {Rix}, {Sand ford}, {Buder},
  {Lind}, {Liu}, {Shi}, \& {Zhang}}]{Xiang2019}
{Xiang}, M., {Ting}, Y.-S., {Rix}, H.-W., {et~al.} 2019, \apjs, 245, 34

\bibitem[{{Xiang} {et~al.}(2022){Xiang}, {Rix}, {Ting}, {Kudritzki}, {Conroy},
  {Zari}, {Shi}, {Przybilla}, {Ramirez-Tannus}, {Tkachenko}, {Gebruers}, \&
  {Liu}}]{Xiang2021}
{Xiang}, M., {Rix}, H.-W., {Ting}, Y.-S., {et~al.} 2022, \aap, 662, A66

\bibitem[{{Yan} {et~al.}(1998){Yan}, {Tambasco}, \&
  {Drake}}]{1998PhRvA..57.1652Y}
{Yan}, Z.-C., {Tambasco}, M., \& {Drake}, G.~W.~F. 1998, \pra, 57, 1652

\bibitem[{{Yong} {et~al.}(2013){Yong}, {Mel{\'e}ndez}, {Grundahl}, {Roederer},
  {Norris}, {Milone}, {Marino}, {Coelho}, {McArthur}, {Lind}, {Collet}, \&
  {Asplund}}]{Yong2013}
{Yong}, D., {Mel{\'e}ndez}, J., {Grundahl}, F., {et~al.} 2013, \mnras, 434,
  3542

\bibitem[{{Zhao} {et~al.}(2012){Zhao}, {Zhao}, {Chu}, {Jing}, \&
  {Deng}}]{Zhao2012}
{Zhao}, G., {Zhao}, Y.-H., {Chu}, Y.-Q., {Jing}, Y.-P., \& {Deng}, L.-C. 2012,
  Research in Astronomy and Astrophysics, 12, 723

\bibitem[{{Zinn} {et~al.}(2020){Zinn}, {Stello}, {Elsworth}, {Garc{\'\i}a},
  {Kallinger}, {Mathur}, {Mosser}, {Bugnet}, {Jones}, {Hon}, {Sharma},
  {Sch{\"o}nrich}, {Warfield}, {Luger}, {Pinsonneault}, {Johnson}, {Huber},
  {Silva Aguirre}, {Chaplin}, {Davies}, \& {Miglio}}]{Zinn2020}
{Zinn}, J.~C., {Stello}, D., {Elsworth}, Y., {et~al.} 2020, \apjs, 251, 23

\bibitem[{{Zinn} {et~al.}(2022){Zinn}, {Stello}, {Elsworth}, {Garc{\'\i}a},
  {Kallinger}, {Mathur}, {Mosser}, {Hon}, {Bugnet}, {Jones}, {Reyes}, {Sharma},
  {Sch{\"o}nrich}, {Warfield}, {Luger}, {Vanderburg}, {Kobayashi},
  {Pinsonneault}, {Johnson}, {Huber}, {Buder}, {Joyce}, {Bland-Hawthorn},
  {Casagrande}, {Lewis}, {Miglio}, {Nordlander}, {Davies}, {Silva}, {Chaplin},
  \& {Silva Aguirre}}]{Zinn2022}
---. 2022, \apj, 926, 191

\end{thebibliography}

\appendix

\section{Initial parameters}

\begin{figure*}[ht]
 \centering
 \includegraphics[width=\textwidth]{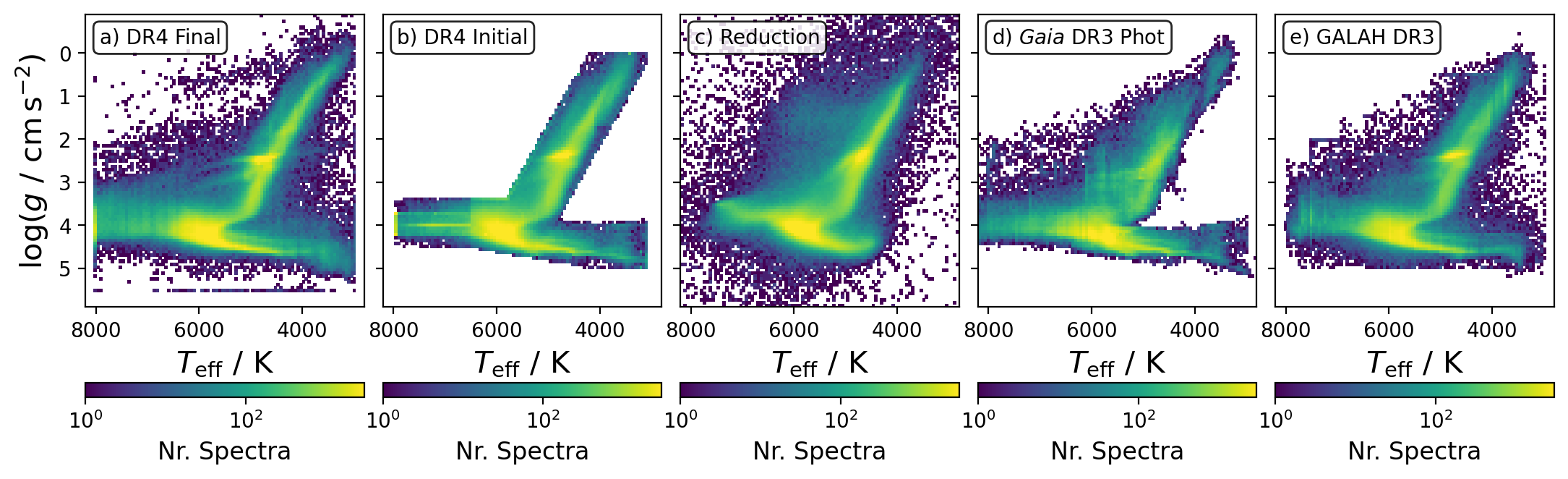}
 \includegraphics[width=\textwidth]{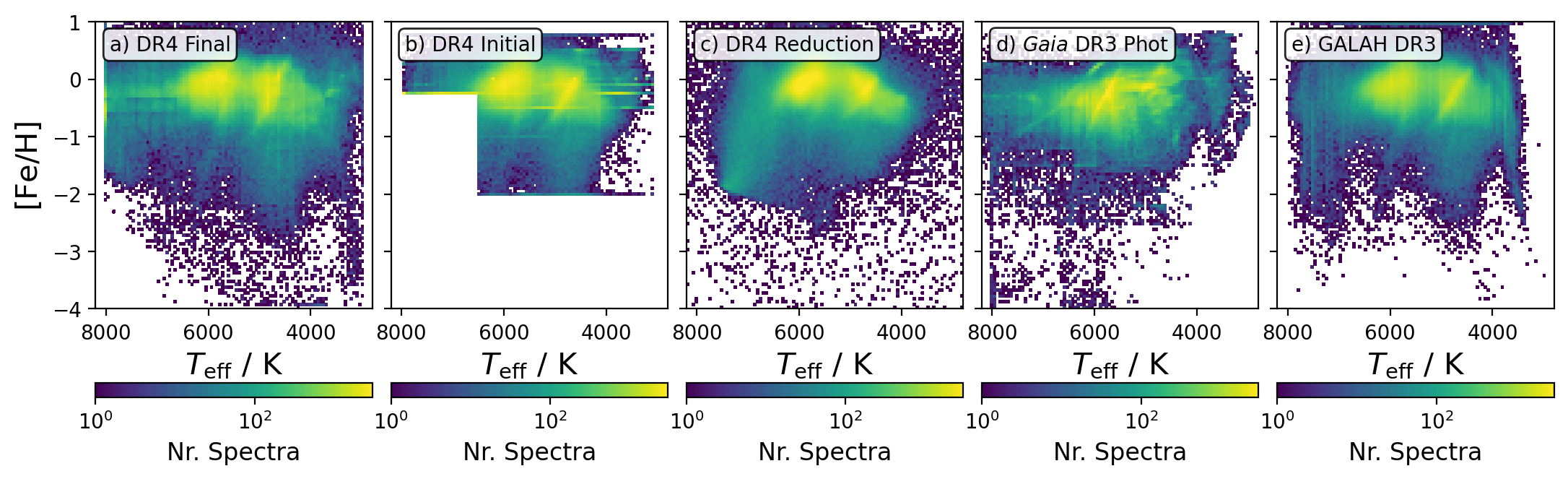}
 \includegraphics[width=\textwidth]{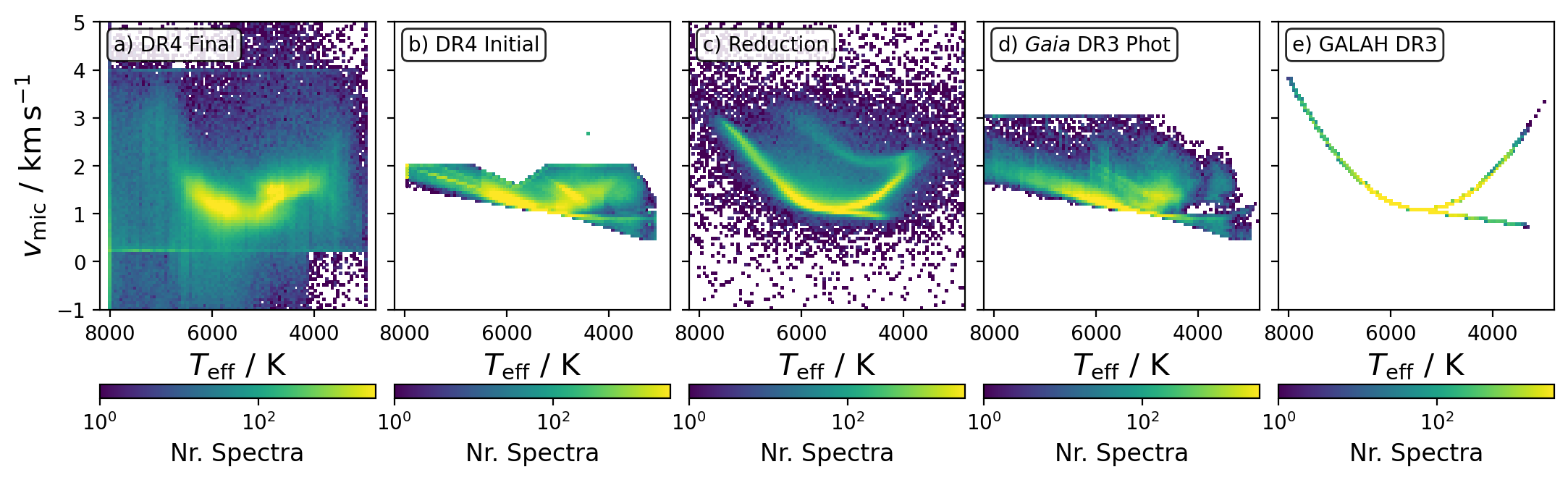}
 \includegraphics[width=\textwidth]{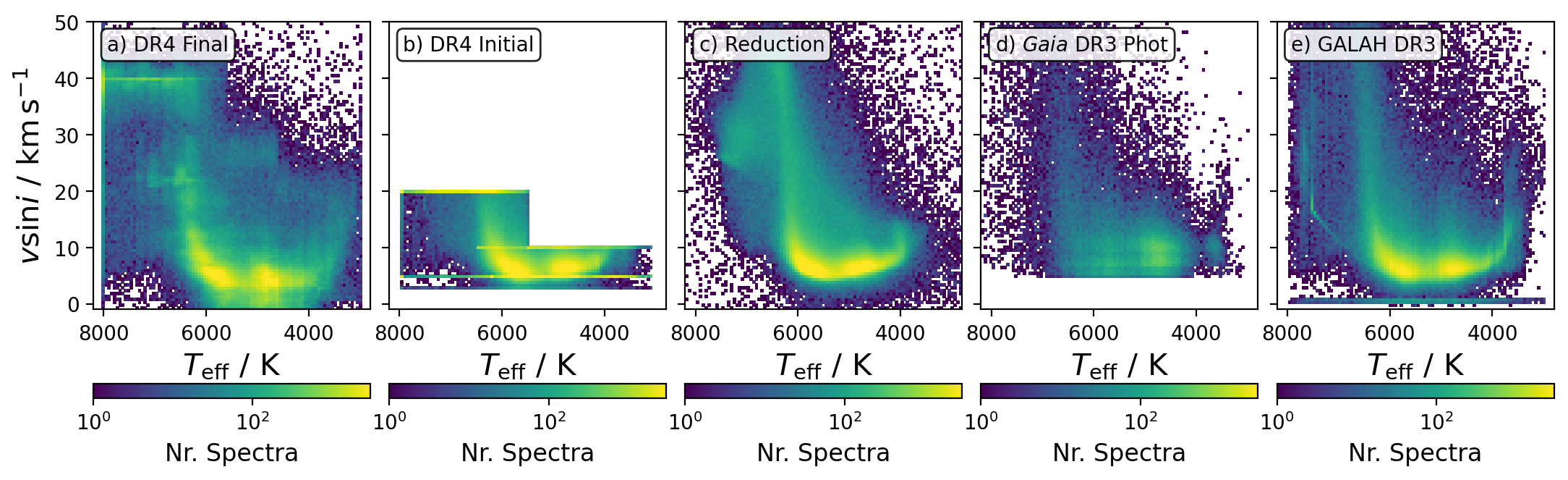} \caption{\textbf{Comparison of final GALAH DR4 stellar parameters (first column) against the initial parameters used in the \textit{allstar} analysis (second column), estimates from the GALAH DR4 reduction pipeline (third column), \Gaia DR3 \citep[fourth column with \vmic based on the adjusted formula from ][]{DutraFerreira2016}, and GALAH DR3 (fifth column).}} \label{fig:initial_parameters}
\end{figure*}

We append the overview of the initial and final stellar parameters of GALAH DR4 in Figure~\ref{fig:initial_parameters}. We show the density distribution of \logg, \feh, \vmic, and \vsini in each row as a function of \Teff.

\section{Data Products}

\begin{figure*}[ht]
 \centering
 \includegraphics[width=0.875\textwidth]{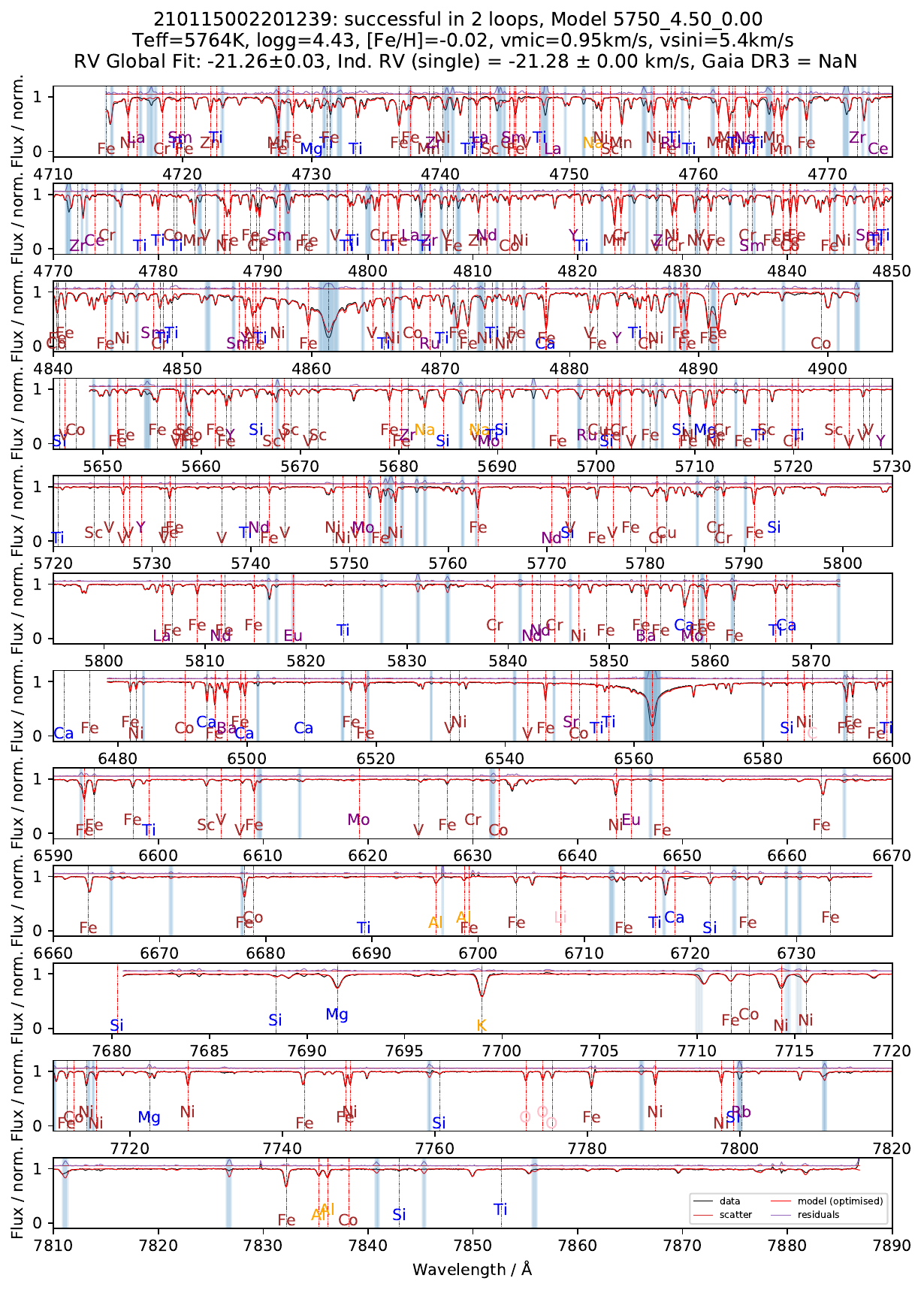} \caption{\textbf{ Example output of the \texttt{allstar} analysis for Vesta (210115002201239).} The observed flux (black) is compared with the fitted model flux (red), and the residuals (purple) show the difference between the observed and modelled spectra. Important spectral lines are annotated with their corresponding elements, with element groups colour-coded for clarity. Blue-shaded regions represent the 5\% of the spectrum that was masked and excluded from the fit to avoid contamination from outliers or poorly modelled lines.}
\label{fig:210115002201239_allstar_fit_comparison}
\end{figure*}

\begin{figure}[ht]
 \centering
 \includegraphics[width=\textwidth]{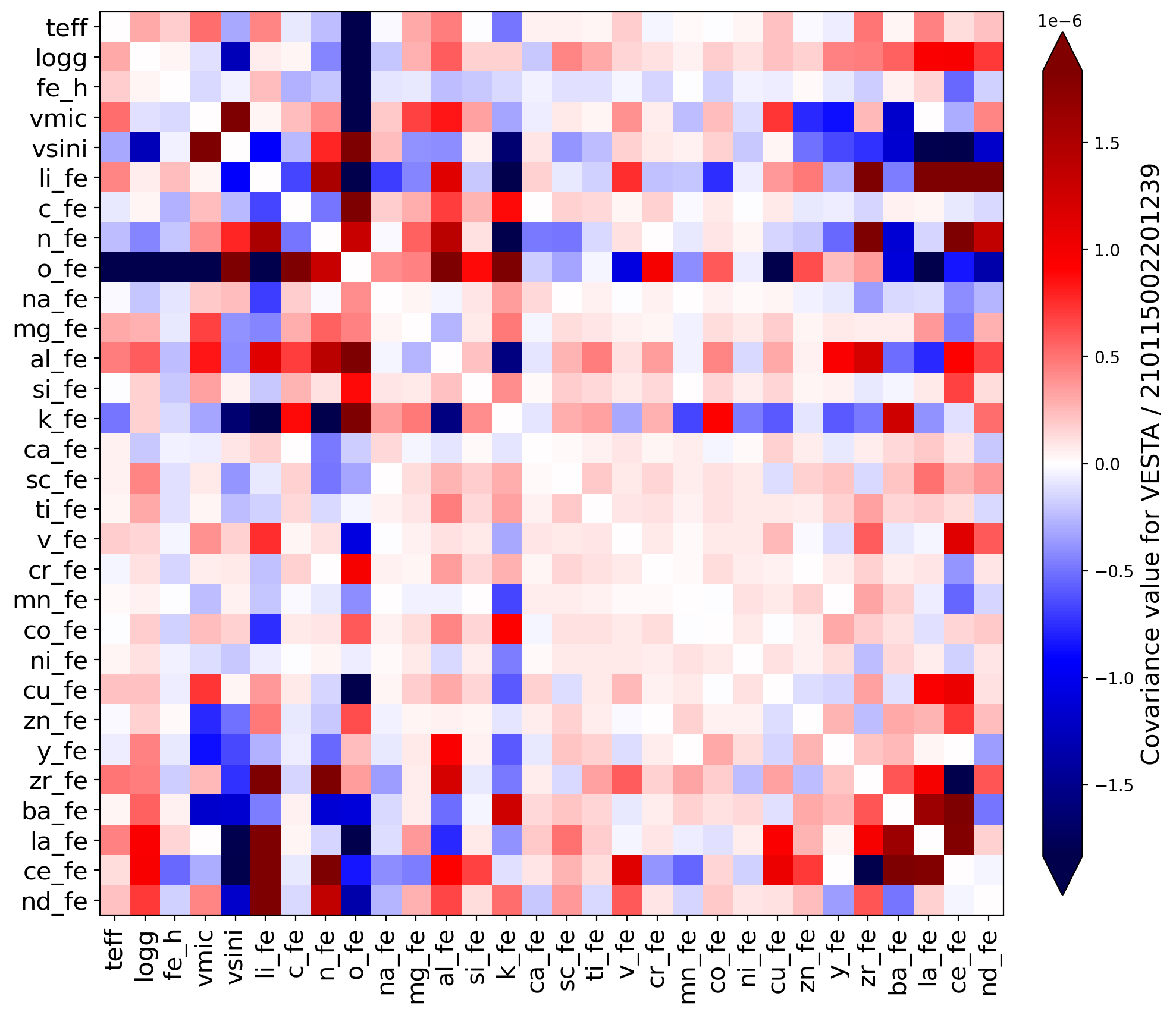}
 \hfill
 \includegraphics[width=\textwidth]{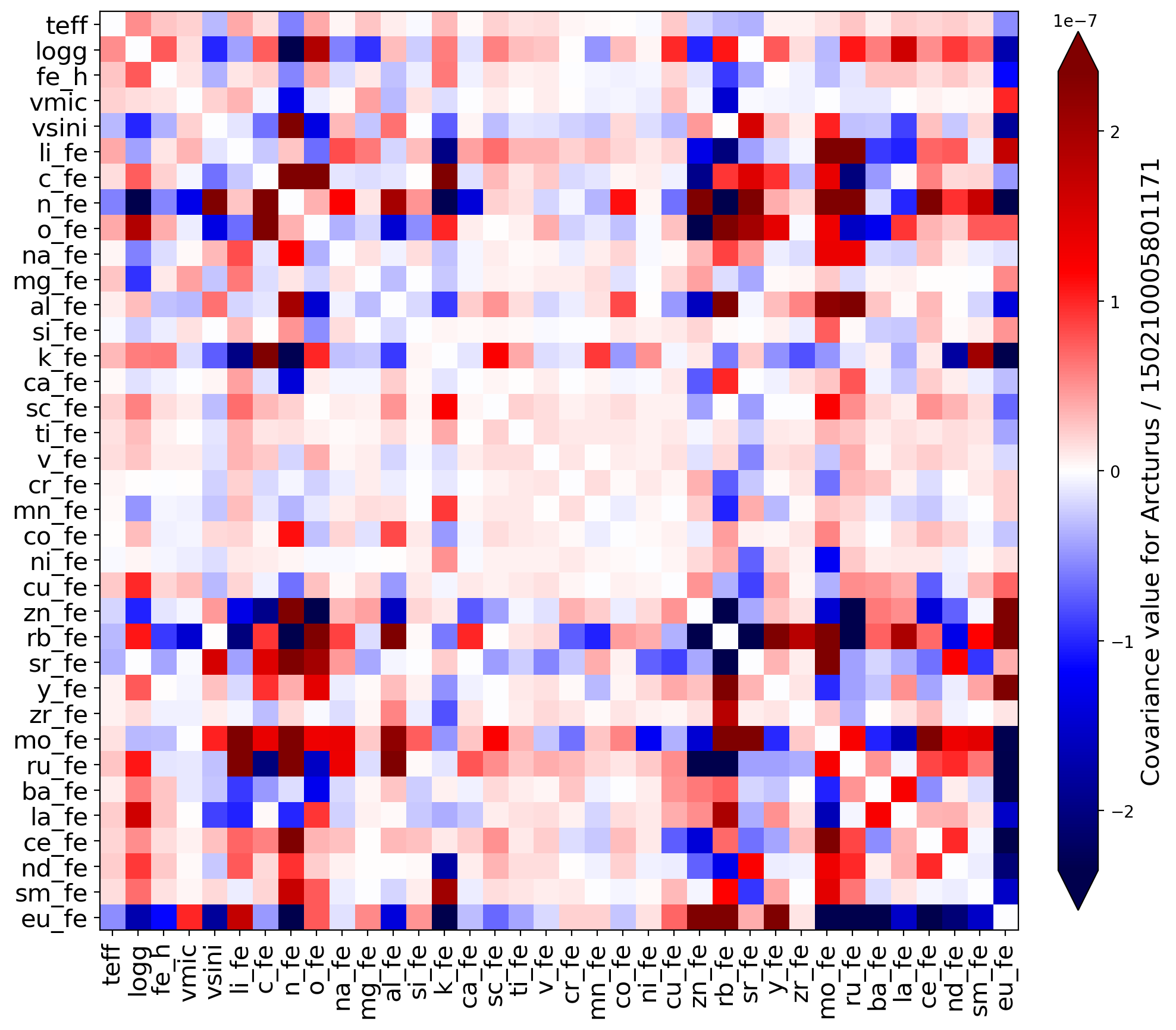}
 \caption{\textbf{Covariance matrices for labels for Vesta (panel a) and Arcturus (panel b).}}
 \label{fig:covariance_vesta_arcturus}
\end{figure}

We append examples of data products of GALAH DR4 that were not already shown in the main manuscript. Table~\ref{tab:main_catalog_schema} shows a shortened table schema of the \texttt{allstar} and \texttt{allspec} catalogues. Figure~\ref{fig:210115002201239_allstar_fit_comparison} shows the automatically created fit comparison of the \texttt{allstar} module for Vesta (210115002201239). Figure~\ref{fig:covariance_vesta_arcturus} shows examples of the covariance matrices for Vesta and Arcturus, as representative examples for main-sequence and giant stars.

\begin{table*}[ht]
\centering
\caption{Table schema of the GALAH DR4 main catalogues. Columns that are part of \texttt{allspec}, but not \texttt{allstar} are listed below the middle line. For compactness, we have combined repetitive columns (for example with integers $N$). Detailed table schemas are available in the FITS headers of each catalogue file.}
\label{tab:main_catalog_schema}
\begin{tabular}{llll}
\hline \hline
Column & Description & Column & Description \\
\hline
sobject\_id & GALAH identifier & tmass\_id & 2MASS identifier \\ 
gaiadr3\_source\_id & \Gaia DR3 source\_id & survey\_name & 2dF-HERMES Program \\ 
field\_id & GALAH Field ID & setup & allspec/allstar \\ 
mjd & Modified Julian Date & ra & propagated from \Gaia DR3 \\ 
dec & propagated from \Gaia DR3 & flag\_sp & Major spectroscopic quality bitmask flag \\ 
flag\_sp\_fit & Major fitting quality flag & opt\_loop & Nr of optimisation loops used for fitting \\ 
flag\_red & Quality bitmask flag of reduction pipeline & snr\_px\_ccdN & Average SNR for CCD N \\ 
chi2\_sp & Chi2 value of spectroscopic fitting & px\_used\_perc & Percentage of spectrum used for fit \\ 
model\_name & Used neural network for synthesis & closest\_model & Closest neural network for synthesis \\ 
comp\_time & Computation time spent on spectrum & fit\_global\_rv & RV fitted or fixed after co-adding? \\ 
rv\_comp\_1 & Radial velocity of primary source & e\_rv\_comp\_1 & Uncertainty of rv\_comp\_1 \\ 
rv\_comp\_2 & Radial velocity of potential secondary source & e\_rv\_comp\_2 & Uncertainty of rv\_comp\_1 \\ 
rv\_gaia\_dr3 & Radial velocity in \Gaia DR3 & e\_rv\_gaia\_dr3 & Uncertainty of rv\_gaia\_dr3 \\ 
v\_bary\_eff & Barycentric velocity correction & teff & Spectr. effective temperature \\ 
e\_teff & Uncertainty teff & logg & Photometric surface gravity \\ 
e\_logg & Uncertainty logg\_plx & fe\_h & Abundance of Fe as pseudo-metallicity \\ 
e\_fe\_h & Uncertainty fe\_h & flag\_fe\_h & Quality flag fe\_h \\ 
vmic & Microturbulence velocity (fitted) & e\_vmic & Uncertainty vmic \\ 
vsini & Broadening velocity & e\_vsini & Uncertainty of vsini \\ 
nn\_li\_fe & Elemental abundance for [Li/Fe] & nn\_e\_li\_fe & Uncertainty nn\_li\_fe \\ 
nn\_flag\_li\_fe & Quality bitmask flag of Li\_fe & x\_fe & Elemental abundance for [X/Fe] \\ 
e\_x\_fe & Uncertainty of elemental abundance [X/Fe] & flag\_x\_fe & Quality bitmask flag of [X/Fe] \\ 
mass & Mass used for calculating $\log g$ (plx) & age & Age estimated when calculating mass \\ 
bc\_ks & Bolometric Correction of $K_S$ band & a\_ks & Attenuation in $K_S$-band $A(K_S)$ \\ 
lbol & Bolometric Luminosity & r\_med & Median Distance \\ 
r\_lo & Lower Limit Distance & r\_hi & Higher Limit Distance \\ 
sb2\_rv\_N & Nth perc. of RV residual signal & ew\_h\_beta & Equivalent Width of observed Hbeta core \\ 
ew\_h\_alpha & Equivalent Width of observed Halpha core & res\_h\_beta & Residual EW in Hbeta core \\ 
res\_h\_alpha & Residual EW in Halpha core & ew\_k\_is & EW of K7699 Interstellar Line \\ 
sigma\_k\_is & Gaussian sigma of K7699 Interstellar Line & rv\_k\_is & RV of K7699 Interstellar Line \\ 
ew\_dib5780 & Equivalent width of DIB NNNN & sigma\_dib5780 & Gaussian sigma of DIB NNNN \\ 
rv\_dib5780 & RV of DIB NNNN & ebv & Extinction $E(B-V)$ \\ 
phot\_g\_mean\_mag & Mean \Gaia DR3 G-band apparent magnitude & bp\_rp & Color of $BP-RP$ bands \\ 
j\_m & 2MASS $J$-band magnitude & j\_msigcom & Uncertainty of j\_m \\ 
h\_m & 2MASS $H$-band magnitude & h\_msigcom & Uncertainty of h\_m \\ 
ks\_m & 2MASS $K_S$-band magnitude & ks\_msigcom & Uncertainty of ks\_m \\ 
W2mag & AllWISE W2-band magnitude & e\_W2mag & uncertainty of W2mag \\ 
ruwe & RUWE reported by \Gaia DR3 & parallax & Astrometric parallax used for GALAH DR4 \\ 
parallax\_error & Uncertainty of astrometric parallax & ew\_li & Eqiuvalent width of Lithium 6708 LiI line \\ 
e\_ew\_li\_low & Lower uncertainty ew\_li & e\_ew\_li\_upp & Upper uncertainty ew\_li \\ 
a\_li & Absolute 3D NLTE Li abundance & a\_li\_upp\_lim & Upper limit of absolute 3D NLTE A(Li) \\ 
e\_a\_li\_low & Lower uncertainty of a\_li & e\_a\_li\_upp & Upper uncertainty of a\_li \\ 
e\_a\_li\_teff & Uncertainty of A(Li) due to temperature & flag\_a\_li & Flag for a\_li measurement \\ 
\hline
rv\_comp\_nr & Nr RV cross-correlation function peaks & rv\_comp\_1\_p & Prominence of rv\_comp\_1 in CCF \\ 
rv\_comp\_2\_h & Height of rv\_comp\_1 in CCF & rv\_comp\_2\_p & Prominence of rv\_comp\_1 in CCF \\ 
logg\_spec & Spectroscopic surface gravity estimate & e\_logg\_spec & Uncertainty logg\_spec \\ 
\hline
\end{tabular}
\end{table*}

\section{Stellar parameter and abundance validation} \label{sec:stellar_param_abund_valid}

\begin{figure*}[ht]
 \centering
 \includegraphics[width=\textwidth]{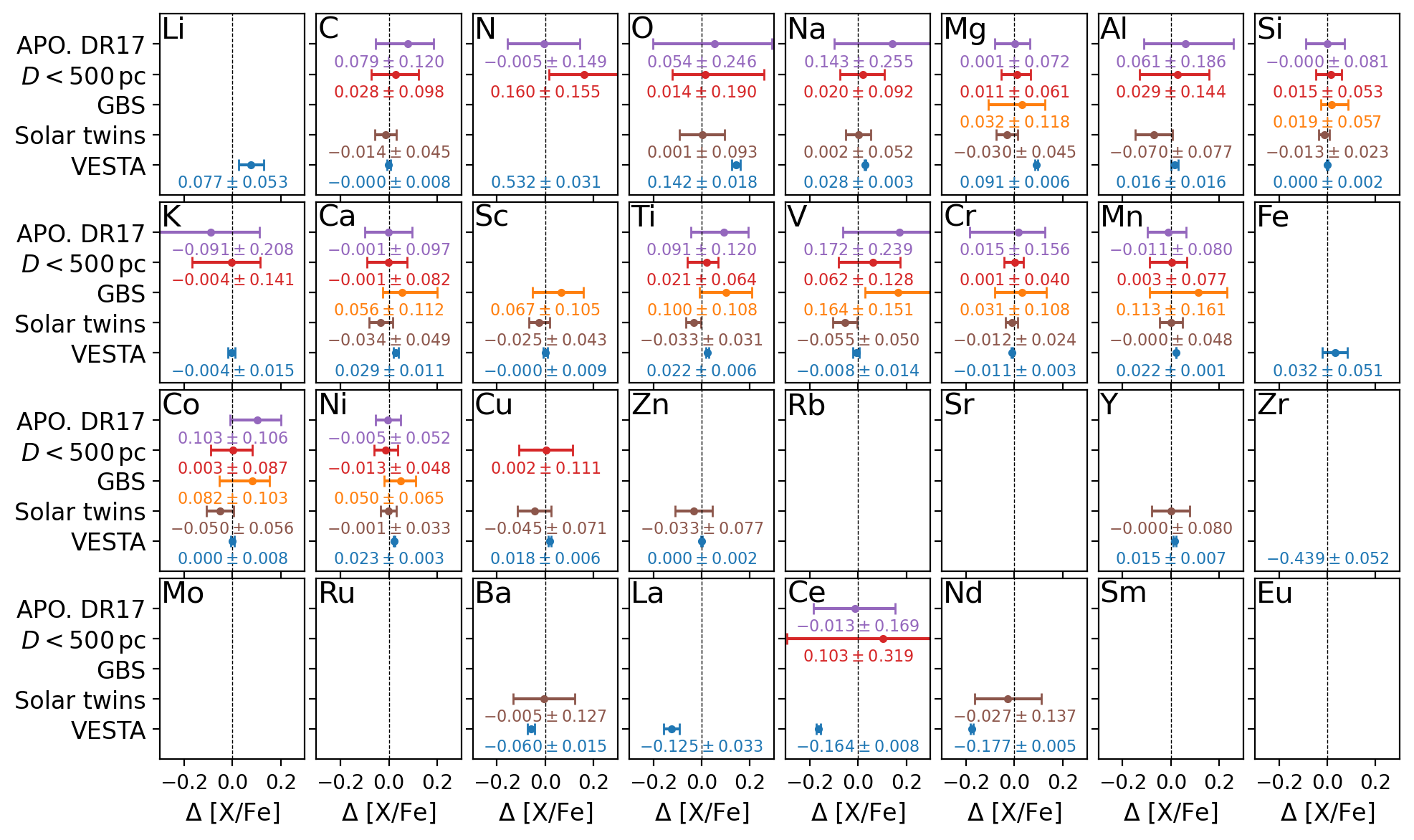}
 \caption{\textbf{zero-point estimates of elemental abundances for GALAH DR4.} Each panel shows the comparison to literature (DR4 - literature) for Vesta (blue), \Gaia FKG Benchmark Stars (orange), Stars with $\vert \mathrm{[Fe/H]} \vert \leq 0.1$ closer than $D_\varpi < 0.5\,\mathrm{kpc}$ (red), as well as stars that were also observed by APOGEE DR17 (purple).}
 \label{fig:galah_dr4_zeropoint_checks_allstar}
\end{figure*}
\begin{figure*}[ht]
 \centering
 \includegraphics[width=\textwidth]{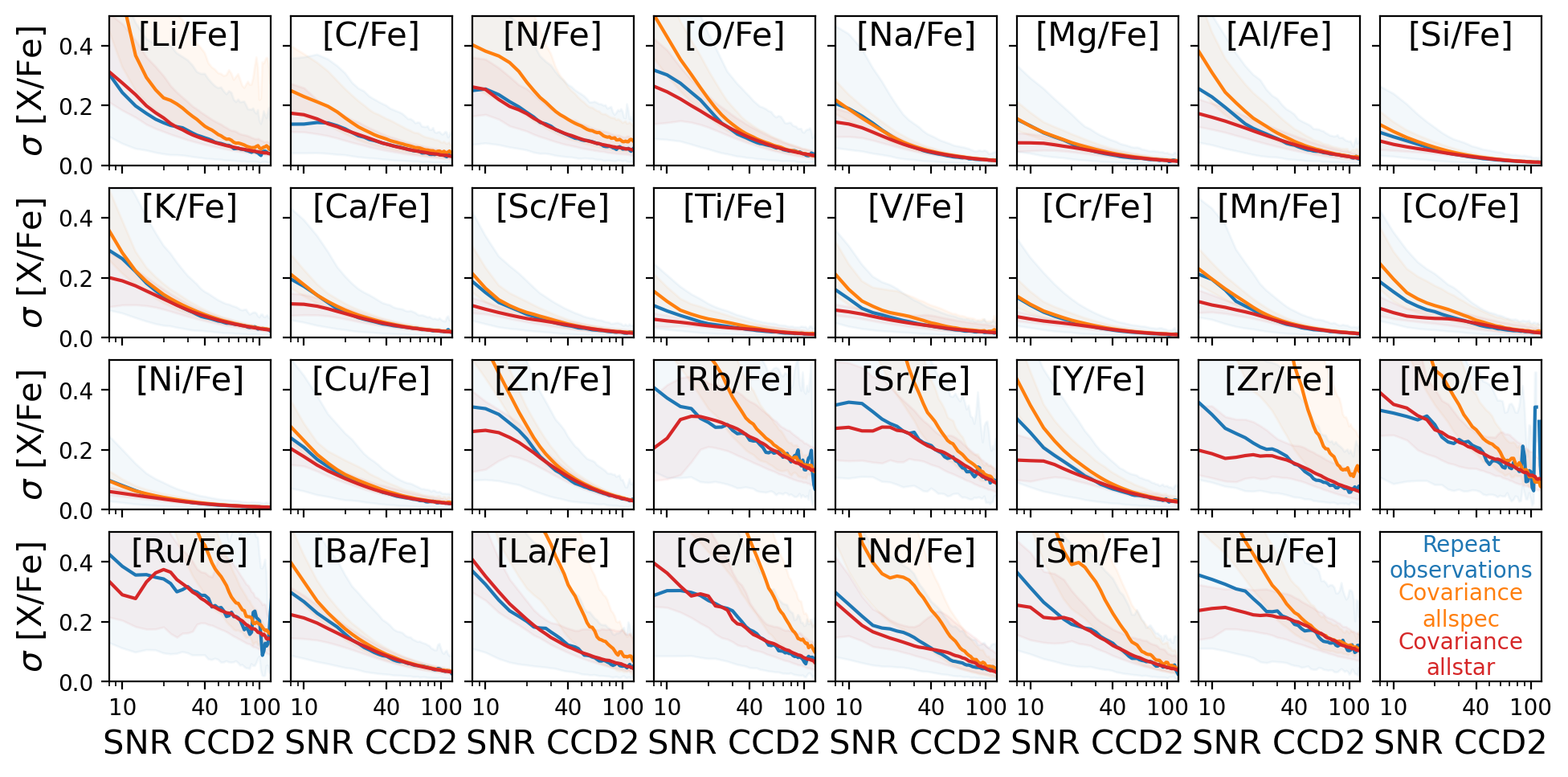}
 \caption{\textbf{Precision monitoring (with a median line and standard deviation shading) of elemental abundances as a function of SNR for the green CCD2 across for GALAH DR4.} Each panel shows the behaviour for bins of width 10 for the scatter of repeat observations of the \texttt{allspec} runs (blue) as well as covariance uncertainties of \texttt{allspec} (orange) and \texttt{allstar} (red) setups.}
 \label{fig:galah_dr4_precision_abundances}
\end{figure*}
\begin{figure*}
\includegraphics[width=\textwidth]{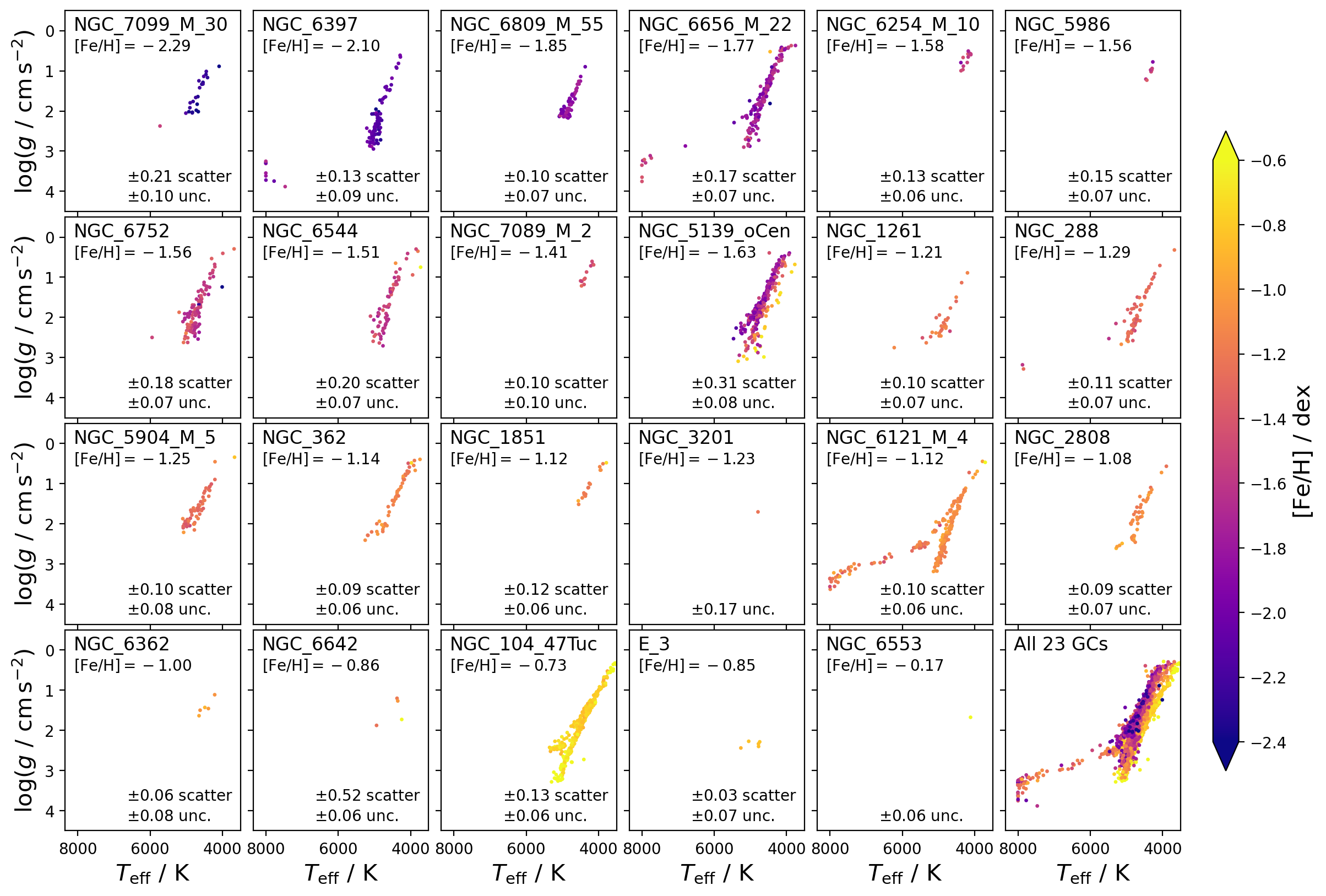}
\caption{
\textbf{Collage of globular clusters in the \Teff-\logg space, coloured by stellar metallicity \feh.} There are only minor trends between \feh and \Teff, even for the horizontal branch stars in NGC 288, NGC 6656 (M22), and NGC 6121 (M4). NGC 5139 ($\upomega$Cen) shows a significant range in \feh. RMS scatter and median metallicity uncertainties for each cluster are given in the lower right of each panel.}
\label{fig:galah_dr4_gcs_teff_logg}
\end{figure*}
\begin{figure*}[ht]
 \centering
 \includegraphics[width=\textwidth]{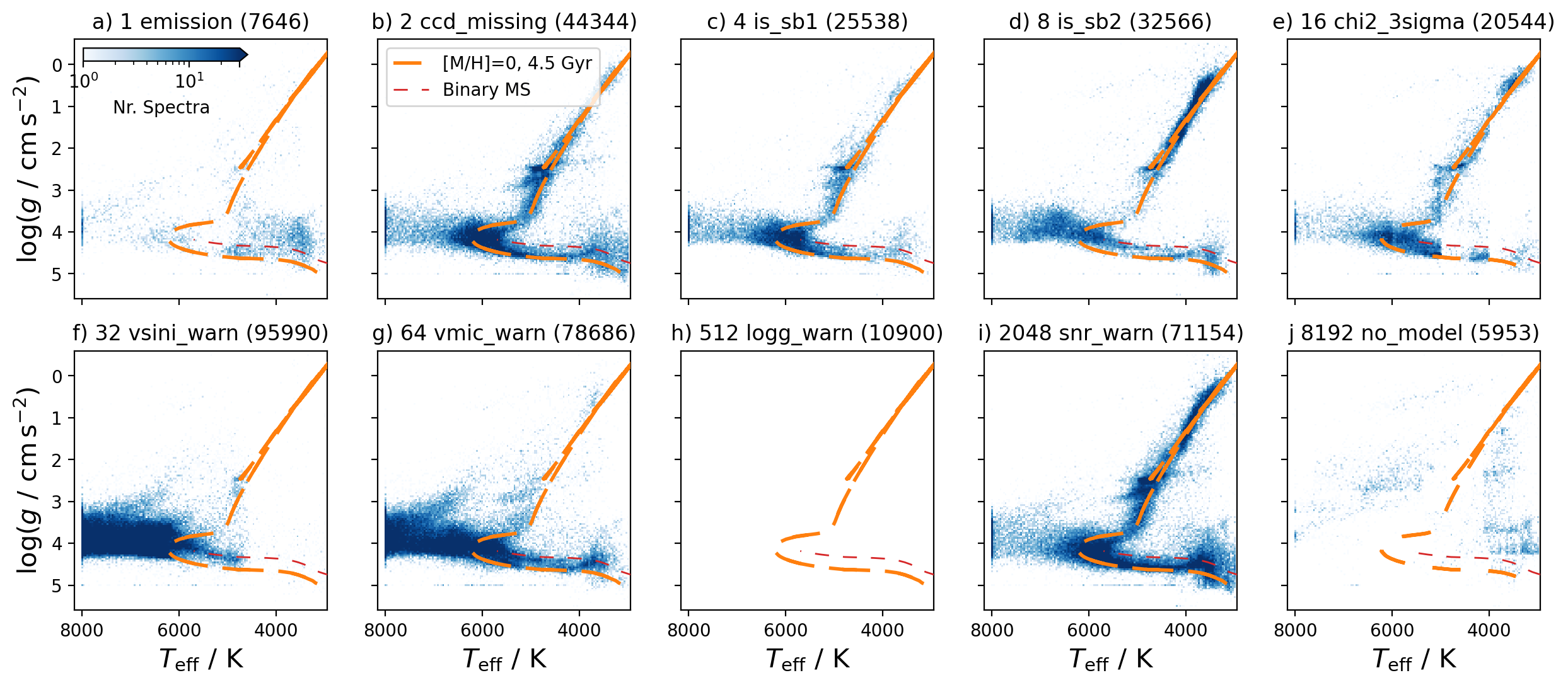}
 \caption{\textbf{Parameter overview of stars with raised major quality flag \texttt{flag\_sp} for \texttt{allstar}.}
 \textbf{Each panel} shows the logarithmic density distribution of stars in the \Teff and \logg plane with blue colourmaps. A PARSEC isochrone with $\mathrm{[M/H]}=0$ and $\tau = 4.5\,\mathrm{Gyr}$ is overplotted in orange and the same mass binary main-sequence (shifted from the single star one by $\Delta \log g = -0.3\,\mathrm{dex}$) is shown in red. Panel heads denote the bit mask and its description as well as how many times the flag was raised. We neglect distributions with no flag (0), for flags which have not been raised (8,9,11), and for which no results were available (15).} \label{fig:flag_sp_overview_allstar}
\end{figure*}

Stellar parameter and abundance zero-points of the \texttt{allstar} module are listed in Table~\ref{tab:zeropoints}. A complete table, including the zero-points for the \texttt{allspec} module can be found as FITS file in the online \href{https://github.com/svenbuder/GALAH_DR4/blob/main/catalogs/galah_dr4_zeropoints_240705.fits}{repository}. A compromise between the different accuracy abundance indicators is shown in Figure~\ref{fig:galah_dr4_zeropoint_checks_allstar} for the \texttt{allstar} module. Figure~\ref{fig:galah_dr4_precision_abundances} shows the precision of individual abundances. Figures~\ref{fig:comparison_dr3_dr4_apo17_rest} and \ref{fig:comparison_dr3_dr4_apo17_rest2} show the remaining comparisons with APOGEE DR17 in addition to Figures~\ref{fig:comparison_dr4_apo17} and \ref{fig:comparison_dr4_dr3_apo17}.
Although many of the observed globular clusters are expected to show an abundance spread, including for iron, we show a collage of globular clusters with ascending iron abundance in Figure~\ref{fig:galah_dr4_gcs_teff_logg}, with each panel indicating the median iron abundance per cluster as well as the spread (scatter) of the iron abundance distribution and the average measurement uncertainty. A more comprehensive analysis of the globular clusters will be presented in upcoming work (McKenzie et al., in preparation). Finally, Figure~\ref{fig:flag_sp_overview_allstar} shows the distribution of flagged stars in the Kiel diagram.

\begin{table}[ht]
\centering
\caption{Zero point estimates and corrections applied to the \texttt{allstar} measurements. We used \citet{Prsa2016} as reference for Solar parameters and \citet{Grevesse2007}, consistent with the \marcs model atmosphere composition \citep{Gustafsson2008}, as reference for Solar abundances. For reference, we also show the combined rotational and macroturbulence as well as microturbulence velocities from \citet{Jofre2014}. Values for Vesta indicate our uncorrected measurements for the Vesta spectrum.}
\label{tab:zeropoints}
\begin{tabular}{cccccc}
\hline \hline
Property & Reference & Zeropoint & Shift & Vesta & $\Delta$Vesta \\
& $R$ & $Z$ & $Z-R$ & $V$ & $V - R$ \\
\hline
\Teff & 5772.0 & 5772.0 & 0.0 & 5752.261 & -19.739 \\
\logg & 4.438 & 4.438 & 0.0 & 4.429 & -0.009 \\
\feh & 0.0 & 0.049 & 0.049 & -0.019 & -0.068 \\
A(Fe) & 7.45 & 7.499 & 0.049 & 7.431 & -0.068 \\
\vmic & 1.06 & 1.06 & 0.0 & 1.0 & -0.06 \\
\vsini & 4.5 & 4.5 & 0.0 & 5.552 & 1.052 \\
A(Li) & 1.05 & 1.05 & 0.0 & 1.108 & 0.058 \\
A(C) & 8.39 & 8.393 & 0.003 & 8.348 & -0.045 \\
A(N) & 7.78 & 7.705 & -0.075 & 8.368 & 0.663 \\
A(O) & 8.66 & 8.659 & -0.001 & 8.784 & 0.125 \\
A(Na) & 6.17 & 5.999 & -0.171 & 6.35 & 0.351 \\
A(Mg) & 7.53 & 7.445 & -0.085 & 7.687 & 0.242 \\
A(Al) & 6.37 & 6.185 & -0.185 & 6.552 & 0.367 \\
A(Si) & 7.51 & 7.486 & -0.024 & 7.515 & 0.029 \\
A(K) & 5.08 & 5.029 & -0.051 & 5.108 & 0.079 \\
A(Ca) & 6.31 & 6.287 & -0.023 & 6.361 & 0.074 \\
A(Sc) & 3.17 & 3.167 & -0.003 & 3.12 & -0.047 \\
A(Ti) & 4.9 & 4.876 & -0.024 & 4.882 & 0.006 \\
A(V) & 4.0 & 4.124 & 0.124 & 3.849 & -0.275 \\
A(Cr) & 5.64 & 5.64 & 0.0 & 5.61 & -0.03 \\
A(Mn) & 5.39 & 5.289 & -0.101 & 5.494 & 0.205 \\
A(Co) & 4.92 & 5.05 & 0.13 & 4.771 & -0.279 \\
A(Ni) & 6.23 & 6.228 & -0.002 & 6.236 & 0.008 \\
A(Cu) & 4.21 & 4.418 & 0.208 & 4.002 & -0.416 \\
A(Zn) & 4.6 & 4.651 & 0.051 & 4.53 & -0.121 \\
A(Rb) & 2.6 & 2.6 & 0.0 & -- & -- \\
A(Sr) & 2.92 & 2.92 & 0.0 & -- & -- \\
A(Y) & 2.21 & 2.204 & -0.006 & 2.152 & -0.052 \\
A(Zr) & 2.58 & 2.58 & 0.0 & 2.122 & -0.458 \\
A(Mo) & 1.92 & 1.92 & 0.0 & -- & -- \\
A(Ru) & 1.84 & 1.84 & 0.0 & -- & -- \\
A(Ba) & 2.17 & 2.108 & -0.062 & 2.113 & 0.005 \\
A(La) & 1.13 & 1.19 & 0.06 & 0.986 & -0.204 \\
A(Ce) & 1.7 & 1.77 & 0.07 & 1.447 & -0.323 \\
A(Nd) & 1.45 & 1.328 & -0.122 & 1.276 & -0.052 \\
A(Sm) & 1.0 & 1.0 & 0.0 & -- & -- \\
A(Eu) & 0.52 & 0.52 & 0.0 & -- & -- \\
\hline
\end{tabular}
\end{table}

\begin{figure*}[ht]
 \centering
 \includegraphics[width=0.87\textwidth]{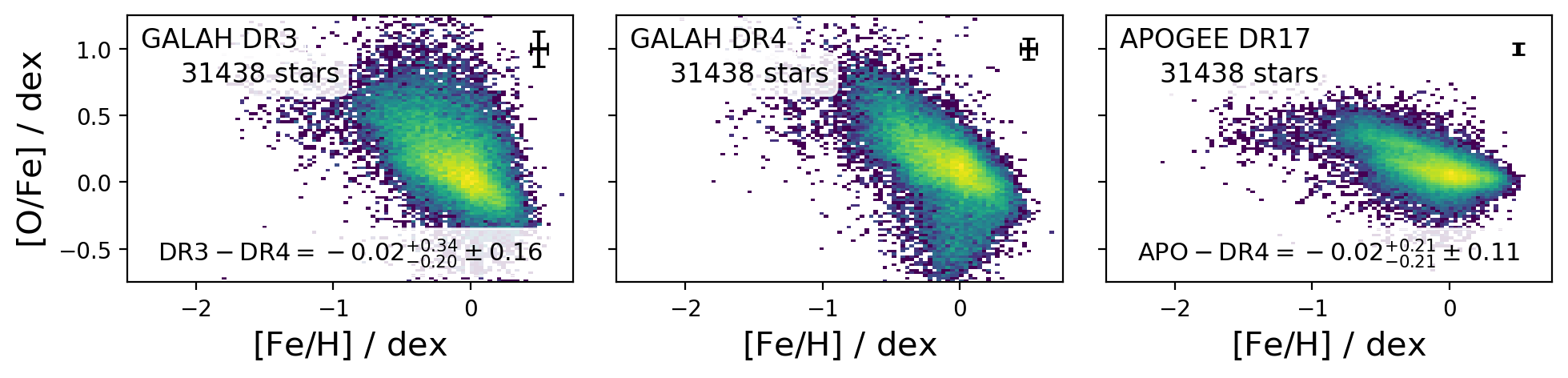}
 \includegraphics[width=0.87\textwidth]{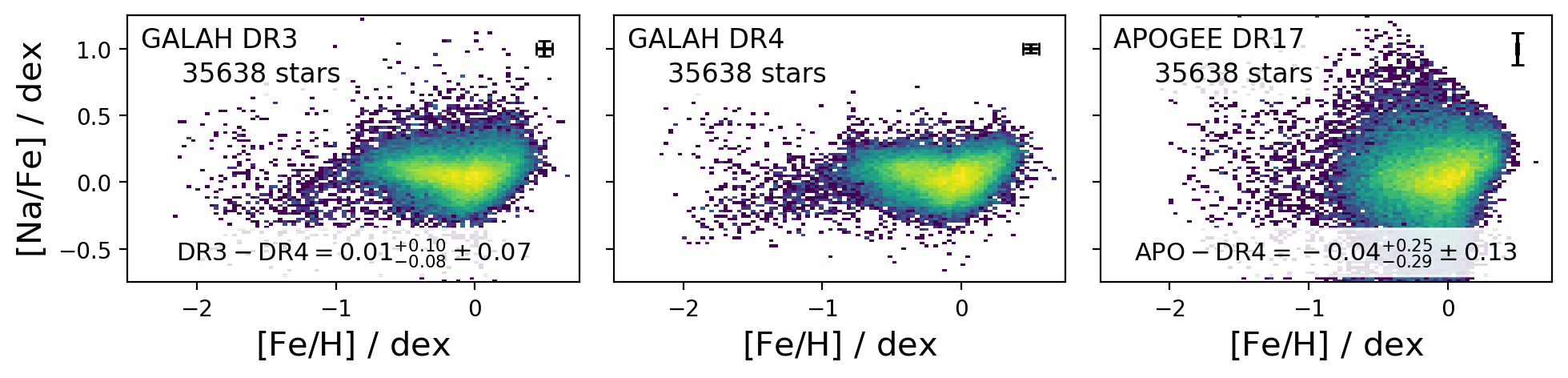}
 \includegraphics[width=0.87\textwidth]{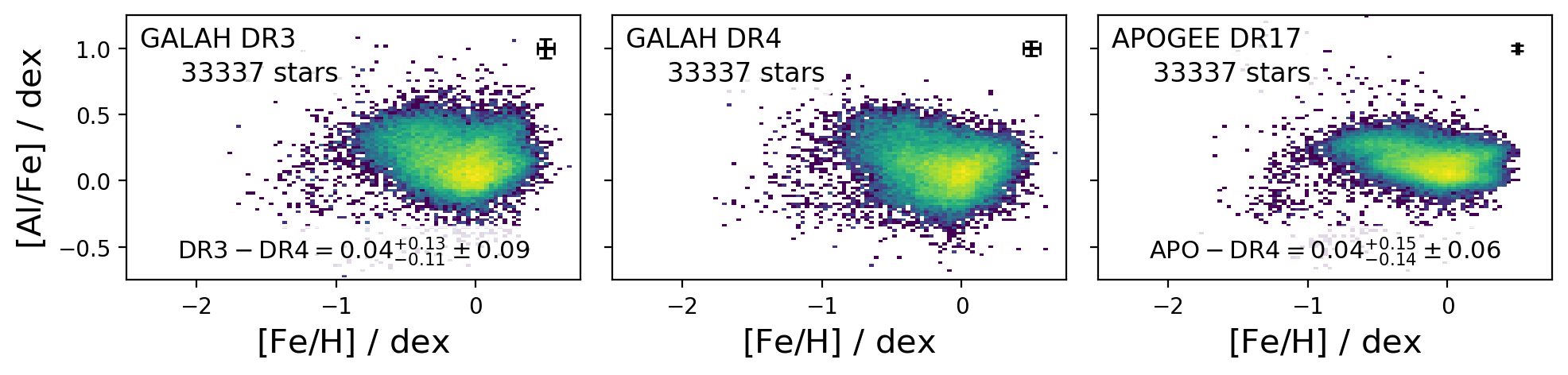}
 \includegraphics[width=0.87\textwidth]{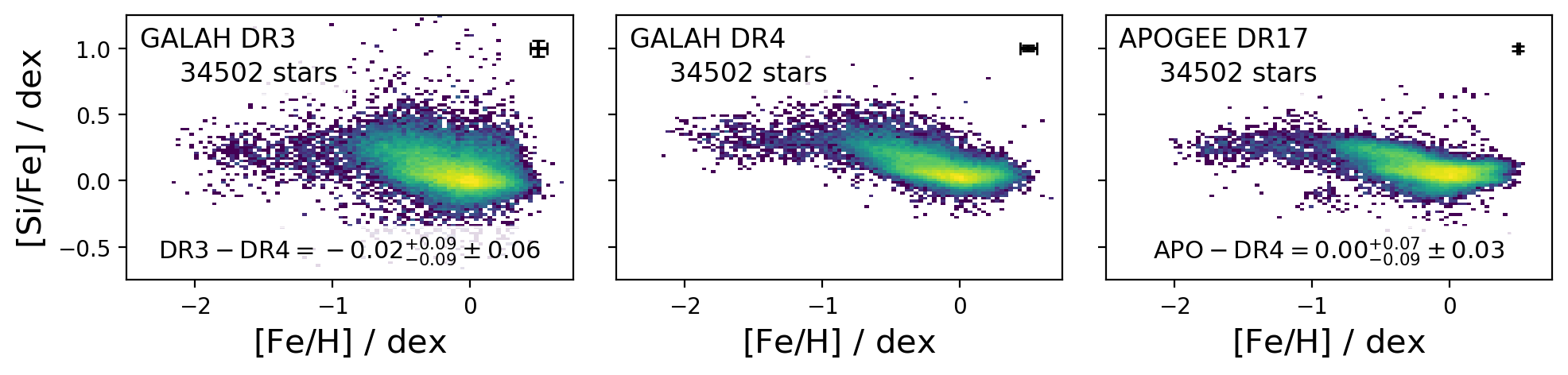}
 \includegraphics[width=0.87\textwidth]{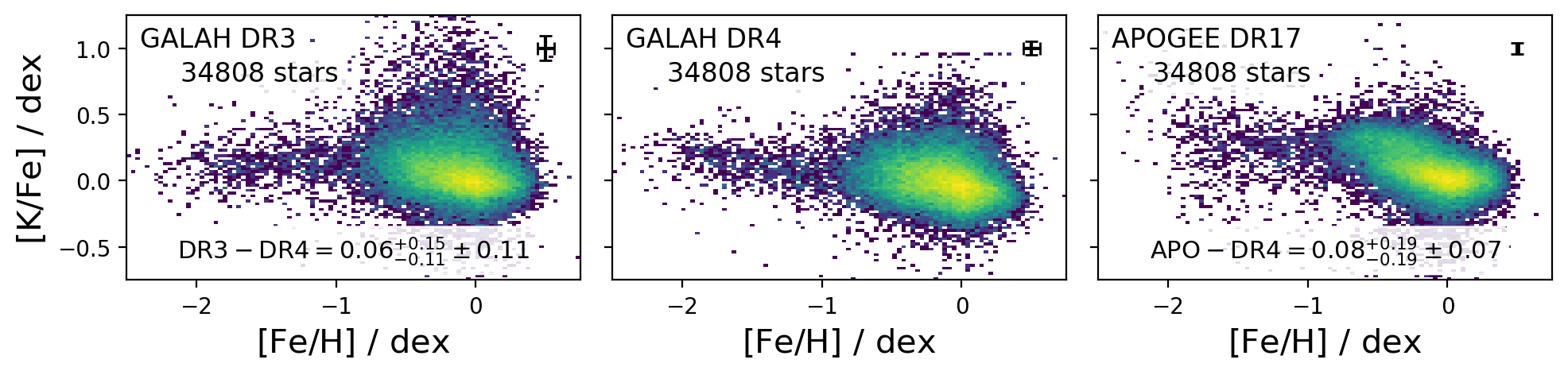}
 \includegraphics[width=0.87\textwidth]{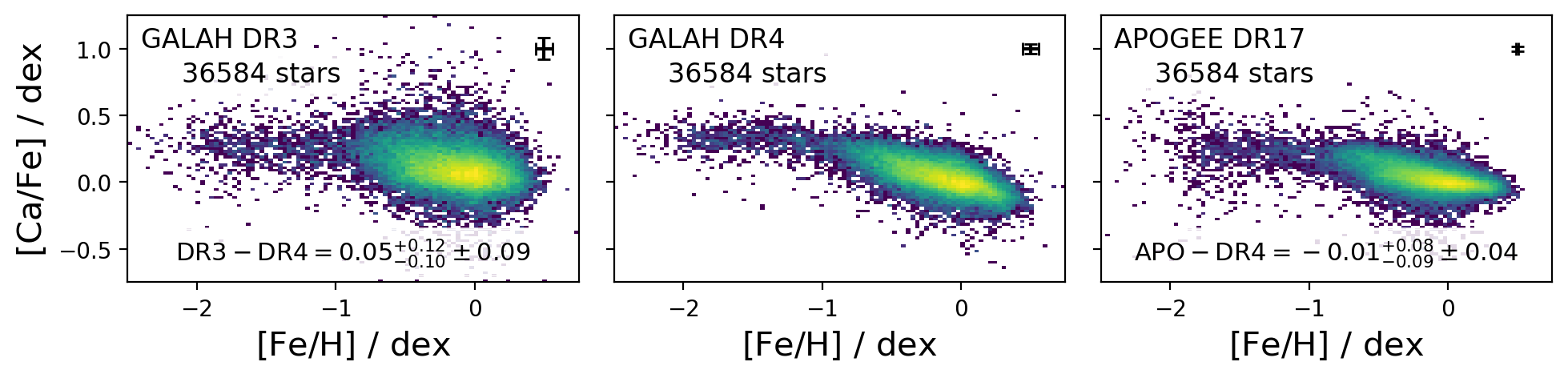}
 \caption{\textbf{Comparison of stars with available measurements in GALAH DR3 (left column), GALAH DR4 (middle column) and APOGEE DR17 (right) for O, Na, Al, Si, K, and Ca}.}
 \label{fig:comparison_dr3_dr4_apo17_rest}
\end{figure*}

\begin{figure*}[ht]
 \centering
 \includegraphics[width=0.87\textwidth]{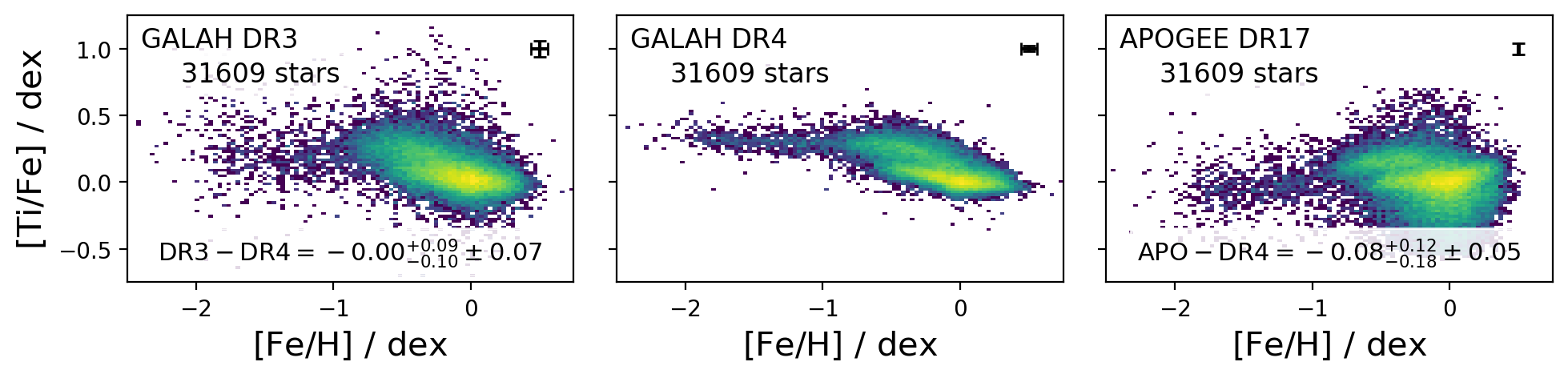}
 \includegraphics[width=0.87\textwidth]{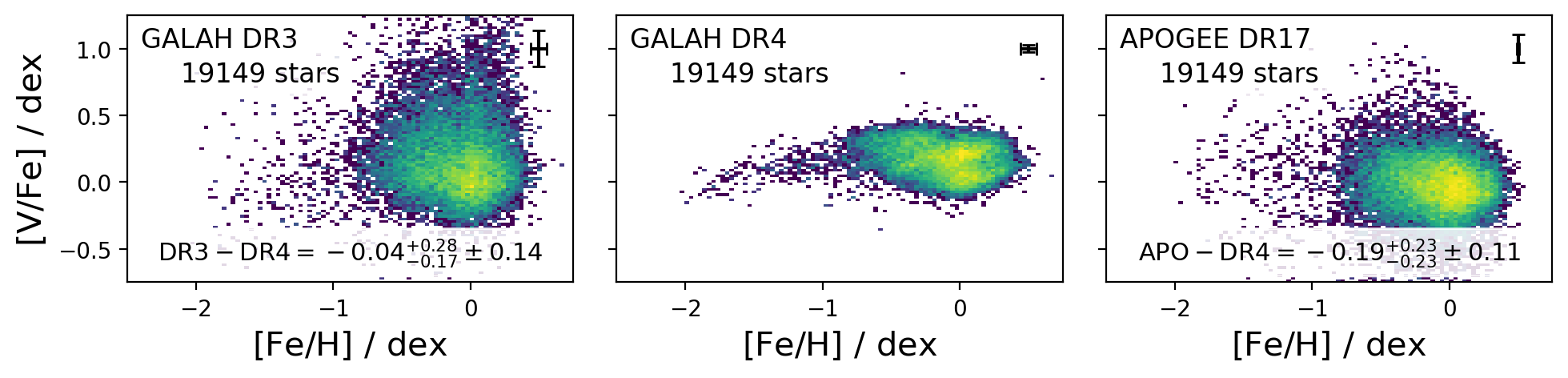}
 \includegraphics[width=0.87\textwidth]{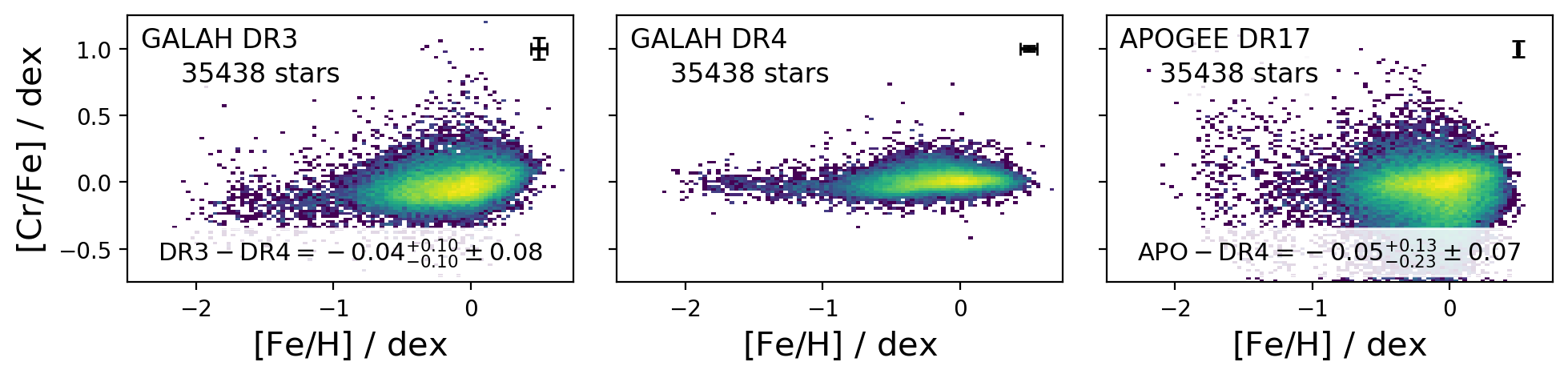}
 \includegraphics[width=0.87\textwidth]{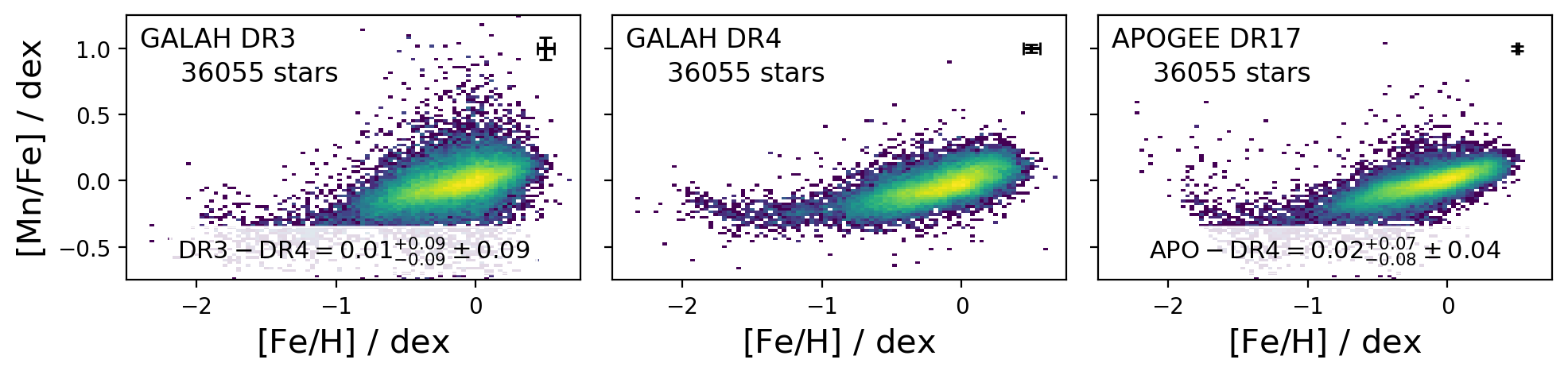}
 \includegraphics[width=0.87\textwidth]{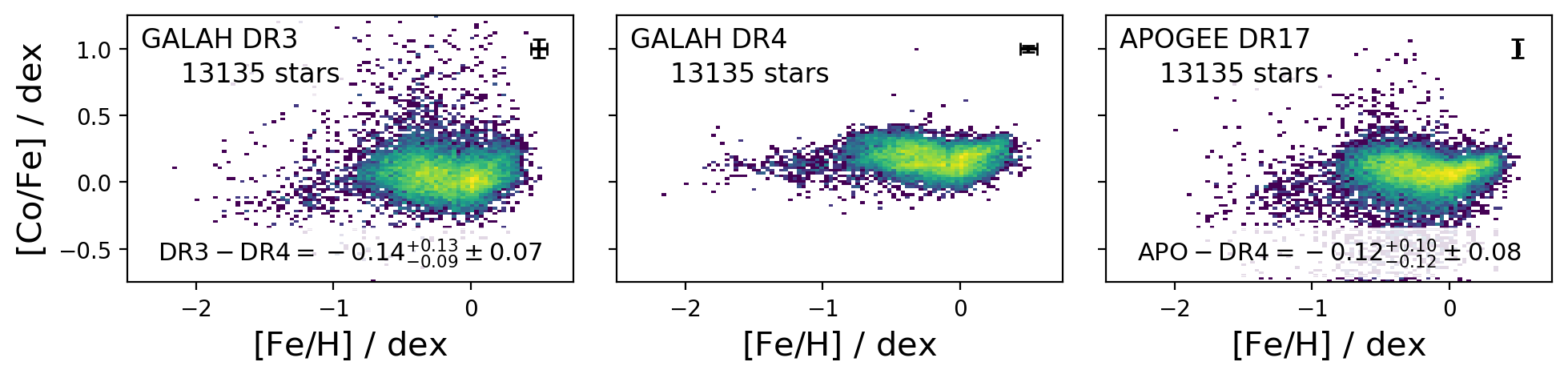}
 \includegraphics[width=0.87\textwidth]{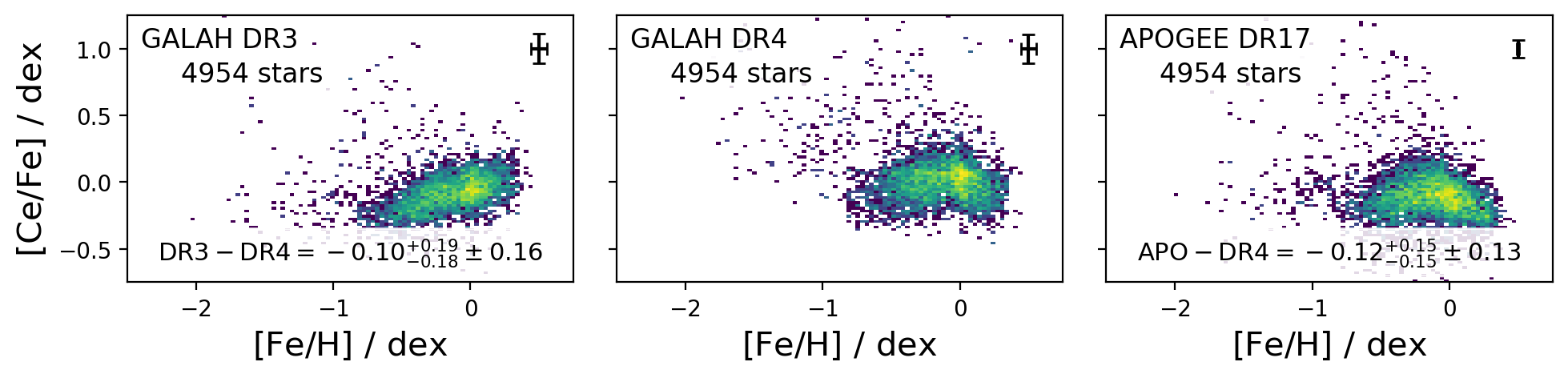}
 \caption{\textbf{Continuation of Figure~\ref{fig:comparison_dr3_dr4_apo17_rest} for Ti, V, Cr, Mn, Co, and Ce}.}
 \label{fig:comparison_dr3_dr4_apo17_rest2}
\end{figure*}

\end{document}